\documentclass[twoside,12pt]{article}
\usepackage{epsfig} 
\usepackage[english]{babel}
\usepackage{amsmath,amssymb}
\usepackage{amsfonts}
\usepackage{bbm}
\newcommand{\be}{\begin{equation}}
\newcommand{\ee}{\end{equation}}
\newcommand{\bea}{\begin{eqnarray}}
\newcommand{\eea}{\end{eqnarray}}

\newcommand {\ignore}[1]{}
\newcommand{\lsim}{
\mathrel{\hbox{\rlap{\hbox{\lower4pt\hbox{$\sim$}}}\hbox{$<$}}}}
\newcommand{\gsim}{
\mathrel{\hbox{\rlap{\hbox{\lower4pt\hbox{$\sim$}}}\hbox{$>$}}}}

\let\vev\VEV
\def\ZP{$Z^\prime$ }
\def\e6{$E(6)$}
\def\10{$SO(10)$}
\def\21{$SU(2) \otimes U(1) $}
\def\lr{$SU(3) \otimes SU(2)_L \otimes SU(2)_R \otimes U(1)_{B-L}$ }
\def\422{$SU(4) \otimes SU(2) \otimes SU(2)$ }
\def\321{$SU(3) \otimes SU(2) \otimes U(1)$ }
\def\O{\hbox{$\cal O$ }}
\def\meff{\langle m_{\nu} \rangle}
\def\lfv{lepton flavour violation }
\def\lnv{lepton number violation }
\newcommand{\ed}{\end{document}}
\DeclareMathAlphabet{\mathsc}{OT1}{cmr}{m}{sc}

\newcommand{\CL}   {C.L.}
\newcommand{\dof}  {d.o.f.}

\newcommand{\eVq}  {\rm{eV}^2}
\newcommand{\Sol}  {\mathsc{sol}}
\newcommand{\Atm}  {\mathsc{atm}}

\newcommand{\Dms}  {\Delta m^2_\Sol}
\newcommand{\Dma}  {\Delta m^2_\Atm}

\def \nbb {$\beta\beta_{0\nu}$ }

\def\meff{\langle m_{\nu} \rangle}

\newcommand{\AddrAHEP}{%
 AHEP Group, Instituto de F\'{\i}sica Corpuscular,
  C.S.I.C. -- Universitat de Val{\`e}ncia \\
  Edificio de Institutos de Paterna, Apartado 22085,
  E--46071 Val{\`e}ncia, Spain\\}
\begin{document}
\title{CP Violation and Neutrino Oscillations}
\author{Hiroshi Nunokawa $^{1}$, Stephen Parke $^2$ and Jos\'e W. F. Valle$^3$ \\
$^1$ Departamento de F\'isica, Pontif\'icia Universidade Cat\'olica do Rio de Janeiro,\\
C. P. 38071, 22452-970, Rio de Janeiro, Brazil\\
$^2$ Theoretical Physics Department, Fermi National Accelerator Laboratory,\\
P.O. Box 500, Batavia, IL 60510, USA\\
$^3$ \AddrAHEP}
\maketitle
\begin{abstract} 
  We review the basic mechanisms of neutrino mass generation and the
  corresponding structure of the lepton mixing matrix.
  We summarize the status of three-neutrino oscillation parameters as
  determined from current observations, using state-of-the-art solar
  and atmospheric neutrino fluxes, as well as latest experimental data
  as of September 2007. We also comment on recent attempts to account
  for these results and to understand flavour from first principles.
  We discuss extensively the prospects for probing the strength of CP
  violation in two near term accelerator neutrino oscillation
  experiments, T2K and NO$\nu$A, as well as possible extensions such as
  T2KK and a second large off-axis detector near the NO$\nu$A
  detector.
  We also briefly discuss the possibility of probing the effect of
  Majorana phases in future neutrinoless double beta decay searches
  and discuss other implications of leptonic CP violation such as
  leptogenesis.  
  Finally we comment on the issue of robustness of the current
  oscillation interpretation and possible ways of probing for
  non-standard neutrino interactions in precision oscillation studies.

\vfill
\end{abstract}

\tableofcontents

\section{Introduction}
\label{sec:introduction}

CP violation has been discussed phenomenologically since the sixties.
Gauge theories naturally account for its presence, although leave its
ultimate origin unanswered~\cite{Kobayashi:1973fv}.  Understanding the
origin of CP violation from first principles constitutes one of the
central challenges of theoretical elementary particle physics.  With
the historic discovery of neutrino
oscillations~\cite{fukuda:2002pe,ahmad:2002jz,araki:2004mb,Kajita:2004ga,ahn:2002up}
the issue of leptonic CP violation has also come to the center of the
agenda of the particle and nuclear physics communities.
The existence of CP violation in the lepton sector is expected in
gauge theories of neutrino mass.
The main difference with respect to CP violation in the quark mixing
matrix is the appearance of new phases associated to the Majorana
nature of neutrinos and/or with the admixture of \321 singlet leptons
in the charged current weak
interaction~\cite{schechter:1980gr,schechter:1982cv}. The latter also
leads to the effective violation of unitarity in the lepton mixing
matrix describing neutrino oscillations.
Irrespective of what the underlying origin of neutrino mass may turn
out to be, Majorana phases are a generic feature of gauge theories
that account for the smallness of neutrino mass through the feebleness
of lepton (or B-L) number violation. This includes two alternative
classes of gauge theories, that differ by the scale at which $L$ is
broken and neutrino masses arise.
In Sec.~\ref{sec:origin-neutrino-mass} we give a very sketchy summary
of the various ways to endow neutrinos with mass, and in
Sec.~\ref{sec:struct-lept-mixing} we describe the basic structure of
the lepton mixing matrix that follows from theory.

The analysis of current neutrino oscillation experiments is given in
Sec.~\ref{sec:stat-neutr-oscill} within the simplest CP-conserving
three-neutrino mixing pattern, leaving aside the LSND and Mini-Boone
data~\cite{Zimmerman:2002xj,Aguilar:2001ty,Aguilar-Arevalo:2007it}.
We summarise the status of neutrino mass and mixing
parameters~\cite{Maltoni:2004ei} as determined from current neutrino
oscillation
data~\cite{fukuda:2002pe,ahmad:2002jz,araki:2004mb,Kajita:2004ga,ahn:2002up}
~\cite{apollonio:1999ae,boehm:2001ik}.
In addition to a determination of the solar angle $\theta_{12}$, the
atmospheric angle $\theta_{23}$ and the corresponding mass squared
splittings, the data place a constraint on the last angle in the
three--neutrino leptonic mixing matrix, $\theta_{13}$.
Necessary inputs of such interpretation are the solar and atmospheric
neutrino fluxes~\cite{Bahcall:2004fg,Honda:2004yz}, the neutrino cross
sections and response functions, as well as the accurate description
of neutrino propagation in the Sun and the Earth, taking into account
matter effects~\cite{wolfenstein:1978ue,mikheev:1985gs}.
%


It is well known that CP violation should manifest itself in neutrino
oscillation experiments~\cite{Cabibbo:1977nk}.
In Sec.~\ref{sec:cp-viol-neutr} we review the basic theoretical
features of CP violation in neutrino oscillation probabilities, while
in Sec.~\ref{sec:near-term-exper} we proceed to a summary of the
prospects for probing $\theta_{13}$ and CP violation in neutrino
oscillations.  We focus on the case of accelerator neutrino
experiments and give a detailed discussion of the potential of the T2K
and NO$\nu$A experiments and their possible extensions such as T2KK
and a second large off-axis detector near the NO$\nu$A detector.


Lepton number violating processes such as neutrinoless double beta
decay~\cite{elliott:2002xe,doi:1985dx} are briefly discussed in Sec.
\ref{sec:cp-violation-lepton}.
In particular, searching for \nbb constitutes a very important
milestone for the future, as this will probe the fundamental nature of
neutrinos, irrespective of the nature of the neutrino mass-generation
mechanism~\cite{Schechter:1982bd}.  Within the simplest mass mechanism
the search for \nbb probes not only the absolute scale of neutrino
mass but is also sensitive to CP violation induced by the so-called
Majorana phases~\cite{schechter:1980gr}, inaccessible in conventional
oscillations~\cite{bilenky:1980cx,Schechter:1981gk,doi:1981yb,deGouvea:2002gf}.

In Sec.~\ref{sec:robustn-neutr-oscill} we also briefly comment on the
robustness of the current oscillation interpretation and point out the
interest in probing for non-standard neutrino interactions in future
precision oscillation studies.

\section{Kinematics of neutrino mass}
\label{sec:kinem-neutr-mass}

\subsection{Majorana and Dirac masses}
\label{sec:major-dirac-mass-1}

Massive fermions can either be Dirac or Majorana. If they carry
electric charge, we have no choice, they must be Dirac. Electrically
neutral fermions, like neutrinos (or supersymmetric `` inos''), are
expected to be Majorana-type on general grounds, irrespective of how
they acquire their mass (see Sec.~\ref{sec:origin-neutrino-mass}).
Phenomenological differences between Dirac and Majorana neutrinos are
tiny for most processes, such as neutrino oscillations: first because
neutrinos are known to be light and, second, because the weak
interaction is chiral, well described by the V-A form.  Nevertheless
it is basic, in particular, for the issue of CP violation and, for
this reason, it will be reviewed here.

The most basic spin $1/2$ fermion corresponding to the lowest
representation of the Lorentz group is given in terms of a 2-component
spinor $\rho$, with the following free
Lagrangean~\cite{schechter:1980gr}
\begin{equation}
{\cal L_M}=-i\rho^{\dagger} \sigma_{\mu} \partial_{\mu} \rho
-\frac{m}{2}\rho^T \sigma_2 \rho + H.C.
\label{eq:LM}
\end{equation}
where we use the $2 \times 2$ $\sigma$ matrices, with $\sigma_i$ being
the usual Pauli matrices and $\sigma_4=-i \: I$, $I$ being the
identity matrix, in Pauli's metric conventions, where $a.b \equiv
a_{\mu} b_{\mu} \equiv \vec{a} \cdot \vec{b} + a_4 b_4$, $a_4=ia_0$.
Under a Lorentz transformation, $x \to \Lambda x$, the spinor $\rho$
transforms as $\rho \to S(\Lambda)\rho(\Lambda^{-1}x)$ where $S$ obeys
\begin{equation}
S^{\dagger} \sigma_{\mu} S = \Lambda_{\mu \nu}\sigma_{\nu}~.
\label{eq:HOMO}
\end{equation}
The kinetic term in Eq.~(\ref{eq:LM}) is clearly invariant, and so is
the mass term, as a result of unimodular property $det\:S=1$.
However, the mass term is not invariant under a phase transformation
\begin{equation}
\rho \to  e^{i \alpha} \rho.
\label{eq:repha0}
\end{equation}
The equation of motion following from Eq.~(\ref{eq:LM}) is
\begin{equation}
-i \sigma_{\mu} \partial_{\mu} \rho = m \sigma_2 \rho^*.
\label{eq:EQM}
\end{equation}
As a result of the conjugation and Clifford properties of the
$\sigma$-matrices, one can verify that each component of the spinor
$\rho$ obeys the Klein-Gordon wave-equation.
 
In order to display clearly the relationship between our theory,
Eq.~(\ref{eq:LM}) and the usual theory of a massive spin $1/2$ Dirac
fermion, defined by the familiar Lagrangean
\begin{equation}
{\cal L_D}= - \bar{\Psi} \gamma_{\mu} \partial_{\mu} \Psi -
m\:\bar{\Psi} \Psi,
\label{eq:LD}
\end{equation}
where by convenience we use the chiral representation of the Dirac
algebra $\gamma_{\mu} \gamma_{\nu} + \gamma_{\nu} \gamma_{\mu} =
2\:\delta_{\mu \nu}$ in which $\gamma_5$ is diagonal, \bea \gamma_i =
\left(\begin{array}{ccccc}
    0 & -i\sigma_i\\
    i\sigma_i & 0\\
\end{array}\right), &
\gamma_4 = \left(\begin{array}{ccccc}
                        0 & I\\
                        I & 0\\
\end{array}\right),&
\gamma_5 = \left(\begin{array}{ccccc}
                        I & 0\\
                        0 & -I\\
\end{array}\right).
\label{eq:GAMMAS}
\eea
In this representation the charge conjugation matrix $C$ obeying
\bea
C^T = - C\\
C^\dagger = C^{-1}\\
C^{-1}\:\gamma_{\mu}\:C = -\:\gamma_{\mu}^T\\
\label{eq:conj}
\eea is simply given in terms of the basic conjugation matrix
$\sigma_2$ as
\begin{equation}
C = \left(\begin{array}{ccccc}
                        -\sigma_2 & 0\\
                        0 & \sigma_2 \\
\end{array}\right).
\label{eq:CHCONJ}
\end{equation}
A Dirac spinor can then be written as
\begin{equation}
\Psi_D = \left(\begin{array}{ccccc}
                \chi\\
                \sigma_2\:\phi^*\\
                \end{array}\right),
\label{eq:PSID}
\end{equation}
so that the corresponding charge-conjugate spinor
$\Psi_D^c = C\:\bar{\Psi}_D^T$ is the same as $\Psi_D$ but
exchanging $\phi$ and $\chi$, i.~e.
\begin{equation}
\Psi_D^c = \left(\begin{array}{ccccc}
                \phi\\
                \sigma_2\:\chi^*\\
                \end{array}\right).
\label{eq:PSIDCONJ}
\end{equation}
A 4-component spinor is said to be Majorana or self-conjugate if $\Psi
= C \bar{\Psi}^T$ which amounts to setting $\chi = \phi$.  Using
Eq.~(\ref{eq:PSID}) we can rewrite Eq.~(\ref{eq:LD}) as
\begin{equation}
{\cal L_D}=-i\sum_{\alpha=1}^2 \rho_{\alpha}^{\dagger} \sigma_{\mu}
\partial_{\mu} \rho_{\alpha}
-\frac{m}{2} \sum_{\alpha=1}^2 \rho_{\alpha}^T \sigma_2 \rho_{\alpha} + H.C.
\label{eq:LD2}
\end{equation}
where
\bea
\chi = \frac{1}{\sqrt2} (\rho_2 + i\rho_1), \nonumber \\
\phi = \frac{1}{\sqrt2} (\rho_2 - i\rho_1),
\label{eq:DECOMP}
\eea are the left handed components of $\Psi_D$ and of the
charge-conjugate field $\Psi_D^c$, respectively. In this way the Dirac
fermion is shown to be equivalent to two Majorana fermions of equal
mass. The $U(1)$ symmetry of the theory described by Eq.~(\ref{eq:LD})
under $\Psi_D \to e^{i \alpha} \Psi_D$ corresponds to continuous
rotation symmetry between $\rho_1$ and $\rho_2$
\begin{eqnarray*}
\rho_1 \to \cos \theta \rho_1 + \sin \theta \rho_2 \\
\rho_2 \to -\sin \theta \rho_1 + \cos \theta \rho_2
\label{eq:repha1}
\end{eqnarray*}
which result from the mass degeneracy between the $\rho$'s, showing
that, indeed, the concept of fermion number is not basic.

\subsection{Quantization }
\label{sec:quantization}

The mass term in Eq.~(\ref{eq:LM}) vanishes unless $\rho$ and $\rho^*$
are anti-commuting, so we consider the Majorana fermion, right from
the start, as a quantized field. The solutions to Eq.~(\ref{eq:LM})
can easily be obtained in terms of those of Eq.~(\ref{eq:LD}), which
are well known. The answer is
\bea
\Psi_M = (2\pi)^{-3/2} \int d^3k \sum_{r=1}^2 (\frac{m}{E})^{1/2} [
e^{i k.x} A_r(k) u_{Lr}(k) + e^{-i k.x} A_r^{\dagger}(k) v_{Lr}(k)],
\label{eq:PSIM}
\eea 
where $u=C\:\bar{v}^T$ and $E(k) = (\vec{k}^2 + m^2)^{1/2}$ is the
mass-shell condition.  The creation and annihilation operators obey
canonical anti-commutation rules and, like the $u$'s and $v$'s, depend
on the momentum $k$ and helicity label $r$. 
The expression in Eq.~(\ref{eq:PSIM}) describes the basic Fourier
expansion of a massive Majorana fermion. It differs from the usual
Fourier expansion for the Dirac spinor in Eq.~(\ref{eq:PSID2}) in two
ways,
\begin{itemize}
\item the spinor is two-component, as there is a chiral
projection acting of the $u$'s and $v$'s
\item there is only one Fock space, particle and anti-particle
  coincide, showing that a massive Majorana fermion corresponds to one
  half of a conventional massive Dirac fermion.
\end{itemize}
The $u$'s and $v$'s are the same wave functions that appear in the
Fourier decomposition the Dirac field
\bea 
\Psi_D = (2\pi)^{-3/2} \int d^3k
\sum_{r=1}^2 (\frac{m}{E})^{1/2} [ e^{i k.x} a_r(k) u_r(k) + e^{-i
  k.x} b_r^{\dagger}(k) v_r(k)]\;.
\label{eq:PSID2}
\eea

Using the helicity eigenstate wave-functions,
\bea \vec{\sigma} \cdot \vec{k} \: u_L^{\pm}(k) = \pm \mid \vec{k}
\mid u_{L}^{\pm}(k),\\
\vec{\sigma} \cdot \vec{k} \: v_L^{\pm}(k) = \mp \mid \vec{k} \mid
v_{L}^{\pm}(k),
\label{eq:limit}
\eea 
one can show that, out of the $4$ linearly independent wave functions
$u_{L}^{\pm}(k)$ and $v_{L}^{\pm}(k)$, only two survive as the mass
approaches zero, namely, $u_{L}^{-}(k)$ and $v_{L}^{+}(k)$
~\cite{schechter:1981hw}. This way the Lee-Yang two-component massless
neutrino theory is recovered as the massless limit of the Majorana
theory.

Two independent propagators follow from Eq.~(\ref{eq:LM}),
\bea
<0 \mid \rho(x) \: \rho^*(y) \mid 0>
=i \sigma_{\mu} \partial_{\mu} \Delta_F(x-y;m),\\
\label{eq:NORMAL}
<0 \mid \rho(x) \: \rho (y) \mid 0>
= m \: \sigma_2 \: \Delta_F(x-y;m),
\label{eq:LNV}
\eea 
where $\Delta_F(x-y;m)$ is the usual Feynman function.  The first one
is the ``normal'' propagator that intervenes in total lepton number
conserving ($\Delta L = 0 $) processes, while the one in
Eq.~(\ref{eq:LNV}) describes the virtual propagation of Majorana
neutrinos in $\Delta L = 2 $ processes such as neutrinoless
double-beta decay.
 
\subsection{CP properties}
\label{sec:CP-properties}

The Lagrangean in Eq.~(\ref{eq:LM}) can be easily generalized for a
system of an arbitrary number of Majorana neutrinos, giving
\begin{equation} 
{\cal L_M}=-i \sum_{\alpha=1}^{n}
  \rho_{\alpha}^{\dagger} \sigma_{\mu} \partial_{\mu} \rho_{\alpha}
  -\frac{1}{2} \sum_{\alpha,\beta=1}^{n}
  M_{\alpha\beta}\rho_{\alpha}^T \sigma_2 \rho_{\beta} + H.C.
\label{eq:LM2}
\end{equation}
where the sum runs over the ``neutrino-type'' indices $\alpha$ and
$\beta$. By Fermi statistics the mass coefficients $M_{\alpha \beta}$
must form a symmetric matrix, in general complex. This matrix can
always be diagonalized by a complex $n \times n$ unitary matrix
$U$~\cite{schechter:1980gr}
\begin{equation}
M_{diag} = U^T M U \:.
\end{equation}
When $M$ is real (CP conserving) its diagonalizing matrix $U$ may be
chosen to be orthogonal and, in general, the mass eigenvalues can have
different signs. These may be assembled as a signature matrix
\begin{equation}
\eta = diag(+,+,...,-,-,..)
\end{equation}
For two neutrino types there are two classes of models, one with $\eta
= diag(+,-)$ and another characterized by $\eta = diag(+,+)$.  The
class with $\eta = diag(+,-)$ contains as a limit the case where the
two fermions make up a Dirac neutrino.
Depending on whether the two neutrinos are active-sterile or
active-active the limit is called
quasi-~\cite{valle:1983yw,valle:1983dk} or pseudo-Dirac
neutrino~\cite{wolfenstein:1981kw}. Oscillations between the
quasi-Dirac neutrino components potentially change the predictions of
Big Bang Nucleosynthesis while oscillations between the components of
a pseudo-Dirac neutrino do not.

Note that one can always make all masses positive by introducing
appropriate phase factors in the wave functions, such as the factors
of $i$ in Eq.~(\ref{eq:DECOMP}).
When interactions are added (see Sec.~\ref{sec:struct-lept-mixing})
these signs become physical. As emphasized by Wolfenstein, these signs
and the corresponding CP phases play an important role in the
discussion of neutrinoless double beta
decay~\cite{Wolfenstein:1981rk}. 

\section{The origin of neutrino mass}
\label{sec:origin-neutrino-mass}

The fifteen basic building blocks of matter listed in Table
\ref{tab:sm} are all 2-component sequential ``left-handed'' chiral
fermions, one set for each generation.  Parity violation in the weak
interaction is accounted for ``effectively'' by having ``left'' and
``right'' fermions which behave differently with respect to the \321
gauge group. In contrast to charged fermions, neutrinos come only in
one chiral species.
It has been long noted by Weinberg~\cite{Weinberg:1980bf} that one can
add to the Standard \321 Model (SM) an effective dimension-five
operator $\O = \lambda \ell \ell \Phi \Phi$, where $\ell$ denotes a
lepton doublet for each generation and $\Phi$ is the SM scalar doublet.
\begin{figure}[h] \centering
    \includegraphics[height=3.5cm,width=.4\linewidth]{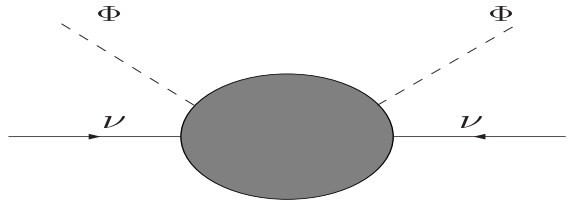}
    \caption{\label{fig:d5} 
    Dimension five  operator responsible for neutrino mass.}
\end{figure}

\begin{table*}
  \centering
\begin{math}
\begin{array}{|c|c|} \hline
& \ \ \ {\mbox SU(3)\otimes SU(2)\otimes U(1)} \\
\hline
\ell_a = (\nu_a, l_a)^T & (1,2,-1)\\
e_a^c   & (1,1,2)\\
\hline
Q_a = (u_a, d_a)^T    & (3,2,1/3)\\
u_a^c   & (\bar{3},1,-4/3)\\
d_a^c   & (\bar{3},1,2/3)\\
\hline
\Phi  & (1,2,1)\\
\hline
\end{array}
\end{math}
\caption{Matter and scalar multiplets of the Standard Model}
\label{tab:sm}
\end{table*}

Once the electroweak symmetry breaks through the nonzero vacuum
expectation value (vev) $\vev{\Phi}$ Majorana neutrino masses $\propto
\vev{\Phi}^2$ are induced, in contrast to the masses of the charged
fermions which are linear in $\vev{\Phi}$.
This constitutes the most basic definition of neutrino mass, in which
its smallness relative to the masses of the SM charged fermions is
ascribed to the fact that $\O$ violates lepton number by two units
($\Delta L=2$) whereas the other fermion masses do not. Note that this
argument is totally general and holds irrespective of the underlying
origin of neutrino mass. From such general point of view the emergence
of Dirac neutrinos would be a surprise, an ``accident'', justified
only in the presence of a fundamental lepton number symmetry, in
general absent. For example, neutrinos could naturally get very small
Dirac masses via mixing with a bulk fermion in models involving extra
dimensions~\cite{Dienes:1998sb,Arkani-Hamed:1998vp,Ioannisian:1999sw}.
Barring such very special circumstances, gauge theories expect
neutrinos to be Majorana.

Little more can be said from first principles about the {\sl
  mechanism} giving rise to the operator in Fig.~\ref{fig:d5}, its
associated mass {\sl scale} or its {\sl flavour structure}.  For
example, the strength $\lambda$ of the operator $\O$ may be suppressed
by a large scale $M_X$ in the denominator (top-down) scenario, leading
to
$$ m_{\nu} = \lambda_0 \frac{\vev{\Phi}^2}{M_X}, $$
where $\lambda_0$ is some unknown dimensionless constant.
Gravity, which in a sense "belongs" to the SM, could induce the
dimension-five operator $\O$, providing the first example of a
top-down scenario with $M_X = M_P$, the Planck scale. In this case the
magnitude of the resulting Majorana neutrino masses are too small to
be relevant in current searches.

Alternatively, the strength $\lambda$ of the operator $\O$ may be
suppressed by small parameters (e.g. scales, Yukawa couplings) in the
numerator and/or loop-factors (bottom-up scenario).
Both classes of scenarios are viable and allow for many natural
realizations. While models of the top-down type are closer to the idea
of unification, bottom-up schemes are closer to experimental
verification.

Models of neutrino mass may also be classified according to whether or
not additional neutral heavy states are present, in addition to the
three isodoublet neutrinos. As an example, such leptons could be \321
singlet ``right-handed'' neutrinos.
In what follows we classify models according to the mass scale at
which $\O$ is induced, namely bottom-up and top-down scenarios.

\subsection{Seesaw-type neutrino masses}
\label{sec:top-down-scenario}

The most popular top-down scenario is the seesaw. The idea is to
generate the operator $\O$ by the exchange of heavy states. The
smallness of its strength is understood by ascribing it to the
violation of lepton number at a high mass scale, namely the scale at
which the heavy states acquire masses.

\subsubsection{The  majoron seesaw}
\label{sec:majoron-seesaw}

The simplest possibility for the seesaw is to have ungauged lepton
number~\cite{schechter:1982cv}. It is also the most general, as it can
be studied in the framework of just the \321 gauge group. Such
``1-2-3'' scheme is characterized by \321 singlet, doublet and triplet
mass terms, described by the
matrix~\cite{schechter:1980gr,schechter:1982cv}
\be
\label{ss-matrix} {\mathcal M_\nu} = \left(\begin{array}{cc}
    Y_3 v_3 & Y_\nu \vev{\Phi} \\
    {Y_\nu}^{T} \vev{\Phi}  & Y_1 v_1 \\
\end{array}\right) 
\ee 
in the basis $\nu_{L}$, $\nu^{c}_{L}$, corresponding to the three
``left'' and three ``right'' neutrinos, respectively. Note that,
though symmetric, by the Pauli principle, ${\mathcal M_\nu}$ is
complex, so that its Yukawa coupling sub-matrices \(Y_\nu\) as well as
\(Y_3\) and \(Y_1\) are complex matrices denoting the relevant Yukawa
couplings, the last two symmetric.

\begin{figure}[ht] \centering
\hglue 3.2cm
   \includegraphics[height=3.8cm,width=.15\linewidth]{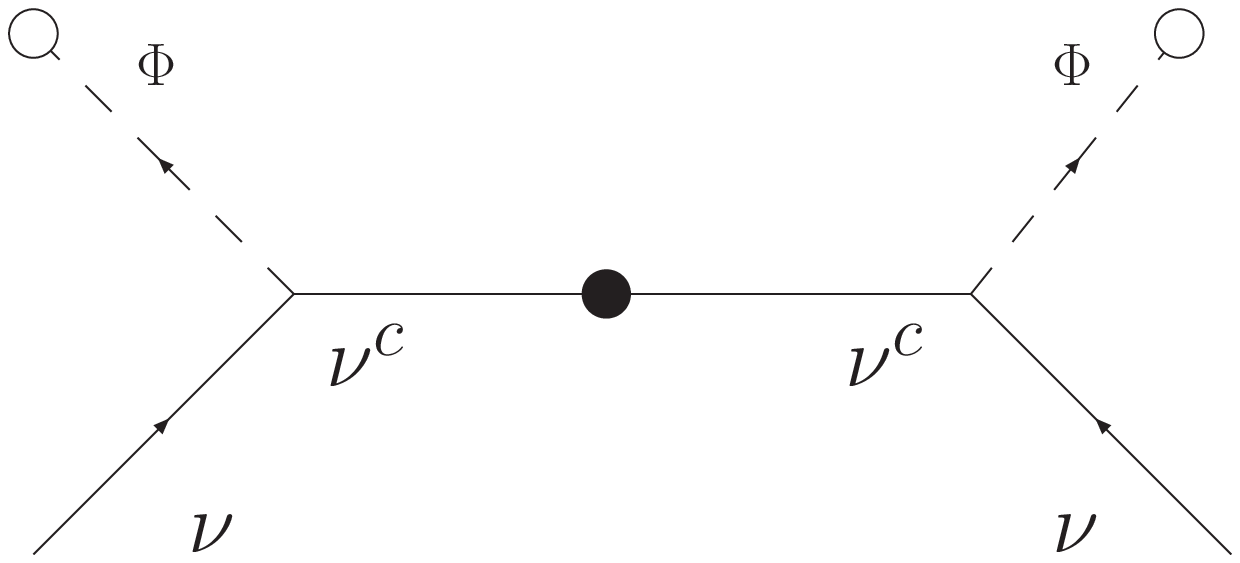} \hskip 3cm
    \includegraphics[height=4cm,width=.5\linewidth]{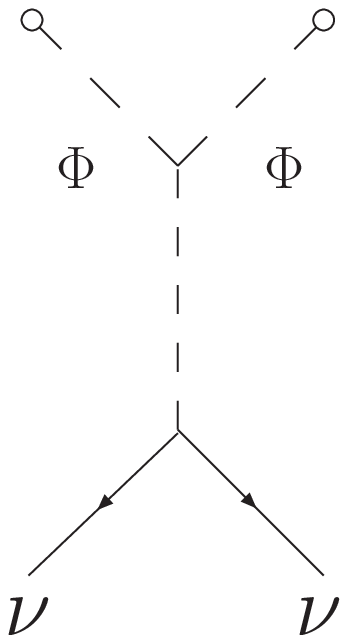}
    \caption{\label{fig:seesaw} %
      Two types of seesaw mechanism. Left: right-handed neutrino
      exchange (type-I), right: heavy \321 triplet exchange
      (type-II).}
\end{figure}

Such \321 seesaw contains singlet, doublet and triplet scalar
multiplets, obeying a simple ``1-2-3'' vev seesaw relation of the type
 \begin{equation}
   v_3 v_1 \sim {v_2}^2 \:\:\: \mathrm{with} \:\:\: v_1 \gg v_2 \gg v_3 
 \label{eq:123-vev-seesaw}
 \end{equation}
 This follows simply from the minimization condition of the \321
 invariant scalar potential and arises in a wide variety of seesaw
 type models, as reviewed in \cite{Orloff:2005nu,Valle:2006vb}. It
 implies that the triplet vev $v_3 \to 0$ as the singlet vev $v_1$
 grows.  Here $v_2 \equiv \vev{\Phi}$ denotes the SM Higgs doublet
 vev.
 Small neutrino masses are induced either by heavy \321 singlet
 ``right-handed'' neutrino exchange (type I) or the smallness of the
 induced triplet vev that follows from heavy scalar exchange (type
 II), as illustrated in Fig.~\ref{fig:seesaw}.
The matrix \(\mathcal{M_\nu}\) is diagonalized by a unitary mixing
matrix \(U_\nu\),
\begin{equation}
   U_\nu^T {\mathcal M_\nu} U_\nu = \mathrm{diag}(m_i,M_i),
\end{equation}
yielding 6 mass eigenstates, including the three light neutrinos with
masses \(m_i\), and three two-component leptons.
The light neutrino mass states \(\nu_i\) are given in terms of the
flavour eigenstates via the unitary matrix
$U_\nu$~\cite{schechter:1980gr,schechter:1982cv}
\begin{equation}
   \nu_i = \sum_{a=1}^{6}(U_\nu)_{ia} n_a.
\end{equation}
The effective light neutrino mass, obtained this way is of the
form
\begin{equation}
  \label{eq:ss-formula0}
  m_{\nu} \simeq Y_3 v_3 -
Y_\nu {Y_1}^{-1} {Y_\nu}^T \frac{{\vev{\Phi}}^2}{v_1}
\end{equation}
The diagonalization matrices can be worked out explicitly as a
perturbation series, see Ref.~\cite{schechter:1982cv}.

There is an important feature of such ``1-2-3'' seesaw namely, that
since lepton number is ungauged, there is a physical Goldstone boson
associated with its spontaneous breakdown, the
Majoron~\cite{chikashige:1981ui}.  
It is often argued that, due to quantum gravity effects the associated
majoron will pick up a mass. It has been shown that, a massive majoron
with mass in the fraction of keV range can provide the observed dark
matter of the Universe~\cite{Lattanzi:2007ux}. Such majoron decaying
dark matter offers an alternative to the supersymmetric dark matter
scenario and may lead to interesting phenomenological implications.

 \subsubsection{Left-right symmetric seesaw}
 \label{sec:left-right-symmetric}

 A more elegant setting for the seesaw is a gauge theory containing
 B-L as a generator, such as \lr or the unified models based on \10 or
 \e6~\cite{Minkowski:1977sc,Orloff:2005nu,Lazarides:1980nt}.
 For example in \10 each matter generation is naturally assigned to a
 {\bf 16} (spinorial in \10) so that there are {\bf 16} . {\bf 16} .
 {\bf 10} as well as {\bf 16} .  $\overline{\bf 126}$ . {\bf 16} terms
 generating Dirac and Majorana neutrino mass terms, respectively,
 leading to the neutrino mass matrix \be
\label{ss-matrix} {\mathcal M_\nu} = \left(\begin{array}{cc}
    Y_L \vev{\Delta_L} & Y_\nu \vev{\Phi} \\
    {Y_\nu}^{T} \vev{\Phi}  & Y_{R} \vev{\Delta_R} \\
\end{array}\right) 
\ee 
in the basis $\nu_{L}$, $\nu^{c}_{L}$, corresponding to the ``left''
and ``right'' neutrinos, respectively, where \(Y_L\) and \(Y_R\)
denote the Yukawas of the {\bf 126} of \10, whose vevs
$\vev{\Delta_{L,R}}$ give rise to the Majorana terms, while ${Y_\nu}$
is the Dirac Yukawa coupling in {\bf 16} . {\bf 16} .  {\bf 10}. The
matrix \(Y_\nu\) is an arbitrary complex matrix in flavour space,
while \(Y_L\) and \(Y_R\) are complex symmetric \(3\times3\) matrices
that correspond to \(Y_1\) and \(Y_3\) of the simplest \321 model.
Small neutrino masses are induced either by heavy \321 singlet
``right-handed'' neutrino exchange (type I) or heavy scalar boson
exchange (type II) as illustrated in Fig.~\ref{fig:seesaw}.
The matrix \(\mathcal{M_\nu}\) is diagonalized by a unitary mixing
matrix \(U_\nu\) as before. The diagonalization matrices can be worked
out explicitly as a perturbation series, using the same method of
Ref.~\cite{schechter:1982cv}. 
This means that the explicit formulas for the \(6\times6\) unitary
diagonalizing matrix \(U\) given explicitly in
Ref.~\cite{schechter:1982cv} also hold in the left-right case,
provided one takes into account that $v_1 \to \vev{\Delta_R}$ and $v_3
\to \vev{\Delta_L}$.
The effective light neutrino mass, obtained this way is of the
form
\begin{equation}
  \label{eq:ss-formula}
  m_{\nu} \approx Y_L \vev{\Delta_L} -
Y_\nu {Y_R}^{-1} {Y_\nu}^T \frac{{\vev{\Phi}}^2}{\vev{\Delta_R}}.
\end{equation}
We have the new vev seesaw relation
 \begin{equation}
  \vev{\Delta_{L}} \vev{\Delta_{R}} \sim \vev{\Phi}^2,
 \label{eq:vev-seesaw}
 \end{equation}
 which naturally follows from minimization of the left-right symmetric
 scalar potential, together with the vev
 hierarchy~\cite{Orloff:2005nu}
 \begin{equation}
  \vev{\Delta_{L}} \ll  \vev{\Phi} \ll \vev{\Delta_{R}}.
 \label{eq:vev-hierarchy}
 \end{equation}
%
 This implies that both type I and type II contributions vanish as
 $\vev{\Delta_{R}} \to \infty$.
 The structure of the seesaw is exactly the same as before, enabling
 one to employ the same perturbative diagonalization method in
 \cite{schechter:1982cv}.

 The new insight that is obtained this way is that now one can arrange
 for the breakdown of parity invariance to be spontaneous, so that
 smallness of neutrino masses gets correlated to the observed
 maximality of parity violation in low-energy weak interactions, as
 stressed by Mohapatra and Senjanovic~\cite{Orloff:2005nu}. However
 this is hardly relevant phenomenologically in view of the large value
 of the B-L scale needed both to fit the neutrino masses, as well as
 unify the gauge couplings.
 Another important difference with the \321 seesaw case is the absence
 of the Majoron, now absorbed as the longitudinal mode of the gauge
 boson corresponding to the B-L generator.

\subsubsection{Double seesaw}
\label{sec:double-seesaw}

Nothing is sacred about the number of (anomaly-free) gauge singlet
leptons $S_i$ in the SM~\cite{schechter:1980gr} or of singlets outside
the {\bf 16} in \10 or the {\bf 27} in \e6~\cite{mohapatra:1986bd}.
New important features may emerge when the seesaw is realized with
non-minimal lepton content. Here we mention the seesaw scheme
suggested in Ref.~\cite{mohapatra:1986bd} with \e6
motivations~\cite{Witten:1985xc}.
The model extends minimally the particle content of the SM by the
sequential addition of a pair of two-component \321 singlet leptons,
$\nu_i^c, S_i$, with \(i\) a generation index running over \(1,2,3\).
In the \(\nu,\nu^c,S\) basis, the \(9\times9\) neutral leptons
mass matrix $\mathcal{M_\nu}$ is given as
\begin{equation}
\label{eqn:doubleSeesaw}
{\mathcal M_\nu}=\left(
   \begin{array}{ccc}
      0   & Y_\nu^T \vev{\Phi} & 0   \\
      Y_\nu  \vev{\Phi} & 0     & M^T \\
      0   & M     & \mu
   \end{array}\right),
\end{equation}
in the basis $\nu_{L}$, $\nu^{c}_{L}$, $S_{L}$, where \(Y_\nu\) is an
arbitrary \(3\times3\) complex Yukawa matrix, \(M\) and \(\mu\) are
\321 singlet complex mass matrices, \(\mu\) being symmetric. 
Notice that it has zeros in the $\nu_{L}$-$\nu_{L}$ and
$\nu^{c}_{L}$-$\nu^{c}_{L}$ entries, a feature of several
string models~\cite{Witten:1985xc}. 

\begin{figure}[h] \centering
    \includegraphics[height=3cm,width=.45\linewidth]{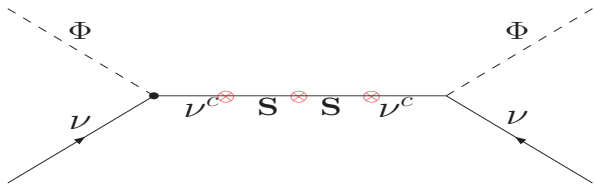}
    \caption{\label{fig:iss-mass} 
    ``Double'' and ``inverse'' seesaw mechanism.}
\end{figure}

For $\mu \gg M$ one has to first approximation that the $S_i$ decouple
leaving the simpler seesaw at scales below that. In such a ``double''
seesaw scheme the three light neutrino masses are determined from
\begin{equation}
\label{eqn:lightNu}
    m_\nu \approx  {\vev{\Phi}^2} Y_\nu^T { M^{T}}^{-1} \mu M^{-1}  Y_\nu~.
\end{equation}
The mass generation is illustrated in Fig.~\ref{fig:iss-mass}.
This formula can be readily derived from the method given in
Ref.~\cite{schechter:1982cv} and decoupling the heavy states in two
steps, first $S$ then $\nu^c$ since $\mu \gg M$.
A new feature is that there are two independent scales, \(\mu\) and
\(M\), with the B-L symmetry broken only by the largest scale \(\mu\).
It is completely natural also to consider the case where \(\mu\)
instead of large is very small~\cite{mohapatra:1986bd}, even smaller
than the electroweak vev.  This low \(\mu\) case will be considered in
Sec.~\ref{sec:inverse-seesaw}, and opens new phenomenological
possibilities, discussed in Sec.~\ref{sec:phen-neutr-mass}.

Irrespective of what sets its scale, the entry \(\mu\) may be
proportional to the vev of an \321 singlet scalar, in which case the
model contains a singlet majoron.  

 \subsubsection{Unconventional seesaw}
 \label{sec:low-b-l}

 There are many types of seesaw. More important than keeping track of
 the taxonomy of schemes (type
 I~\cite{Minkowski:1977sc,Orloff:2005nu,Lazarides:1980nt}, type
 II~\cite{schechter:1980gr,schechter:1982cv}, type
 III~\cite{Akhmedov:1995vm,Barr:2005ss,Fukuyama:2005gg}, etc.) is
 understanding that the seesaw is not a theory but a mechanism that
 allows for many possible realizations. Schemes leading to the same
 pattern of neutrino masses may differ in many other respects.
%

\begin{figure}[h] \centering
    \includegraphics[height=3cm,width=.45\linewidth]{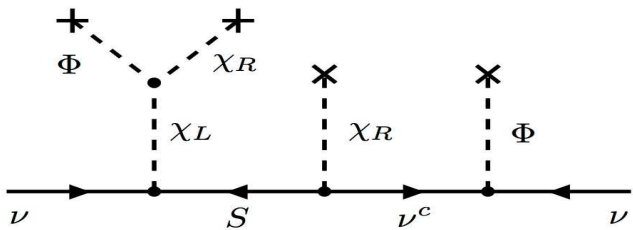}
    \caption{\label{fig:new-seesaw} 
    Unconventional seesaw mechanism.}
\end{figure}

 As an example of extended seesaw models, let us consider one that has
 recently been suggested~\cite{Malinsky:2005bi}. It belongs to the
 class of supersymmetric \10 models with broken D-parity.  In addition
 to the three right-handed neutrinos of the standard seesaw model, it
 contains three sequential gauge singlets $S_{iL}$ with the following
 mass matrix
\be \label{ess-matrix}
{\mathcal M_\nu} =
\left(\begin{array}{ccc}
0 & Y_\nu \vev{\Phi} & F \vev{\chi_L} \\
{Y_\nu}^{T} \vev{\Phi} & 0 & \tilde F \vev{\chi_R}     \\
F^{T} \vev{{\chi}_L}    & \tilde F^{T} \vev{\chi_R} & 0
\end{array}\right) 
\ee 
in the basis $\nu_{L}$, $\nu^{c}_{L}$, $S_{L}$. Notice that it has
zeros along the diagonal, specially in the $\nu_{L}$-$\nu_{L}$ and
$\nu^{c}_{L}$-$\nu^{c}_{L}$ entries, thanks to the fact that there is
no {\bf 126}, a feature of several string-inspired
models~\cite{Witten:1985xc,mohapatra:1986bd}. The resulting neutrino
mass is
\begin{eqnarray}
m_{\nu} & \simeq &
\frac{\vev{\Phi}^2}{M_\mathrm{unif}} 
\left[Y_\nu ( F \tilde F^{-1})^{T}+( F \tilde F^{-1}) {Y_\nu}^{T}\right],
\end{eqnarray}
where $M_\mathrm{unif}$ is the unification scale, $F$ and $\tilde F$
denote independent combinations of Yukawa couplings of the $S_{iL}$.
One can see that the neutrino mass is suppressed by the unification
scale $M_{\mathrm{unif}}$ \emph{irrespective of how low is the B-L
  breaking scale}.
In contrast to all familiar seesaws, see e.g.
Eq.~(\ref{eq:ss-formula}), this new seesaw mechanism is {\sl linear}
in the Dirac Yukawa couplings $Y_\nu$, as illustrated in
Fig.~\ref{fig:new-seesaw}.
It is rather remarkable that one can indeed take the B-L scale as low
as TeV without generating inconsistencies with gauge coupling
unification~\cite{Malinsky:2005bi}.

\subsection{Bottom-up neutrino masses}
\label{sec:bottom-up-scenario}

There is a variety of models of neutrino mass where the operator $\O$
is induced from physics at accessible scales, TeV or less.  The
smallness of its strength is naturally achieved due to loop and Yukawa
couplings suppression. Note also that lepton number violating
parameters may appear in the numerator instead of the denominator of
$\O$ with the smallness of its strength natural in t'Hofft's
sense~\cite{'tHooft:1979bh}. For example, in the inverse seesaw
scheme of Sec.~\ref{sec:inverse-seesaw}, one can consistently take
\(\mu\) to be small as the symmetry of the theory increases in the
limit of vanishing \(\mu\), namely B-L is restored.

\subsubsection{Radiative models}
\label{sec:radiative-models}

\begin{figure}[h] \centering
    \includegraphics[height=3cm,width=.45\linewidth]{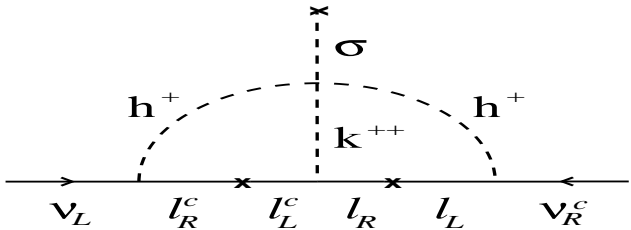}
    \caption{\label{fig:neumass} 
    Two-loop origin for neutrino mass.}
\end{figure}

The first possibility is that neutrino masses are induced by
calculable radiative corrections~\cite{zee:1980ai}. For example, they
may arise at the two-loop level~\cite{babu:1988ki} as illustrated in
Fig.~\ref{fig:neumass}. Up to a logarithmic factor one has,
schematically,
\begin{equation}
   \label{eq:babu}
{\mathcal M_\nu} \sim \lambda_0 \left(\frac{1}{16\pi^2}\right)^2 
f Y_l h Y_l f^T \frac{\vev{\Phi}^2}{(m_k)^2} \vev{\sigma}
 \end{equation}
 in the limit where the doubly-charged scalar $k$ is much heavier than
 the singly charged one. Here $l$ denotes a charged lepton, $f$ and
 $h$ are their Yukawa coupling matrices and $Y_l$ denotes the SM Higgs
 Yukawa couplings to charged leptons and $\vev{\sigma}$ is an \321
 singlet vev introduced in Ref.~\cite{Peltoniemi:1993pd}. Clearly,
 even if the proportionality factor $\lambda_0$ is large, the neutrino
 mass can be made naturally small by the presence of a product of five
 small Yukawas and the appearance of the two-loop factor. A remarkable
 feature of the model is that, thanks to the anti-symmetry of the $f$
 Yukawa coupling matrix, one of the neutrinos is massless.

\subsubsection{Supersymmetric neutrino masses}
\label{sec:supersymm-as-orig}

Another very interesting alternative are models where low energy
supersymmetry is the origin of neutrino mass~\cite{Hirsch:2004he}.
The intrinsically supersymmetric way to break lepton number is to
break the so-called R parity. This could happen spontaneously, driven
by a nonzero vev of an \321 singlet
sneutrino~\cite{Masiero:1990uj,romao:1992vu,romao:1997xf}.  This way
one is led to a very simple reference model which can be regarded as
the minimal way to include neutrino masses into the MSSM. In this
model R parity is violated only through an effective bilinear
term~\cite{Diaz:1998xc}.
Neutrino mass generation takes place in a hybrid scenario, with one
scale generated at tree level by the mixing of neutralinos and
neutrinos, induced by the sneutrino vevs, and the other induced by
``calculable'' radiative corrections~\cite{Hirsch:2000ef}.
\begin{figure}[h] \centering
    \includegraphics[height=3.5cm,width=.5\linewidth]{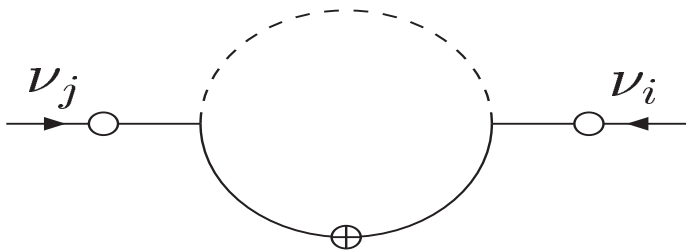}
  \caption{\label{fig:d-5} Loop origin of solar mass scale.
    Atmospheric scale arises from a type-I weak-scale seesaw-like
    graph involving exchange of sleptons and squarks.}
\end{figure}
Here the two blobs in each graph denote $\Delta L=1$ insertions, while
the crossed blob accounts for chirality flipping.  The general form of
the expression is quite involved but the approximation
\begin{equation}
   \label{eq:bilinear}
{\mathcal M_\nu} \sim \left(\frac{1}{16\pi^2}\right) {\vev{\Phi}^2} \frac{A}{m_0} Y_d Y_d 
 \end{equation}
 holds in some regions of parameters. Here $A$ denotes the trilinear
 soft supersymmetry breaking coupling (not shown in the diagram),
 $m_0$ the scalar supersymmetry breaking mass, and $Y_d$ the suitable
 Yukawa coupling.
The neutrino mass spectrum naturally follows a normal hierarchy, with
the atmospheric scale generated at the tree level and the solar mass
scale arising from calculable loops.

\subsubsection{Inverse seesaw}
\label{sec:inverse-seesaw}

Before closing this section we mention that there are also tree level
neutrino mass schemes with naturally light neutrinos. One is the
inverse seesaw scheme suggested in Ref.~\cite{Witten:1985xc} from
heterotic string motivations which led to the \e6 gauge group.
The model extends minimally the particle content of the SM by the
sequential addition of a pair of two-component \321 singlet leptons,
$\nu_i^c, S_i$, with \(i\) denoting a generation index running over
\(1,2,3\).

In the \(\nu,\nu^c,S\) basis, the \(9\times9\) neutral leptons mass
matrix $\mathcal{M_\nu}$ is the formally exactly the same as given in
Eq.~(\ref{eqn:doubleSeesaw})~\cite{mohapatra:1986bd}. 
Again the diagonalization follows the method in
Ref.~\cite{schechter:1982cv} but, in contrast to what is done for the
large \(\mu\) regime, now one separates the full heavy sector
consisting of three Quasi-Dirac states (six two-component leptons) at
once.
One obtains exactly the same light neutrino mass formula despite the
important difference that now the entry \(\mu\) is taken very small,
e.~g.  \(\mu \ll Y_\nu \vev{\Phi} \ll M\).

As before \(Y_\nu\) and \(M\) are arbitrary \(3\times3\) complex
Yukawa matrices, \(\mu\) being symmetric due to the Pauli principle.
Notice that for small \(\mu\) neutrino masses vanish with \(\mu\), as
we saw in Eq.~(\ref{eqn:lightNu}), illustrated in
Fig.~\ref{fig:iss-mass}.  The fact that the neutrino mass vanishes as
\(\mu\ \to 0\) is just the opposite of the behaviour of the seesaw
formulas in Eqs.~(\ref{eq:ss-formula0}) and (\ref{eq:ss-formula}) with
respect to $v_3$ and $\vev{\Delta_R}$, respectively; thus this is
sometimes called inverse seesaw model of neutrino masses.
The entry \(\mu\) may be proportional to the vev of an \321 singlet
scalar, in which case spontaneous B-L violation leads to the existence
of a majoron~\cite{gonzalez-garcia:1989rw}. This would be
characterized by a relatively low scale, so that the corresponding
phase transition could take place after the electroweak
transition~\cite{Peltoniemi:1993pd}.
The naturalness of the model stems from the fact that the limit when
$\mu \to 0$ increases the symmetry of the theory.  One possible
phenomenological implication would be the phenomenon of invisibly
decaying Higgs
boson~\cite{Joshipura:1993hp,romao:1992zx,Hirsch:2004rw}.  In such
schemes it will be crucial to take into account the existence of
sizeable invisible Higgs boson decay channels in the analysis of
experimental data on Higgs
searches~\cite{deCampos:1997bg,Abdallah:2003ry}.
Another possible implication is the existence of novel neutrino decay
and annihilation processes that may be relevant in dense supernova
media~\cite{kachelriess:2000qc}. Other aspects of the phenomenology
are mentioned in Sec.~\ref{sec:phen-neutr-mass}.

\subsection{Phenomenology of neutrino masses and mixings}
\label{sec:phen-neutr-mass}

Clearly the first phenomenological implication of neutrino mass models
is the phenomenon of neutrino oscillations, required to account for
the current data. Before turning to the prospects for probing neutrino
oscillations with high precision at the future generation of long
baseline oscillation experiments discussed in
Sec.~\ref{sec:cp-viol-neutr}, let us comment on other manifestations
of neutrino mass. 

First comes the issue of the absolute scale of neutrino masses. This
will be tested directly in searches for tritium beta
decay~\cite{Drexlin:2005zt,Masood:2007rc} and
cosmology~\cite{Lesgourgues:2006nd,Hannestad:2006zg}.
Some models do indeed suggest a lower bound on the absolute neutrino
mass. For example the model in Ref.~\cite{babu:2002dz} gives
$m_{\nu}\gsim 0.3$ eV and therefore will be tested soon.  Lastly, the
nature (Dirac or Majorana) of neutrinos and the new sources of CP
violation associated to its Majorana character can be scrutinized in
neutrinoless double beta decay searches and other \lnv processes (see
Sec.~\ref{sec:cp-violation-lepton}).
What else is there?


If neutrino masses arise {\sl a la seesaw} this may be all. The
simplest \10 seesaw has a drawback that the dynamics responsible for
generating the small neutrino masses seems most likely untestable. In
other words, beyond neutrino masses and oscillations, the model can
not be probed phenomenologically at low energies, due to the large
scale involved.  

\begin{figure}[h] \centering
  \includegraphics[height=6cm,width=.8\linewidth]{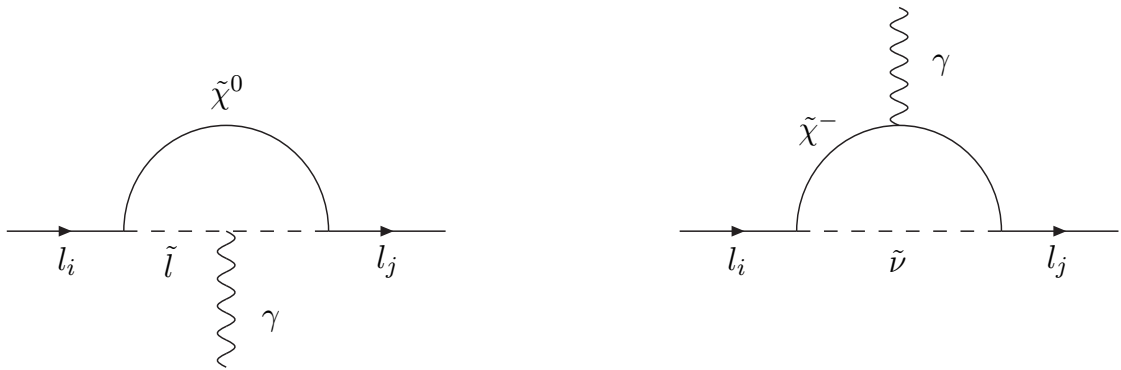}
\vglue -1cm
\caption{\label{fig:mueg1} Supersymmetric Feynman diagrams for
  \(l_{i}^{-} \to l_{j}^{-}\gamma\). They involve the exchange of
  charginos(neutralinos) and sneutrinos (charged sleptons).}
\end{figure}

\begin{figure}[h]
\centering
\includegraphics[clip,height=6cm,width=0.8\linewidth]{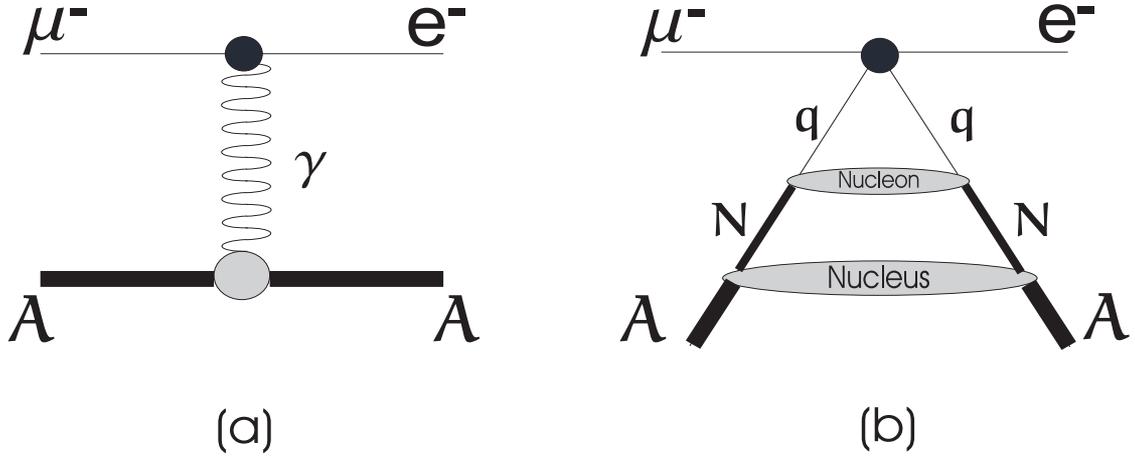}
\caption{Contributions to the nuclear $\mu^-- e^-$ conversion: (a)
  long-distance and (b) short-distance.}
     \label{fig:Diagrams}
\end{figure}

\begin{figure}[!h] \centering
  \includegraphics[height=6cm,width=.45\linewidth]{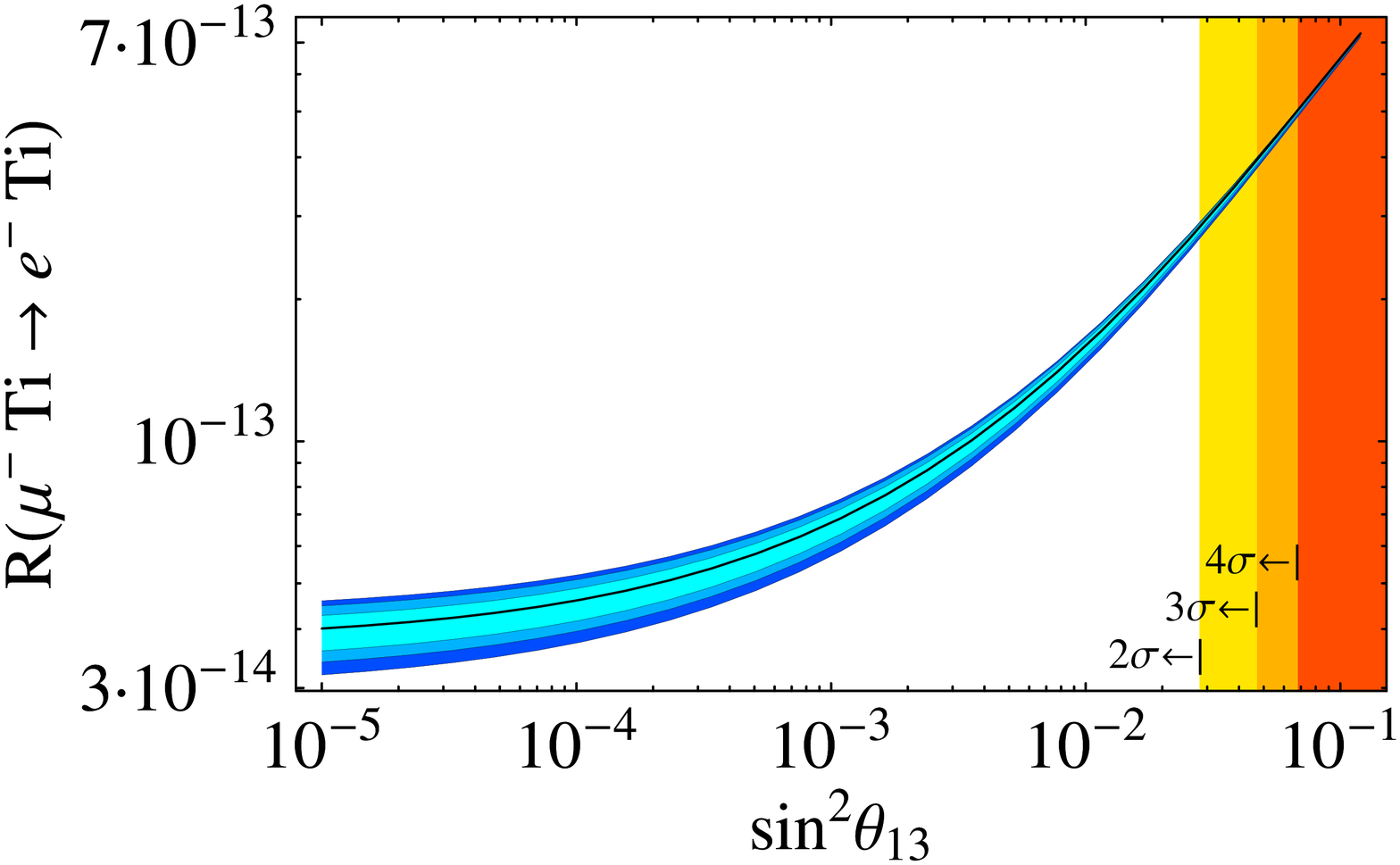}
 \includegraphics[height=6cm,width=.45\linewidth]{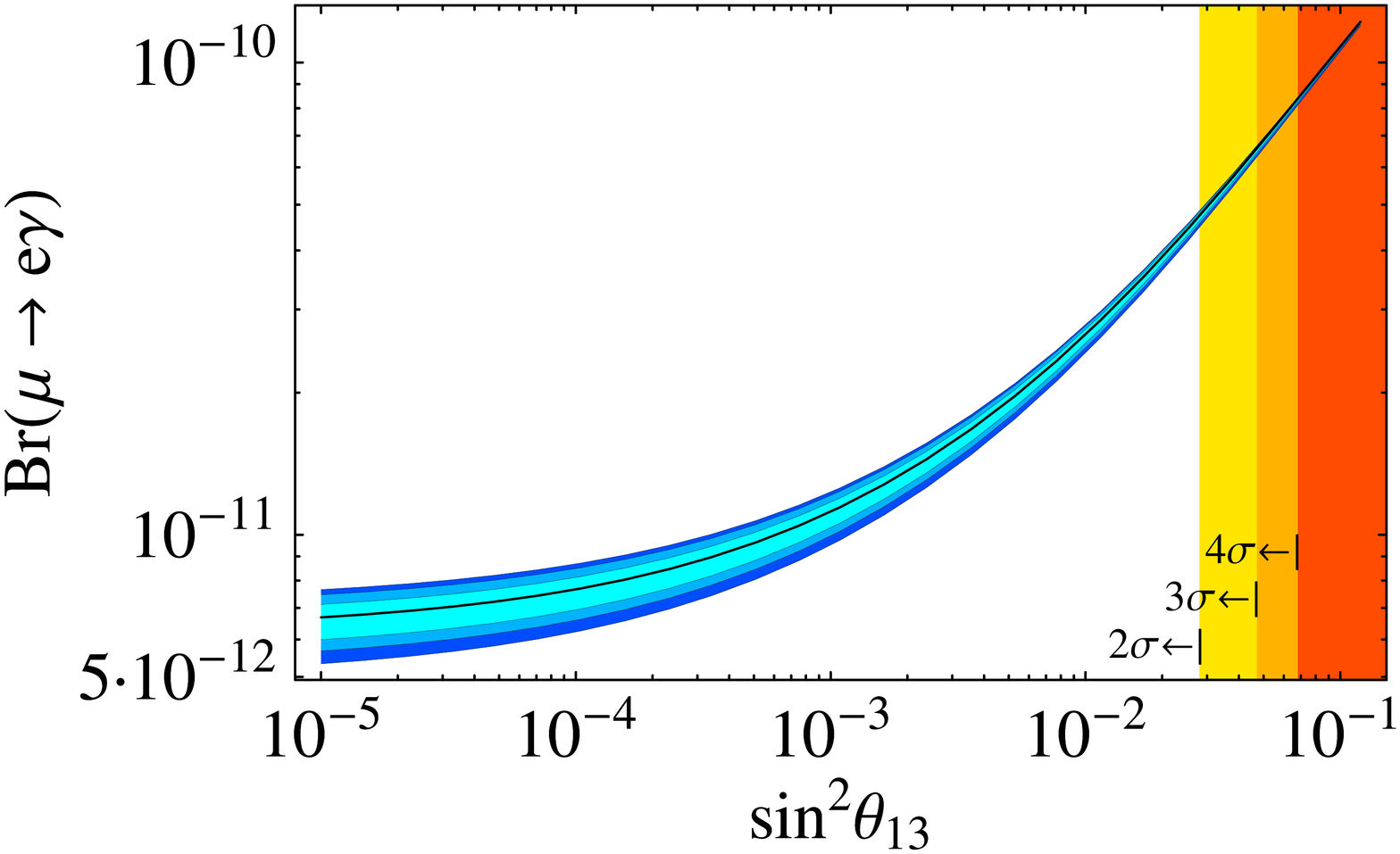}
 \caption{\label{fig:mueg2} LFV branching ratios in the supersymmetric
   inverse seesaw model of neutrino mass (see text).}
\end{figure}

\begin{figure}[!h] \centering
  \includegraphics[height=6cm,width=.45\linewidth]{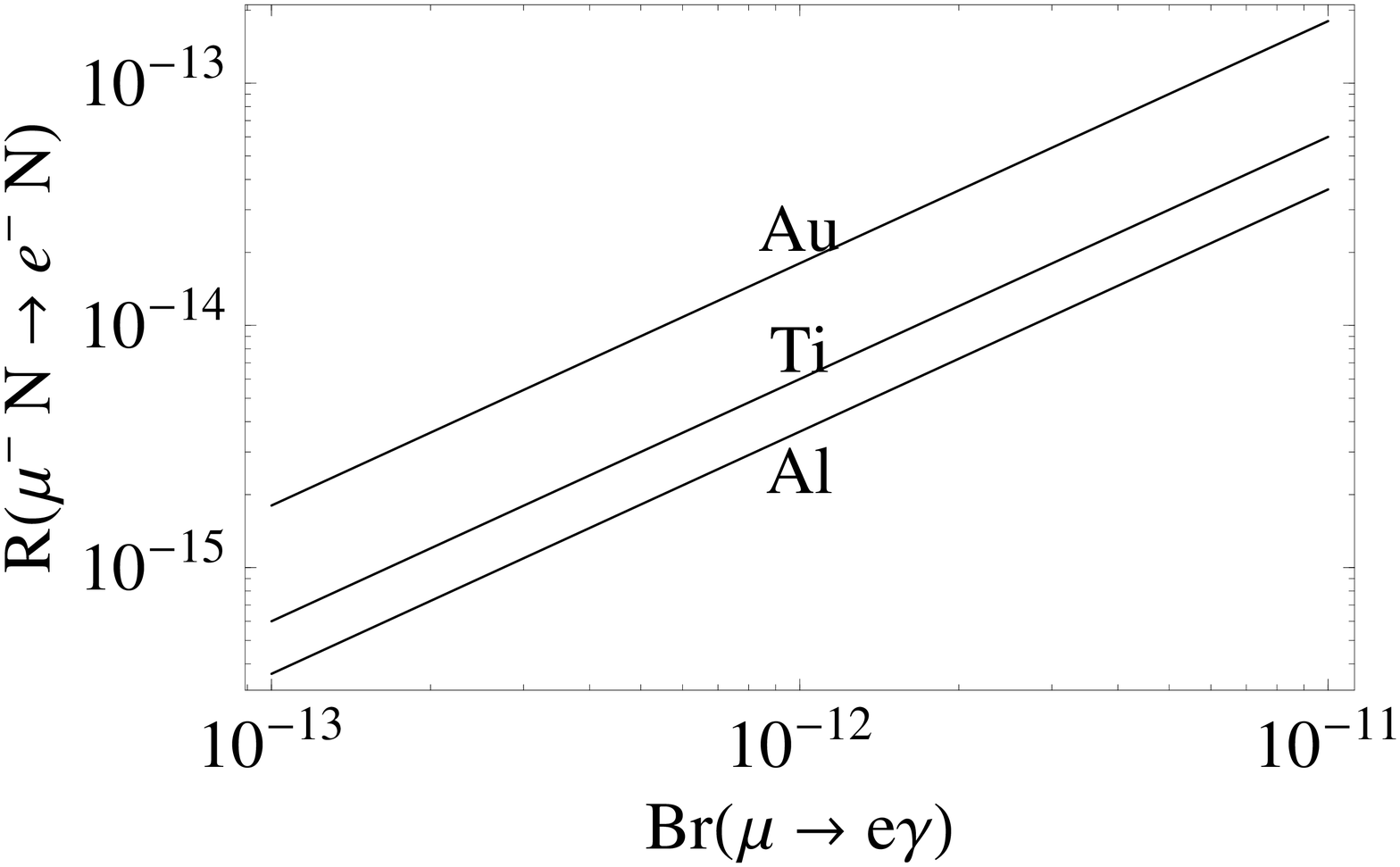}
\caption{\label{fig:mueg3} Correlation between \(Br(\mu\to e\gamma)\)
  and muon-electron conversion in nuclei from Ref.~\cite{Deppisch:2005zm}.}
\end{figure}

However, in the presence of supersymmetry, the phenomenology of the
seesaw mechanism can be richer.
A generic feature of supersymmetric seesaw models is the existence
\lfv decays like $\mu^-\to e^-\gamma$, flavour violating tau decays
(Fig.~\ref{fig:mueg1}) and nuclear $\mu^--e^-$ conversion
(Fig.~\ref{fig:Diagrams}). These can have accessible rates
(Figs.~\ref{fig:mueg2} and \ref{fig:mueg3}) which depend not only on
the seesaw mechanism, but also on the details of supersymmetry
breaking and on a possible theory of flavour.
The way these LFV processes arise through
supersymmetry~\cite{hall:1986dx} is illustrated in
Fig.~\ref{fig:mueg1}. 
The existence of such loop effects leads to enhanced rates for flavour
violating
processes~\cite{borzumati:1986qx,Hisano:2001qz,casas:2001sr,Antusch:2006vw,Calibbi:2006nq,Joaquim:2006uz}
\cite{Deppisch:2004fa,Deppisch:2005zm}. 
We note that such LFV and/or CP violating effects may arise at the
one--loop level from the exchange of relatively light neutral heavy
leptons, even in the absence of
supersymmetry~\cite{bernabeu:1987gr,branco:1989bn,rius:1990gk,gonzalez-garcia:1992be,Ilakovac:1994kj}.

As an illustration of the interplay of supersymmetry with heavy lepton
exchange in engendering \lfv processes we consider the rates for the
$\mu^-\to e^-\gamma$ decay in the framework of the supersymmetric
inverse seesaw model~\cite{Deppisch:2004fa,Deppisch:2005zm}.
Fig.~\ref{fig:mueg2} displays the dependence of the branching ratios
for $\mu^--e^-$ conversion in Ti (left) and $\mu^-\to e^-\gamma$
(right) with the small neutrino mixing angle $\theta_{13}$, for
different values of $\theta_{12}$ (black curve: $\theta_{12}$ best fit
value, blue bands denote $2\sigma,~3\sigma,~4\sigma$ confidence
intervals for the solar mixing angle $\theta_{12}$).  The inverse
seesaw parameters are given by: $M=1$~TeV and $\mu=30$~eV.  The light
neutrino parameters used are from \cite{Maltoni:2004ei}, except for
$\theta_{13}$ which is varied as shown in the plots.  The vertical
coloured bands denote the exclusion areas from the 2$\sigma$,
3$\sigma$ and $4\sigma$ limits on $\sin^2\theta_{13}$, respectively.
These \lfv rates may be testable in the new generation of upcoming
experiments.  For large \(M\) the estimates are similar to those of
the standard supersymmetric seesaw.

The novel feature present in this model and not in the minimal seesaw
is the possibility of enhancing \(Br(\mu\to e\gamma)\) and other tau
decays with \lfv even in the {\sl absence} of supersymmetry in the
case where \(M\) is low, around TeV or so.
In this region of parameters the model also gives rise to large
estimates for the nuclear $\mu^--e^-$ conversion, depicted in
Fig.~\ref{fig:Diagrams}.  The latter fall within the sensitivity of
future experiments such as PRISM~\cite{Kuno:2000kd}. Note that LFV
happens even in the absence of supersymmetry and even in the massless
neutrino limit. The allowed rates are hence unsuppressed by the
smallness of neutrino
masses~\cite{bernabeu:1987gr,branco:1989bn,rius:1990gk,gonzalez-garcia:1992be,Ilakovac:1994kj}.
Finally, for low enough \(M\) the corresponding quasi-Dirac heavy
leptons could be searched directly at
accelerators~\cite{Dittmar:1990yg,Abreu:1997pa}.


{\bf Supersymmetric neutrino mass and collider tests}

We now turn to the case of low-scale models of neutrino mass,
considered in Sec.~\ref{sec:bottom-up-scenario}. As an example we
consider the case of models where supersymmetry is the origin of
neutrino mass~\cite{Hirsch:2004he,Hirsch:2000ef}, considered in
Sec.~\ref{sec:supersymm-as-orig}.
A general feature of these models is that, unprotected by any
symmetry, the lightest supersymmetric particle (LSP) is unstable. In
order to reproduce the masses indicated by current neutrino
oscillation data, typically the LSP is expected to decay inside the
detector.  More strikingly, its decay properties correlate with
neutrino mixing angles. For example, if the LSP is the lightest
neutralino, it is expected to have the same decay rate into muons and
taus, since the observed atmospheric angle, as we will see below, is
close to
$\pi/4$~\cite{Porod:2000hv,romao:1999up,mukhopadhyaya:1998xj}.
\begin{figure}[!h]
\centering
\includegraphics[clip,height=5cm,width=0.42\linewidth]{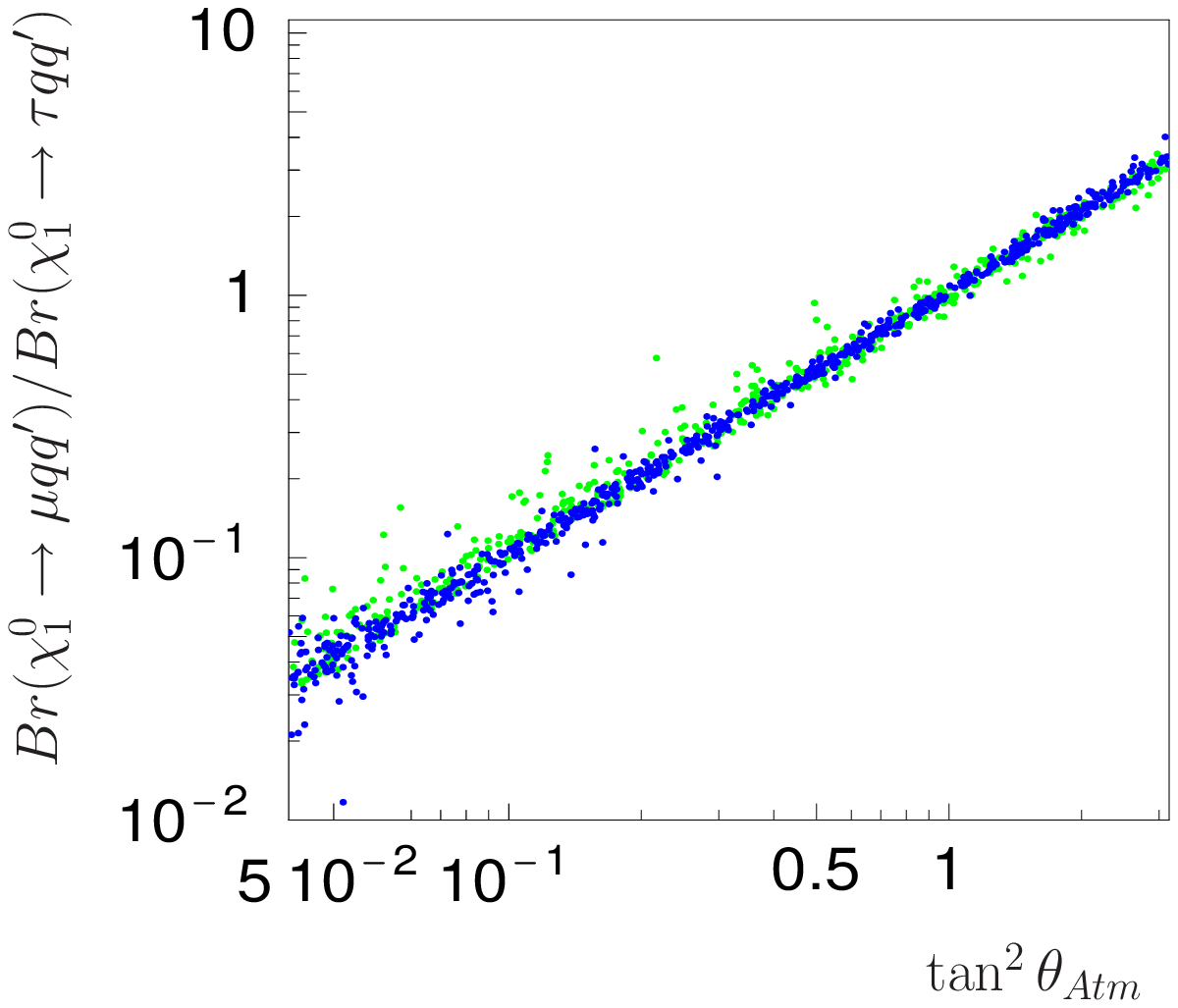}
\includegraphics[clip,height=5cm,width=0.42\linewidth]{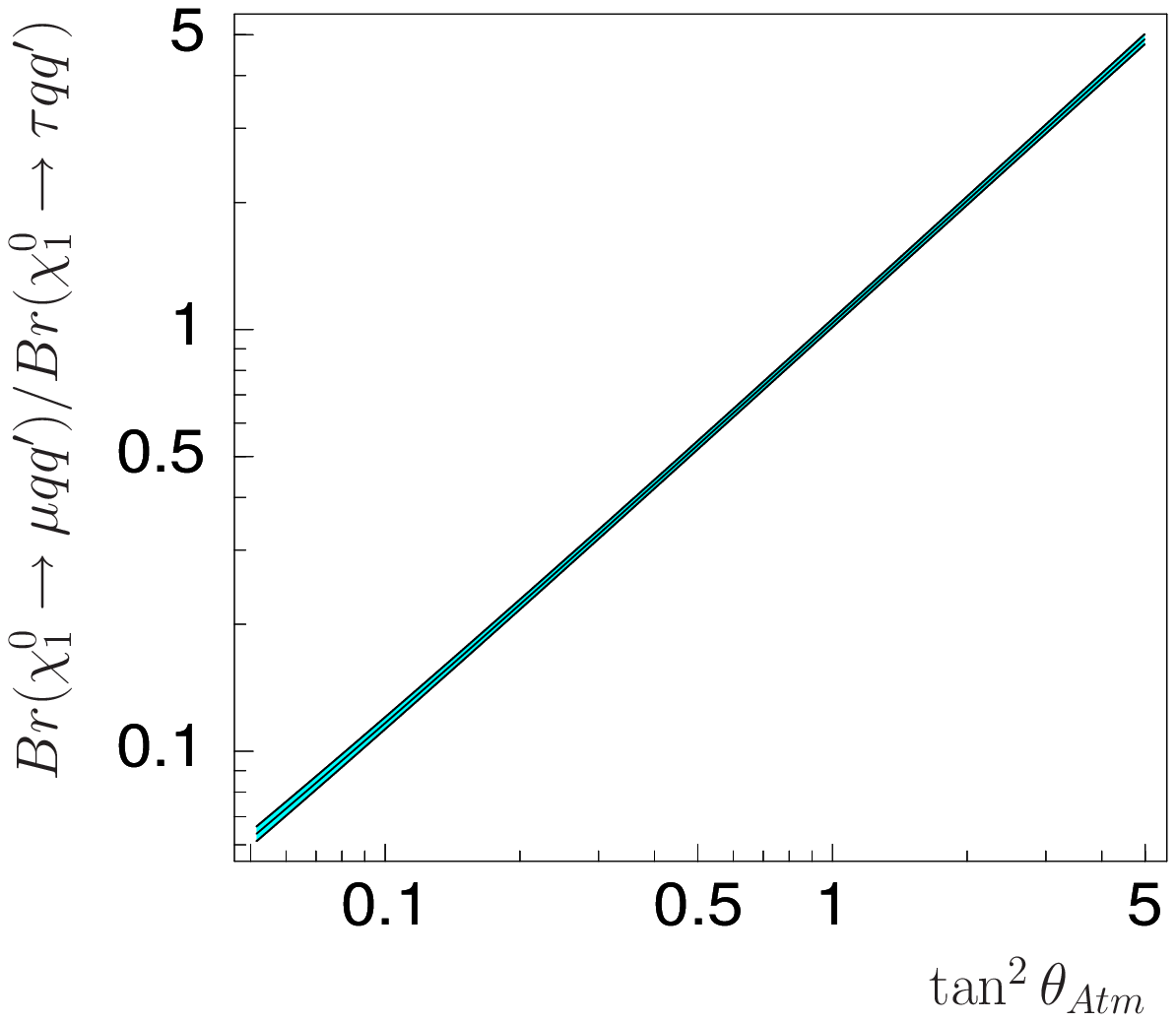}
\caption{LSP decays trace the atmospheric mixing angle~\cite{Porod:2000hv}.}
     \label{fig:correlation}
\end{figure}
This opens the tantalizing possibility of testing neutrino mixing 
at high energy accelerators, like the "Large Hadron Collider" (LHC) 
and the "International Linear Collider" (ILC) and constitutes a smoking 
gun signature of this proposal that for sure will be tested.
This possibility also illustrates the complementarity of accelerator
and non-accelerator approaches in elementary particle physics.

{\bf Majorons and B-L gauge bosons}

Before closing let us also mention that there could be a dynamical
``tracer'' of the mass-generation mechanism. In the case of the
``1-2-3'' majoron seesaw considered in Sec.~\ref{sec:majoron-seesaw}
the existence of the Goldstone boson brings new interactions of
neutrinos with majoron, but the rates are probably too small to be
cosmologically relevant.  There can also be a new neutrino decay mode,
$\nu_3 \to \nu + J$, involving the emission of the majoron, denoted
$J$~\cite{schechter:1982cv}.
However, the strong suppression of the majoron emitting decay rates
obtained in Ref.~\cite{schechter:1982cv} and the smallness of neutrino
masses that follows from current laboratory experiments, as well as
cosmology, indicates that such neutrino decays are purely academic.
However, there is an important exception. In models with
family-dependent lepton numbers the decay is enhanced even in
vacuum~\cite{valle:1983ua,gelmini:1984ea}. Alternatively, if the
neutrino decays in high density media, like supernovae, characterized
by huge matter effects, then they could lead to detectable signals in
underground water Cerenkov experiments~\cite{kachelriess:2000qc} even
in the simplest majoron schemes.
Conversely, in the case of seesaw mechanisms with gauged B-L symmetry,
if the corresponding B-L scale is low, as in the model discussed in
Sec.~\ref{sec:low-b-l}, there will exist a light new neutral gauge
boson, \ZP that could be detected in searches for Drell-Yan processes
at the LHC.

\section{The general structure of the lepton mixing matrix}
\label{sec:struct-lept-mixing}

In any gauge theory in order to identify physical particles one must
diagonalize all relevant mass matrices, which typically result from
gauge symmetry breaking.
Mechanisms giving mass to neutrinos generally imply the need for new
interactions whose Yukawa couplings (like $Y_\nu$) will coexist with
that of the charged leptons, $Y_l$.
The lepton mixing matrix $V$ follows from a mismatch between the
diagonalization of the charged lepton mass matrix and that of the
neutrino mass matrix, which involve these Yukawas. This is similar to
the way the Cabibbo-Kobayashi-Maskawa matrix arises in the quark
sector, namely from a mismatch between up and down-type Yukawa
couplings.
Hence, like quarks, massive neutrinos will generally mix. The
structure of this mixing is not generally predicted from first
principles.
Whatever the ultimate high energy gauge theory may be it must be
broken to the SM at low scales, so one should characterize the
structure of the lepton mixing matrix in terms of the \321 structure.

\subsection{Quark-like lepton mixing matrix}
\label{sec:preliminaries}

From the start, in the absence of a fundamental theory of flavour, the
lepton mixing matrix of massive Dirac neutrinos is a unitary matrix
$V$ which can always be parametrized as
\begin{equation}
  \label{eq:par0}
V = \omega_0 (\gamma) \prod_{i<j}^{n} \omega_{ij} (\eta_{ij})\:
\end{equation}
where $$\omega_0 (\gamma)=\mathrm{exp}\: i(\sum_{a=1}^{n} \gamma_a
A_a^a)$$ is a diagonal unitary matrix described by $n-1$ real
parameters $\gamma_a$. (By choosing an overall relative phase between
charged leptons and Dirac neutrinos we can take $V$ as unimodular,
i.~e.~det $V=1$, so that the phases in the ``Cartan'' matrix
$\omega_0$ obey $\sum_{a=1}^{n} \gamma_a =1$).
On the other hand
$$\omega_{ab} (\eta_{ab})=\mathrm{exp}\: \sum_{a=1}^{n} (\eta_{ab}A_a^b - \eta_{ab}^*
A_b^a)$$ is a complex rotation in ${ab}$ with parameter $\eta_{ab} =
|\eta_{ab}| \mathrm{exp}\:i\theta_{ab}$.  For example,
\begin{equation}
  \label{eq:par12}
\omega_{12} (\eta_{12}) =
\left(\begin{array}{ccccc}
c_{12} & e^{i \phi_{12}} s_{12} & 0 ...\\
-e^{-i \phi_{12}} s_{12} & c_{12} & 0 ...\\
0  & 0 & 1 ...\\
...&....&.......
\end{array}\right)
\end{equation}
However, in order to obtain the parametrization of this matrix, as
presented in Ref.~\cite{schechter:1980gr}, one must take into account
that its structure can be further simplified by taking into account
that, once the charged leptons and Dirac neutrino mass matrices are
diagonal, one can still rephase the corresponding fields by $\omega_0
(\alpha)$ and $\omega_0 (\gamma-\alpha)$, respectively, keeping
invariant the form of the free Lagrangean. This results in the form
\begin{equation}
  \label{eq:par}
V = \omega_0 (\alpha) \prod_{i<j}^{n} \omega_{ij} (\eta_{ij})\:
\omega_0^\dagger (\alpha).
\end{equation}
There are $n-1$ $\alpha$-values associated to Dirac neutrino phase
redefinitions which we are still free to choose.  Using the
conjugation property
\begin{equation}
  \label{eq:par-conj}
\omega_0 (\alpha) \omega_{ab} (|\eta_{ab}| \mathrm{exp}\:i\theta_{ab})\:
\omega_0^\dagger (\alpha)  = \omega_{ab} [|\eta_{ab}| \mathrm{exp}\:i(\alpha_a + \theta_{ab} - \alpha_b )]
\end{equation}
we arrive at the final Dirac lepton mixing matrix which is, of course,
identical in form to that describing quark mixing.
It involves a set of 
\begin{equation}
  \label{eq:count}
n(n-1)/2 \:  \: \:\mathrm{mixing  \: \: angles} \: \: \theta_{ij} \: \: \mathrm{and} \: \: \: n(n-1)/2-(n-1) 
\mathrm{ \: CP  \: \:phases}\,.   
\end{equation}

One sees that $(n-1)$ phases were eliminated by rephasing, a
possibility that would be absent in the case of Majorana neutrinos.
This is the parametrization as originally given in
\cite{schechter:1980gr}, with unspecified factor ordering. It needs to
be supplemented only by an ordering prescription, for this we refer
the reader to the PDG choice~\cite{Yao:2006px}.  

In summary, if neutrinos masses were added {\sl a la Dirac} their
charged current weak interaction would have exactly the same structure
as that of quarks. From Eq.~(\ref{eq:count}) for the case $n=3$ there
are 3 angles and precisely one leptonic CP violating phase, just as in
the Kobayashi-Maskawa matrix describing quark mixing.

However the imposition of lepton number conservation is {\sl ad hoc}
in a gauge theory and hence neutrinos are expected to be
Majorana~\footnote{The same happens for electrically neutral
  supersymmetric fermions such as the gravitino, gluino and
  neutralinos.}.  Note that the argument in favour of neutrinos being
Majorana has nothing to do with the neutrino mass generation
mechanism, it is much more basic.

\subsection{Majorana case: unitary approximation}
\label{sec:unit-appr-lept}

Here we consider the form of the lepton mixing matrix in models where
neutrino masses arise in the absence of right-handed neutrinos.  These
include, for example, the models in Sec.~\ref{sec:radiative-models}
and constitute a good approximation for high-scale seesaw models in
Sec.~\ref{sec:top-down-scenario}.

For $n$ generations of Majorana neutrinos the lepton mixing matrix has
exactly the same form given in Eq.~(\ref{eq:par}). The difference
insofar as the structure of lepton mixing is concerned come about
since in the case of Majorana neutrinos the mass terms are manifestly
not invariant under rephasings of the neutrino fields. As a result,
the parameters $\alpha$ in Eq.~(\ref{eq:par}) can not be used to
eliminate $n-1$ Majorana phases as we just did in
Sec.~\ref{sec:preliminaries}. Consequently these are additional
sources of CP violation in the currents of gauge theories with
Majorana neutrinos. They are sometimes called ``Majorana phases''.
They already exist in a theory with just two generations of Majorana
neutrinos, $n=2$, which is described by
\begin{equation}
   \label{eq:w12}
\omega_{13} = \left(\begin{array}{ccccc}
c_{12} &  e^{i \phi_{12}} s_{12} \\
-e^{-i \phi_{12}} s_{12} &  c_{12}
\end{array}\right)\,,
 \end{equation}
 where $\phi_{12}$ is the Majorana phase.  Such ``Majorana''-type CP
 phase is, in a sense, mathematically more ``fundamental'' than the
 Dirac phase whose existence, as we just saw, requires three
 generations at least.

 For the case of three neutrinos the lepton mixing matrix can be
 parametrized as ($n=3$)~\cite{schechter:1980gr}
\begin{equation}
  \label{eq:2227}
K =  \omega_{23} \omega_{13} \omega_{12},
\end{equation}
where each factor in the product of the $\omega$'s is effectively
$2\times 2$, characterized by an angle and a CP phase. Two of the
three angles are involved in solar and atmospheric oscillations, so we
set $\theta_{12} \equiv \theta_\Sol$ and $\theta_{23} \equiv
\theta_\Atm$.  The last angle in the three--neutrino leptonic mixing
matrix is $\theta_{13}$,
\begin{equation}
   \label{eq:w13}
\omega_{13} = \left(\begin{array}{ccccc}
c_{13} & 0 & e^{i \phi_{13}} s_{13} \\
0 & 1 & 0 \\
-e^{-i \phi_{13}} s_{13} & 0 & c_{13}
\end{array}\right)\,.
 \end{equation}
 Such symmetrical parametrization of the lepton mixing matrix, $K$
 can be written as:
\begin{equation}
K=\left[ \begin{array}{c c c}
c_{12}c_{13}&s_{12}c_{13}e^{i{\phi_{12}}}&s_{13}e^{i{\phi_{13}}}\\
-s_{12}c_{23}e^{-i{\phi_{12}}}-c_{12}s_{13}s_{23}e^{i({\phi_{23}}-{\phi_{13}})}
&c_{12}c_{23}-s_{12}s_{13}s_{23}
e^{i({\phi_{12}}+{\phi_{23}}-{\phi_{13}})}&c_{13}s_{23}e^{i{\phi_{23}}}\\
s_{12}s_{23}e^{-i({\phi_{12}}+{\phi_{23}})}-c_{12}s_{13}c_{23}e^{-i{\phi_{13}}}
&-c_{12}s_{23}e^{-i{\phi_{23}}}-
s_{12}s_{13}c_{23}e^{i({\phi_{12}}-{\phi_{13}})}&c_{13}c_{23}\\
\end{array} \right].
\label{writeout}
\end{equation}
All three CP violating phases are physical~\cite{Schechter:1981gk}:
$\phi_{12},\phi_{23}$ and $\phi_{13}$. Even though the parametrization
is fully ``symmetric'' there is a basic difference between Dirac and
Majorana phases.
The ``invariant'' combination $\delta \equiv
\phi_{12}+\phi_{23}-\phi_{13}$ corresponding to the ``Dirac phase''
exists only beyond $n=3$. At $n=3$ a single phase (say $\phi_{13}$)
may be taken to be non-zero.  This is the phase that corresponds to
the Dirac phase present in the quark sector, and affects neutrino
oscillations involving three neutrinos. The other two phases are
associated to the Majorana nature of neutrinos.  These already exist
for $n=2$ but show up only in lepton-number violating processes, like
neutrinoless double beta decay. They do not affect conventional
neutrino
oscillations~\cite{Schechter:1981gk,doi:1981yb,deGouvea:2002gf} but do
enter in the ``neutrino oscillation thought-experiment'' described in
\cite{Schechter:1981gk}. Needless to say the latter process is of
conceptual interest, only. 


 An important subtlety arises regarding the conditions for CP
 conservation in gauge theories of massive Majorana neutrinos.  Unlike
 the case of Dirac fermions, where CP invariance implies that the
 mixing matrix should be real, in the Majorana case the condition one
 needs is
$$ K^* = K \eta$$
where $\eta = \mathrm{diag}(+,+,..., , ,..)$ is the signature matrix
describing the relative signs of the neutrino mass eigenvalues that
follow from diagonalizing the relevant Majorana mass matrix, if one
insists in using real diagonalizing matrices, like
Wolfenstein~\cite{Wolfenstein:1981rk}.  Consequently the value
$\phi_{12} = \pi/2$ and $\phi_{12} = 0$ are both CP conserving.
These important signs determine the CP properties of the neutrinos and
play a crucial role in \nbb.

Before concluding let us mention that the above parametrization of the
lepton mixing matrix was originally given in \cite{schechter:1980gr},
but with unspecified factor ordering. In what follows we tacitly
employ the ordering prescription now adopted by the
PDG~\cite{Yao:2006px}.
Further discussion on the advantages of this ``symmetrical
parametrization'' and additional references can be found in
Ref.~\cite{Masood:2004dq}.

\subsection{General  seesaw-type lepton mixing matrix}
\label{sec:general-form-lepton}

What we now present is a brief summary of the original systematic
study of the effective form of the lepton mixing matrix originally
given in~\cite{schechter:1980gr}, applicable to any scheme of Majorana
neutrino masses where isosinglet and isodoublet mass terms coexist.
In addition to the existence of Majorana phases, which do not require
the existence of \321 singlets, one has doublet-singlet mixing
parameters, in general complex. As a result one finds that leptonic
mixing as well as CP violation may take place even in the massless
neutrino limit~\cite{branco:1989bn,rius:1990gk}.

The most general effective model is described by $(n,m)$, $n$ being
the number of \321 isodoublet and $m$ the number of \321 isosinglet
leptons. In this Section we assume $m \neq 0$, the $(n,0)$ has just
been considered in Sec.~\ref{sec:unit-appr-lept}. The two-component
\321 singlet leptons present in the theory have in general a gauge and
Lorentz invariant Majorana mass term, breaking total lepton number
symmetry.  Their number, $m$, is completely arbitrary, as \321
singlets carry no anomaly.

The resulting structure of the weak currents can be substantially more
complex, since the heavy isosinglets will now mix with the ordinary
$SU(2)$ doublet neutrinos in the charged current weak interaction. As
a result, the mixing matrix describing the charged leptonic weak
interaction is a rectangular matrix, called $K$.

Typically the charged weak interactions of the light (mass-eigenstate)
neutrinos in $(n,m)$ models are effectively described by a mixing
matrix which is non-unitary. For example, one sees that the coupling
of a given light neutrino to the corresponding charged lepton is
decreased by a certain factor. The existence of these neutral heavy
leptons could therefore be inferred from low energy weak decay
processes, where the neutrinos that can be kinematically produced are
only the light ones. There are constraints on the strength of the such
mixing matrix elements that follow from low energy weak decay
measurements as well as from LEP.

An explicit parametrization of the weak charged current mixing matrix
$K$ that covers the most general situation present in these $(n, m)$
models has also been given in Ref.~\cite{schechter:1980gr} so that
here we only highlight the main points. It involves in general 
\begin{equation}
  \label{eq:angle}
n(n+2m-1)/2  
\end{equation}
mixing angles $\theta_{ij}$ and 
\begin{equation}
  \label{eq:phase}
n(n+2m- 1)/2  
\end{equation}
CP violating phases $\phi_{ij}$.

This number far exceeds the corresponding number of parameters
describing the charged current weak interaction of quarks
Eq.~(\ref{eq:par}).  The reason is twofold: (i) neutrinos are Majorana
particles, their mass terms are not invariant under rephasings, and
(ii) the isodoublet neutrinos in general mix with the \321 singlets
and, in addition, so CP may also be violated in this mixing.
As a result, there are far less CP phases that can be eliminated by
field redefinitions. They may play a role in leptogenesis as well as
neutrino oscillations.

Another important feature which arises in any theory based on \321
where isosinglet and isodoublet lepton mass terms coexist is that the
leptonic neutral current is non-trivial~\cite{schechter:1980gr}: there
are non diagonal couplings of the Z to the mass-eigenstate neutrinos.
They are expressed as a projective Hermitian matrix 
$$P = K^\dagger K$$
This contrasts with the neutral current couplings of mass-eigenstate
neutrinos in theories where there are no isosinglet neutrinos , i.e.,
$m = 0$. In that case, just as in the case of Dirac neutrinos, the
neutral current couplings of mass-eigenstate neutrinos is diagonal.

Before closing this section we mention explicitly that the $(3,3)$
seesaw model is characterized, from Eqs.~(\ref{eq:angle}) and
(\ref{eq:phase}) by 12 mixing angles and 12 CP phases (both Dirac and
Majorana-type)~\cite{schechter:1980gr}.

Before we close, note that, in a scheme with $m < n$, $n - m$
neutrinos will remain massless, while $2m$ neutrinos will acquire
Majorana masses, $m$ light and $m$ heavy.  For example, in a model
with $n = 3$ and $m = 1$ one has one light and one heavy Majorana
neutrino, in addition to the two massless ones. In this case clearly
there will be less parameters than present in a model with $m = n$.
Note also that for $m > n$, the case $m = 2n$ corresponds to the
models considered in Secs.~\ref{sec:low-b-l} and
\ref{sec:inverse-seesaw}.

\subsection{Leptonic CP violation and leptogenesis}
\label{sec:lept-neutr-mass}

\begin{figure}[h]
\centering
\includegraphics[clip,height=3.5cm,width=0.8\linewidth]{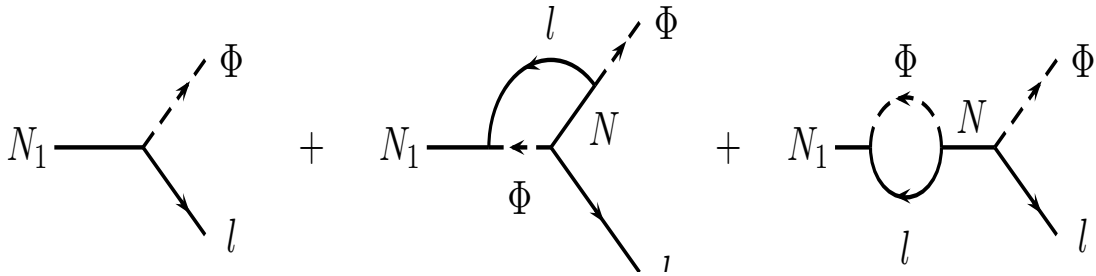}
\caption{Diagrams contributing to heavy neutrino decays whose
  interference leads to leptogenesis.}
     \label{fig:lep-g}
\end{figure}

An interesting cosmological implication of leptonic CP violation is
that it opens an attractive possibility of accounting for the
matter-antimatter asymmetry in the Universe within the so-called
thermal leptogenesis
mechanism~\cite{Fukugita:1986hr,Buchmuller:2005eh}.
In this picture the heavy ``right-handed'' neutrinos of the seesaw
mechanism play a crucial role.
They decay out of equilibrium generating a lepton asymmetry,
through diagrams in Fig.~\ref{fig:lep-g}.
This lepton (or B-L) asymmetry gets converted, through sphaleron
processes, into a baryon asymmetry~\cite{kuzmin:1985mm}.
One of the crucial ingredients is CP violation in the lepton sector.

In the framework of a supersymmetric seesaw scheme the high temperature
needed for leptogenesis leads to an overproduction of gravitinos,
which destroys the standard predictions of Big Bang Nucleosynthesis
(BBN).
In minimal supergravity models, with $m_{3/2} \sim$ 100 GeV to 10 TeV
gravitinos are not stable, decaying during or after BBN. Their rate of
production can be so large that subsequent gravitino decays completely
change the standard BBN scenario.
To prevent such ``gravitino crisis'' one requires an upper bound on
the reheating temperature $T_R$ after inflation, since the abundance
of gravitinos is proportional to the reheating temperature.  
A recent detailed analysis derived a stringent upper bound $T_R \lsim
10^6$ GeV when the gravitino decay has hadronic
modes~\cite{Kawasaki:2004qu}.
This upper bound is in conflict with the temperature required for
leptogenesis, $T_R > 2 \times 10^9$ GeV~\cite{Buchmuller:2004nz}.
Therefore, thermal leptogenesis in the usual hierarchical case seems
difficult to reconcile with low energy supersymmetry if gravitino
masses lie in the range suggested by the simplest minimal supergravity
models.
Their required mass is typically too large in order for them to be
produced after inflation, implying that the minimal type I
supersymmetric seesaw schemes may be in trouble~\footnote{A loophole
  is the possibility of resonant
  leptogenesis~\cite{Pilaftsis:2003gt}.}.
\begin{figure}[h]
\centering
\includegraphics[height=8cm,width=.8\linewidth]{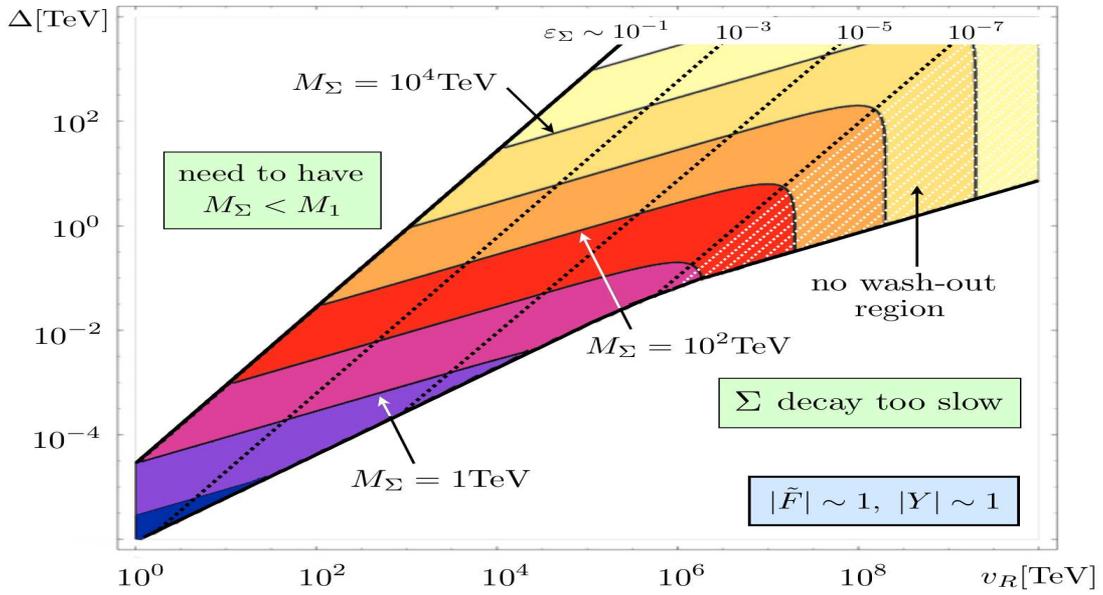}
  \caption{Parameter space in low-scale leptogenesis model of
    Ref.~\cite{Hirsch:2006ft}.}
 \label{fig:correlation1}
\end{figure}

\begin{figure}[!h] \centering
  \includegraphics[width=7.5cm,height=0.28\textheight]{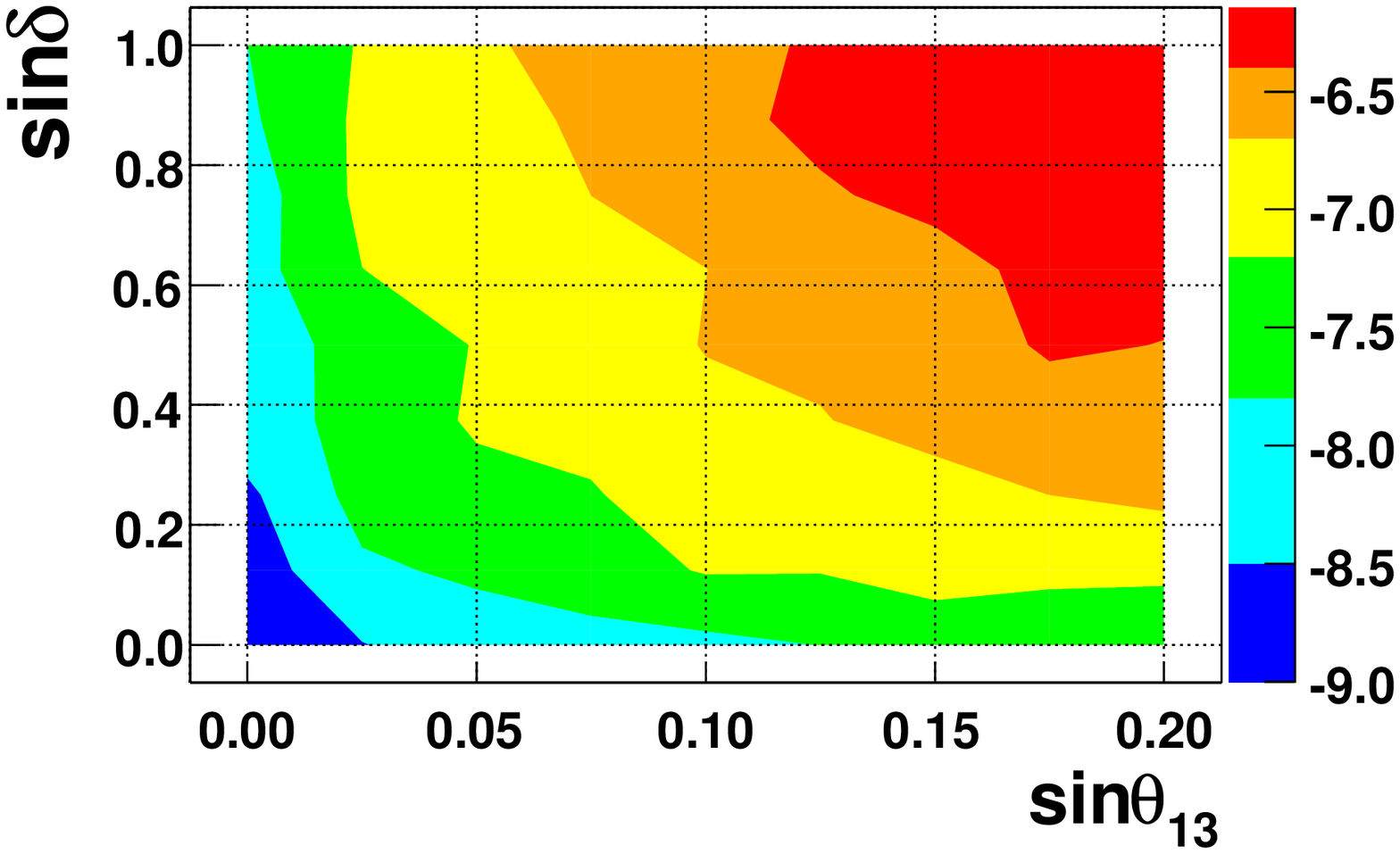}
  \includegraphics[width=7.5cm,height=0.28\textheight]{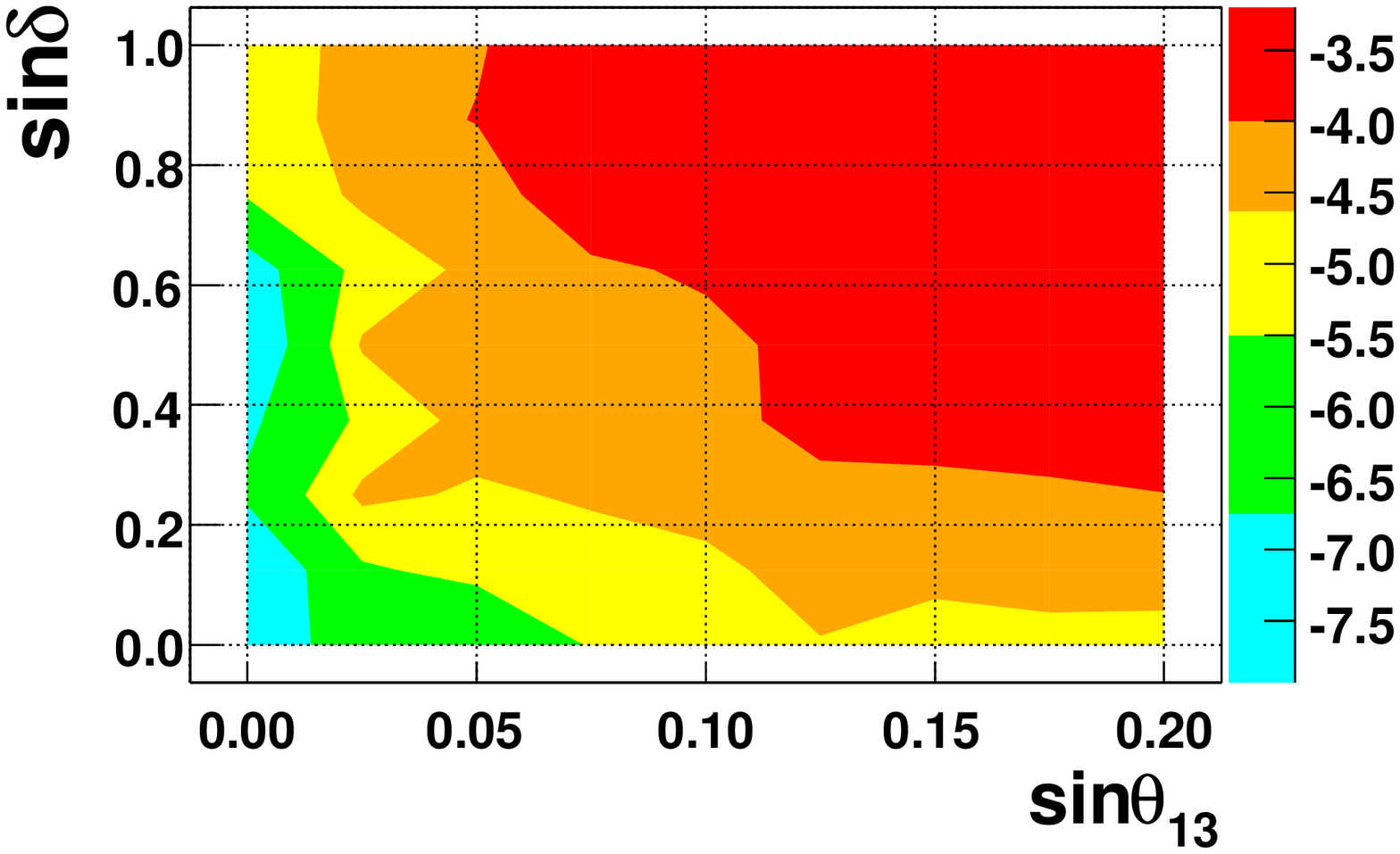}
 \caption{Contours of the maximum value of the asymmetry $\epsilon_1$
    obtained using the Fritzsch texture~\cite{Romao:2007jr}.}
\label{fig:larger}
\end{figure}

One way to cure this inconsistency has been suggested in
Ref.~\cite{Hirsch:2006ft} is to add a singlet $\Sigma$ to the
supersymmetric \10 seesaw model discussed in \cite{Malinsky:2005bi}.
Leptogenesis can occur at the TeV scale through the decay of $\Sigma$,
thereby avoiding the gravitino crisis.  Washout of the asymmetry is
suppressed by the absence of direct couplings of $\Sigma$ to leptons.
The leptogenesis parameter region is illustrated in
Fig.~\ref{fig:correlation1} and the reader is addressed to the
original paper for details. Here we simply comment that in this model
successful leptogenesis can occur for $M_\Sigma = 1$ TeV and low $v_R
= 10$ TeV. As an example we consider the $\Sigma$ decay asymmetries
$\epsilon_1$ that can be generated in this way. The contours of the
maximum $\epsilon_1$ values obtained in this model using the Fritzsch
texture for the quark masses are illustrated in Fig. \ref{fig:larger}.
The left and right panels correspond to two alternative choices of the
relevant model parameters, explained in Ref.~\cite{Romao:2007jr}.
Taking into account washout effects, acceptable values of the baryon
asymmetry $\mathcal O(10^{-10})$ require $\epsilon_1$ above $\mathcal
O(10^{-7})$ or so. From Fig.~\ref{fig:larger} one sees that this is
clearly possible to achieve within this model even if the only source
of CP violation is that provided by the three-flavour Dirac phase that
enters in neutrino oscillations.

An alternative viable leptogenesis mechanism based on a modified
type-I seesaw was suggested in Ref.~\cite{Farzan:2005ez}.  It consists
in adding a small R-parity violating $\lambda_i \hat{\nu^c}_i
\hat{H}_u \hat{H}_d$ term in the superpotential, where $\hat{\nu^c}_i$
are right-handed neutrino supermultiplets. One can show that in the
presence of this term, the produced lepton-antilepton asymmetry can be
enhanced.

Last, but not least, leptogenesis may also be induced by the type-II
seesaw mechanism, with or without supersymmetry.
In the non-supersymmetric case this requires the addition of two heavy
Higgs scalar triplets~\cite{PhysRevLett.80.5716}, while in
supersymmetry even with a minimal triplet content, leptogenesis can be
naturally accommodated thanks to the resonant interference between
superpotential and soft supersymmetry breaking
terms~\cite{D'Ambrosio:2004fz}.

\section{Neutrino oscillations}
\label{sec:stat-neutr-oscill}

\subsection{The formalism}
\label{sec:formalism}

We start with a very general discussion aimed at fixing the notation
used in this review. As we have already seen in
Sec.~\ref{sec:general-form-lepton}, in general models of neutrino mass
the lepton mixing matrix $K$ contain both Dirac and Majorana-type CP
phases and is in general rectangular as it couples the charged leptons
also to the heavy (mainly isosinglet) neutrinos postulated, e.~g. in
type-I seesaw models, in order to produce neutrino
masses~\cite{schechter:1980gr}.  Such states are too heavy to
participate in neutrino oscillations which are effectively described
by a non-unitary mixing matrix.  Such deviations from unitarity are
the origin of gauge-induced neutrino non-standard interactions and
may, in some cases, be sizeable, see
Sec.~\ref{sec:robustn-neutr-oscill}.
In the discussion of neutrino oscillations that we give in
Secs.~\ref{sec:stat-neutr-oscill},~\ref{sec:cp-viol-neutr} and
\ref{sec:near-term-exper} we will tacitly assume that $K$ is strictly
unitary, so that the three active neutrinos are mixed as follows,
\begin{equation}
\nu_{\alpha L}  = \sum_{i=1}^3 U_{\alpha i}
\nu_{i L}, \ \ (\alpha = e,\mu, \tau), 
\label{eq:mixing1}
\end{equation}
where we have now denoted $K$ by $U$, to highlight that $U^\dagger
U=UU^\dagger=1$.  Here $\nu_{\alpha L}~ (\alpha = e,\mu, \tau)$
describe the left handed neutrino fields with definite flavor whereas
$\nu_{i L} ~(i = 1,2,3)$ describe the fields with definite masses.
Here, the matrix $U$ is the leptonic analogue of the quark mixing
matrix~\cite{Kobayashi:1973fv}.  In this review, whenever it is
necessary to use the explicit parametrization, we use the following
standard parametrization~\cite{Yao:2006px},
\begin{eqnarray}
U & = &
\left[
\begin{matrix}
1  &  0       & 0       \cr
0  &  c_{23}  & s_{23}  \cr
0  & -s_{23}  & c_{23}  
\end{matrix}
\right]
~\left[
\begin{matrix}
c_{13} & 0  &  s_{13} \text{e}^{-i\delta}\\ 
0  & 1 & 0 \\ 
-s_{13}\text{e}^{i\delta}  & 0 & c_{13} 
\end{matrix}
\right]
~\left[
\begin{matrix}
 c_{12} & s_{12} & 0 \\ 
-s_{12} & c_{12} & 0 \\ 
0  & 0 & 1 
\end{matrix}
\right] \\
 & = & 
\left[
\begin{matrix}
c_{12} c_{13} & s_{12}c_{13}  &  s_{13} \text{e}^{-i\delta}\\ 
-s_{12}c_{23}-c_{12}s_{23}s_{13}\text{e}^{i\delta}  
& c_{12}c_{23}-s_{12}s_{23}s_{13}\text{e}^{i\delta}  & s_{23}c_{13} \\ 
s_{12}s_{23}-c_{12}c_{23}s_{13}\text{e}^{i\delta}  & 
-c_{12}s_{23}-s_{12}c_{23}s_{13}\text{e}^{i\delta}  & c_{23}c_{13} 
\end{matrix}
\right]
\label{eq:cc_pdg}
\end{eqnarray}
\vglue 0.05cm
\noindent
where $c_{ij} \equiv \cos\theta_{ij}$, $s_{ij} \equiv \sin\theta_{ij}$
and $\delta$ is the CP violating phase. This form is the same as
Eq.~(\ref{eq:2227})~\cite{schechter:1980gr}, taking
$\phi_{12}=0=\phi_{23}$ in the ``invariant'' combination $\delta
\equiv \phi_{12}+\phi_{23}-\phi_{13}$ that corresponds to the ``Dirac
phase'' relevant for neutrino oscillations.

Eq.~(\ref{eq:mixing1}) implies that in terms of state vectors
$| \nu_\alpha \rangle\, (\alpha = e,\mu, \tau) $ for flavor 
and $| \nu_i \rangle (i = 1,2,3) $ for mass eigenstates, 
they are related by $U^\ast$ instead of $U$ as 
(see e.g.\cite{Giunti:2003qt}), 
\begin{equation}
| \nu_\alpha \rangle = \sum_{i=1}^3 U^\ast_{\alpha i}
| \nu_i \rangle, \ \ (\alpha = e,\mu, \tau), 
\end{equation}
where for simplicity, we omitted the indices $L$ which indicates the
left handed chirality.  Then it is straightforward to compute the
$\nu_\alpha \to \nu_\beta$ oscillation probability, which is given,
for ultra-relativistic neutrinos, by,
\begin{eqnarray}
P(\nu_\alpha \to  \nu_\beta)  & =  &
\left|~\sum_j U^*_{\alpha j} ~U_{\beta j} 
\text{e}^{-i\frac{m^2_j}{2E}L}~\right|^2\nonumber \\
& = & \delta_{\alpha \beta} 
-4 \sum_{i>j} {\Re}(U^*_{\alpha i} U_{\alpha j} U_{\beta i} U^*_{\beta j} )
\sin^2\left(  \frac{\Delta m^2_{ij}}{4E}L \right) \nonumber \\
& & \quad \quad \quad +2 \sum_{i>j} 
{\Im}(U^*_{\alpha i} U_{\alpha j} U_{\beta i} U^*_{\beta j} )
\sin \left( \frac{\Delta m^2_{ij}}{2E}L \right),
\label{eq:prob_vac}
\end{eqnarray}
where $E$ is the neutrino energy, $L$ is the distance traveled by
neutrino, and $\Delta m_{ij} \equiv m_i^2 - m_j^2$ ($m_i$ being mass
eigenvalues) are the mass squared differences. Here $\Re$ and $\Im$
denote real and imaginary parts.


Let us now describe the neutrino evolution equation in matter.  It is
given in terms of flavor eigenstates as,
\begin{equation}
i\frac{d}{dx}
\left[
\begin{matrix}
\nu_e \cr 
\nu_\mu \cr 
\nu_\tau 
\end{matrix}
\right]
= H(x) 
\left[
\begin{matrix}
\nu_e \cr 
\nu_\mu \cr 
\nu_\tau 
\end{matrix}
\right],
\label{eqn:nu-evol}
\end{equation}
where $\nu_\alpha \, (\alpha = e,\mu,\tau)$ is the amplitude 
for the $\alpha$-flavor. 
The Hamiltonian matrix $H$ is given by 
\begin{equation}
H(x) = U 
\left[
\begin{matrix}
 \frac{m_1^2}{2E}  & 0 & 0 \\ 
0  &   \frac{m_2^2}{2E} & 0 \\ 
0  & 0 &  \frac{m_3^2}{2E}
\end{matrix}
\right]
U^{\dagger}
+ 
\left[
\begin{matrix}
V(x)  & 0 & 0 \\ 
0  & 0 & 0 \\ 
0  & 0 & 0
\end{matrix}
\right],
\label{eq:hamil}
\end{equation}
where $x$ is the position along the neutrino trajectory and $V(x)$ is
the matter potential which is given, for unpolarized 
medium~\footnote{Neutrino evolution in polarized media was treated in
  \cite{Nunokawa:1997dp}.}, as
\begin{equation}
V(x) = \sqrt{2} G_F N_e(x)~,
\end{equation}
where $G_F$ is the Fermi constant, $N_e(x)$ is the electron number
density at $x$.  The Hamiltonian matrix in Eq.~(\ref{eq:hamil}) can be
replace by,
\begin{equation}
H(x) = U 
\text{diag}\left[0, \frac{\Delta m_{21}^2}{2E}, \frac{\Delta m_{31}^2}{2E}
\right]
U^{\dagger}
+ 
\text{diag}[V(x), 0,0],
\label{eqn:Hmatter2}
\end{equation}
by re-phasing all of the neutrino flavors by $\exp[{-im^2_1x/(2E)}]$.
For anti-neutrinos, the same Eq. holds with the change $V(x) \to
-V(x)$ and $U \to U^{\ast}$ (or equivalently $\delta \to - \delta$).

For a given $x$, Hamiltonian can be diagonalized by using the effective 
mixing matrix in matter $U(N)$, 
\begin{equation}
H(x) = U(N) 
\text{diag}\left[0, \frac{\Delta m_{21}^2(N)}{2E}, 
\frac{\Delta m_{31}^2(N)}{2E}
\right]
U^{\dagger}(N),
\end{equation}
where $\Delta m_{ij}^2(N)$ are the effective mass 
squared differences in matter (see Eq.~\ref{eq:deltam2_matter}). 

\subsection{The current data}
\label{sec:neutr-oscill-data}

Current neutrino oscillation data have no sensitivity to CP violation.
Thus we neglect all phases in the analysis and take, moreover, the
simplest unitary 3-by-3 form of the lepton mixing matrix in
Eq.~(\ref{eq:2227}).
In this approximation oscillations depend on the three mixing
parameters $\sin^2\theta_{12}, \sin^2\theta_{23}, \sin^2\theta_{13}$
and on the two mass-squared splittings $\Dms \equiv \Delta m^2_{21}
\equiv m^2_2 - m^2_1$ and $\Dma \equiv \Delta m^2_{31} \equiv m^2_3 -
m^2_1$ characterizing solar and atmospheric neutrinos.  The hierarchy
$\Dms \ll \Dma$ implies that one can set $\Dms = 0$, to a good
approximation, in the analysis of atmospheric and accelerator data.
Similarly, one can set $\Dma$ to infinity in the analysis of solar and
KamLAND data.
Apart from the data already mentioned, the analysis also includes the
constraints from "negative" searches at reactor experiments.
                             
\subsubsection{Solar and reactor data}
\label{sec:solar-+-kamland}
                                                                            
The solar neutrino data includes the rates of the chlorine experiment
($2.56 \pm 0.16 \pm 0.16$~SNU), the gallium results of SAGE
($66.9~^{+3.9}_{-3.8}~^{+3.6}_{-3.2}$~SNU) and GALLEX/GNO ($69.3 \pm
4.1 \pm 3.6$~SNU), as well as the 1496--day Super-K data (44 bins: 8
energy bins, 6 of which are further divided into 7 zenith angle bins).
The SNO sample includes the 2002 spectral day/night data (17 energy
bins for each day and night period)~\cite{ahmad:2002jz} and the most
recent data (391-day data) from the salt phase in the form of the
neutral current (NC), charged current (CC) and elastic scattering (ES)
fluxes~\cite{Aharmim:2005gt}.  The analysis includes both statistical
errors, as well as systematic uncertainties such as those of the eight
solar neutrino fluxes.  Taking into account new radiative opacities,
Bahcall et al (for references see Appendix C in hep-ph/0405172-v5)
obtain new solar neutrino fluxes, neutrino production distributions
and solar density profile, included in the analysis.

Reactor anti-neutrinos from a network of power stations in Japan are
detected by the KamLAND collaboration at the Kamiokande site through
the process $\bar\nu_e + p \to e^+ + n$, where the delayed coincidence
of the prompt energy from the positron and a characteristic gamma from
the neutron capture allows an efficient reduction of backgrounds.
Most of the incident $\bar{\nu}_e$'s come from nuclear plants at
distances of $80-350$ km from the detector, far enough to probe large
mixing angle (LMA) oscillations.
To avoid large uncertainties associated with geo-neutrinos an energy 
cut at 2.6~MeV prompt energy is applied for the oscillation analysis.

The first KamLAND data corresponding to a 162 ton-year exposure gave
54 anti-neutrino events in the final sample, after cuts, whereas $86.8
\pm 5.6$ events are predicted for no oscillations with $0.95\pm 0.99$
background events~\cite{eguchi:2002dm}. This is consistent with the
no--disappearance hypothesis at less than 0.05\% probability, giving
the first terrestrial confirmation of oscillations with $\Dms$.
Additional KamLAND data with a larger fiducial volume of the detector
corresponding to an 766.3~ton-year exposure have been presented
in~\cite{araki:2004mb}.  In total 258 events have been observed,
versus $356.2\pm 23.7$ reactor neutrino events expected in the case of
no disappearance and $7.5\pm 1.3$ background events. This leads to a
confidence level of 99.995\% for $\bar\nu_e$ disappearance, in
addition to evidence for spectral distortion consistent with
oscillations.

Finally, recent data from the KamLAND experiment have been presented
at the 10th International Conference on Topics in Astroparticle and
Underground Physics, TAUP 2007~\cite{KamLAND:2007}.  These data
correspond to a total exposure of 2881 ton-year.  They provide a
better measurement of the solar neutrino oscillation parameters,
especially the ``solar'' mass-squared splitting. This is due to the
reduction of systematic uncertainties thanks to the full volume
calibration.
These data correspond to a total exposure of 2881 ton-year, almost 4
times larger than 2004 data.  They provide a very precise measurement
of the solar neutrino oscillation parameters, mainly the mass
splitting. This is due to the reduction of systematic undertainties
thanks to the full volume calibration.

Various systematic errors associated to the neutrino fluxes,
backgrounds, reactor fuel composition and individual reactor powers,
small matter effects, and improved $\bar{\nu}_e$ flux parametrization
are included in the analysis~\cite{Maltoni:2004ei}.  Assuming CPT
invariance one can directly compare the information obtained from
solar neutrino experiments with the KamLAND reactor results.  

One finds that a strong evidence for spectral distortion in the
KamLAND data, leading to a much improved $\Dms$ determination,
substantially reducing the allowed region of oscillation parameters.
Altogether the KamLAND data single out the LMA solution from the
previous ``zoo'' of alternatives~\cite{Maltoni:2003da}.  
As discussed in Sec.~\ref{sec:robustn-neutr-oscill}, more than just
cornering the oscillation parameters~\cite{Maltoni:2002aw}, KamLAND
has eliminated all previously viable non-oscillation
solutions~\cite{pakvasa:2003zv} playing a key role in the resolution
of the solar neutrino problem.
                                
Last, but not least, the Borexino collaboration has also recently
presented their first data~\cite{Borexino:2007}.  While they provide
the first real time detection of $^7$Be solar neutrinos, and an
important confirmation of the Standard solar model and the large
mixing oscillations, these data currently do not affect the
determination of neutrino oscillation parameters.
                                         
\subsubsection{Atmospheric and accelerator data}
\label{sec:atmospheric-+-k2k}
                                                                               
The first evidence for neutrino conversions was the zenith angle
dependence of the $\mu$-like atmospheric neutrino data from the
Super-K experiment in 1998, an effect also seen in other atmospheric
neutrino experiments.  At that time there were equally good
non-oscillation solutions, involving non-standard neutrino
interactions~\cite{Gonzalez-Garcia:1998hj}.  Thanks to the
accumulation of up-going muon data, and the observation of the dip in
the $L/E$ distribution of the atmospheric $\nu_\mu$ survival
probability, the signature for atmospheric neutrino oscillations has
now become convincing.  We have used the Super-K charged-current
atmospheric neutrino events, with the $e$-like and $\mu$-like data
samples of sub- and multi-GeV contained events grouped into 10
zenith-angle bins, with 5 angular bins of stopping muons and 10
through-going bins of up-going muons~\cite{Kajita:2004ga}.  Multi-ring
$\mu$ and neutral-current events and $\nu_\tau$ appearance are not
used, since an efficient Monte-Carlo simulation of these data would
require further details of the Super-K experiment, in particular of
the way the neutral-current signal is extracted from the data. As far
as atmospheric neutrino fluxes are concerned, the analysis of
\cite{Maltoni:2004ei} employs state--of--the--art three--dimensional
calculations given in ~\cite{Honda:2004yz}.
\clearpage
                                                                  
The disappearance of $\nu_\mu$'s over a long-baseline probing the same
$\Delta m^2$ region relevant for atmospheric neutrinos is now
available from accelerator neutrino oscillation experiments, such as
the KEK to Kamioka (K2K) neutrino oscillation experiment and the MINOS
Experiment using the NuMI Beamline facility at Fermilab.

Neutrinos produced by a 12~GeV proton beam from the KEK proton
synchrotron consist of 98\% muon neutrinos with a mean energy of
1.3~GeV. The beam is controlled by a near detector 300~m away from the
proton target.  Comparing these near detector data with the $\nu_\mu$
content of the beam observed by the Super-K detector at a distance of
250~km gives information on neutrino oscillations.

                 
The K2K collaboration has published details of the analysis of their
full data sample (K2K-I and K2K-II)~\cite{Ahn:2006zz}. The data have
been taken in the period from June 1999 to November 2004 and
correspond to $0.922\times 10^{20}$~p.o.t. Without oscillations
$158^{+9.2}_{-8.6}$ events are expected whereas only 112 events have
been observed. Out of these, 58 events are single--ring events where
the reconstruction of the neutrino energy is possible. These events
have been used in order to perform a spectral analysis of the K2K
data, as described in Ref.~\cite{Maltoni:2004ei}.


The first MINOS results were released in 2006.
MINOS is a long--baseline experiment that searches for $\nu_\mu$
disappearance in a neutrino beam with a mean energy of 3~GeV produced
at Fermilab. It consists of a near detector, located at 1 km from the
neutrino source and a far detector located at the Soudan Mine, at
735~km from Fermilab. First data corresponding to $0.93\times
10^{20}$~p.o.t.\ ~\cite{Tagg:2006sx}, have comparable weight as the
final K2K data sample. In the absence of oscillations $177\pm11$
$\nu_\mu$ events with $E<10$~GeV are expected, whereas 92 have been
observed, which provides a $5.0\sigma$ evidence for disappearance.
New experimental data have been released by the MINOS Collaboration.
These have been collected from June 2006 to July 2007 (Run-IIa), and
they have been analyzed together with the first data sample (Run-I),
with a total exposure of 2.5$\times$10$^{20}$ p.o.t.  In total, 563
$\nu_\mu$ events have been observed at the far detector, while
738$\pm$30 events were expected for no oscillation.

One finds that the values of the oscillation parameters from the
$\nu_\mu$ disappearance results at MINOS are consistent with the ones
from K2K, as well as from Super-K atmospheric data, providing strong
terrestrial confirmation of oscillations with $\Dma$ with accelerator
neutrinos. Statistics of the current data sample still does not
strongly constrain the mixing angle.
However, although the determination of $\sin^2\theta_\Atm$ is
completely dominated by atmospheric data, K2K and MINOS data give an
important restriction on the allowed $\Dma$
values~\cite{Maltoni:2004ei}. 
They constrain $\Dma$ from below, which is important for further
future long-baseline experiments, since their sensitivities are
drastically affected if $\Dma$ lies in the lower part of the 3$\sigma$
range indicated by current atmospheric data.
Note that there is currently no sensitivity to the sign of the
atmospheric mass-squared splitting, this being one of the challenges
for upcoming experiments.
                   
\subsection{Status of three-neutrino oscillations}
\label{sec:stat-three-neutr}
                                                                             
\begin{figure}[t] \centering
\includegraphics[width=.45\linewidth,height=6.5cm]{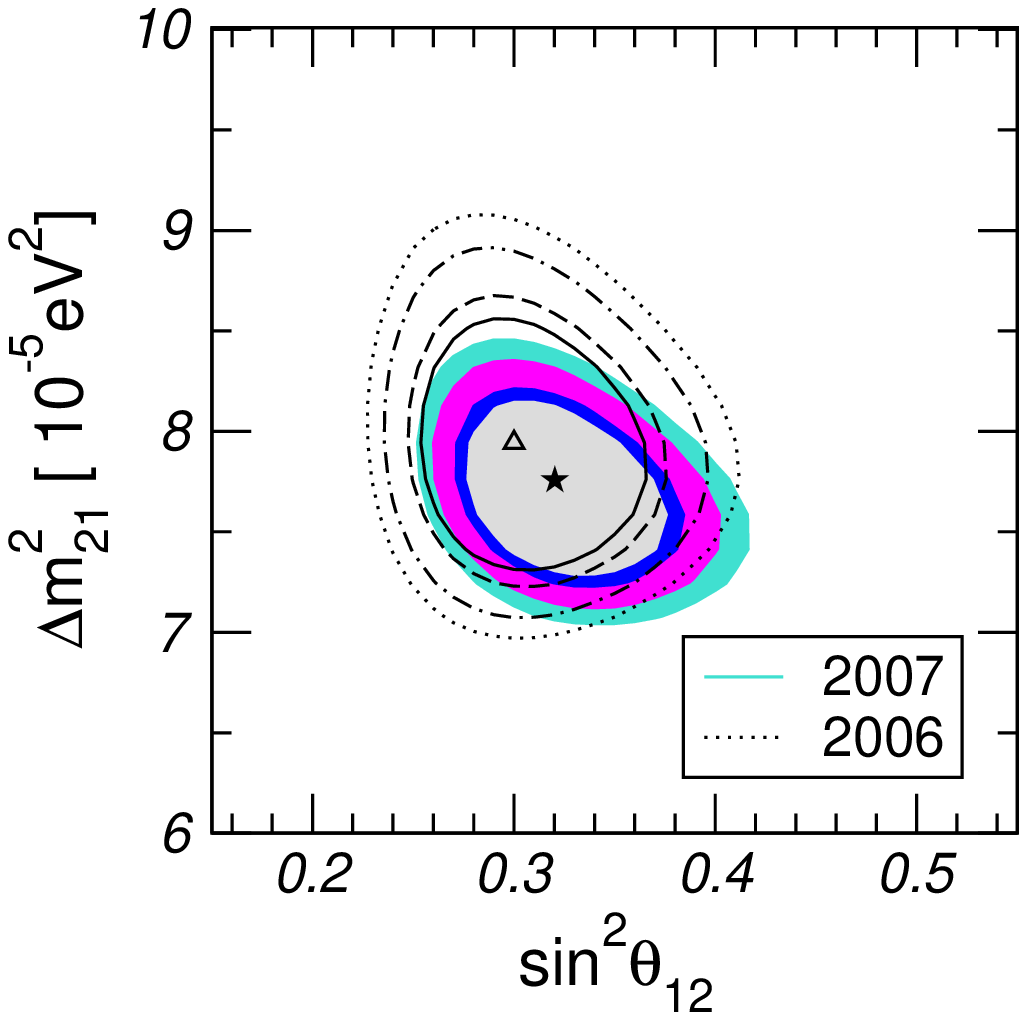}
\includegraphics[width=.45\linewidth,height=6.5cm]{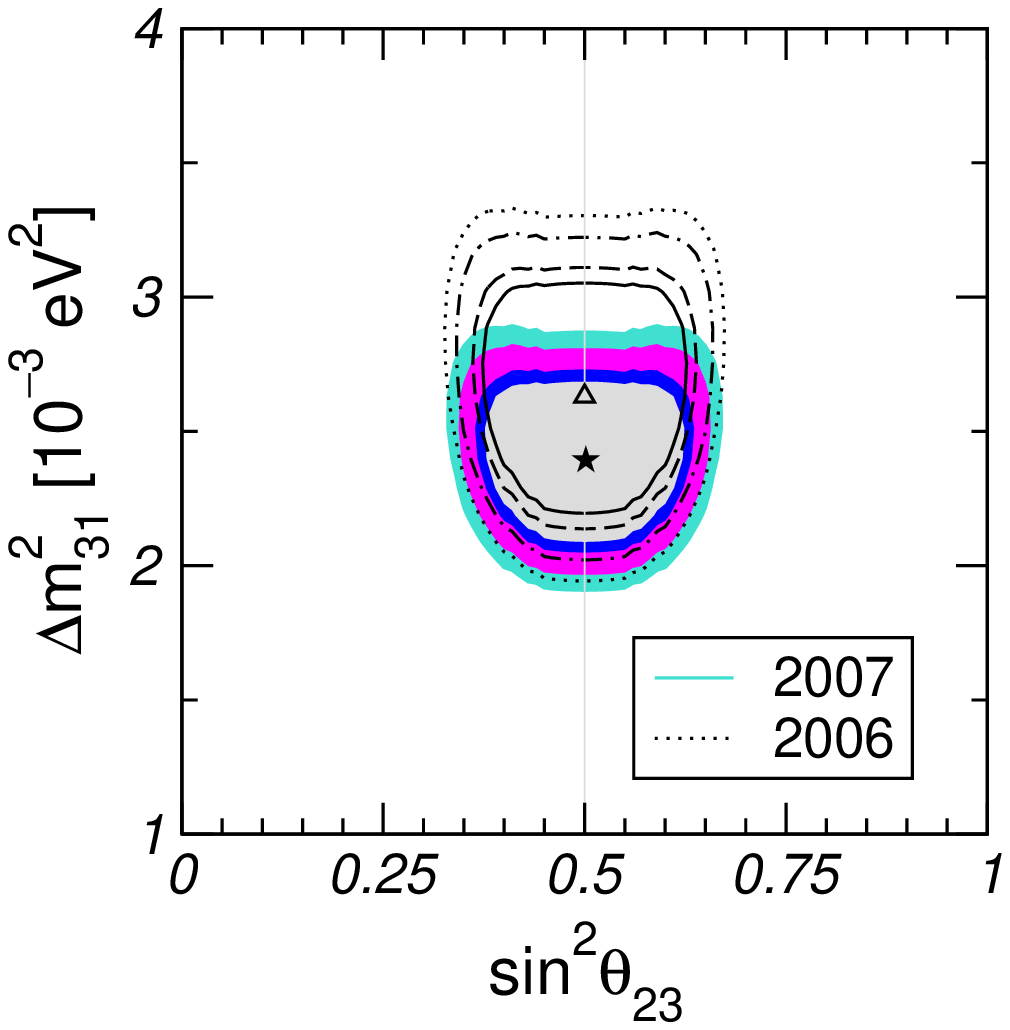}
\caption{\label{fig:global} %
  Current regions of neutrino oscillation parameters allowed by the
  world's neutrino oscillation data at 90\%, 95\%, 99\%, and 3$\sigma$
  \CL\ for 2 \dof\, from Ref.~\cite{Maltoni:2004ei}. Shaded regions
  include 2007 data, while the empty regions correspond to results
  before the latest update. }
\end{figure}

\begin{figure}[!h] \centering
\includegraphics[height=6.2cm,width=.46\linewidth]{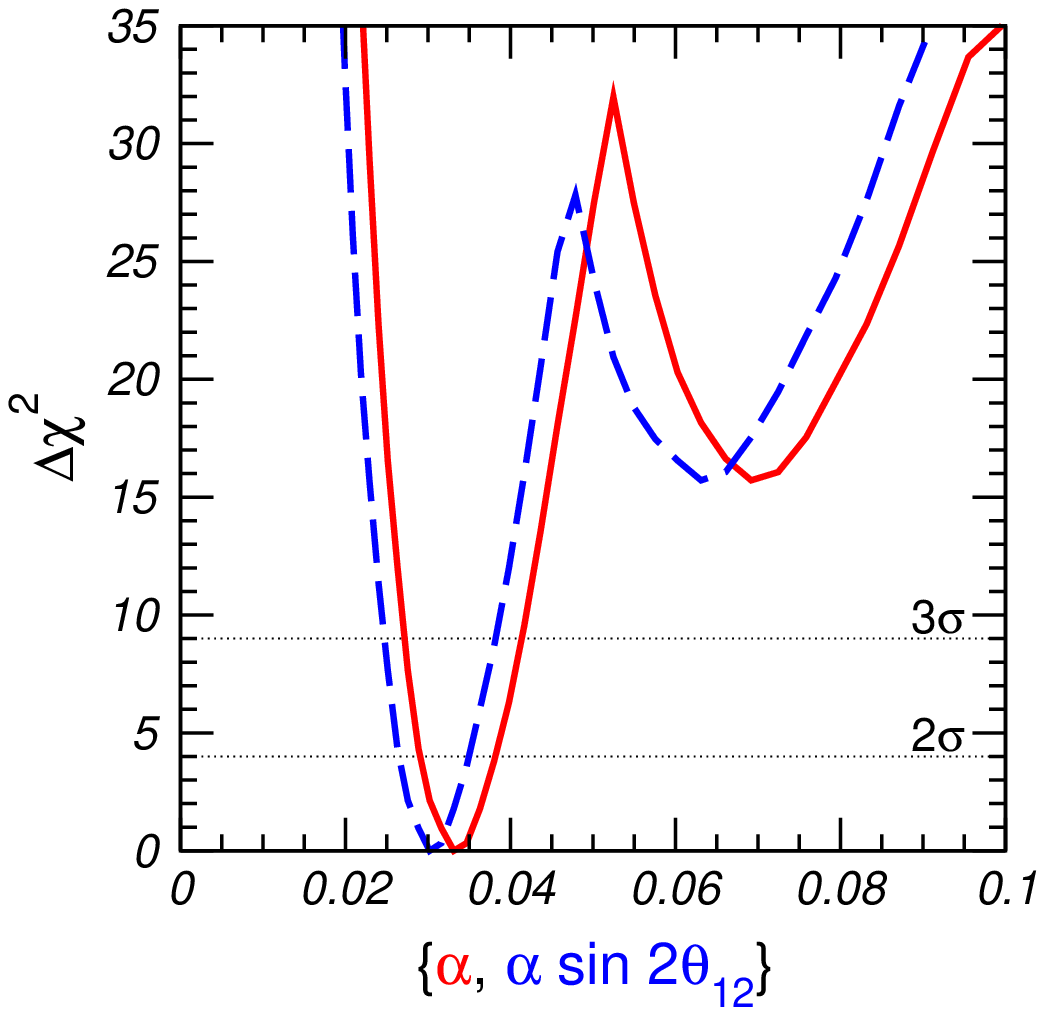}
\includegraphics[height=6.5cm,width=.46\linewidth]{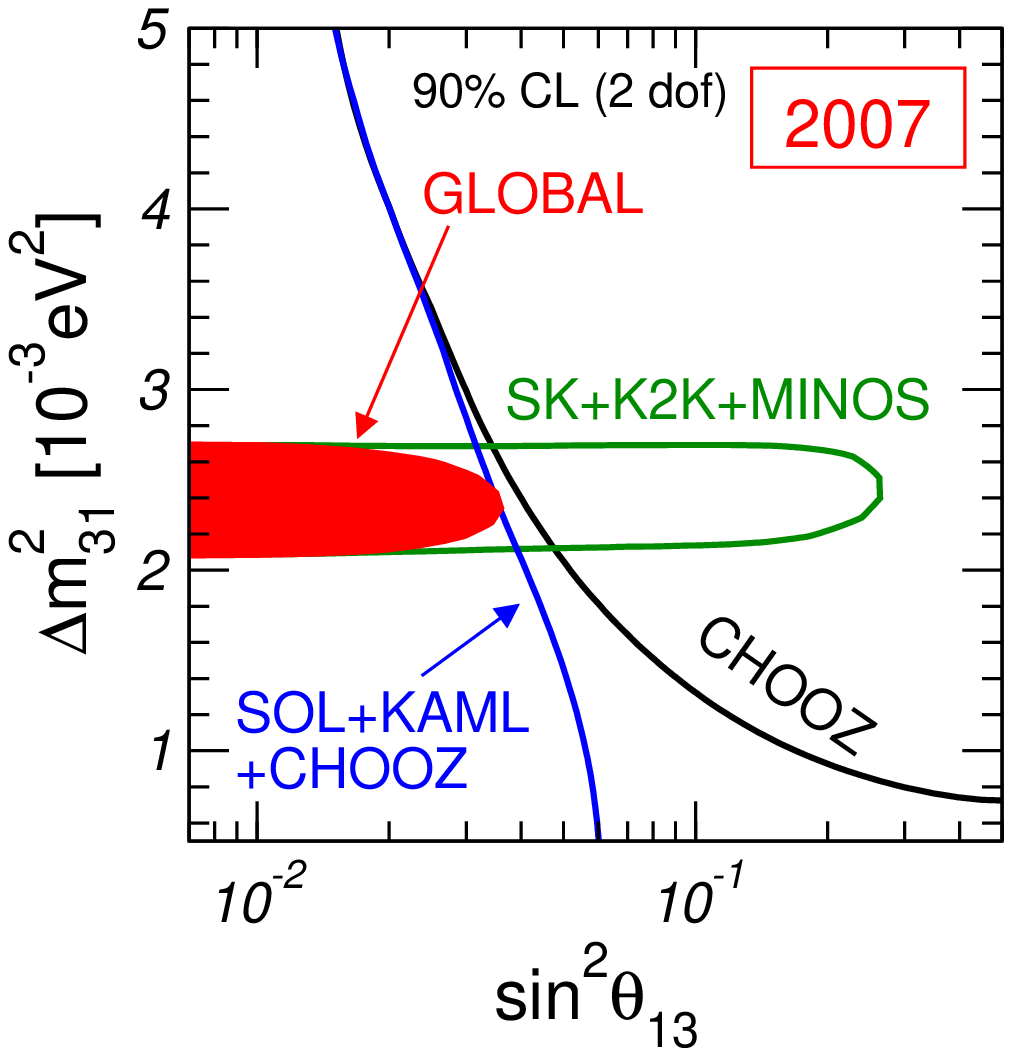}
\caption{\label{fig:alpha}%
  Current determination of $\alpha \equiv \Dms / \Dma$ and bound on
  $\sin^2\theta_{13}$ from the world's neutrino oscillation data, from
  Ref.~\cite{Maltoni:2004ei}.}
\end{figure}

The three--neutrino oscillation parameters that follow from the global
analysis of neutrino oscillation data have been reviewed by Maltoni et
al in Ref.~\cite{Maltoni:2004ei}. An update of the results of this
analysis can be found in the arXiv as version 6 of hep-ph/0405172.
This update includes the latest relevant experimental data, as well as
the latest update of the SSM of Bahcall et al (for references see
Appendix C in hep-ph/0405172-v6). The results are summarized in
Figs.~\ref{fig:global} and \ref{fig:alpha} as well as in the Table.

The plots shown in Fig.~\ref{fig:global} are two-dimensional
projections of the allowed regions in the five-dimensional parameter
space.
The left panel in Fig.~\ref{fig:global} gives the allowed regions of
the ``solar'' neutrino oscillation parameters ($\sin^2\theta_{12}$,
$\Delta m_{21}^2$). One sees the important role of the 2007 KamLAND
data in providing an improved determination of both parameters,
especially the mass squared splitting, now displayed on a linear
scale.
In the right panel of Fig.~\ref{fig:global} we give the allowed
regions of the ``atmospheric'' parameters ($\sin^2\theta_{23}, \Delta
m^2_{31}$) also on a linear scale.  In addition to a confirmation of
oscillations with $\Dma$ with accelerator neutrinos, these data
provide a better determination of $\Dma$ with little impact on the
other parameters. Indeed, one can see from the figure that the
inclusion of the MINOS data is crucial in the determination of $\Dma$.
We see how accelerator data are starting to play the main role in the
determination of the ``atmospheric'' splitting, a trend which will
become stronger in the future.
The best fit values and the allowed 3$\sigma$ ranges of the
oscillation parameters from the global data are summarized in the
Table.
\begin{figure}[!h] \centering
\includegraphics[scale=1.05]{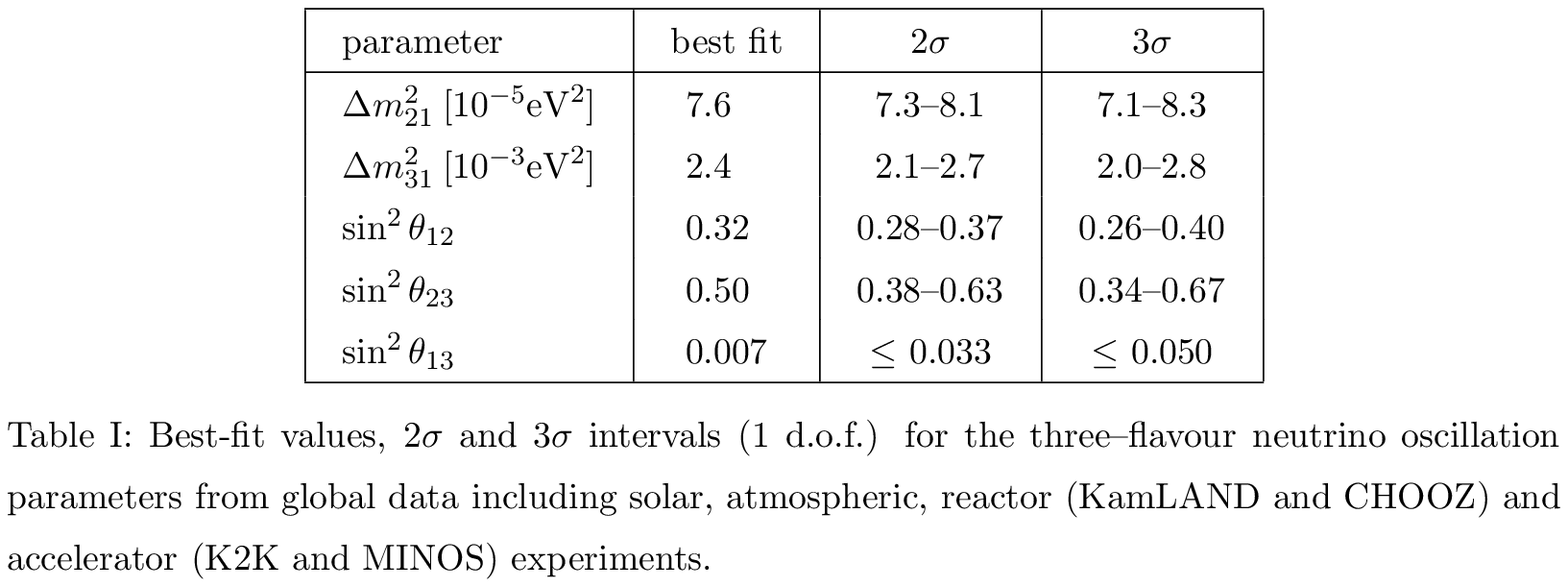}
\end{figure}

The left panel in Fig.~\ref{fig:alpha} gives the parameter $\alpha$,
namely the ratio of solar over atmospheric splittings, as determined
from the global $\chi^2$ analysis.
The right panel in Fig.~\ref{fig:alpha} illustrates how the bound on
$\sin^2\theta_{13}$ emerges from the interplay of different data
samples.
The plot shows the upper bound on $\sin^2\theta_{13}$ as a function of
$\Dma$ from CHOOZ data alone and compares to the bound obtained from
an analysis including also solar and reactor neutrino data. One sees
that, although for larger $\Dma$ values the bound on
$\sin^2\theta_{13}$ is dominated by CHOOZ, this bound deteriorates
quickly as $\Dma$ decreases, so that for $\Dma \lsim 2 \times 10^{-3}
\eVq$ the solar and KamLAND data become relevant.

In summary, we find the following bounds at 90\% \CL\ (3$\sigma$) for
1 \dof:
\begin{equation}\label{eq:th13}
    \sin^2\theta_{13} \le \left\lbrace \begin{array}{l@{\qquad}l}
    0.051~(0.084) & \text{(solar+KamLAND)} \\
    0.028~(0.059) & \text{(CHOOZ+atmospheric+K2K+MINOS)} \\
    0.028~(0.050) & \text{(global data)}
\end{array} \right.
\end{equation}

\subsection{Predicting  neutrino oscillation parameters}
\label{sec:pred-lept-mixing}

As we saw in Sec.~\ref{sec:stat-neutr-oscill} only five of the basic
parameters of the lepton sector are currently probed in neutrino
oscillation
studies~\cite{fukuda:2002pe,ahmad:2002jz,araki:2004mb,Kajita:2004ga,ahn:2002up}
~\cite{apollonio:1999ae,boehm:2001ik}: the angle $\theta_{12}$ and the
splitting $\Dms$, which are determined from solar and KamLAND data,
together with the angle $\theta_{23}$ and the corresponding mass
squared splitting $\Dma$ determined by atmospheric and accelerator
data.
\clearpage
Out of the five neutrino oscillation parameters in the three--neutrino
leptonic mixing matrix, two play a key role for future CP violation
studies in neutrino oscillation experiments: the third ``small'' angle
$\theta_{13}$, together with the ratio of solar to atmospheric
splittings, derived from the oscillation analysis.  These are given in
Fig.~\ref{fig:alpha}.

As we have seen, current neutrino data point towards a well defined
pattern of neutrino mixing angles, quite distinct from that of quarks.
The data seem to indicate an intriguing complementarity between the
angles that characterize the quark and lepton mixing
matrices~\cite{Raidal:2004iw,Minakata:2004xt,Ferrandis:2004vp,Dighe:2006zk}.
It is not clear whether this numerological coincidence has a deeper
meaning.

Some ``flavour--blind'' gauge models have the interesting feature of
``accidentally'' predicting a zero neutrino mass. For example, due to
the anti-symmetry of the one of the Yukawa coupling matrices, the
radiative models~\cite{zee:1980ai,babu:1988ki} considered in
Sec.~\ref{sec:radiative-models} predict that one of the neutrinos is
massless. Similarly for the model considered in
Ref.~\cite{valle:1983dk}. These are rather exceptional examples and
the predictions obtained are incomplete and/or problematic, such as
those of the simplest Zee model~\cite{zee:1980ai} which also include a
nearly maximal solar mixing angle, now in conflict with observation.
This illustrates that, typically, gauge symmetry by itself is not
sufficient to predict masses and mixings, neither for quarks, nor
for leptons.  


\begin{figure}[h] \centering
\includegraphics[height=5.6cm,width=.45\linewidth]{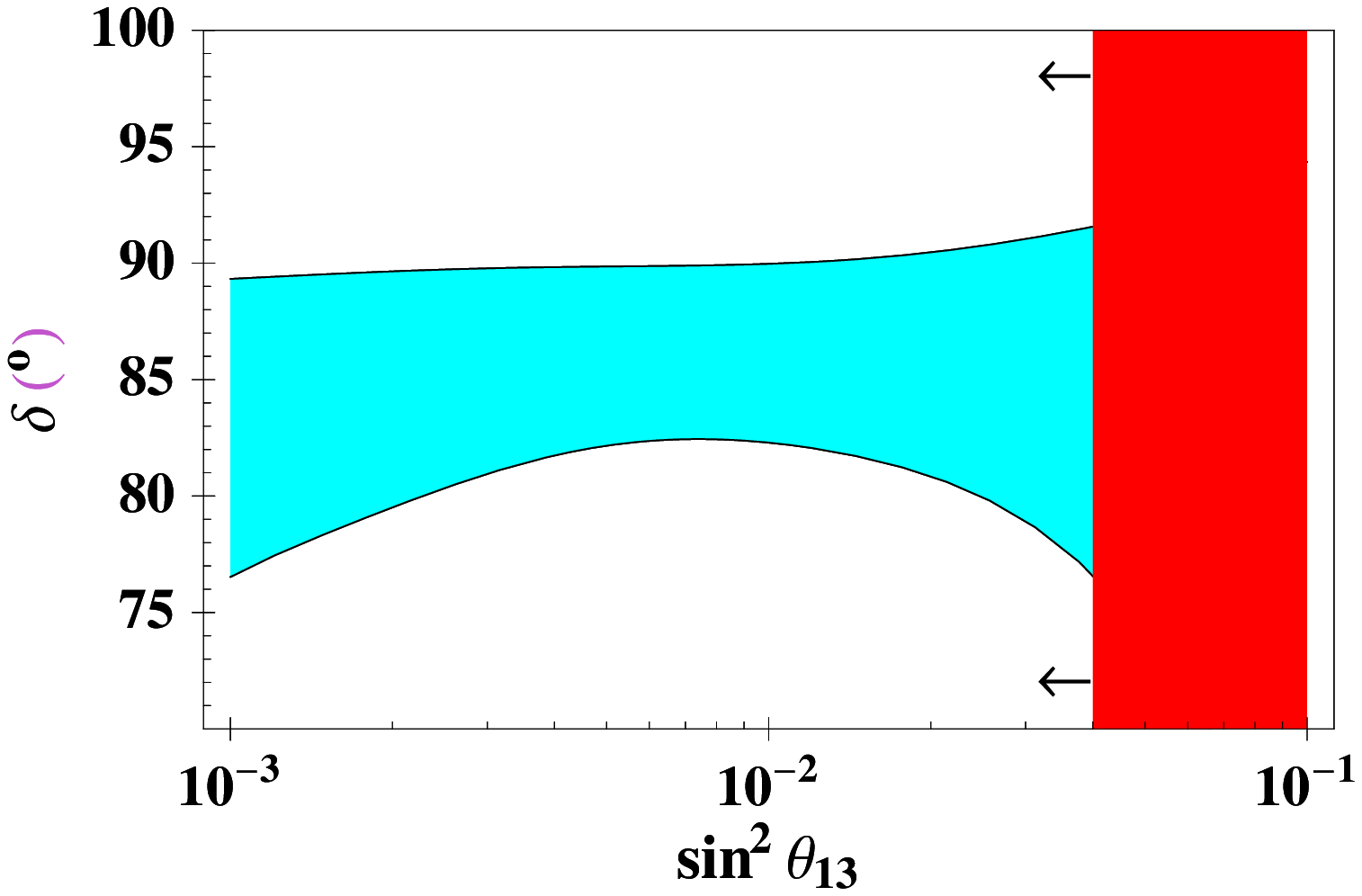}
\includegraphics[height=5.6cm,width=.45\linewidth]{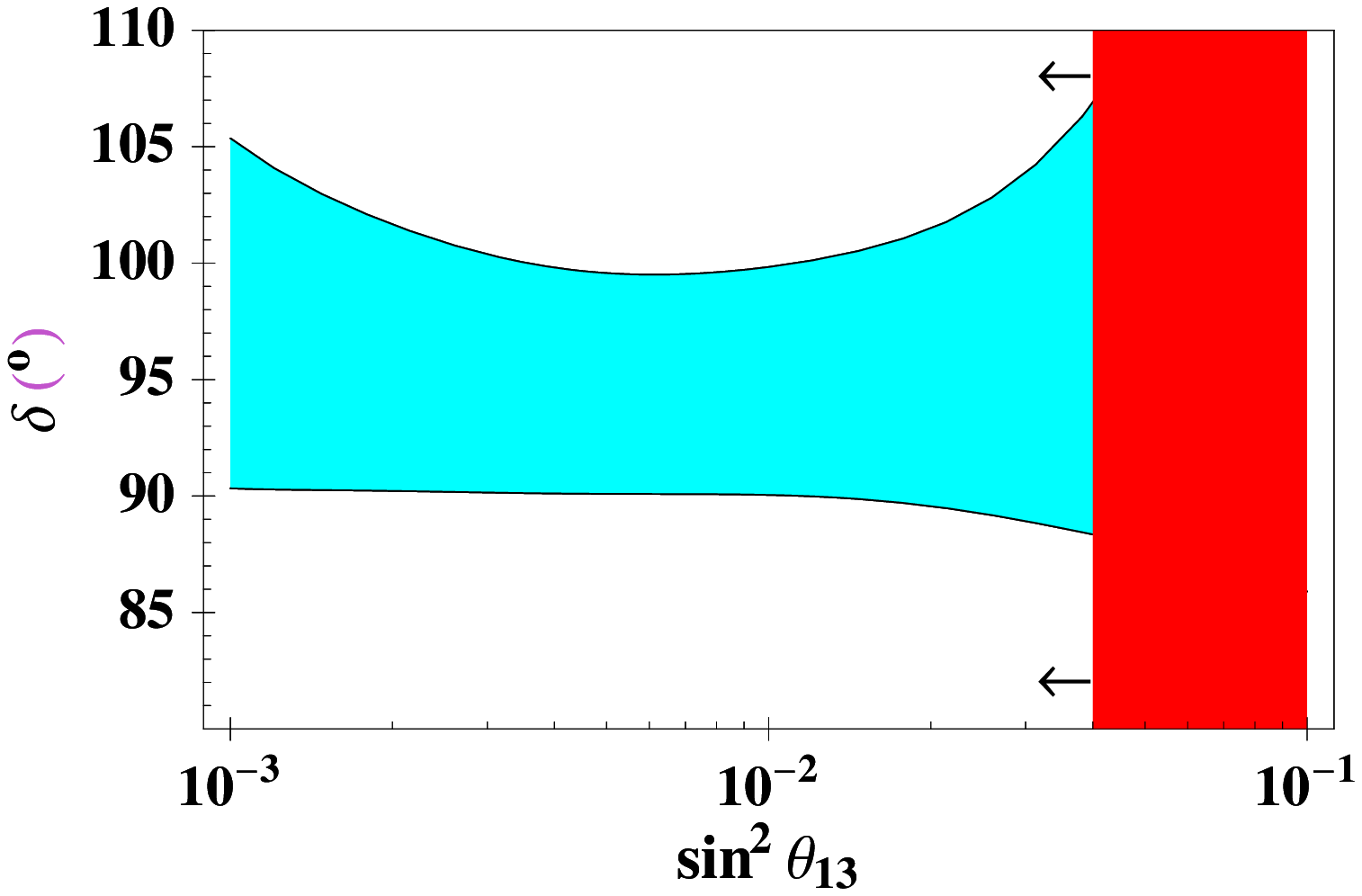}
\caption{\label{fig:dbd-lower-bound}%
Near maximal CP violation in neutrino oscillations, as predicted in 
Ref.~\cite{Hirsch:2007kh}.}
\end{figure}
\begin{figure}[h] \centering
\includegraphics[height=6cm,width=.5\linewidth]{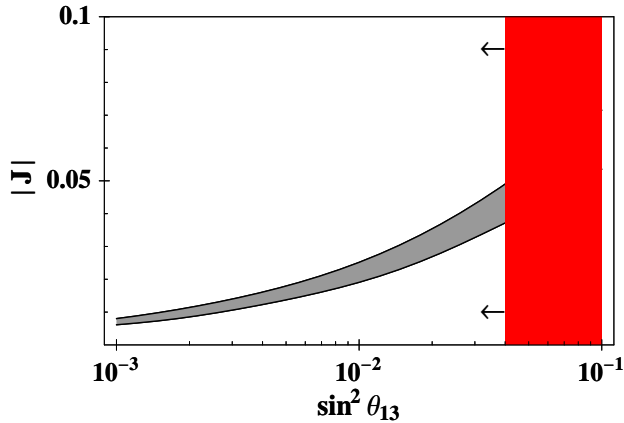}
\caption{\label{fig:max-cpv}%
  Magnitude of CP violation invariant in neutrino oscillations, as
  predicted in Ref.~\cite{Hirsch:2007kh}.}
\end{figure}

There has been a rush of papers attempting to understand the values of
the leptonic mixing angles from underlying symmetries at a fundamental
level.  
For example one can account for the maximum mixing in atmospheric
neutrino oscillation, as well as vanishing of the $\theta_{13}$ 
\begin{equation}
  \label{eq:pred}
\theta_{23}=\pi/4 \:\:\: \rm{and} \:\:\: \theta_{13}=0  
\end{equation}
in terms of a mu-tau symmetry, under which the neutrino mass matrix is
invariant under the interchange of second and third generation
neutrinos~\cite{Lam:2001fb}.  Other models with $Z_2$ symmetries have
been considered~\cite{Grimus:2005jk}. For example,
in~\cite{Grimus:2004cc} one such symmetry has been considered leading
to a maximal atmospheric mixing angle $\theta_{23}$, its breaking
inducing a nonzero value of $\theta_{13}$ in a way that strongly
depends on the neutrino mass hierarchy.

More generally, it has been argued that the form of the mixing at high
energies could be~\cite{Harrison:2002kp}
\begin{align}
\label{eq:hps}
\tan^2\theta_{\Atm}&=\tan^2\theta_{23}^0=1\\ \nonumber
\sin^2\theta_{\textrm{Chooz}}&=\sin^2\theta_{13}^0=0\\
\tan^2\theta_{\Sol}&=\tan^2\theta_{12}^0=0.5 \nonumber
\end{align}
Such Harrison-Perkins-Scott pattern of lepton
mixing~\cite{Harrison:2002er} could result from some kind of flavour
symmetry, valid at a very high energy scale where the dimension-five
neutrino mass operator arises.

One idea is that neutrino masses arise from a common seed at some
``neutrino mass unification'' scale $M_X$~\cite{chankowski:2000fp},
very similar the merging of the gauge coupling constants at high
energies due to supersymmetry~\cite{amaldi:1991cn}.
Unfortunately in its simplest (CP conserving) form this very simple
theoretical ansatz is inconsistent with the significantly non-maximal
value of the solar mixing angle $\theta_{12}$ inferred from current
data~\cite{Maltoni:2004ei}.


A more satisfactory and fully viable alternative realization of the
``neutrino mass unification'' idea employs an $A_4$ flavour symmetry
in the context of a seesaw scheme~\cite{babu:2002dz}. Starting from
three-fold degeneracy of the neutrino masses at a high energy scale, a
viable low energy neutrino mass matrix can indeed be obtained.
The model predicts maximal atmospheric angle and vanishing
$\theta_{13}$, as in Eq.~(\ref{eq:pred}).  Moreover, if CP is violated
$\theta_{13}$ becomes arbitrary but the Dirac CP violation phase is
maximal~\cite{Grimus:2003yn}.
The solar angle $\theta_{12}$ is unpredicted. However, one expects it
to be large,
$$\theta_{12}=\O(1).$$ 
(There have been variant realizations of the $A_4$ symmetry that enable
one to predict the solar angle, see, for example
Ref.~\cite{Hirsch:2005mc}).
Within such $A_4$ flavour symmetry seesaw scheme one can show that the
lepton and slepton mixings are intimately related. It was shown that
the resulting slepton spectrum must necessarily include at least one
mass eigenstate below 200 GeV, which can be produced at the LHC. The
prediction for the absolute Majorana neutrino mass scale $$m_0 \geq
0.3 \mathrm{eV}$$  ensures that the model will be tested by future cosmological
tests and $\beta\beta_{0\nu}$ searches.  Rates for lepton flavour
violating processes $\ell_j \to \ell_i + \gamma$ in the range of
sensitivity of current experiments are typical in the model, with
BR$(\mu \to e \gamma) \gsim 10^{-15}$ and the lower bound BR$(\tau \to
\mu \gamma) > 10^{-9}$.


A variant $A_4$ flavour symmetry seesaw model has been
suggested~\cite{Hirsch:2007kh} which predicts a direct correlation
between the expected magnitude of CP violation in neutrino
oscillations and the value of $\sin^2\theta_{13}$. The model leads to
nearly maximal leptonic CP violation in neutrino oscillations through
the CP phase $\delta$. These predictions are illustrated in
Figs.~\ref{fig:dbd-lower-bound} and \ref{fig:max-cpv}.
For a discussion of various schemes to derive the neutrino mixing
matrix using the tetrahedral group $A_4$ was given in
Ref.~\cite{Zee:2005ut}.

Also discrete flavour groups of the $D_n$ series have been used.  For
example, seesaw models for the lepton sector using $D_4$ and $S_3$
family symmetries have been suggested in \cite{Grimus:2004rj}.  The
model predicts a normal hierarchy neutrino mass spectrum with the
mixing angle $\theta_{13}=0$ and an unpredicted solar mixing angle
$\theta_{12}$.  It employs an enlarged Higgs scalar sector, so that
the atmospheric mixing angle $\theta_{23}$ is given as a ratio of
Higgs scalar vevs, and it is maximal if the full Lagrangian is
$D_4$-invariant. The deviation of $\theta_{23}$ from $\pi / 4$ is
governed by the strength of the soft breaking of the $D_4$ symmetry.


Other non-Abelian finite subgroups of SU(3) have also been used as
family symmetries and shown to generate tri-bimaximal mixing in the
neutrino sector, while allowing for quark and charged lepton
hierarchies~\cite{Luhn:2007sy}.
Many attempts at predicting lepton mixing angles employ the
Harrison-Perkins-Scott mixing pattern in Eq.~(\ref{eq:hps}) at some
high energy scale~\cite{Carr:2007qw}.
Typically one must correct predictions made at a high scale by
renormalization group evolution~\cite{Luo:2005fc}. In this connection
the idea of tri-bimaximal neutrino mixing has also been considered in
the context of discrete symmetries in models with extra
dimensions~\cite{Altarelli:2005yp}.

In short, progress made in the experimental determination of neutrino
parameters has already ruled out many of the proposed models.  For
example, many interesting attempts have failed simply because the
solar mixing angle has been shown to be large but significantly
non-maximal~\cite{Maltoni:2004ei}.


Note that understanding the observed pattern of masses and mixings in
the quark and lepton sector separately in the context of a non-unified
model by appealing to suitable flavour symmetries is a relatively
feasible challenge.
However, in a unified model where quarks and leptons sit in the same
multiplets, there is a tendency to correlate the corresponding mixing
angles.
In a ``generic'' flavour-blind unified model the lepton and quark
mixing angles can always be reconciled thanks to the large number of
parameters involved.
For example, we have seen how in general seesaw-type models the
leptonic mass matrices involves couplings which are absent for the
quarks.  Even in the simplest type-I seesaw mechanism discussed in
Sec.~\ref{sec:top-down-scenario} the B-L violating terms bring in
their own independent flavor structure, distinct from that of the
Dirac mass term which is restricted by the quark sector.
Acceptable mixing patterns can certainly be accommodated, but not
predicted.  To achieve some degree of predictivity one must appeal to
additional symmetries, beyond the gauge symmetry, realized at the
unified level and suitably chosen so as to restrict the flavour
``textures''.
This constitutes a rather demanding challenge, so far not achieved
with full success. Despite some progress on how to obtain successful
unified models of flavour~\cite{Altarelli:2004za} using discrete
non-abelian groups, we currently lack a fully satisfactory unified
theory of flavour, and it is likely that the ``flavour problem'' will
indeed remain with us for a while.

\section{CP violation in neutrino oscillations}
\label{sec:cp-viol-neutr}

In gauge theories of neutrino masses, the lepton mixing matrix
typically contains a number of CP violating phases that may affect
neutrino oscillations~\cite{schechter:1980gr}.
In this section, we discuss theoretical aspects of CP violation in
neutrino oscillations and the possibility of probing it experimentally
at future searches.

It has been recognized for a long time that the Dirac CP violating
phase present in the simplest three-neutrino model could in principle
be observed in neutrino oscillation experiments~\cite{Cabibbo:1977nk}.
These are likely to be the most promising way to probe directly the
Dirac CP phase present in the neutrino mixing matrix, unless
$\theta_{13}$ is too small.  If CPT is conserved violation of CP
implies that of T.  In this review, we do not consider the possible
violation of CPT in the neutrino sector~\footnote{See
  e.g.~\cite{Kostelecky:2003cr,Gonzalez-Garcia:2003jq} for such a
  possibility.}.  Earlier works around 1980 on CP and/or T violation
can be found in
Refs.~\cite{schechter:1980gr},~\cite{bilenky:1980cx,Schechter:1981gk,doi:1981yb}
and \cite{Barger:1980jm,Pakvasa:1980bz}. It has now been understood
that the difference between oscillation probabilities for neutrino and
anti-neutrino is proportional to the leptonic analogue of the
CP-invariant factor of the quark sector~\cite{Jarlskog:1985ht}.
In the last decade, CP violation in neutrino oscillation has received
enormous amount of attention in the community~\cite{Arafune:1996bt,
  Tanimoto:1996ky, Tanimoto:1996by, Arafune:1997hd,
  Minakata:1998bf,Minakata:1997td,Bilenky:1997dd,Kuo:1987km,Krastev:1988yu,
  Minakata:2000ee,Donini:1999jc,Romanino:1999zq}.  Here we consider
three active neutrinos without sterile
neutrino~\cite{peltoniemi:1993ec,peltoniemi:1993ss,caldwell:1993kn}
which have been invoked in connection with LSND and MiniBoone
data~\cite{Aguilar:2001ty,Aguilar-Arevalo:2007it}.

In this review, in order keep the discussion as much general as
possible, we do not show any kind of sensitivity plots for specific
planned experiment or project, but try to keep the discussion at the
level of neutrino oscillation probabilities. 

\subsection{Preliminaries}

The simplest measure of CP violation, which is equivalent to T
violation if CPT is conserved, would be the difference 
of oscillation probabilities between neutrinos and anti-neutrinos, 
$P(\nu_\alpha \to  \nu_\beta)$ and $P(\bar\nu_\alpha
\to  \bar\nu_\beta)$, which is given 
by~\cite{Barger:1980jm,Pakvasa:1980bz},
\begin{equation}
\Delta P_{\nu \bar\nu\, \alpha\beta} \equiv P(\nu_\alpha \to  \nu_\beta)
- P(\bar\nu_\alpha \to  \bar\nu_\beta)
= -16 J_{\alpha \beta} \sin \Delta_{12}
\sin \Delta_{23} \sin \Delta_{31},
\label{eq:DeltaP}
\end{equation}
\begin{figure}[h]
\begin{center}
\includegraphics[width=.9\textwidth,height=9cm]{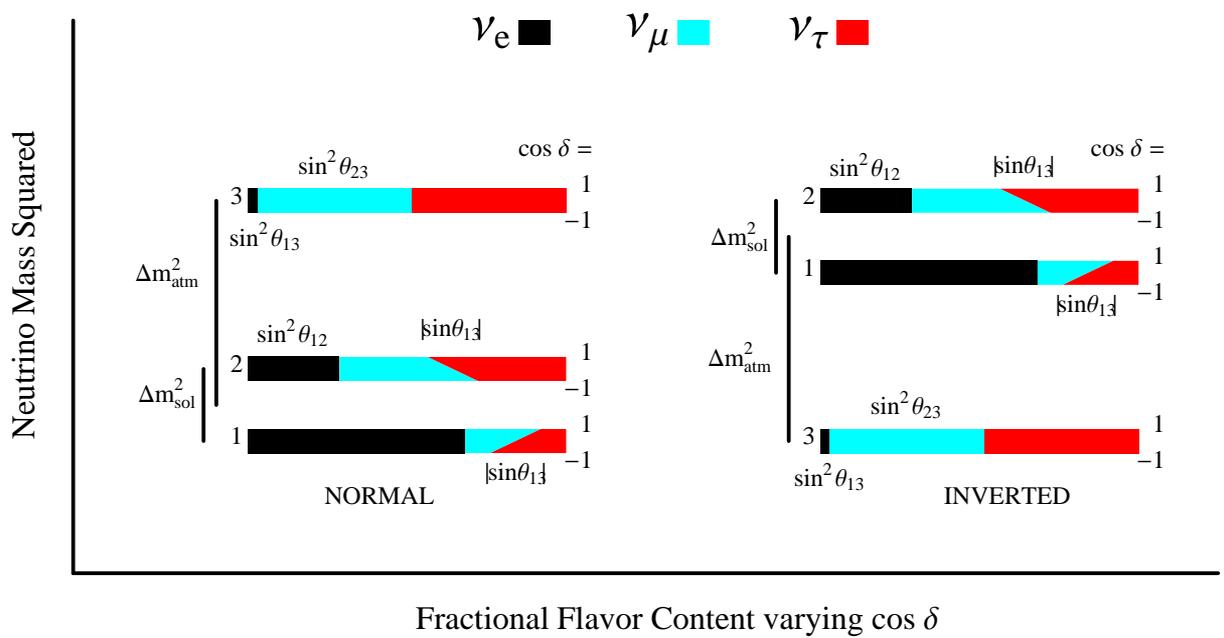}
\label{fig:mass-hierarchy}
\end{center} \vglue -.5cm
\caption[]{The range of probability of finding the $\alpha$-flavor in
  the i-th mass eigenstate as indicated for the two different mass
  hierarchies for the best fit values of the solar and atmospheric
  mixing parameters, $\sin^2 2 \theta_{13}=0.1$ as the CP-violating
  phase, $\delta$ is varied, extracted from \cite{Mena:2003ug}.}
\end{figure}
where we used the notation, $\Delta_{ij} \equiv \Delta m^2_{ij}L/4E$ and 
\begin{equation}
J_{\alpha \beta} \equiv 
{\Im}(U_{\alpha 1}U^*_{\alpha 2} U^*_{\beta 1} U_{\beta 2} )
= \pm J, \ \ J \equiv s_{12}c_{12}s_{23}c_{23}s_{13}c_{13}^2\sin\delta
\label{eqn:J} 
\end{equation}
with positive (negative) sign for (anti-)cyclic permutation of the
flavor indices $e$, $\mu$ and $\tau$.  The parameter $J$ is the
leptonic analogue of the CP-invariant factor for quarks, the unique
and phase-convention-independent measure for CP
violation~\cite{Jarlskog:1985ht}.

Since we do not yet know the value of $\delta$ and $\theta_{13}$, the
current neutrino data give only an upper bound on this quantity, as $J
\lsim 0.04$.  Although we do not yet know whether the mass hierarchy
is normal or inverted this will not cause an ambiguity in the vacuum
$\Delta P$, as it does not depend on the hierarchy~\footnote{Strictly
  speaking, even in vacuum, the oscillation probability for normal and
  inverted hierarchy differ and give a tiny effect, but we do not
  consider such a effect in this review, see
  ~\cite{Nunokawa:2005nx}.}.

Where we can expect to observe the effect of CP violation?  In
contrast with CP violation induced by Majorana phases, which occurs
even for two generations of neutrinos~\cite{schechter:1980gr}, CP
violation in neutrino oscillations is a genuine three (or more) flavor
effect, so it can be observed only when there is an interference
between flavor oscillations involving at least two different $\Delta
m^2$ and three mixing angles, as we will see below.  We observe that
the accelerator based neutrino oscillation experiments will provide
the most promising opportunities to observe such CP violation. This
will be extensively discussed in what follows.

From the expression in Eq.~(\ref{eq:DeltaP}), it is clear that there
is no CP (or T) violation for three generation neutrino mixing in
vacuum if $\delta = 0 ~{\rm or} ~\pi$ and
\begin{itemize}
\item
 one (or more) of the mixing angles is zero,
\item two or more of the masses are degenerate. 
\end{itemize}
Moreover, note that it is impossible to observe CP violation in the
disappearance channels ($\alpha = \beta$) since $\nu_\alpha \to
\nu_\alpha$ is related to $\bar{\nu}_\alpha \to \bar{\nu}_\alpha$ by
CPT. Finally, there is no CP violation if the solar oscillations are
averaged out.

While $\Delta P$ defined in Eq.(\ref{eq:DeltaP}) vanishes for
the disappearance channels,
\begin{eqnarray}
P(\nu_\alpha \rightarrow \nu_\alpha) & = & P(\bar{\nu}_\alpha \rightarrow \bar{\nu}_\alpha) \nonumber \\
& = & 1 -4\vert U_{\alpha 1}\vert^2 \vert U_{\alpha 2}\vert^2 \sin^2 \Delta_{21}
-4\vert U_{\alpha 2}\vert^2 \vert U_{\alpha 3}\vert^2 \sin^2 \Delta_{32}
-4\vert U_{\alpha 3}\vert^2 \vert U_{\alpha 1}\vert^2 \sin^2 \Delta_{31},
\end{eqnarray}
by CPT (hence no direct CP violation), the $\nu_\mu \to \nu_\mu$ and
$\nu_\tau \to \nu_\tau$ disappearance probabilities depends on $\cos
\delta$, using our parametrization of the mixing matrix\footnote{Since
  $\vert U_{\mu 1}\vert^2$, $\vert U_{\mu 2}\vert^2$, $\vert U_{\tau
    1}\vert^2$ and $\vert U_{\tau 2}\vert^2$ depend on $\cos \delta$,
  e.g. $\vert U_{\mu 2}\vert^2 = c^2_{23}c^2_{12} +s^2_{13}s^2_{23}
  s^2_{12} - 2c_{12}s_{12}c_{23} s_{23} s_{13} c_{13} \cos \delta$.}.
Given that if one knows $\cos \delta$ one also knows the magnitude of
$\sin \delta$ up to a sign, can this be used to say anything about
leptonic CP violation?  In principle, the answer is yes.  However it
requires precision measurements in two out of the three disappearance
channels so that one can determine precisely the magnitude of four
independent elements of the mixing matrix.

\begin{figure}[hbt]
\begin{center}
\includegraphics[width=.6\textwidth]{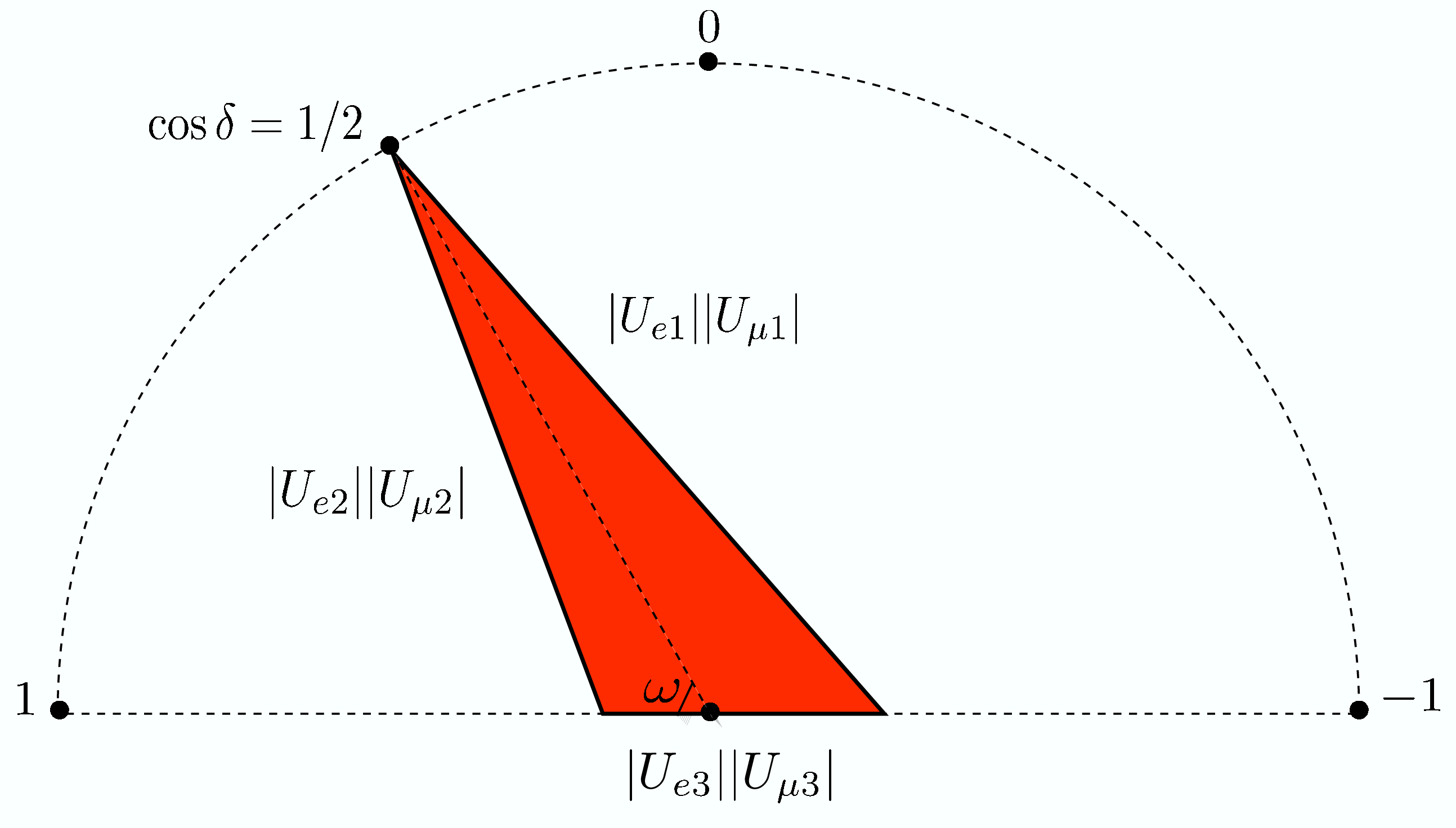}
\end{center}
\caption[]{ The unitarity triangle using the first and second rows of 
  the mixing matrix.  The lengths of each side are as labeled and twice
  the area of this triangle is the absolute value of the CP-invariant
  factor, $\vert J \vert$.  The $|U_{e1}||U_{\mu1}|$ and
  $|U_{e2}||U_{\mu2}|$ vertex moves in a circle as the CP violating
  phase is changed. }
 \label{fig:utri}
\end{figure}

Suppose that one has measured $|U_{e2}|, |U_{e3}|,|U_{\mu2}| ~{\rm
  and} |U_{\mu3}|$ precisely, then from unitarity one knows the moduli
of all the elements of the mixing matrix.  Construct the triangle
which has sides with length $|U_{e1}||U_{\mu1}|$, $|U_{e2}||U_{\mu2}|$
and $|U_{e3}||U_{\mu3}|$, as shown in Fig.~\ref{fig:utri}. This
triangle must close due to the following unitarity relation
\begin{eqnarray}
U^*_{e1}U_{\mu1}+U^*_{e2}U_{\mu2}+U^*_{e3}U_{\mu3}=0.
\end{eqnarray}
Now, twice the area of this triangle is the
absolute value of the CP-invariant factor~\footnote{All other unitarity 
triangles have the same area.}.  Using the rule relating the length
of the sides to the area of the triangle, we have
\begin{eqnarray}
J^2 
& =  &   \frac{1}{4} 
[|U_{e1}||U_{\mu1}| + |U_{e2}||U_{\mu2}| + |U_{e3}||U_{\mu3}|] \times
[-|U_{e1}||U_{\mu1}| + |U_{e2}||U_{\mu2}| + |U_{e3}||U_{\mu3}|] 
\nonumber \\
&  & ~\times[|U_{e1}||U_{\mu1}| - |U_{e2}||U_{\mu2}| + |U_{e3}||U_{\mu3}|]
\times [ |U_{e1}||U_{\mu1}| + |U_{e2}||U_{\mu2}| - |U_{e3}||U_{\mu3}|].
\label{eqn:Jtriangle} 
\end{eqnarray}
The maximum possible difference in the lengths of the two sides,
$|U_{e1}||U_{\mu1}|$ and $ |U_{e2}||U_{\mu2}|$ is
$|U_{e3}||U_{\mu3}|$.  At these extrema, the triangle fits on a line
and has zero area and CP is conserved.  Thus, to claim CP violation
one must show at the required confidence level that $|
|U_{e1}||U_{\mu1}| - |U_{e2}||U_{\mu2}| | < |U_{e3}||U_{\mu3}|$.
Given that $|U_{e3}|$ is known to be small~\cite{Maltoni:2004ei} and
is yet unobserved, this will be a formidable challenge.  The sign of
$J$ or equivalently the sign of CP violation cannot be determined from
vacuum disappearance measurements.

Note that the center of the circle in the triangle diagram
bi-sects the $|U_{e3}||U_{\mu3}|$ side in the ratio of
$|U_{e2}|^2:|U_{e1}|^2$ from left to right. (This is equivalent to
holding the ``(1,2)'' vertex fixed and rotating the ``3'' side about
the same point.)  Using the PDG parametrization of the lepton mixing
matrix, the base of this triangle is $s_{13} c_{13} s_{2}$ and the
height $c_{13} s_{12} c_{12} c_{23} \vert \sin \delta \vert$.  The
angle $\omega$ equals $\delta$ or $2\pi - \delta$.


The CP-invariant factor for any density matter is given by $J(N)
\equiv ( s_{12}c_{12}s_{23}c_{23}s_{13}c_{13}^2\sin\delta)_N$, where
the mixing angles and CP phase are their values in matter, obtained by
writing $U(N)$ in the form of Eq.~(\ref{eq:cc_pdg}).  The matter value
of the CP-invariant factor is related to the vacuum value 
as follows~\cite{Harrison:1999df}:
\begin{equation}
J(N) =  \frac{\Delta m^2_{21} ~\Delta m^2_{32} ~\Delta m^2_{31}}
{\Delta m^2_{21}(N) ~\Delta m^2_{32}(N) ~\Delta m^2_{31}(N)} ~J. 
\end{equation}
This identity guarantees that the $\Delta P$ in Eq.(\ref{eq:DeltaP})
is the same in matter as in vacuum for distances smaller than any of
the matter or vacuum oscillation lengths.
     
\subsection{The oscillation probability {\boldmath $\nu_\mu \to
    \nu$}$_\mathbf{e}$}
\label{introduction}

In this review, we mainly focus on the oscillation channel between
electron and muon neutrinos because it is easier to create and detect
these neutrinos compared to tau neutrinos. 
The drawing below shows schematically the
relation among four possible channels.

\def \al{\mu}
\def \be{e}

{\Large
\begin{center}
\hspace*{+0.7cm}\begin{tabular}{lrclr}
& & {CP} & & \\[0.1in]
&$\nu_\al \to  \nu_\be$ &
$\Longleftrightarrow$ &
$\bar{\nu}_\al \to  \bar{\nu}_\be$  &  \\[0.2in]
{T} &  $\Updownarrow$ 
 \quad \quad
&\quad \quad \quad \quad \quad  & ~~\quad $\Updownarrow$ \quad \quad {T} & \\[0.2in]
&$\nu_\be \to  \nu_\al$ &
$\Longleftrightarrow$ &
$\bar{\nu}_\be \to  \bar{\nu}_\al$ &   \\[0.1in]
& & {CP} & & 
\end{tabular}
\end{center}
}
\noindent 

The horizontal (vertical) processes are related by CP (T) whereas the
processes across the diagonals are related by CPT.  The first row will
be explored in very powerful conventional beams, Superbeams, whereas
the second row could be explored in Neutrino Factories or Beta Beams.

\subsubsection{Vacuum:  {\boldmath $\nu_\mu \to \nu$}$_\mathbf{e}$}

Let us first consider oscillation in vacuum ignoring matter effect.
The transition probability $\nu_\mu \to \nu_e$ can be simply written
as the square of a sum of three amplitudes, one associated with each
neutrino mass eigenstate, as follows~\cite{cervera:2000kp},
\begin{eqnarray}
P(\nu_\mu \to  \nu_e) & = & |~U_{\mu3}^* e^{-im^2_3L/2E} U_{e3} 
+  U_{\mu2}^* e^{-im^2_2L/2E} U_{e2} 
+  U_{\mu1}^* e^{-im^2_1L/2E} U_{e1} ~|^2 \nonumber \\[0.3cm]
&=& | 2U^*_{\mu 3}U_{e3} \sin \Delta_{31} e^{-i\Delta_{32}} + 2 U^*_{\mu2}U_{e2}\sin \Delta_{21}|^2,
\label{pme1}
\end{eqnarray}
\noindent where the unitarity of the lepton mixing
matrix~\cite{Kobayashi:1973fv} has been used to eliminate the
$U^*_{\mu 1}U_{e1}$ term and $\Delta_{jk}$ is used as a shorthand for
the the kinematic phase, $\Delta m^2_{jk}L/4E$.  It is convenient to
rewrite this expression as follows 
\begin{eqnarray}
P(\nu_\mu \to  \nu_e) 
& \approx & |{\sqrt{P_{\text{atm}}}}e^{-i(\Delta_{32}+\delta)} 
+ {\sqrt{P_{\text{sol}}}}|^2  \nonumber \\[0.3cm]
& = & P_{\text{atm}} + 2 \sqrt{P_{\text{atm}}} \sqrt{P_{\text{sol}}} \cos(\Delta_{32}+\delta) 
+ P_{\text{sol}},
\label{eq:pme}
\end{eqnarray}
where as the notation suggests the amplitude $\sqrt{P_{\text{atm}}}
$ depends only on $\Delta m^2_{31}$ and $\sqrt{P_{\text{sol}}} $
depends only on $\Delta m^2_{21}$.  For propagation in the vacuum,
these amplitudes are simply given by
\begin{eqnarray}
\sqrt{P_{\text{atm}}}  & \equiv & 
\sin \theta_{23} \sin 2 \theta_{13} \sin \Delta_{31} \nonumber \\
\sqrt{P_{\text{sol}}} & \equiv  & 
\cos \theta_{23}\cos \theta_{13} \sin 2 \theta_{12} \sin \Delta_{21} 
\approx  
\cos \theta_{23}\cos \theta_{13} \sin 2 \theta_{12} ~\Delta_{21},
\label{eq:patm-psol}
\end{eqnarray}
where $\theta_{13}$ and $\Delta_{21}$ are assumed to be small. 
In the amplitude $\sqrt{P_{\text{sol}}} $, terms proportional to 
$\sin \theta_{13}  \sin \Delta_{21} e^{-i\delta}$ 
have been neglected since they are of second order in 
the small quantities $\sin \theta_{13}$ and $ \Delta_{21}$.

For anti-neutrinos $\delta \to -\delta$. Thus the phase between
$\sqrt{P_{\text{atm}}} $ and $\sqrt{P_{\text{sol}}} $ changes from
$(\Delta_{32}+\delta)$ to $(\Delta_{32}-\delta)$. This changes the
interference term from
 \begin{eqnarray}
 2\sqrt{P_{\text{atm}}} \sqrt{P_{\text{sol}}} \cos(\Delta_{32}+\delta) &  \to  & 2\sqrt{P_{\text{atm}}} \sqrt{P_{\text{sol}}} \cos(\Delta_{32}-\delta).
\end{eqnarray}
Expanding  $\cos(\Delta_{32}\pm\delta)$, one has a CP conserving part,
\begin{eqnarray}
2\sqrt{P_{\text{atm}}} \sqrt{P_{\text{sol}}} \cos\Delta_{32}\cos\delta,
\end{eqnarray}
 and the CP violating part,
\begin{eqnarray}
\mp 2\sqrt{P_{\text{atm}}} \sqrt{P_{\text{sol}}} \sin\Delta_{32}\sin\delta,
\end{eqnarray}
where - (+) sign is for neutrino (anti-neutrino).
This implies that $\Delta P$ given in Eq.~(\ref{eq:DeltaP}) 
can be rewritten as 
\begin{equation}
\label{eq:asym}
\Delta P_{\nu \bar\nu}   \equiv  P(\nu_\mu \to  \nu_e)-
P(\bar\nu_\mu \to  \bar\nu_e) = 
- 4\sqrt{P_{\text{atm}}} \sqrt{P_{\text{sol}}} \sin\Delta_{32}\sin\delta.
\end{equation}
Note that this expression coincides with the exact formula in
Eq.~(\ref{eq:DeltaP}) because terms we neglected in
Eqs.~(\ref{eq:pme}) and (\ref{eq:patm-psol}) do not contribute to the
CP violating term.

Therefore CP violation is maximum when $\Delta_{32} =
(2n+1)\frac{\pi}{2}$ and grows with $n$ since $\sin \Delta_{21}$ grows
with $n$.  Notice also that, as expected, for CP violating term to be
non-zero the kinematical phase $\Delta_{32}$ cannot be $n\pi$.  This
is the neutrino counter part to the non-zero strong phase requirement
for CP violation in the quark sector.

\begin{figure}[hbt]
\includegraphics[width=.48\textwidth]{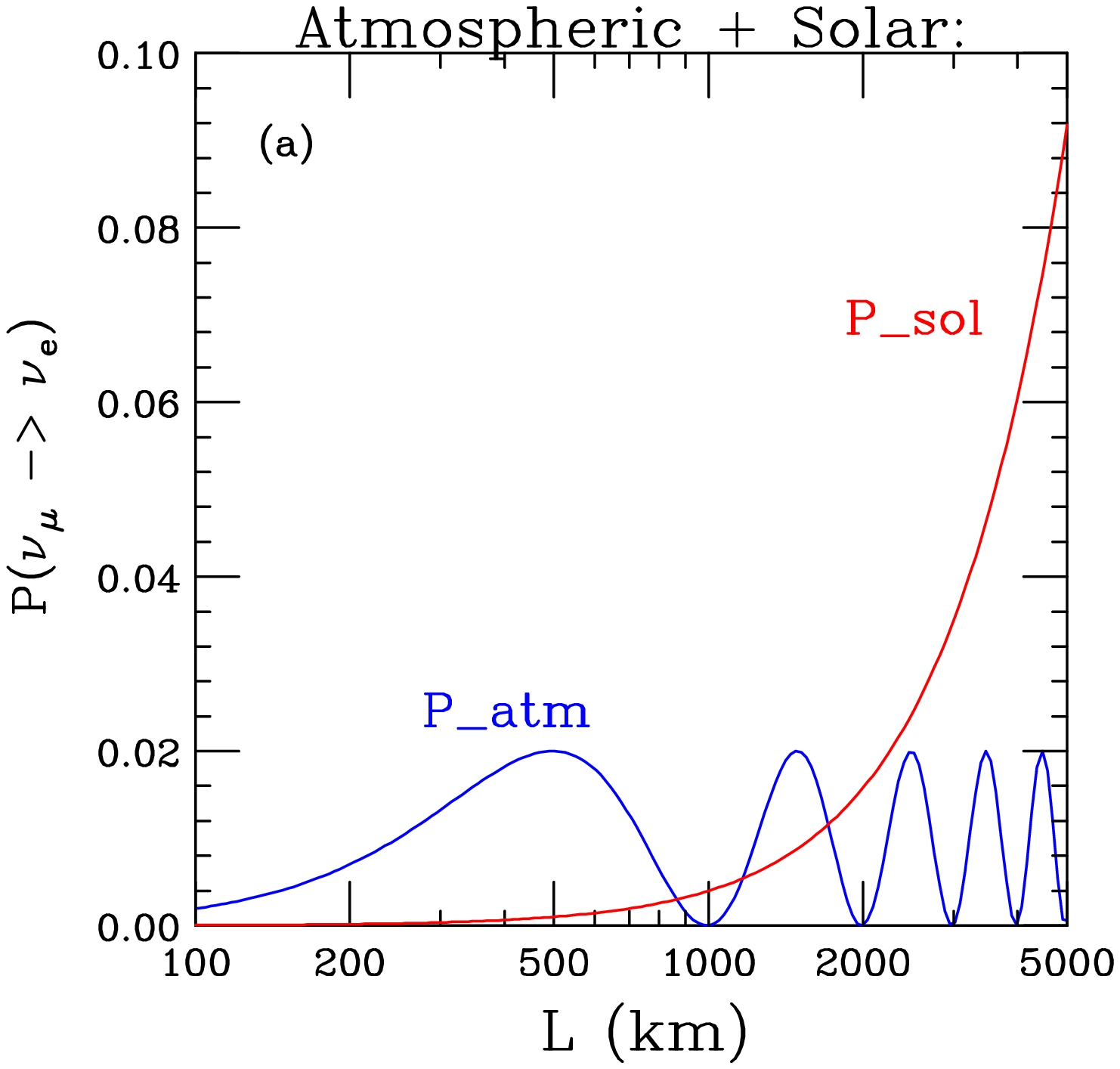}
\includegraphics[width=.48\textwidth]{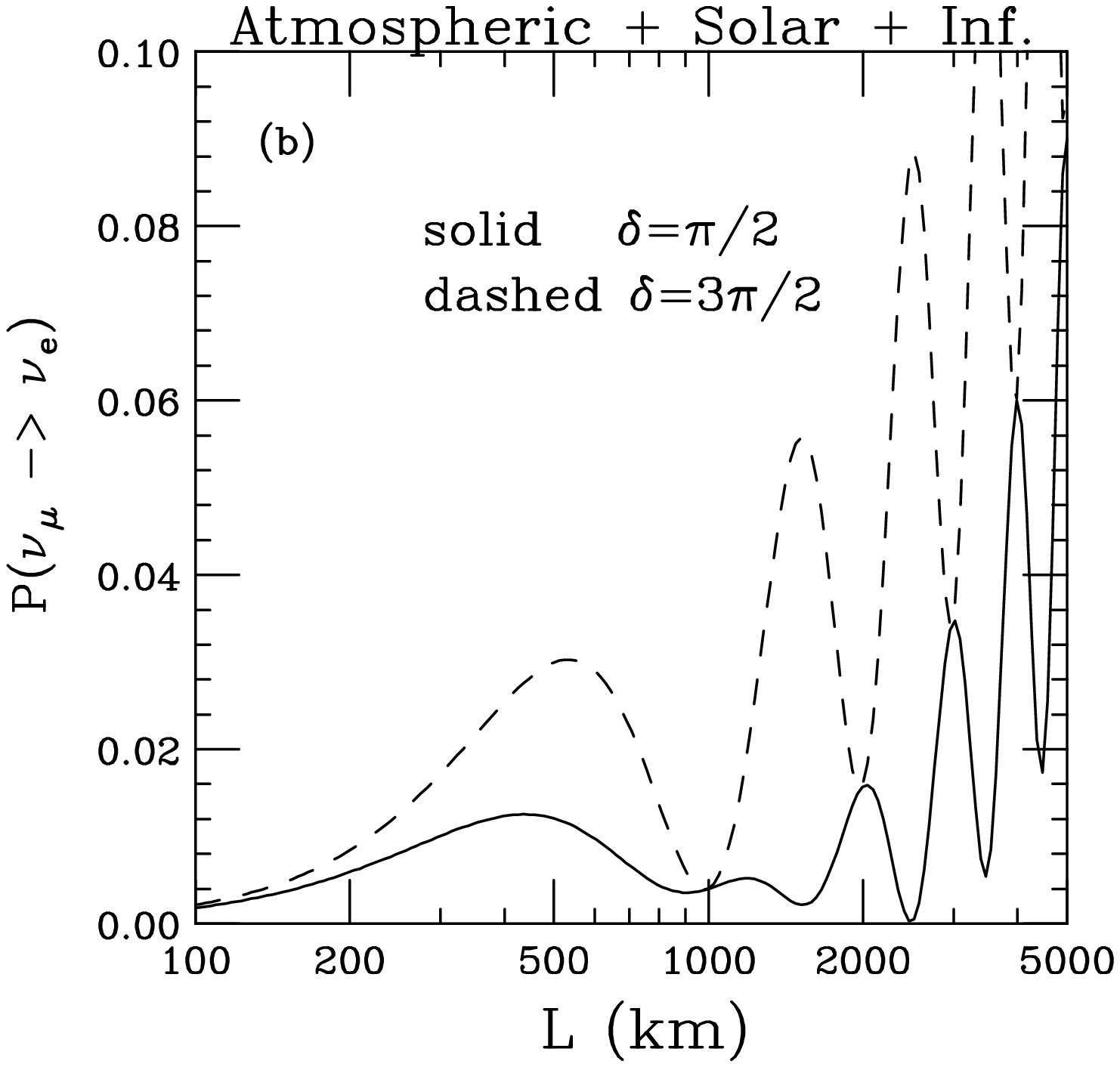}
\caption{The probability $P(\nu_\mu \to \nu_e)$: (a) The components
  $P_{\text{atm}}$ and $P_{\text{sol}}$, (b) The full probability
  including the interference term for $\delta=\frac{\pi}{2} ~{\rm and}
  ~\frac{3\pi}{2}$, solid and dashed respectively.  Since
  $\frac{3\pi}{2} =-\frac{\pi}{2} + 2 \pi$ we have $P(\nu_\mu \to
  \nu_e, \delta=\frac{3\pi}{2}) =P(\bar{\nu}_\mu \to \bar{\nu}_e,
  \delta=\frac{\pi}{2})$ so that these two probabilities can be
  considered to be $P(\nu_\mu \to \nu_e, \delta=\frac{\pi}{2})$
  (solid) and $P(\bar{\nu}_\mu \to \bar{\nu}_e, \delta=\frac{\pi}{2})$
  (dashed).  The difference between these to curves demonstrates CP
  violation for this process.  }
\label{fig:E-pmutoe}
\end{figure}

Fig.~\ref{fig:E-pmutoe} shows the components of $P(\nu_\mu \to 
\nu_e)$ as well as the full probability for selected values of the CP
phase $\delta$ for neutrino energy $E=1$ GeV.
Unless otherwise stated, we fix the absolute value of $\Delta
m_{31}^2$ to be 2.5$\times 10^{-3}$ eV$^2$, $\Delta m_{21}^2 =
8.0\times 10^{-5}$ eV$^2$, $\sin^2\theta_{12} =
0.31$~\cite{Maltoni:2004ei}.  Since we can use CP and T to relate all
the processes discussed earlier in section we have
\begin{equation}
P(\nu_\mu \to  \nu_e, \delta)=
P(\bar{\nu}_\mu \to  \bar{\nu}_e, -\delta) 
 = 
P(\nu_e \to  \nu_\mu, -\delta) = 
P(\bar{\nu}_e \to  \bar{\nu}_\mu, \delta),
\end{equation}
all given from Eq.~(\ref{eq:pme}). 

\begin{figure}[hbt]
\includegraphics[width=0.45\textwidth]{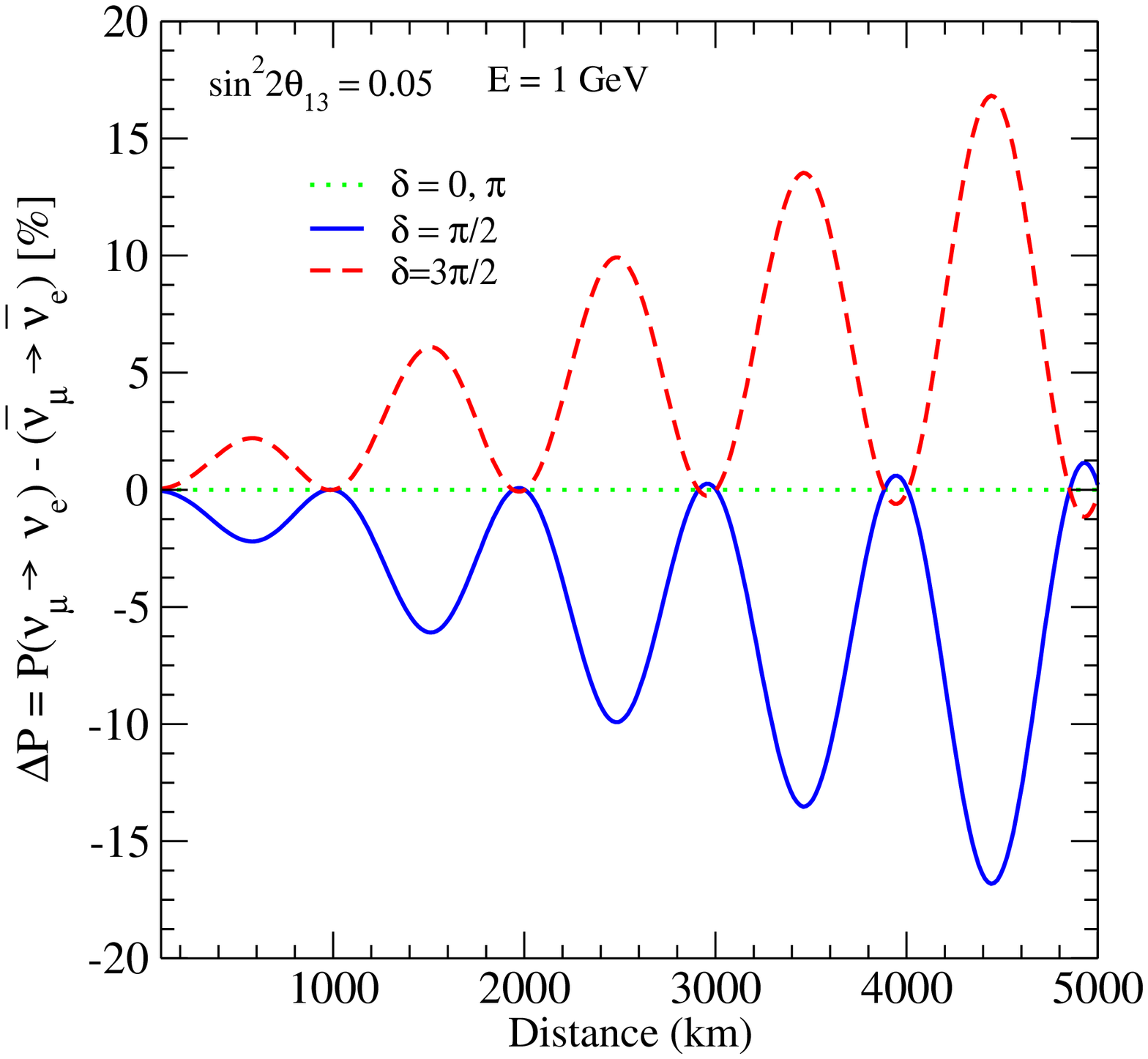}
\includegraphics[width=0.45\textwidth]{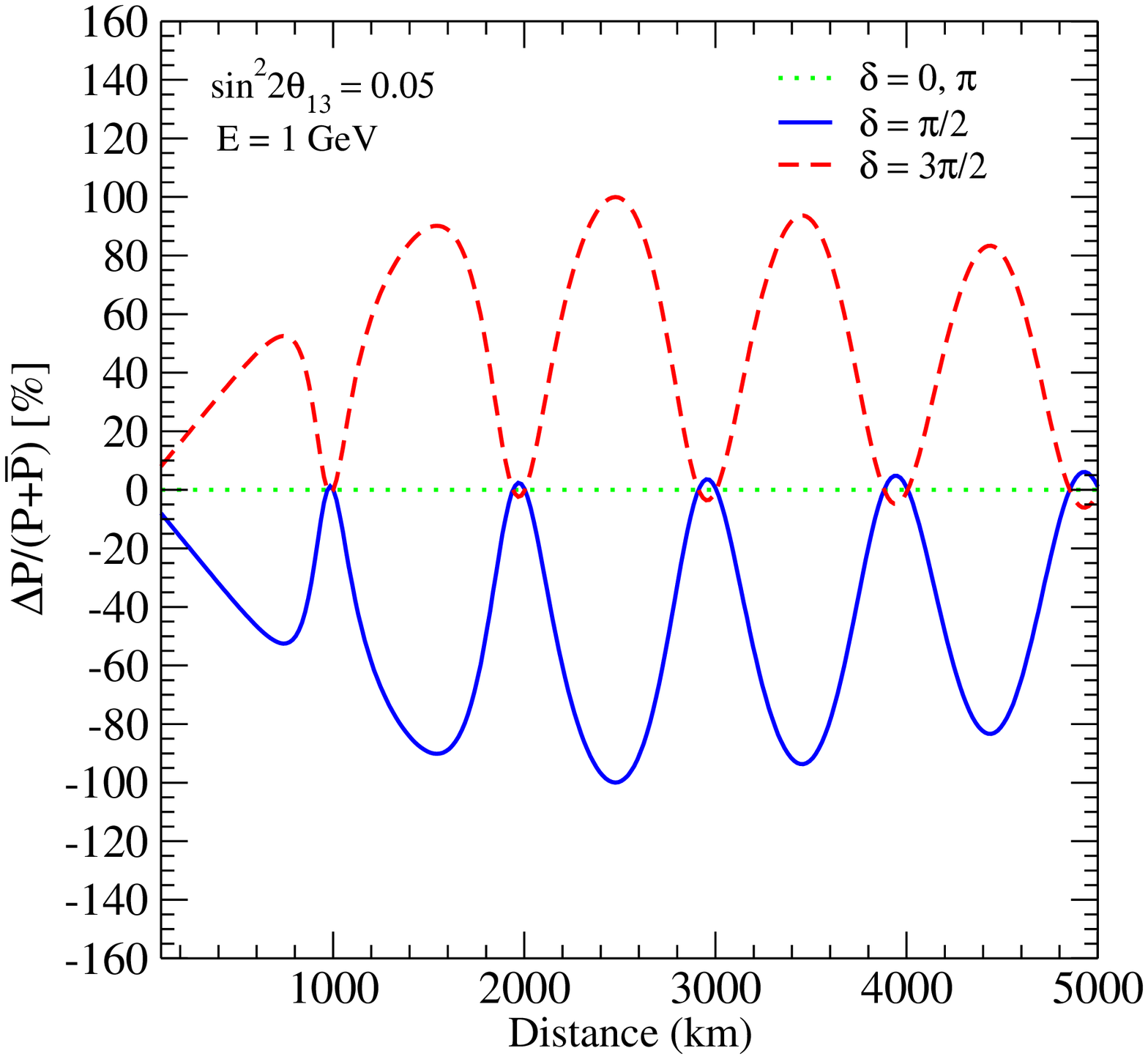}
\caption{Examples of $\Delta P_{\nu \bar\nu} \equiv P(\nu_\mu \to
  \nu_e)-P(\bar\nu_\mu \to \bar\nu_e)$ (left panel) and the asymmetry
  $\Delta P/[P(\nu_\mu \to \nu_e)+ P(\bar\nu_\mu \to \bar\nu_e)]$
  (right panel) in vacuum as a function of distance for fixed value of
  energy, $E=1 $ GeV and $\sin^2 2\theta_{13} = 0.05$.  }
\label{fig:deltap+asym_vac}
\end{figure}

In the left panel of Fig.~\ref{fig:deltap+asym_vac} we show 
$\Delta P_{\nu \bar\nu}$ in vacuum as a function of $L$ for
$\delta = 0, \pi/2, \pi$ and $3\pi/2$ for $\sin^2 2\theta_{13} = 0.05$. 
As mentioned before, peak of $|\Delta P_{\nu \bar\nu}|$ 
occurs at $\Delta_{32} = (2n+1)\frac{\pi}{2}$ growing 
linearly as $n$ or $L$ increases.

\begin{figure}[h]
\includegraphics[width=0.5\textwidth]{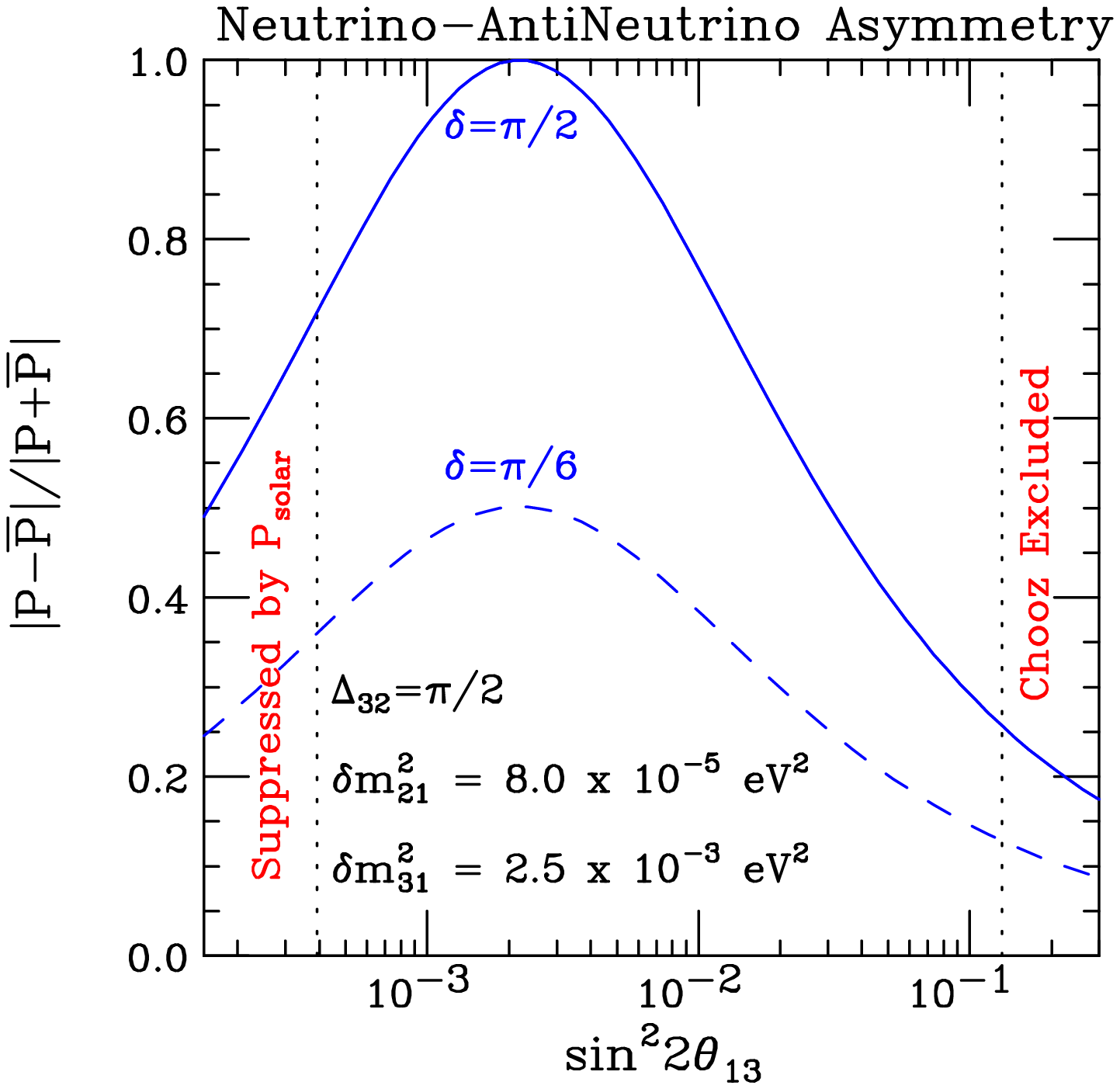}
\includegraphics[width=0.5\textwidth]{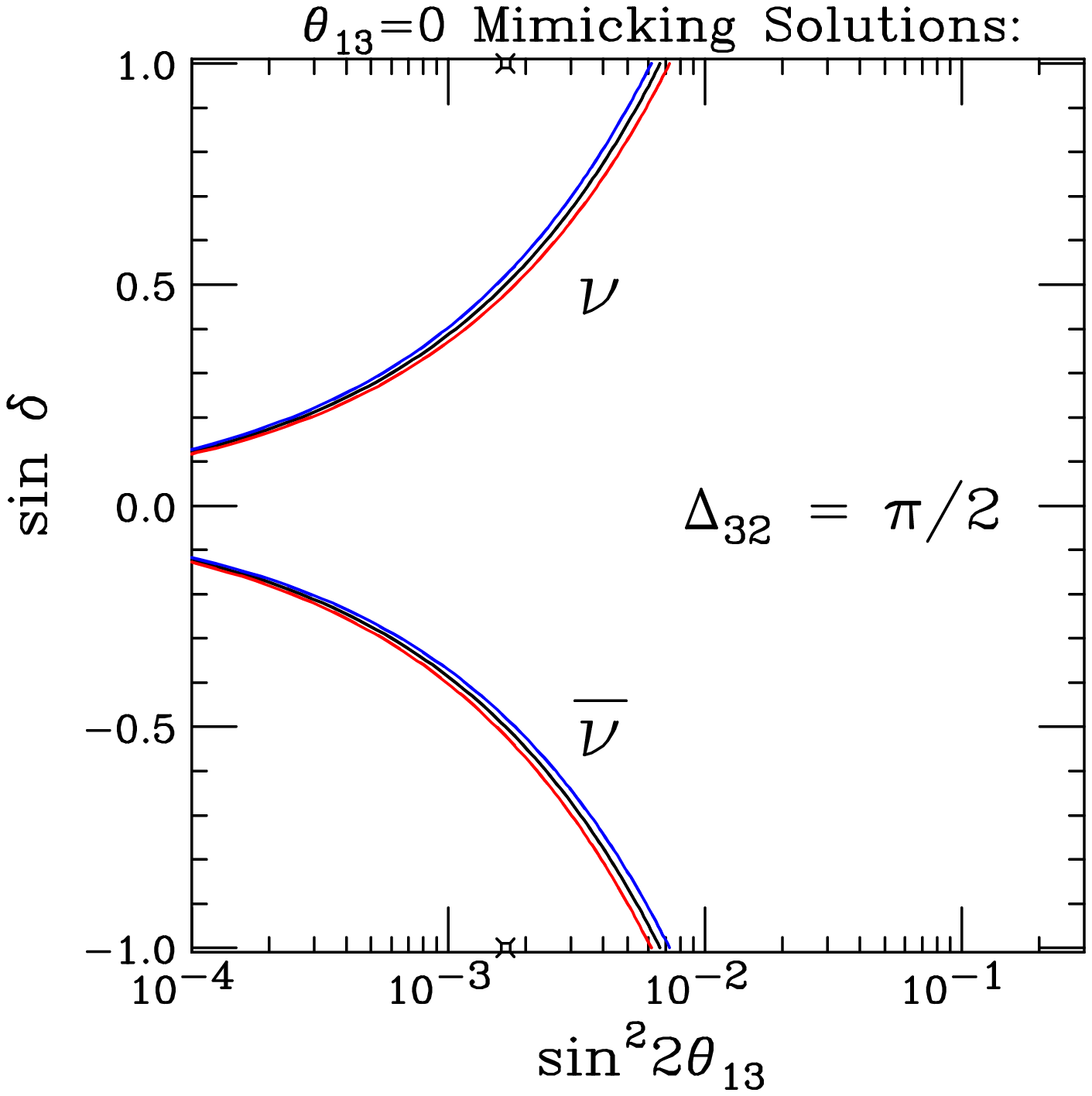}
\caption{The left panel (a) shows the asymmetry between neutrino and
  anti-neutrinos as a function of $\sin^2 2 \theta_{13}$. The peak
  occurs when $\sqrt{P_{\text{atm}}}= \sqrt{P_{\text{sol}}}$ at
  $\sin^2 2 \theta_{13}\approx 0.002$. The right panel (b) shows the
  location of the zero mimicking solutions at the first oscillation
  maximum.  }
\label{fig:asym-zero}
\end{figure}

For a fixed value of $\delta$, as $\theta_{13}$ becomes smaller, 
the absolute value of $\Delta P$ 
become smaller but this does not means that  CP violation gets smaller. 
Consider the  neutrino-anti-neutrino asymmetry, defined as 
\begin{eqnarray}
{\cal A} & \equiv &  \frac{P-\bar{P}}{ P+\bar{P}}
= \frac{2\sqrt{P_{\text{atm}}} \sqrt{P_{\text{sol}}} \sin\Delta_{32}\sin\delta}{ P_{\text{atm}} + 2 \sqrt{P_{\text{atm}}} \sqrt{P_{\text{sol}}} \cos\Delta_{32}\cos\delta + P_{\text{sol}}},
\end{eqnarray}
as a measure of CP violation.
Then, at the vacuum oscillation maxima, $\Delta_{31} =(2n+1) \frac{\pi}{2}$,
\begin{eqnarray}
{\cal A} & = & \frac{2\sqrt{P_{\text{atm}}} \sqrt{P_{\text{sol}}}\sin\delta}{ P_{\text{atm}}  + P_{\text{sol}}}.
\end{eqnarray}
which is maximized when
$\sqrt{P_{\text{atm}}} = \sqrt{P_{\text{sol}}} $. 
This occurs at 
\begin{eqnarray}
\sin^2 2 \theta_{13}= \cot^2 \theta_{23} \sin^2 2\theta_{12} 
\sin^2 \Delta_{21} \approx \cot^2 \theta_{23} \sin^2 2\theta_{12} 
 \left( \frac{\Delta m^2_{21}}{\Delta m^2_{31}}\right)^2~ \Delta^2_{31}.
\label{eqn:Apeak}
\end{eqnarray}
At the first vacuum oscillation maximum, $\Delta_{31} =
\frac{\pi}{2}$, the peak in the asymmetry occurs when $\sin^2 2
\theta_{13} \approx 0.002$.  
In the right panel of Fig.~\ref{fig:deltap+asym_vac} we show the asymmetry 
in vacuum as a function of $L$ for fixed value of $\sin^2 2\theta_{13} = 0.05$ 
and energy $E=1$ GeV.
As function of $\sin^2 2 \theta_{13}$ the
neutrino- anti-neutrino asymmetry is shown in Fig.~\ref{fig:asym-zero}.
At the second vacuum oscillation maximum the asymmetry is maximum when
$\sin^2 2\theta_{13}$ is 9 times larger.

Another feature of the probability that is important is 
the existence of zero mimicking solutions~\cite{Mena:2005sa}.
When,
\begin{equation}
 \sqrt{P_{\text{atm}}} = -2\sqrt{P_{\text{sol}}} \cos(\Delta_{32}\pm\delta)
\end{equation}
we have
\begin{eqnarray}
P(\nu_\mu \to  \nu_e)=P_{\text{sol}}.
\end{eqnarray} 
Thus, even if $\sin^2 2 \theta_{13} \neq 0$, when this condition is
satisfied non-zero $\theta_{13}$ is impossible to observe.  See
Fig.~\ref{fig:asym-zero} for examples of these ``zero mimicking
solutions''.  The maximum value of $\sin^2 2 \theta_{13}$ for which
there are zero mimicking solutions is 4 times the value given in
Eq.~(\ref{eqn:Apeak}).  Luckily, the zero mimicking solutions appear
at different values of $\delta$ if $E/L$ is changed or by switching to
anti-neutrinos.

\subsubsection{Matter:  {\boldmath $\nu_\mu \to  \nu$}$_\mathbf{e}$}
\label{sec:matter}

In describing neutrino propagation in the Earth it is very important
to take into account the effect of
matter~\cite{wolfenstein:1978ue,Barger:1980jm,mikheev:1985gs} because
it can induce a fake CP violating effect even if the CP phase is zero
or $\pi$.  In matter of constant density, the two component appearance
probability, $P^N_{app}$, using the mixing angle and mass squared
difference in matter, $\theta_N$ and $\Delta m^2_N$, is given by
\begin{eqnarray}
P^N_{app}=\sin^2 2\theta_N \sin^2 \Delta_N
\end{eqnarray}
where $\Delta_N \equiv \Delta m^2_N L/4E$.  Since $\Delta m^2_N \sin
2\theta_N$ is independent of the density of matter i.e. it is an
invariant, then
\begin{eqnarray}
P^N_{app}=\sin^2 2\theta_0 \left( \frac{\sin^2 \Delta_N}{\Delta^2_N}\right) \Delta^2_0.
\end{eqnarray}
Note, that this expression does not depend on the mixing angle in
matter, $\theta_N$ and that $\Delta m^2_N$ is given by
\begin{eqnarray}
\Delta m^2_N = \sqrt{(\Delta m_0^2 \cos 2 \theta_0 - 2\sqrt{2}G_F N_eE)^2 + (\Delta m_0^2 \sin 2 \theta_0)^2}
\end{eqnarray}
where $G_F$ is the Fermi constant and $ N_e$ is the number density of
electrons. Except near the resonance, $\Delta m^2_N = \Delta m_0^2 -
2\sqrt{2}G_F N_eE $ is a good approximation.  For matter effects to
significantly alter the appearance probability the following condition
must be satisfied, with the left and right hand side significantly
different
\begin{eqnarray}
    \frac{\sin^2 \Delta_N}{\Delta^2_N} \neq \frac{\sin^2 \Delta_0}{\Delta^2_0}.
\end{eqnarray}
That is, matter must significantly alter the $\Delta m^2$ so that $
\Delta m^2_N \neq \Delta m_0^2$ and the baseline of the experiment
must be a significant fraction of the oscillation length in matter or
vacuum whichever is shorter so that either $\sin \Delta_N \neq
\Delta_N$ and/or $\sin \Delta_0 \neq \Delta_0$.

\begin{figure}[h] 
\centering
\includegraphics[width=0.6\textwidth,height=7cm]{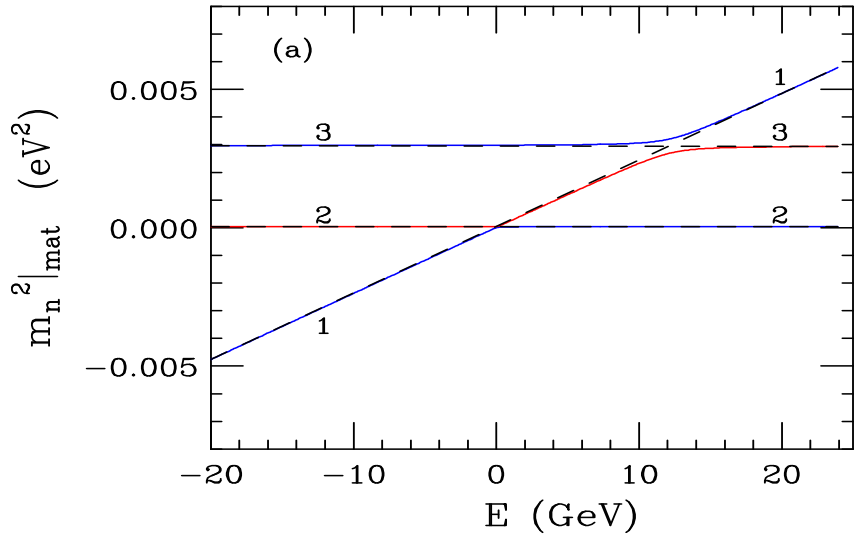}
\caption{The neutrino mass squared differences in matter as a function
  of energy (negative energy is anti-neutrinos).  The density of
  matter is 3 \text{g}$\cdot\text{cm}^{-3}$ and $m^2_1$ at zero energy
  (vacuum) is arbitrarily chosen to be 0 eV$^2$.}
\label{fig:dmsqs}
\end{figure}

For the three neutrino case, nature has chosen two small parameters,
$\sin^2 \theta_{13} \leq 0.04$ and $\Delta m^2_{21}/ \Delta m^2_{31}
\approx 0.03$~\cite{Maltoni:2004ei}; this allows us to factorize the
three neutrino case into a product of two neutrino cases and therefore
the individual $\Delta m^2$ in matter become
\begin{eqnarray}
\Delta m^2_{31} \vert_N & \approx & \Delta m^2_{31} - 2\sqrt{2}G_F N_eE \nonumber \\
\Delta m^2_{21}  \vert_N & \approx &  -2\sqrt{2}G_F N_eE \label{eqn:msqs} \\
\Delta m^2_{32}  \vert_N & \approx & \Delta m^2_{32} . \nonumber 
\label{eq:deltam2_matter}
\end{eqnarray}
In Fig.~\ref{fig:dmsqs} we have plotted the exact mass squared
differences in matter and the approximation given in
Eq.~(\ref{eqn:msqs}) and indeed the approximation is a good one.

\begin{figure}[h]
\includegraphics[width=0.45\textwidth]{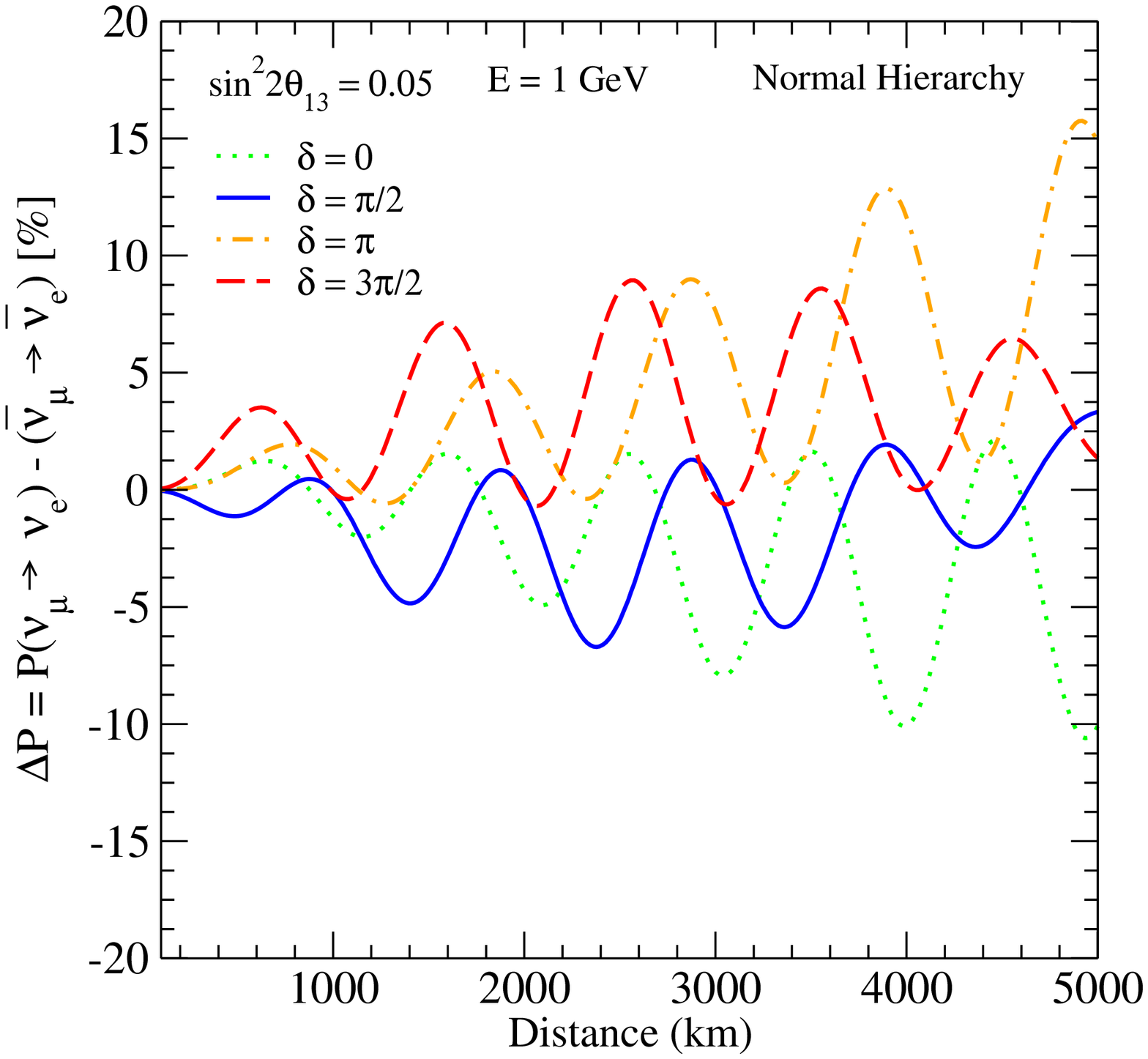}
\includegraphics[width=0.45\textwidth]{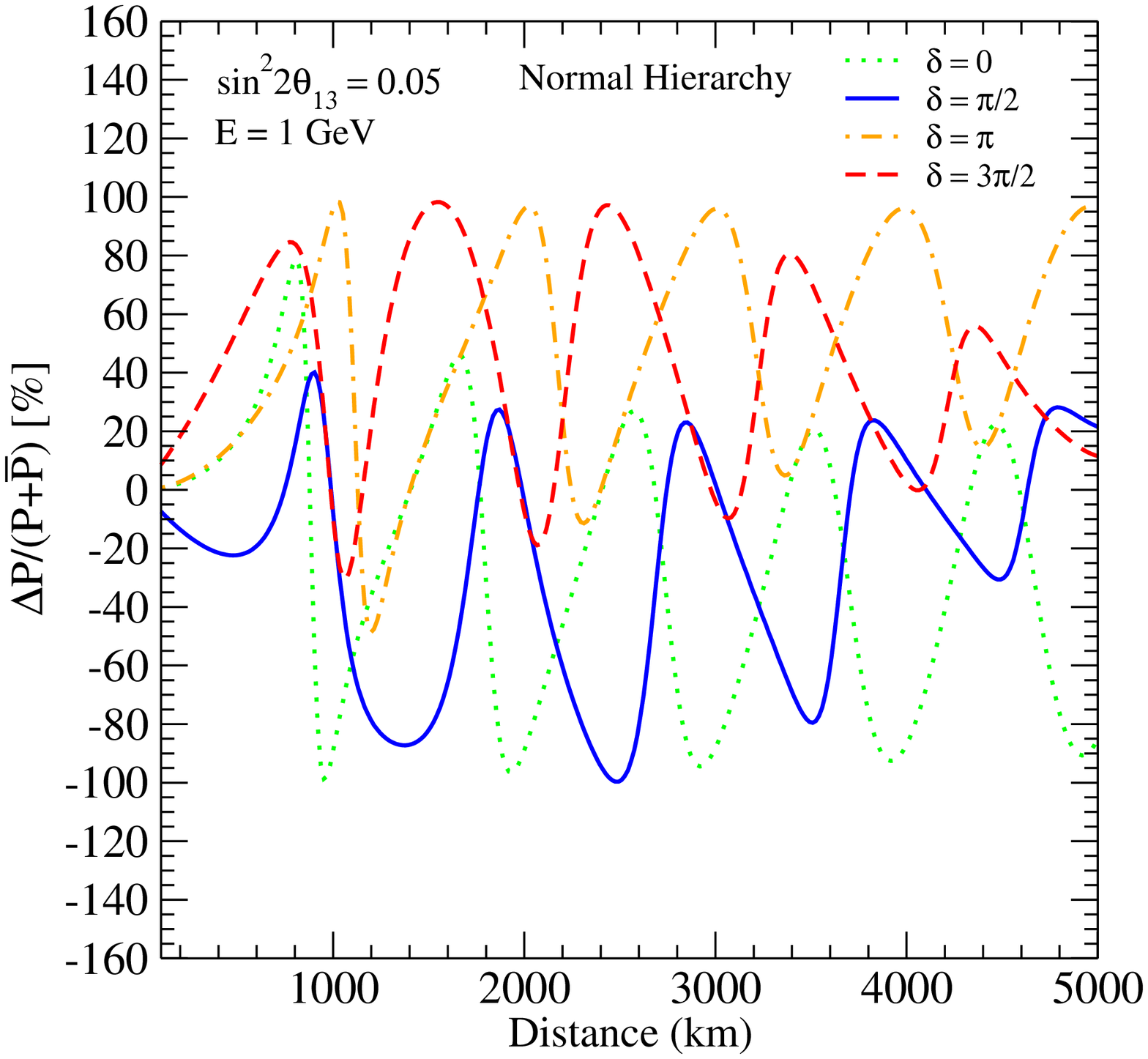}
\caption{Examples of $\Delta P \equiv P(\nu_e \to \nu_\mu)-P(\bar\nu_e
  \to \bar\nu_\mu)$ (left panel) and the asymmetry $\Delta P/[P(\nu_e
  \to \nu_\mu)+ P(\bar\nu_e \to \bar\nu_\mu)]$ (right panel) in matter
  as a function of distance for fixed value of energy, $E=1 $ GeV and
  $\sin^2 2\theta_{13} = 0.05$.  For simplicity, we assume a constant
  electron number density $N_e = 1.5$ mol/cc.  }
\label{fig:deltap+asym_mat}
\end{figure}

Thus the $\sqrt{P_{\text{atm}}} $ and $\sqrt{P_{\text{sol}}} $ in
matter are simply given by
\begin{eqnarray}
\sqrt{P_{\text{atm}}}  & = & \sin\theta_{23} ~\sin 2\theta_{13} 
~\frac{\sin(\Delta_{31} - aL) }{( \Delta_{31} - aL)}~\Delta_{31},\nonumber \\
\sqrt{P_{\text{sol}}} & = &  ~ \cos\theta_{23} ~ \sin 2 \theta_{12}  
~\frac{\sin ( aL)}{(aL)}~\Delta_{21},
\label{eq:pme_matter}
\end{eqnarray}
where $a \equiv G_F N_e/\sqrt{2}$ which is approximately 
$(3500~\text{km})^{-1}$
for $\rho Y_e = 3.0~ \text{g}\cdot\text{cm}^{-3}$.  The relative phase
$(\Delta_{32}+\delta)$ between $\sqrt{P_{\text{atm}}} $ and
$\sqrt{P_{\text{sol}}} $ in Eq.~(\ref{eq:pme}) remains unchanged.  It is
clear from $\sqrt{P_{\text{atm}}}$ that relative size between the
kinematic phase $\Delta_{31}$ compared to $aL$ determines the effects
of matter provided at least one of these is bigger than $\pi/4$.
Since $L$ is common to both $\Delta_{31}$ and $aL$ provided it is a
significant fraction of an oscillation (either in matter or vacuum)
then the comparison is between $\Delta m^2_{31}/E$ and $a=G_F
N_e/\sqrt{2}$.  This implies that the larger the energy of the
neutrino, $E$, the larger the matter effect provided 
the baseline is the same fraction of an oscillation length.
  
\begin{figure}[h]
\hspace{0.5cm}
\includegraphics[height=6.5cm,width=.45\textwidth]{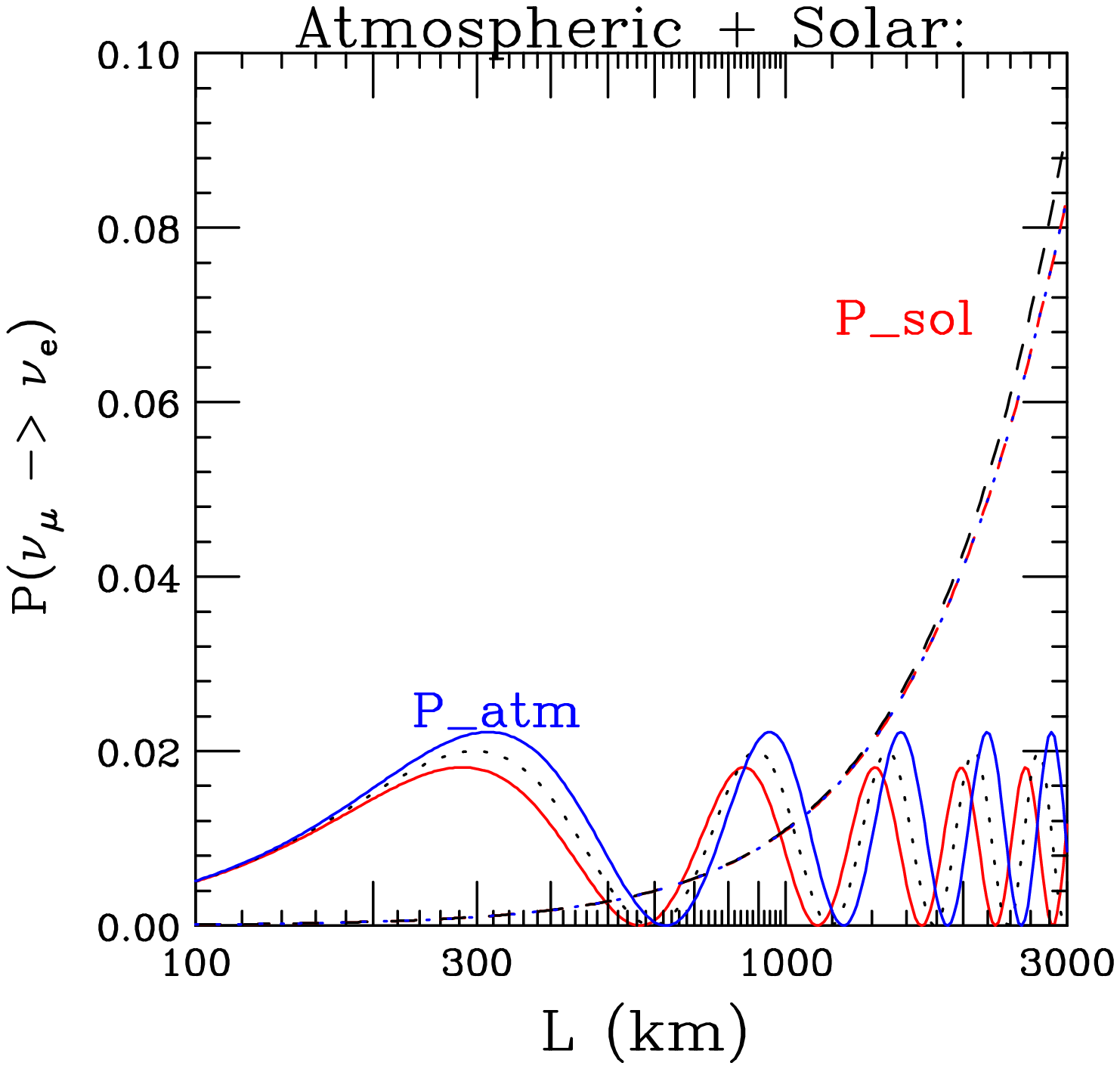}
\includegraphics[height=6.5cm,width=.45\textwidth]{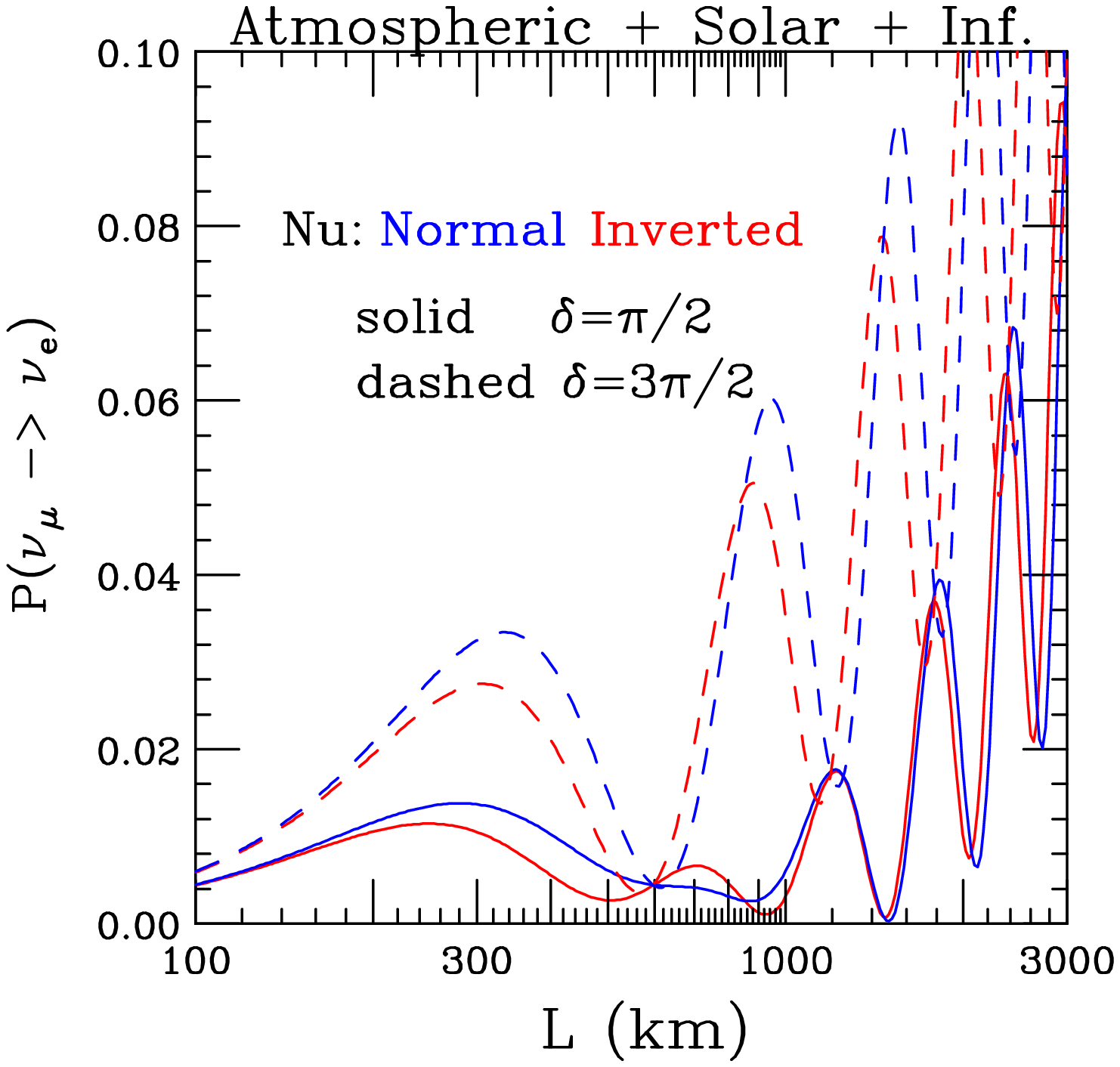}
\vskip 1cm
\hspace{0.5cm}
\includegraphics[height=6.5cm,width=.45\textwidth]{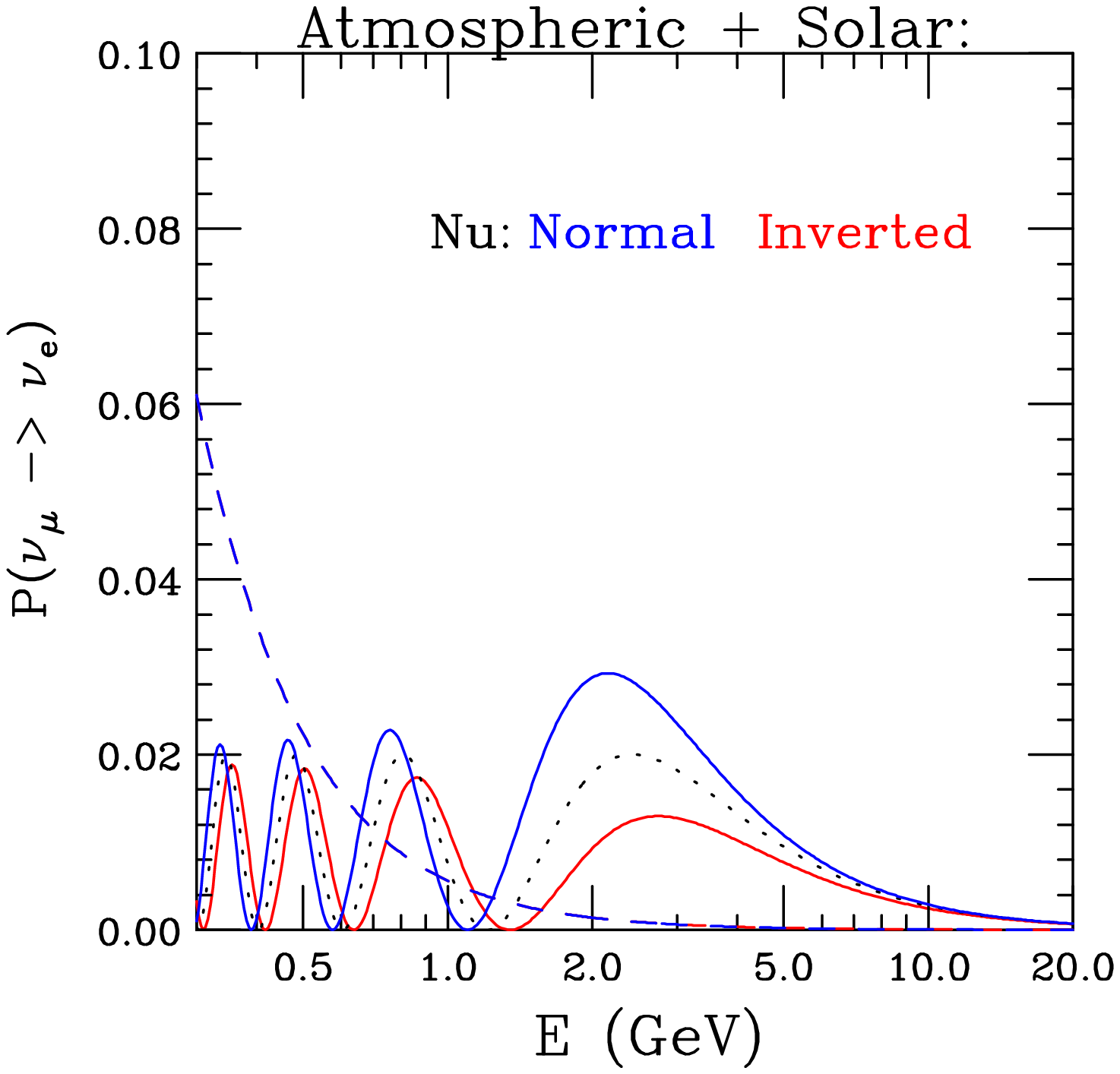}
\includegraphics[height=6.5cm,width=.45\textwidth]{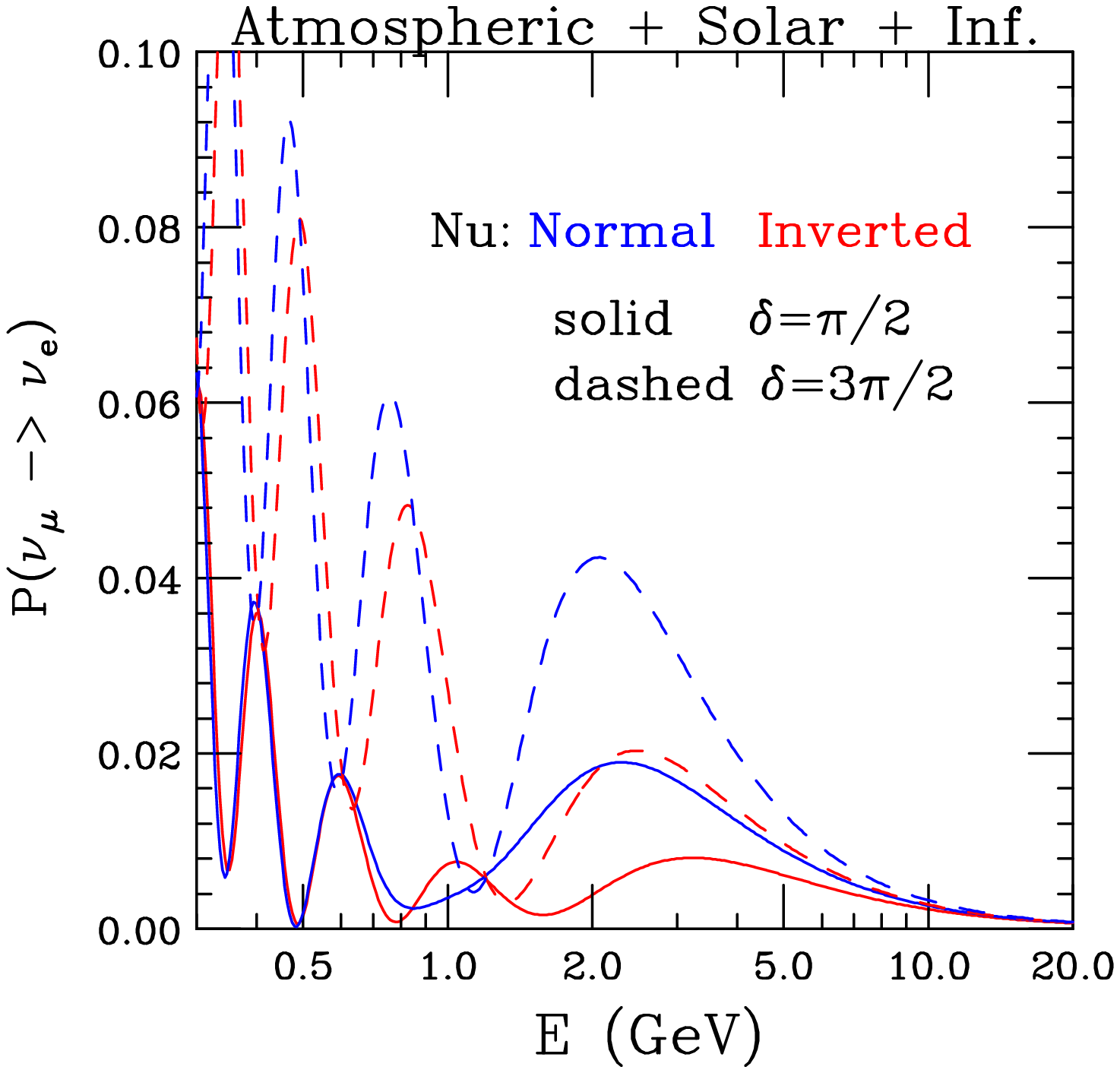}
\caption{The components of the probability  $P(\nu_\mu \to \nu_e)$
in matter: The left column gives $P_{\text{atm}}$ and
$P_{\text{sol}}$, whereas the right column gives the full 
oscillation probability including also the interference term 
between the atmospheric and solar amplitudes. The first row is 
for fixed energy, 0.6 GeV, and varying the baseline, whereas 
the second row is for fixed baseline, 1200 km, and varying 
the energy.}
\label{fig:pmatterE}
\end{figure}
In Fig.~\ref{fig:pmatterE} we show the effects of matter on
$P_{\text{atm}}$ and $P_{\text{sol}}$ for fixed energy and varying the
baseline and for fixed baseline and varying the energy.  For fixed
energy, one sees that the amplitude of the matter effect does not
increase with distance but the shift in the peaks increases in
proportion to the baseline.  For fixed baseline, the amplitude of the
matter effect gets smaller with lower energy, also proportionally.
Thus the biggest matter effect is at the first oscillation peak.

Thus in matter to leading order
 \begin{eqnarray}
 P(\nu_\mu \to  \nu_e) & = &
 \sin^2\theta_{23} ~\sin^2 2\theta_{13} 
~\frac{\sin^2(\Delta_{31} - aL) }
{( \Delta_{31} - aL)^2}~\Delta^2_{31}\nonumber \\
  & &+ \sin 2\theta_{23} ~\sin 2\theta_{13} \sin 2 \theta_{12} ~\frac{\sin(\Delta_{31} - aL) }{( \Delta_{31} - aL)}~\Delta_{31}~\frac{\sin ( aL)}{(aL)}~\Delta_{21}~\cos{(\Delta_{31}+\delta)}
  \nonumber  \\
 & & +~  \cos^2\theta_{23} ~ 
\sin^2 2 \theta_{12}  ~\frac{\sin^2 ( aL)}{(aL)^2}~\Delta^2_{21}, 
 \end{eqnarray}
 where the first (last) term is the atmospheric (solar) probabilities
 and the middle term is the interference between the atmospheric and
 solar contributions.  The phase of the interference has been written
 as $(\Delta_{31}+\delta)$ instead of $(\Delta_{32}+\delta)$ since the
 difference between these two is small. Also, $\sin 2 \theta_{13}$ is
 often written as $2\sin \theta_{13}$ or even $2 \theta_{13}$, again
 the difference is higher order.  We have allowed for $\Delta_{31}$
 and $\Delta_{32}$ to be positive or negative in this derivation, thus
 this expression is valid for both normal, $\Delta m^2_{31} > 0$ and
 inverted, $\Delta m^2_{31} < 0$, hierarchies.
 
 For the CP conjugate process, $\bar{\nu}_\mu \to \bar{\nu}_e$, in
 matter both $a \to -a$ and $\delta \to -\delta$ is required.  Also
 for the T conjugate process, $\nu_e \to \nu_\mu$, replacing $L$ with
 $-L$ including in the kinematic phases, $\Delta$, is necessary.  Note
 also that the CPT conjugate process, $\bar{\nu}_e \to \bar{\nu}_\mu$,
 is obtained by changing the sign of $a$, $\delta$ and $L$.  If one
 changes the sign of $a$ again, since this is equivalent to having
 propagation in anti-matter, the probability is once again the same as
 that for $\nu_\mu \to \nu_e$ as it must from CPT invariance, since
 \begin{eqnarray}
 P(\nu_\mu \to  \nu_e) \vert_{\text{matter}}  \equiv
 P(\bar{\nu}_e \to  \bar{\nu}_\mu) \vert_{\text{anti-matter}}.
\end{eqnarray}
This CPT relationship is true for any symmetric matter profile, $a(x)
= a(L-x)$, where $x$ is the position along the neutrino path and $L$ is
the baseline of the experiment.

\begin{figure}[htb]
\centering
\includegraphics[height=7.5cm]{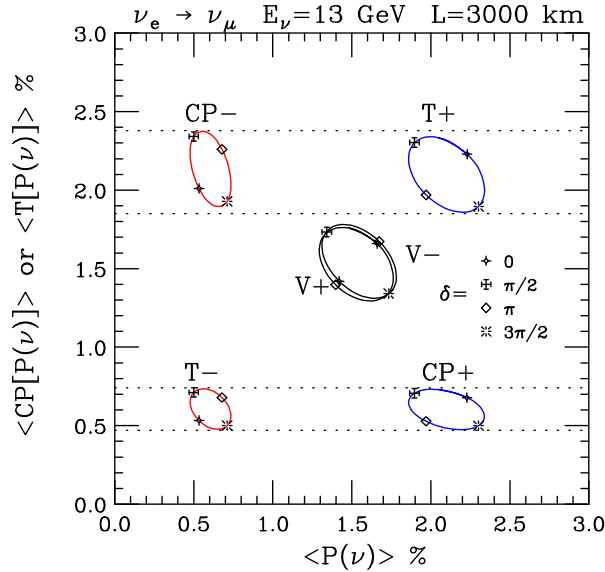}
\caption{$P(\nu_\mu \to \nu_e)$ versus $P(\bar{\nu}_\mu \to \bar{\nu}_e)$
  and $P(\nu_\mu \to \nu_e)$ versus $P(\nu_e \to \nu_\mu)$ in matter.  In
  vacuum the four ellipse overlap, one for CP conjugate pair plus one
  for T conjugate pair times two, one for the normal ($+$) and one for
  the inverted ($-$) hierarchies.  When matter effects are turned on the
  four ellipses split to form a baseball diamond with the vacuum
  ellipse is the pitcher's mound.  All the parameters are held fixed
  except the CP-violating phase, $\delta$ and $\sin^2 2
  \theta_{13}=0.05$. The dotted lines demonstrate the approximate
  relationship of Eq.~\protect{(\ref{T-CP})}.}
\label{fig:gbiprob}
\end{figure}

In fact there are a number of relationships between these
probabilities~\cite{Minakata:2002qe,Blom:2004bk}. Here we discuss
some, especially those that hold for the case of inverting the
hierarchy. The relationship between CP conjugate pairs is given by
\begin{eqnarray}
P(\nu_{e} \to  \nu_{\mu};
\Delta m^2_{31},\Delta m^2_{21}, \delta,a)
&=&
P(\bar{\nu}_{e} \to  \bar{\nu}_{\mu};
-\Delta m^2_{31},-\Delta m^2_{21},\delta, a) \nonumber \\
& \approx &
P(\bar{\nu}_{e} \to  \bar{\nu}_{\mu};
-\Delta m^2_{31},+\Delta m^2_{21},\pi + \delta, a). 
\label{CP-CP}
\end{eqnarray}
The CPT relationship including the effects of switching the hierarchy is
\begin{eqnarray}
P(\nu_{\mu} \to  \nu_{e};
\Delta m^2_{31},\Delta m^2_{21}, \delta, a)
&=&
P(\bar{\nu}_{e} \to  \bar{\nu}_{\mu};
-\Delta m^2_{31},-\Delta m^2_{21},2 \pi - \delta, a) \nonumber  \\
& \approx &
P(\bar{\nu}_{e} \to  \bar{\nu}_{\mu};
-\Delta m^2_{31},+\Delta m^2_{21}, \pi - \delta, a ).
\label{T-CP}
\end{eqnarray}
This relationship is demonstrated in Fig. \ref{fig:gbiprob} by the
dotted horizontal lines.  In fact, there are many such relationships,
e.g., 
\begin{eqnarray}
&P(\nu_{\mu} \to  \nu_{e};
~\Delta m^2_{31},~\Delta m^2_{21}, \delta,a)
&=
P(\nu_{\mu} \to  \nu_{e};-\Delta m^2_{31},-\Delta m^2_{21},
2 \pi - \delta, -a) \nonumber \\
=
&P(\bar{\nu}_{\mu} \to  \bar{\nu}_{e};
-\Delta m^2_{31},-\Delta m^2_{21},\delta, a) \quad
&=
P(\bar{\nu}_{\mu} \to  \bar{\nu}_{e};
~\Delta m^2_{31},~\Delta m^2_{21},2 \pi - \delta, -a).
\label{CP-ids}
\end{eqnarray}
The easiest way to derive such relationships is to look at the
evolution equation, Eqs.~(\ref{eqn:nu-evol}) and (\ref{eqn:Hmatter2}), and
its complex conjugate for neutrinos and anti-neutrinos using a
symmetric matter profile and noting that $U^*(2\pi-\delta)=
U(\delta)$.

We show in Fig.~\ref{fig:deltap+asym_mat} the same quantities shown in
Fig.~\ref{fig:pmatterE} but in the presence of the matter effect. As
we can see, the matter effect can make $\Delta P_{\nu \bar\nu}$ as
well as the asymmetry significantly different from zero even if the CP
phase is zero or $\pi$.
 

When the matter effect is significant the quantity defined in
Eq.~(\ref{eq:asym}) is no longer a good measure of the intrinsic CP
violation, since the matter effect makes this quantity significantly
different from zero even if $\delta=0$ or $\pi$. Therefore we define
\begin{equation}
  \label{eq:cp-asym}
  \delta P_{CP} \equiv \Delta P_{\nu\bar\nu}(\delta=\pi/2) - \Delta P_{\nu\bar\nu}(\delta=0),
\end{equation}
which reduces to Eq.~(\ref{eq:asym}) in vacuum. We note that in this
quantity the matter effect is approximately canceled so that only the
intrinsic CP violation effect remains, to a good approximation.

\begin{figure}[htb]
  \centering
\includegraphics[height=14cm,width=.9\linewidth]{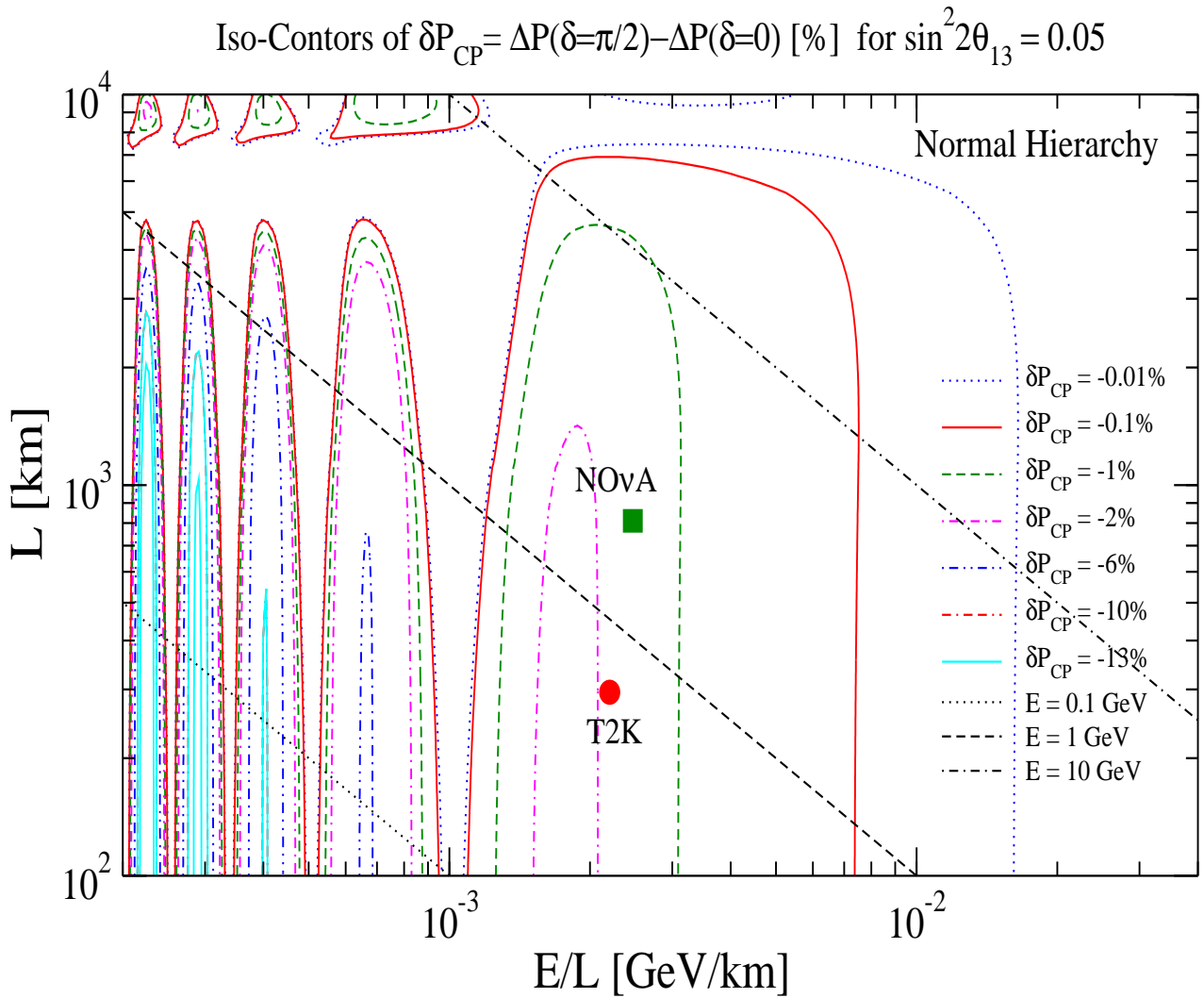}
\caption{Iso-contours of $\delta P_{CP} \equiv \Delta
  P_{\nu\bar\nu}(\delta=\pi/2) - \Delta P_{\nu\bar\nu}(\delta=0)$ for $\sin^2
  \theta_{13}=0.05$ in the $E/L-L$ plane in matter for normal
  hierarchy.}
  \label{fig:T2K-NOVA}
\end{figure} 
In Fig.~\ref{fig:T2K-NOVA} for a given baseline distance the matter
density was assumed constant and estimated from the Preliminary
Reference Earth Model (PREM)~\cite{dziewonski:1981xy} by taking the
average along the neutrino trajectory. The baseline and energy
corresponding to T2K and NO$\nu$A are indicated by the solid circle
and square, respectively. The positions of the peaks of the 5
``peninsula'' regions correspond to, from right to left, 1st, 2nd,
3rd, 4th and 5th oscillation maxima. The best places to search for CP
violation effects in the probabilities are those within the regions
with higher $\delta P_{CP}$ values. Since the neutrino flux decreases
as $1/L^2$ as the baseline gets larger one should, in practice, try to
measure the first or first few oscillation maxima. We note that in the
region with $L\sim 7000-8000$ km no significant CP violation is
expected: at such ``magic'' baselines~\cite{Huber:2003ak} the
dependence of the probability on the CP $\delta$ and the solar
oscillation parameters disappears (see~\cite{Smirnov:2006sm} for its
physical interpretation).

\section{Near Term Experiments}
\label{sec:near-term-exper}

Here we discuss the prospects of observing CP violation by the
following two near term experiments\footnote{At the time of writing
  this review, the beam line for T2K is under construction with first
  beam to the existing Super-Kamiokande detector is expected in 2009.
  The NuMI beam line is operating for the MINOS experiment but the
  NO$\nu$A detector has still to get final approval from DOE to begin
  construction and is expected to come on line after T2K by at least
  one year.}, T2K and NO$\nu$A:

\begin{itemize}
\item T2K~\cite{Itow:2001ee}: Initially this experiment will send a
  beam of $\nu_\mu$'s from the JPARC facility to the existing 50 kton
  Super-Kamiokande (SK) detector.  The beam will be aimed 2.5 degrees
  off the JPARC to SK axis resulting in a mean neutrino energy of 0.63
  GeV with an approximate Gaussian spread of 0.13 GeV at SK.  The
  source detector distance is 295 km with a mean matter density of 2.3
  g cm$^{-3}$.  The beam power will be ramped over time to 0.75 MW.
  Only neutrino running is expect during the initial phase, Phase I.
In the future, there is a possibility to upgrade the beam 
power to as high as 4 MW and to include anti-neutrino running.
\item NO$\nu$A\cite{Ayres:2004js,Ayres:2002nm}: The NO$\nu$A detector
  will be placed 12km off-axis of the existing NUMI neutrino beam line
  at a distance of 810 km.  The mean neutrino energy in this
  configuration is 2.0 GeV with an approximate Gaussian spread of 0.30
  GeV.  By the time the 25kton liquid scintillator detector has been
  completed the NUMI beam power is expected to be greater than or
  equal to 0.4 MW.  This allows for the running in both neutrino or
  anti-neutrino mode.  The mean matter density for this configuration
  is 2.8 g cm$^{-3}$.
In the future, there is a possibility to upgrade the beam 
power up to as high as 2 MW. 
\end{itemize}

Given the current best fit value for $\Delta m^2_{\text{atm}}= 2.5
\times 10^{-3}$ eV$^2$ from the global analysis~\cite{Schwetz:2006dh}
both of these experiments are at a mean energy which is above the
first vacuum oscillation maximum (VOM) energy.  Also, because of the
longer path length for the NO$\nu$A experiment the matter effects are
three times larger for NO$\nu$A than the T2K experiment.

\subsection{Dependence of probabilities on $\theta_{13}$ and $\delta$}
\label{sec:depend-prob-thet}

\begin{figure}[h]
\includegraphics[height=7cm,width=0.44\textwidth]{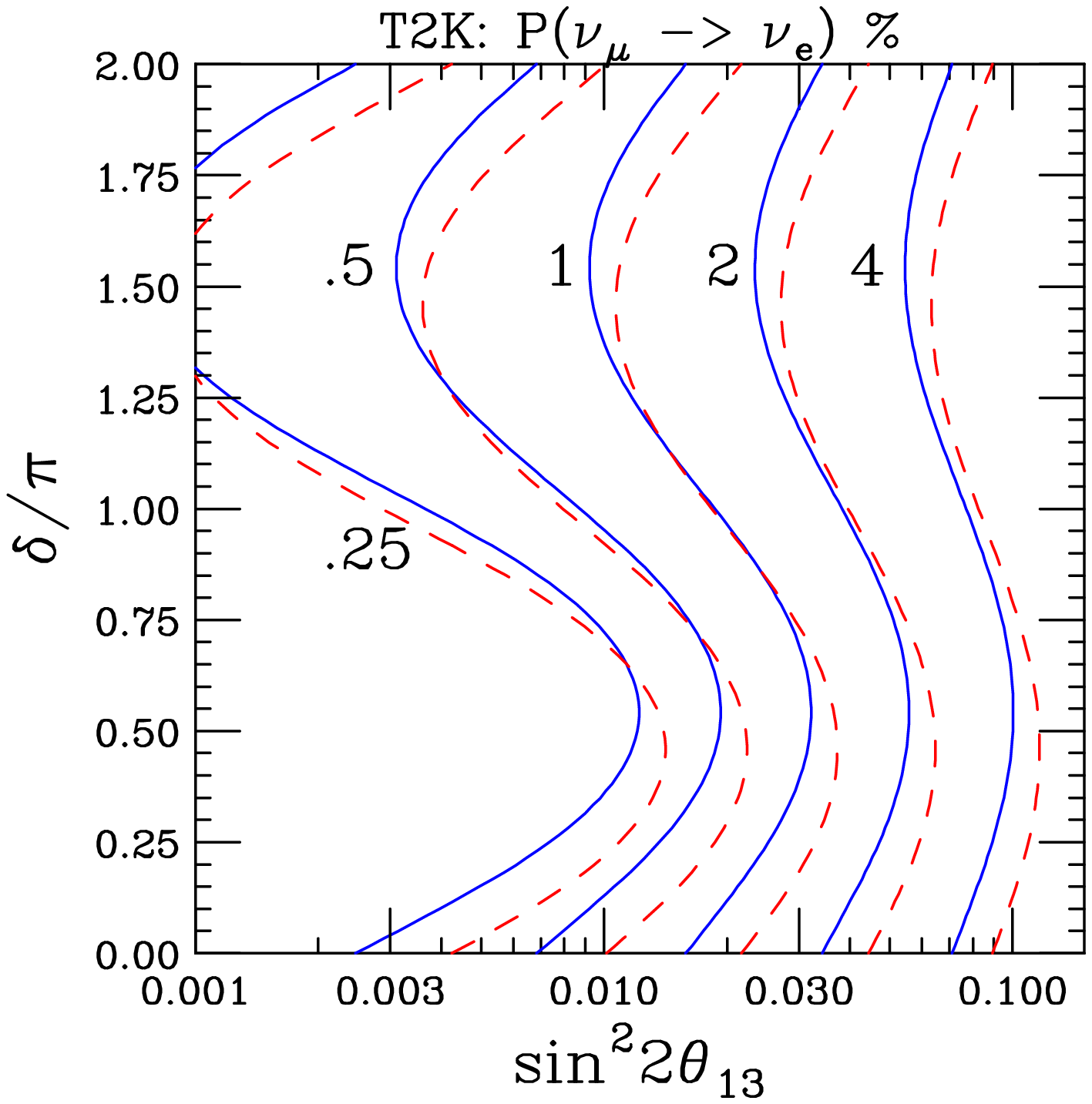}
\vspace{0.5cm}
\includegraphics[height=7cm,width=0.44\textwidth]{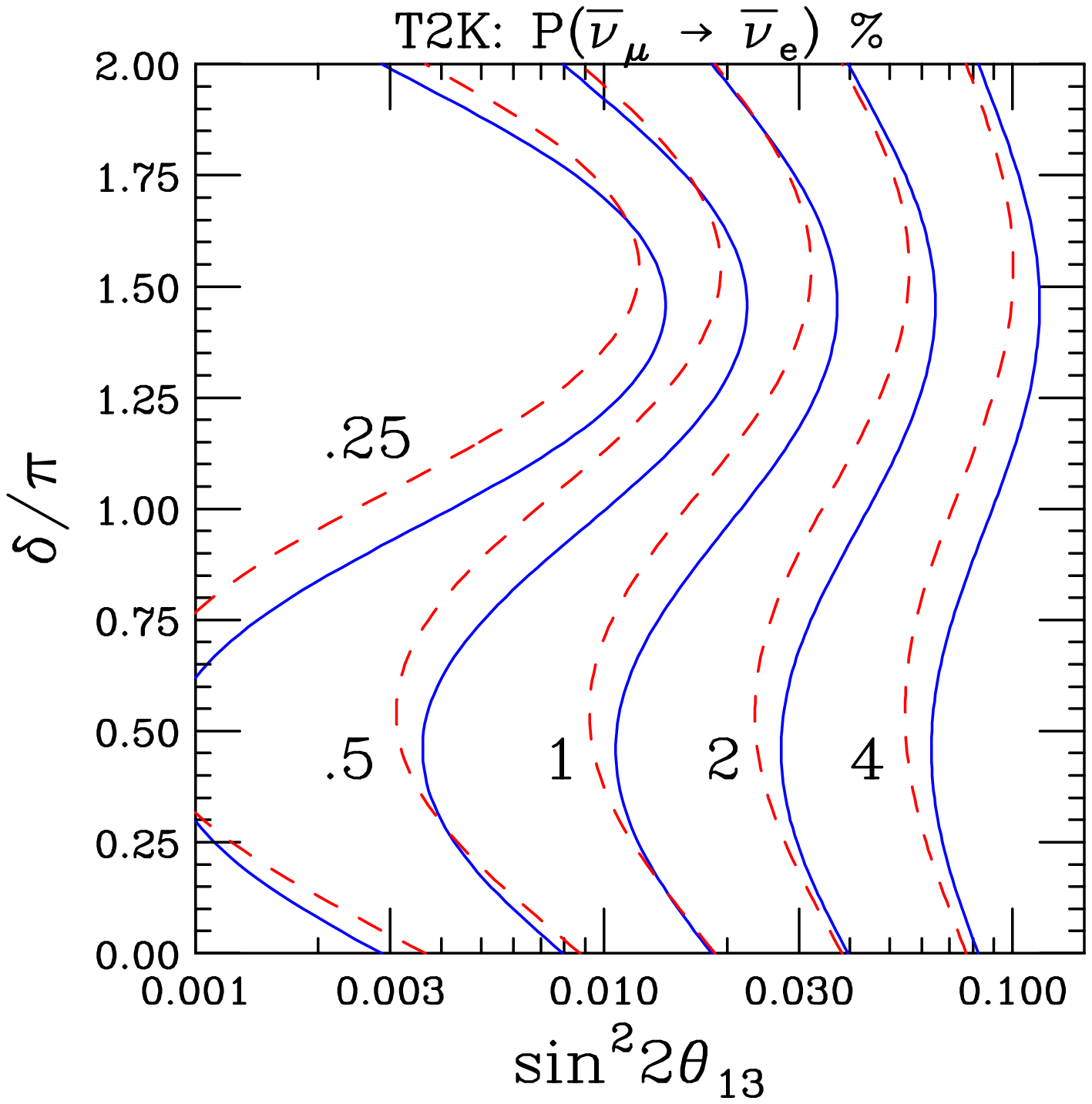}
\vspace{-0.5cm}
\caption{Iso-contours of $P(\nu_\mu \to  \nu_e)$  
and $P(\bar{\nu}_\mu \to  \bar{\nu}_e)$ in matter 
for T2K experiment. The solid (dashed) curves correspond to 
the normal (inverted) hierarchy. In Phase I, T2K is expected to reach a 
sensitivity which is close to the 0.5\% contour for neutrinos.}
\label{fig:T2K}
\end{figure}

First let us look at the how the oscillation probabilities as well as
CP violation (the difference of probabilities) for these two
experiments depends on $\theta_{13}$ and $\delta$.
Fig.~\ref{fig:T2K} shows the iso-probability contours for the T2K
experiment for both hierarchies in neutrino and anti-neutrino running.
The expected sensitivity for the first phase of running of T2K is
approximately equal to the 0.5\% contour in neutrino running.  
\begin{figure}[h]
\includegraphics[height=7cm,width=0.44\textwidth]{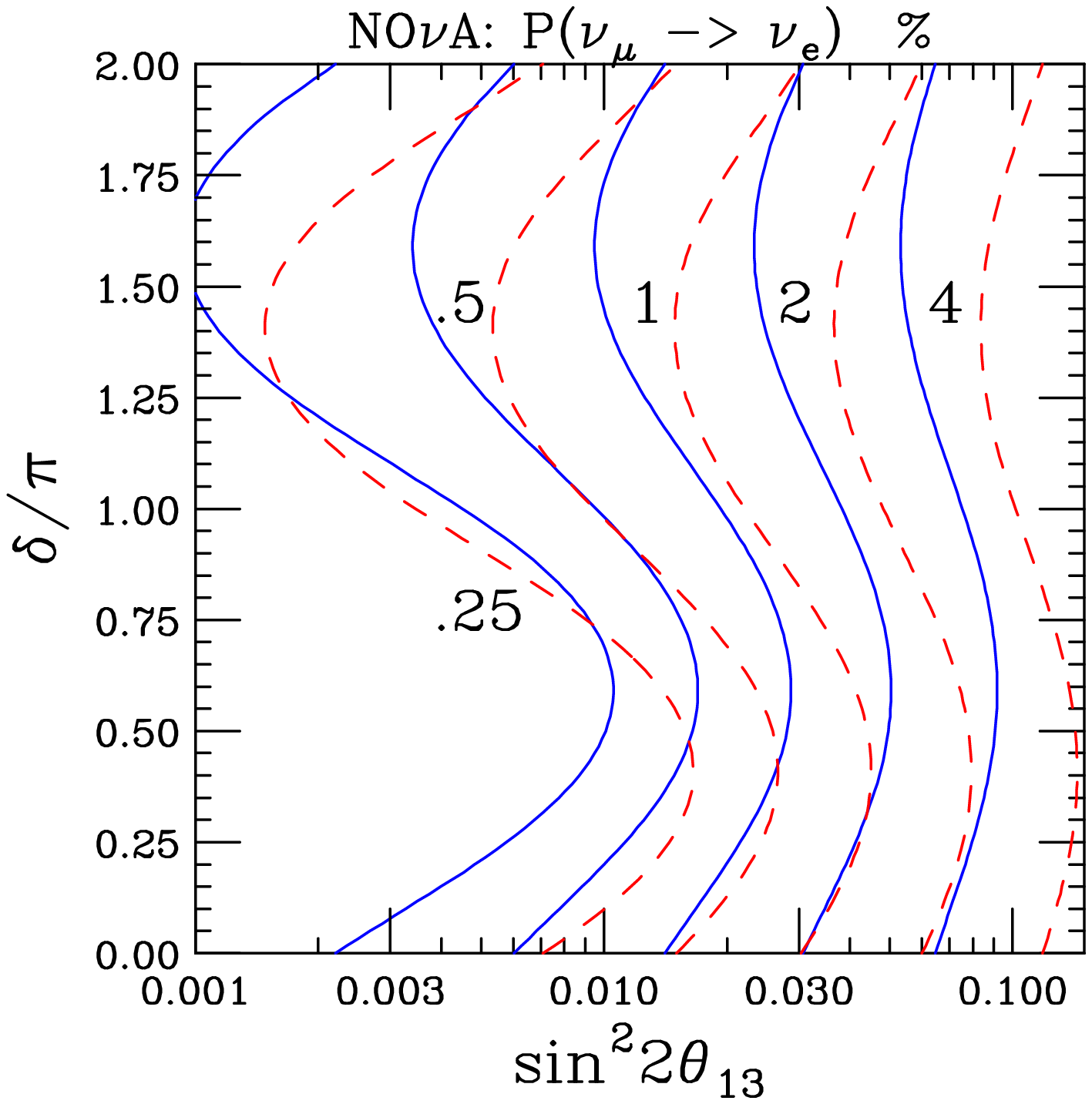}
\includegraphics[height=7cm,width=0.44\textwidth]{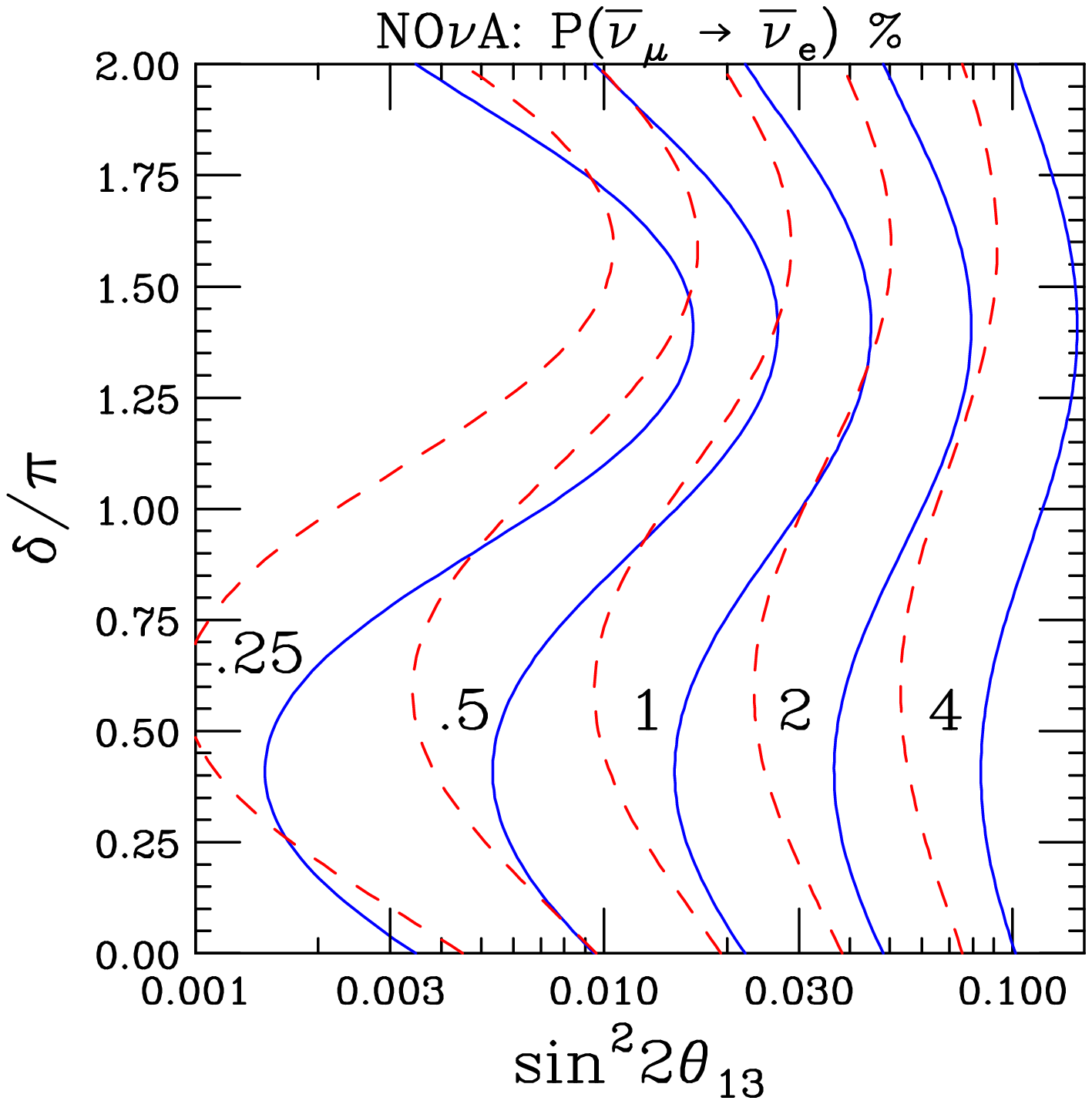}
\vspace{-0.3cm}
\caption{Iso-contours of $P(\nu_\mu \to  \nu_e)$ and
  $P(\bar{\nu}_\mu \to  \bar{\nu}_e)$ in matter for NO$\nu$A
  experiment.  The solid (dashed) curves correspond to the normal
  (inverted) hierarchy. The NO$\nu$A experiment in its initial phase
  is expected to reach a sensitivity close to the 0.5\% contour in
  neutrinos and approximately the 1\% contour in anti-neutrinos. }
\label{fig:nova}
\end{figure}
There is a small difference between the contours for the two
hierarchies due to matter effects and the fact that the mean neutrino
energy is above vacuum oscillation maximum.  Since the beam and fake
backgrounds contribute approximately 1\%, further improvement on the
sensitivity will be challenging.  No anti-neutrino running is expected
in the first phase since the event rate is appreciable smaller than in
neutrino running.  Of course, anti-neutrino will be included as the
beam power is ramped up above the initial phase. Due to the increased
background and smaller statistics, compared to neutrino running, a
reasonable estimate of the sensitivity in anti-neutrino running is the
1\% contour of the anti-neutrino plot in Fig.~\ref{fig:T2K}.
Fig.~\ref{fig:nova} shows the iso-probability contours for the
NO$\nu$A experiment for both hierarchies in neutrino and anti-neutrino
running.  The expected sensitivity for neutrino running is also
approximately 0.5\% and approximately 1\% for anti-neutrino running.
Again, further improvements will be challenging due to backgrounds.
Notice the sizable difference in the sensitivity between the two
hierarchies~\cite{Minakata:2001qm}, primarily due to the matter
effects.  The combined running of neutrinos and anti-neutrinos will
make the combined sensitivity much less dependent on the CP violating
phase, $\delta$.
\begin{figure}[htb]
  \centering
\includegraphics[height=7cm,width=0.46\textwidth]{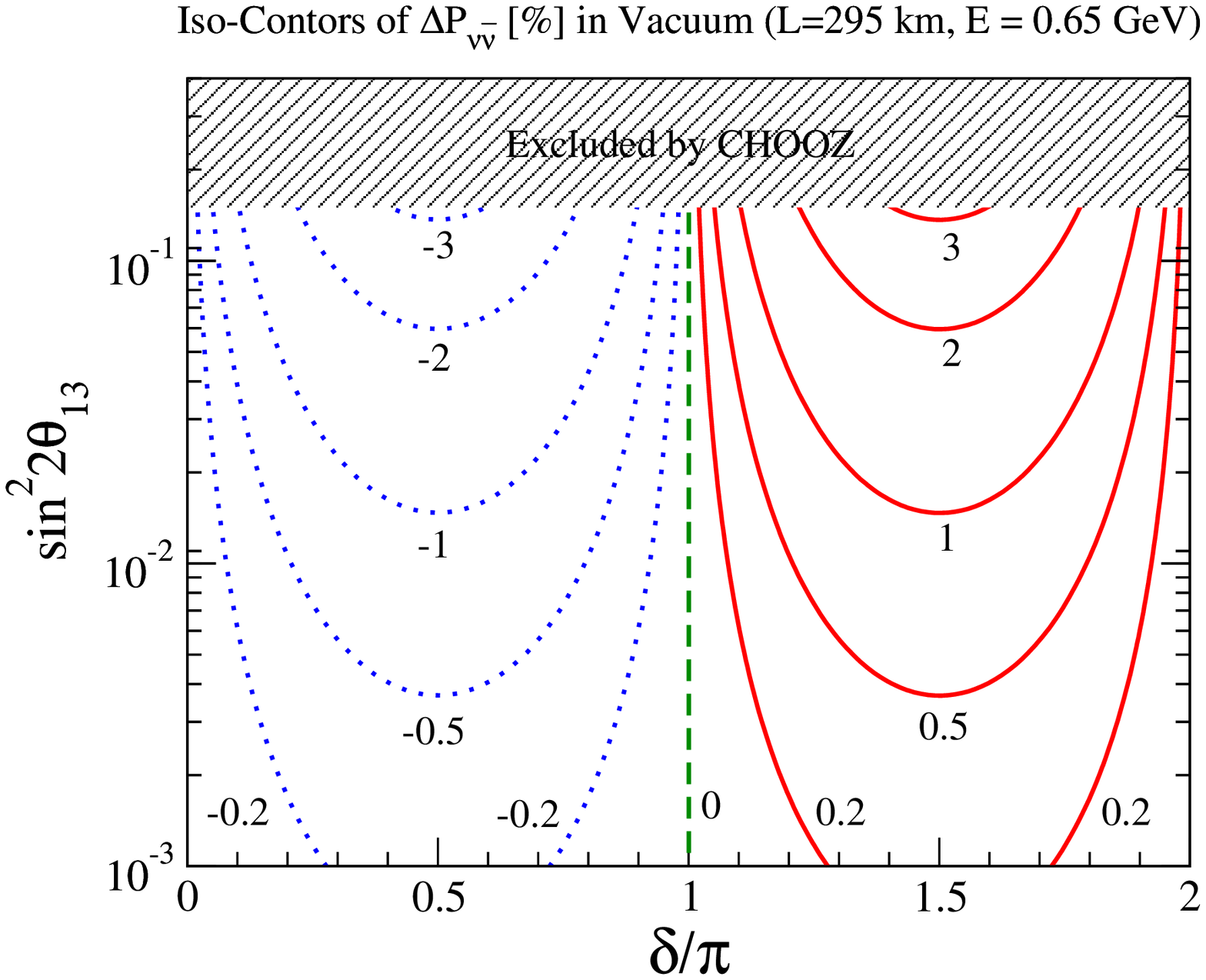}
\includegraphics[height=7cm,width=0.46\textwidth]{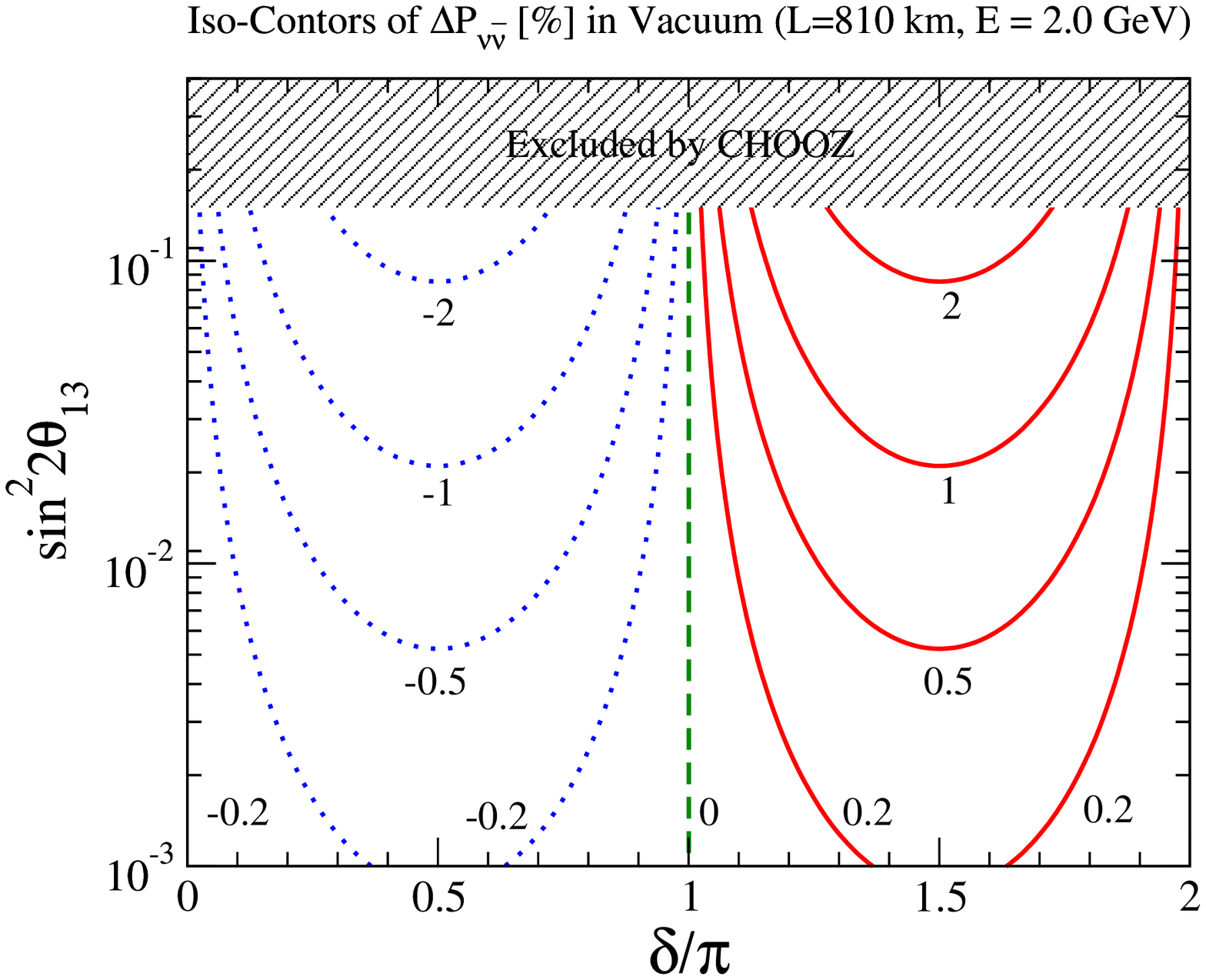}
\vspace{-0.5cm}
\caption{Iso-contours of $\Delta P_{\nu \bar{\nu}} \equiv P(\nu_\mu
  \to  \nu_e)-P(\bar{\nu}_\mu \to  \bar{\nu}_e)$ in the
  $\delta$ and $\sin^2 2\theta_{13}$ in vacuum for T2K (left panel)
  and NO$\nu$A (right panel).  }
\label{fig:DeptaP-cont-vac}
\end{figure}

In order to have some idea about the magnitude of the intrinsic CP
violation (coming from $\delta$) which can potentially 
be observed in these two experiments,
we show in Fig.~\ref{fig:DeptaP-cont-vac}, the iso-contours of $\Delta
P_{\nu \bar{\nu}} \equiv P(\nu_\mu \to  \nu_e)-P(\bar{\nu}_\mu
\to  \bar{\nu}_e)$ in the plane of $\delta$ and $\sin^2
2\theta_{13}$ by switching off the matter effect. Note, that this
quantity is just the CP-invariant factor multiplied by a
constant which depends on the neutrino energy and baseline, see
Eq.~(\ref{eq:DeltaP}).  

\begin{figure}[htb]
\includegraphics[height=7cm,width=0.46\textwidth]{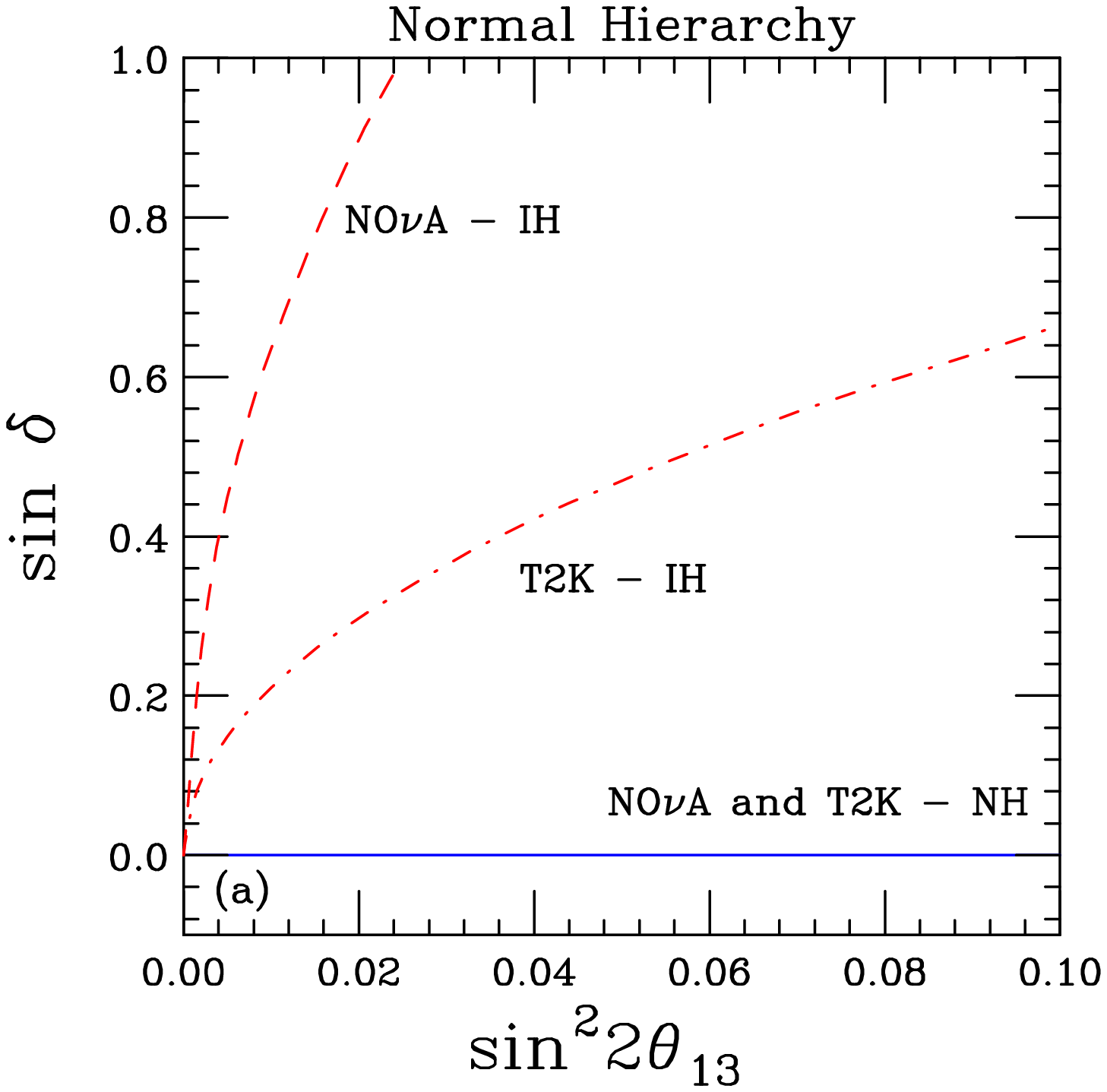}
\includegraphics[height=7cm,width=0.46\textwidth]{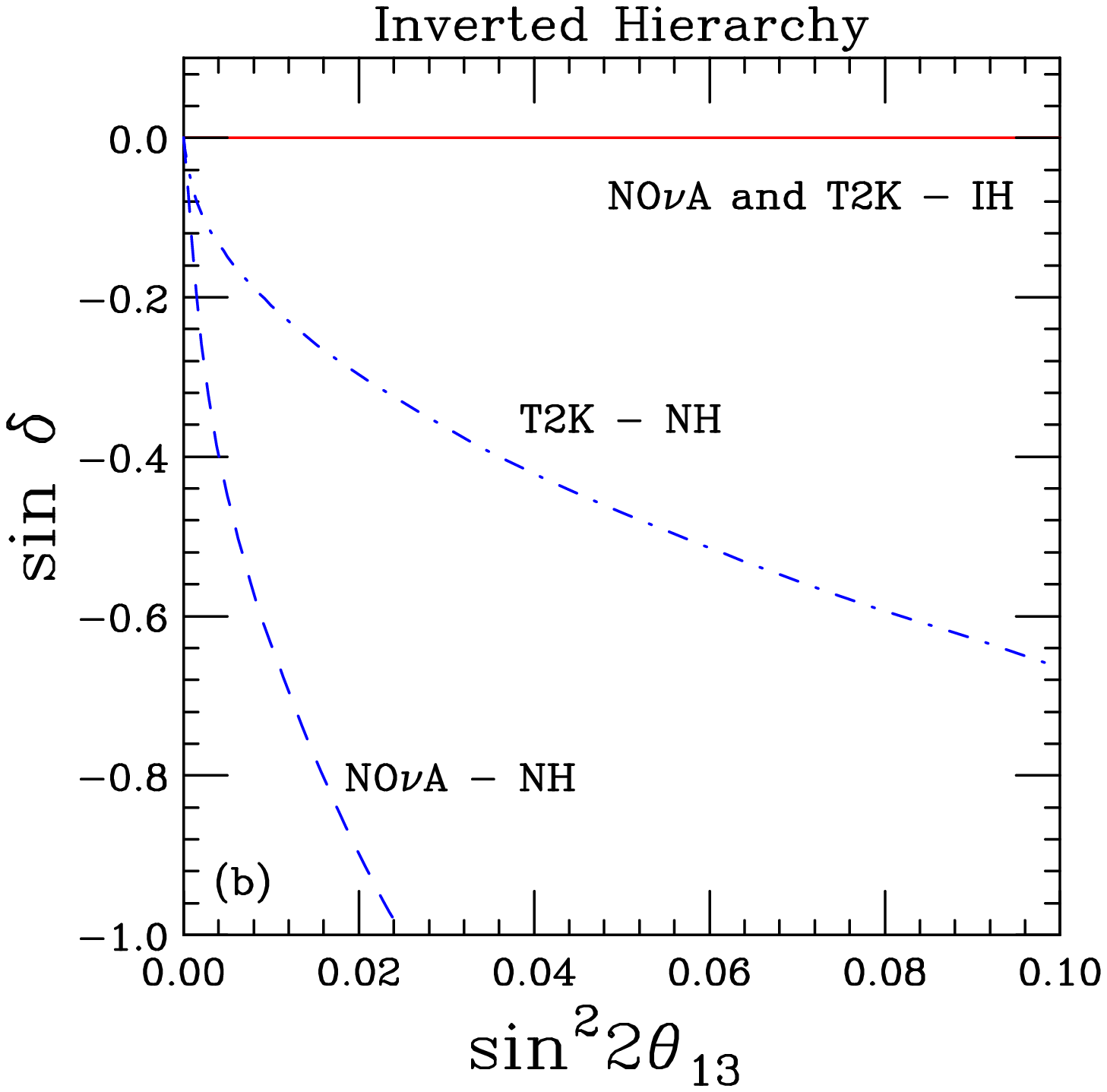}
\vspace{-0.5cm}
\caption{ The positions of the CP conserving solutions in the $(\sin^2
  2\theta_{13}, \sin \delta)$ plane: (a) for the Normal Hierarchy, the
  curves above $\sin \delta =0$ for T2K (dot-dashed) and NO$\nu$A
  (dashed) are the locations of the CP conserving solutions for the
  Inverted Hierarchy.  (b) is the equivalent plot for the Inverted
  Hierarchy. Without determining the hierarchy, a given experiment
  cannot claim the observation of CP violation unless it can exclude
  both of lines with its label at the required confidence level in one
  of these plots.  }
\label{fig:CPCpositions}
\end{figure}

Note also that in vacuum this quantity does not depend on mass
hierarchy so that the contours are symmetric with respect to the line
$\delta = \pi$ except for the sign difference.  The positions of the
CP conserving solutions in the $(\sin^2 2\theta_{13}, \sin \delta)$
plane are shown in Fig.~\ref{fig:CPCpositions}.

\begin{figure}[!b]
\hspace{0.7cm}
\includegraphics[height=7.cm,width=0.45\textwidth]{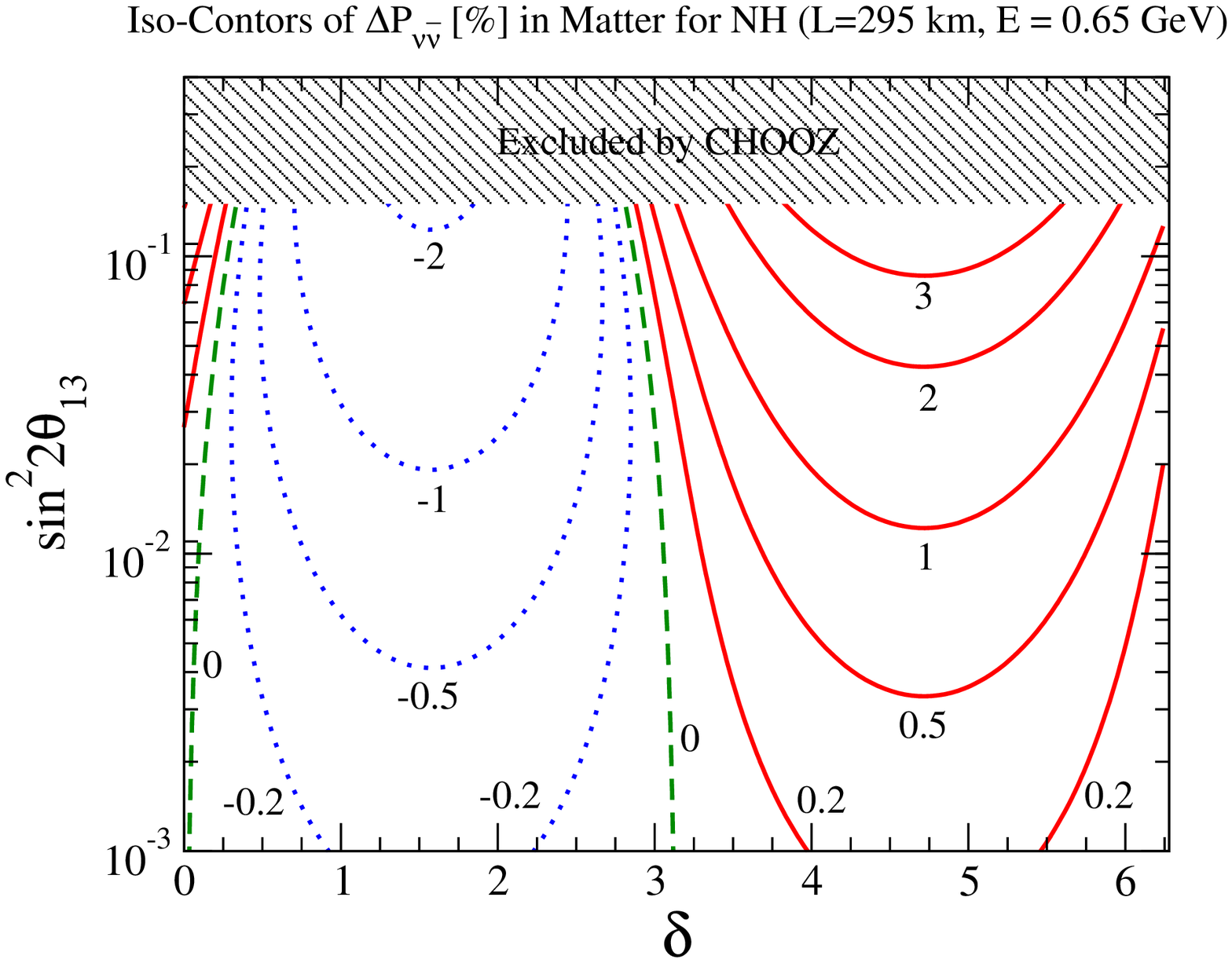}
\hspace{0.2cm}
\includegraphics[height=7.cm,width=0.45\textwidth]{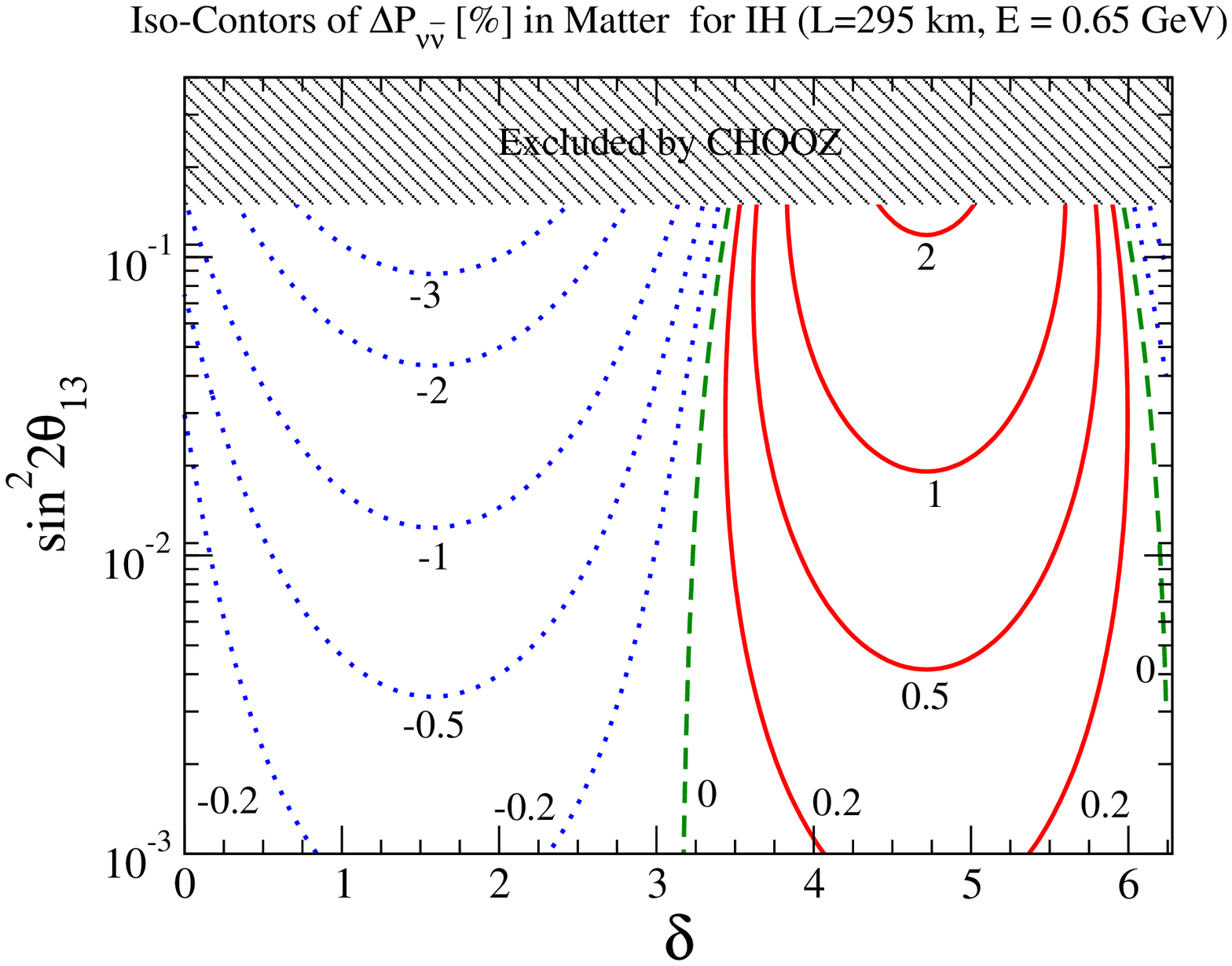}
\caption{Iso-contours of $\Delta P_{\nu \bar{\nu}} \equiv P(\nu_\mu
  \to \nu_e)-P(\bar{\nu}_\mu \to \bar{\nu}_e)$ in the $\delta$ and
  $\sin^2 2\theta_{13}$ for T2K for the normal (left panel) and the
  inverted (right panel) mass hierarchy, with the matter effect
  included.  }
\label{fig:DeptaP-cont-t2k}
\end{figure}

\begin{figure}
\includegraphics[height=7.cm,width=0.45\textwidth]{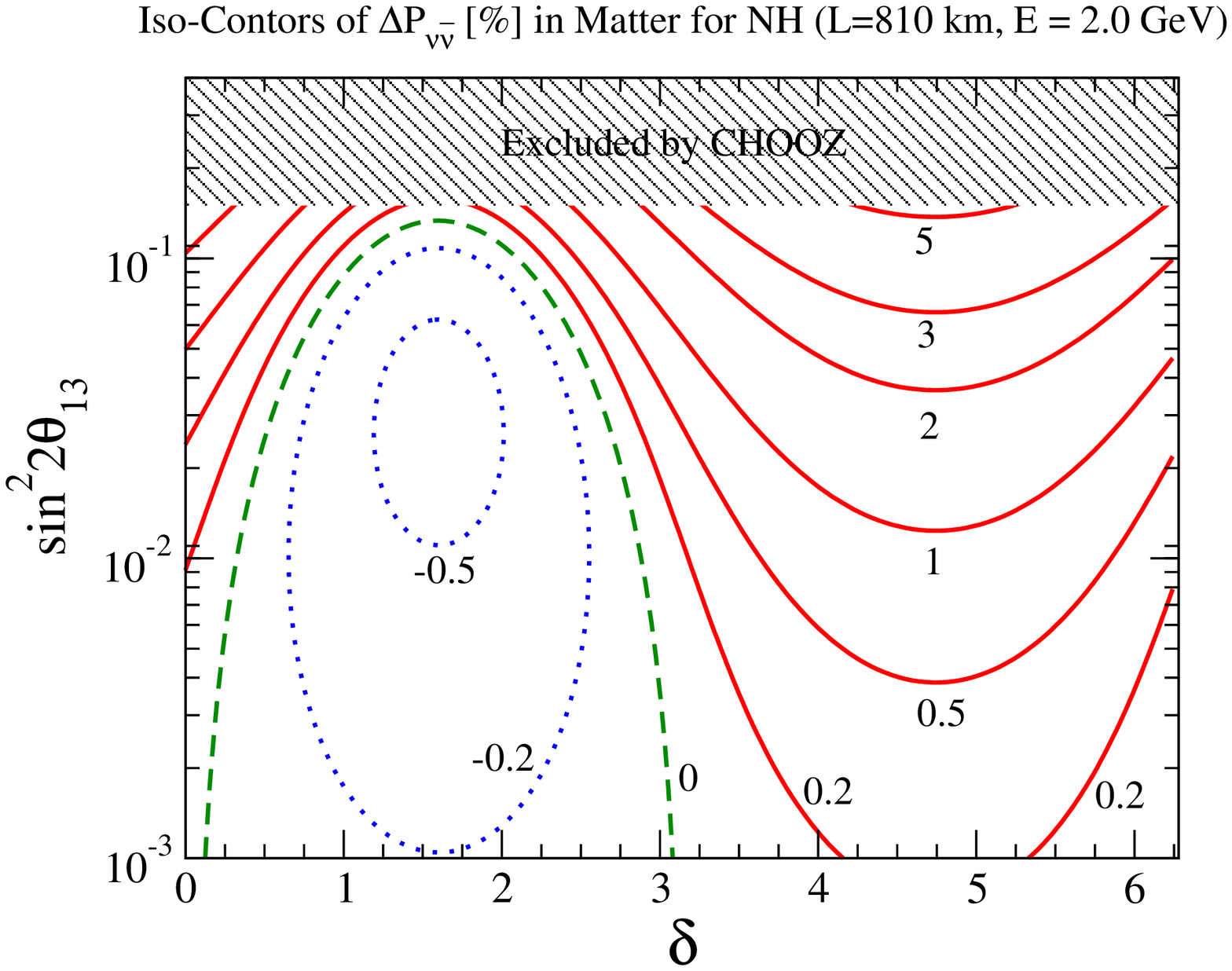}
\includegraphics[height=7.cm,width=0.45\textwidth]{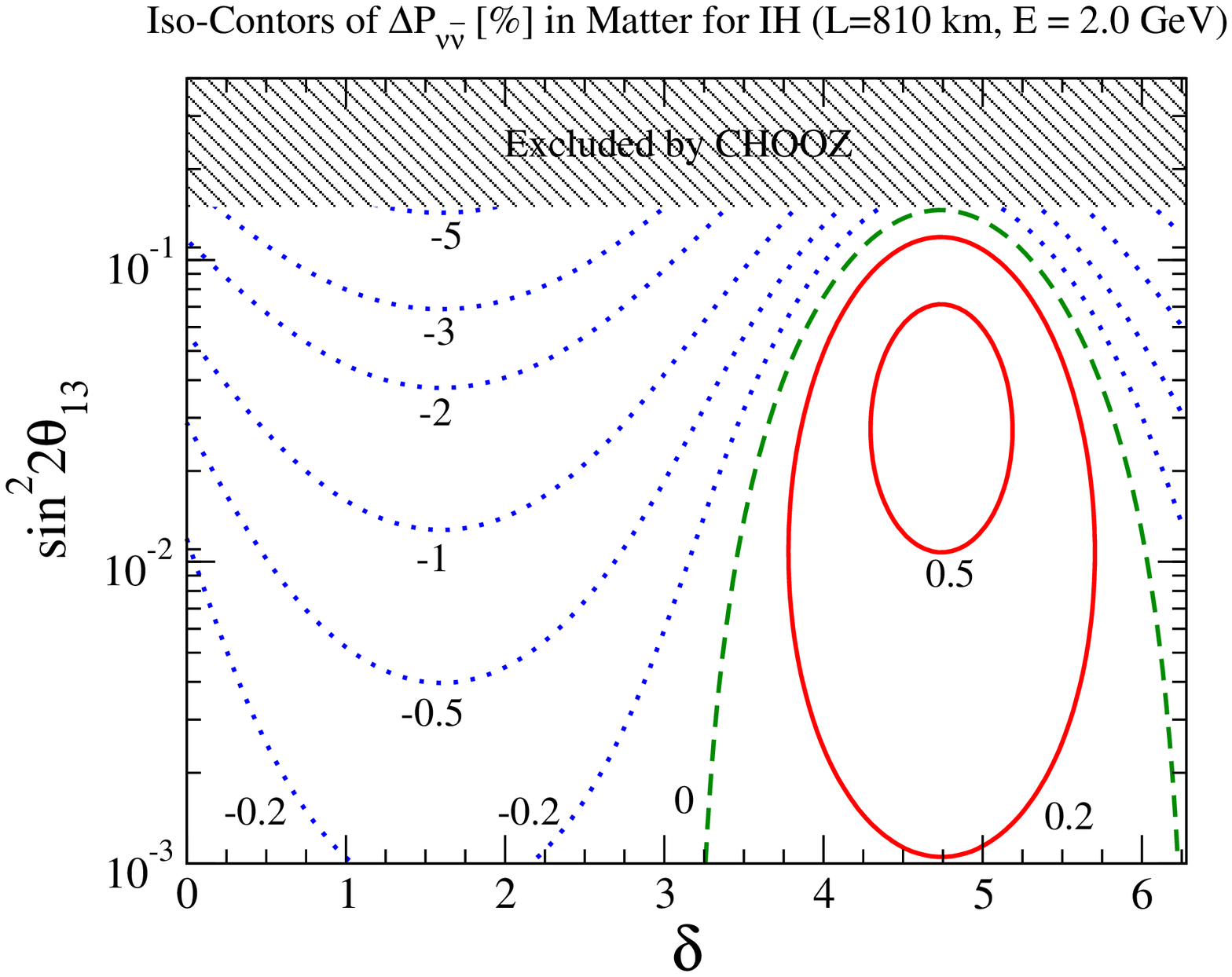}
\caption{Iso-contours of $\Delta P_{\nu \bar{\nu}} \equiv P(\nu_\mu
  \to  \nu_e)-P(\bar{\nu}_\mu \to  \bar{\nu}_e)$ in the
  $\delta$ and $\sin^2 2\theta_{13}$ for NO$\nu$A for the normal (left
  panel) and the inverted (right panel) mass hierarchy, with the
  matter effect included.  }
\label{fig:DeptaP-cont-nova}
\end{figure}

\begin{figure}
\includegraphics[height=8.5cm]{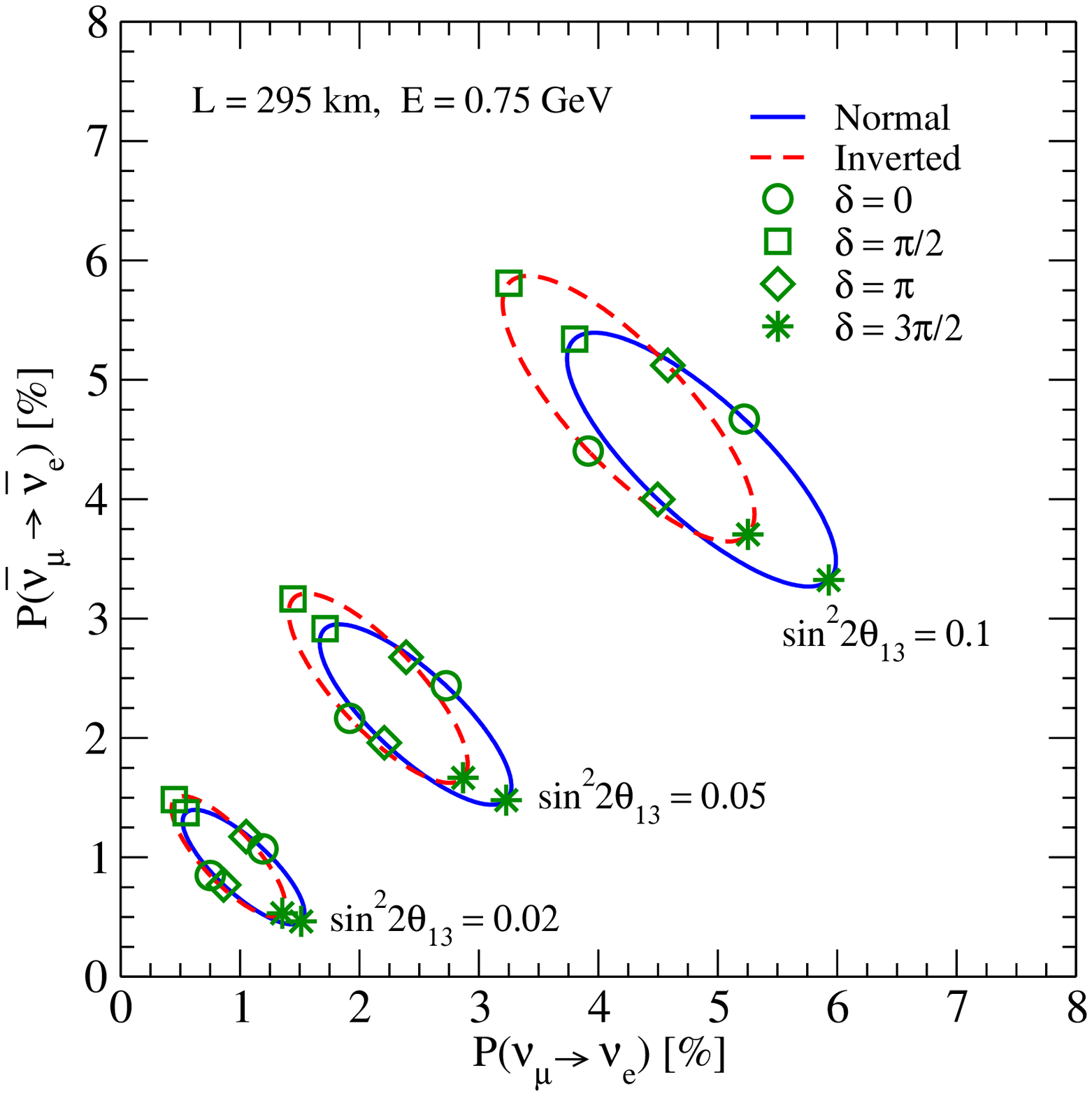}
\includegraphics[height=8.5cm]{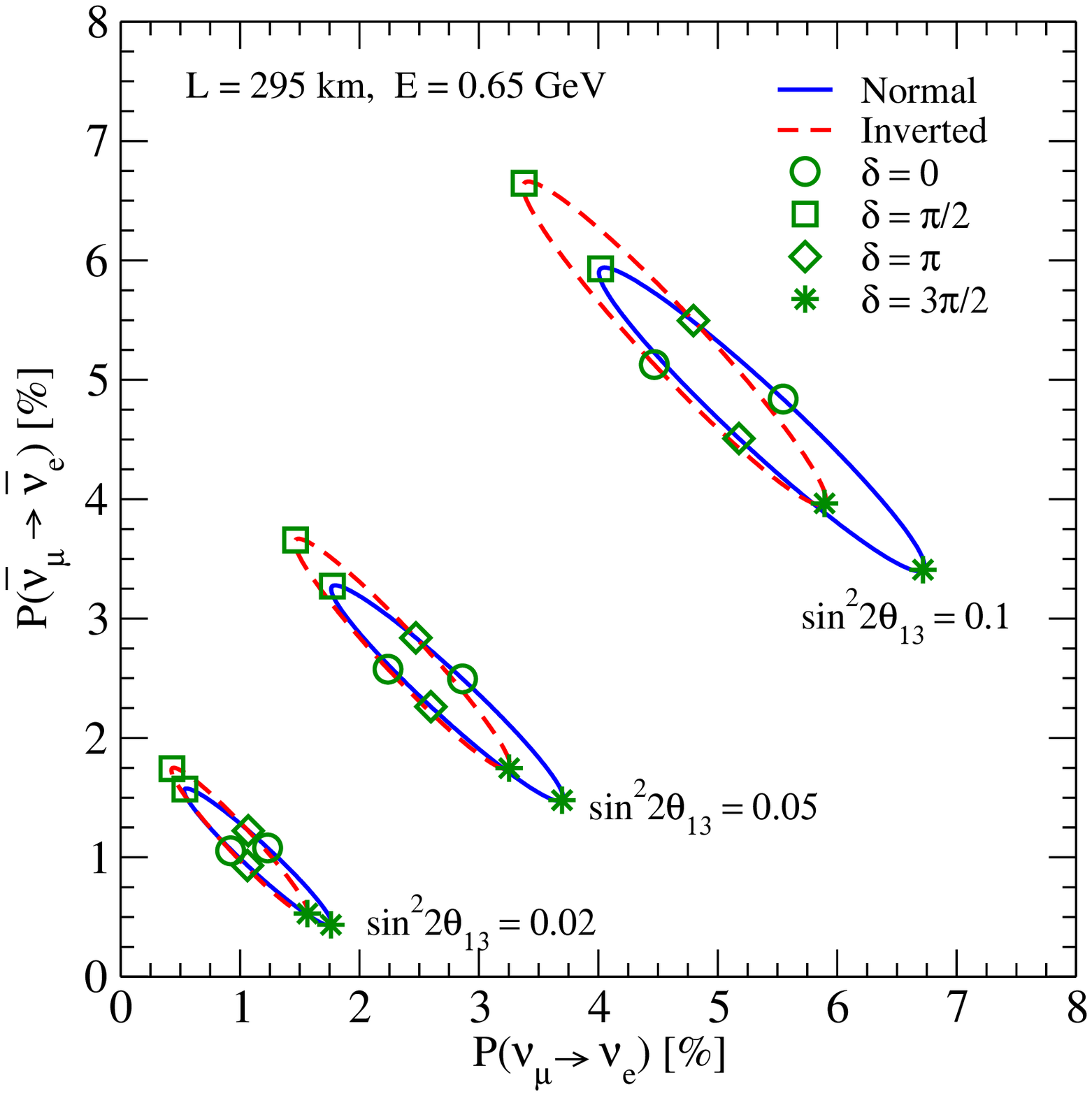}
\caption{The bi-probability $P(\nu_\mu \to \nu_e)$ diagram for T2K
  including matter effects.}
\label{fig:biprob1}
\end{figure}

\begin{figure}
\includegraphics[height=8.5cm]{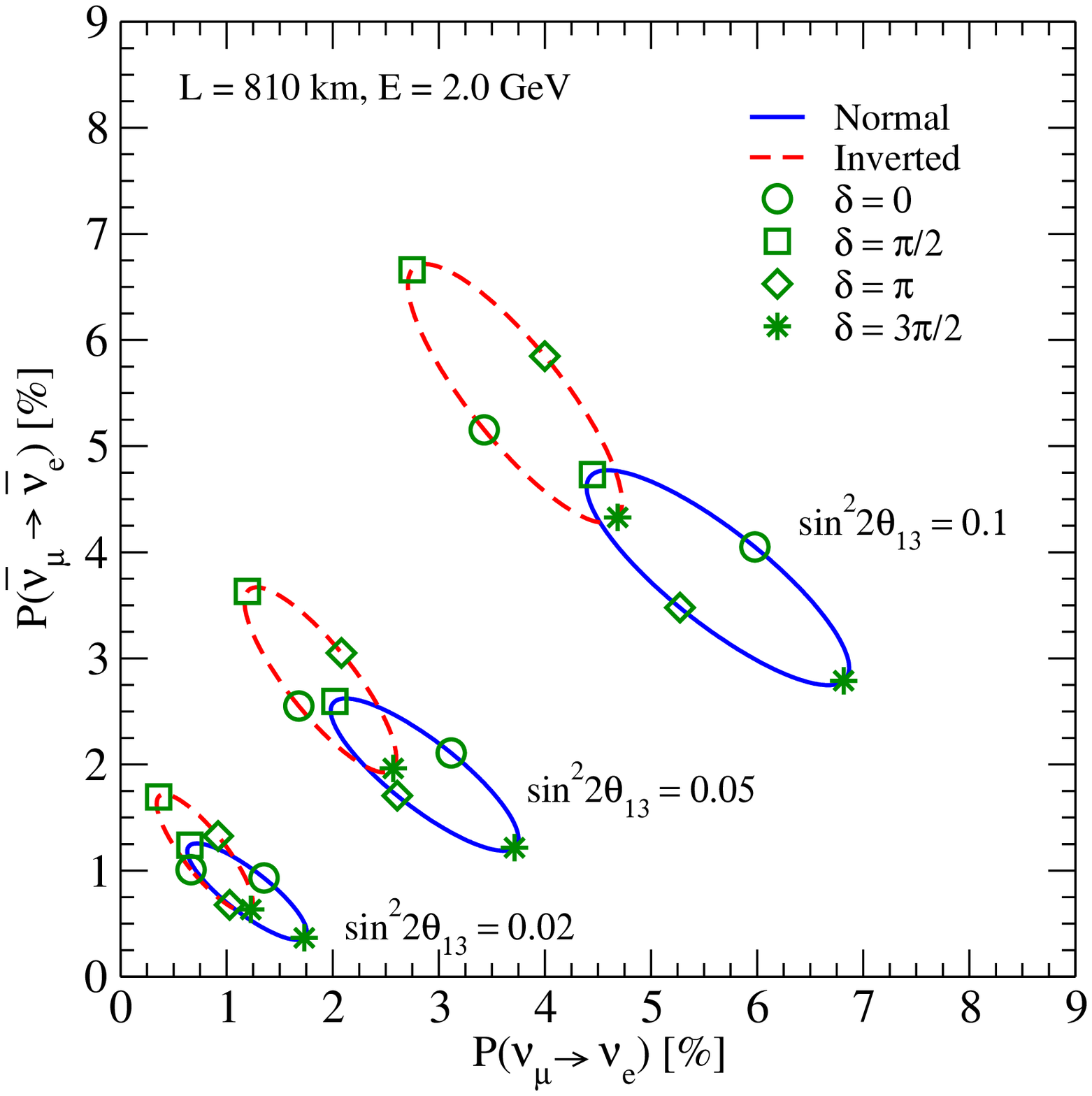}
\includegraphics[height=8.5cm]{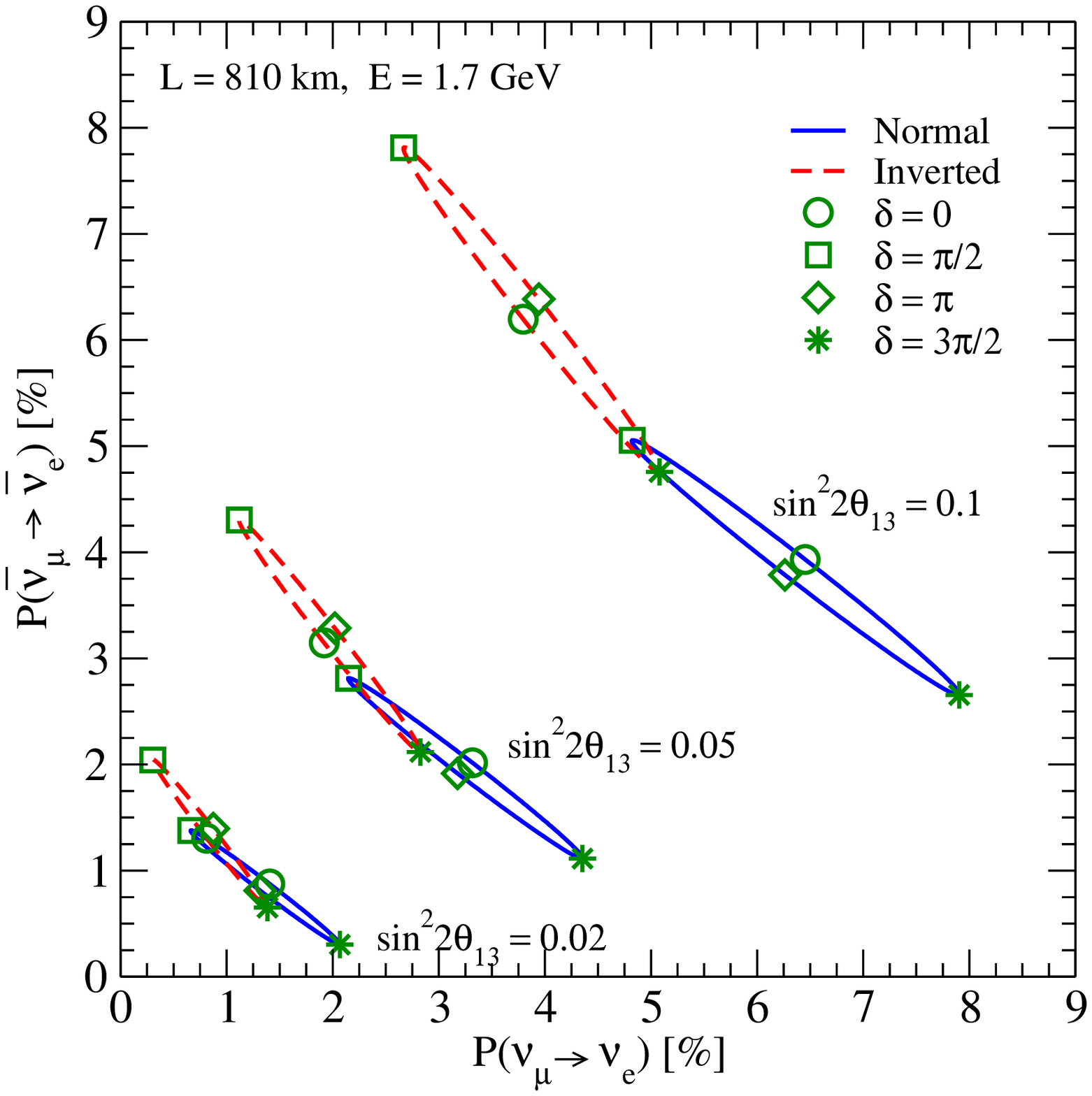}
\caption{The bi-probability $P(\nu_\mu \to  \nu_e)$ diagram 
for NO$\nu$A including matter effects.}
\label{fig:biprob2}
\end{figure}

\begin{figure}[p]
\includegraphics[height=8.5cm]{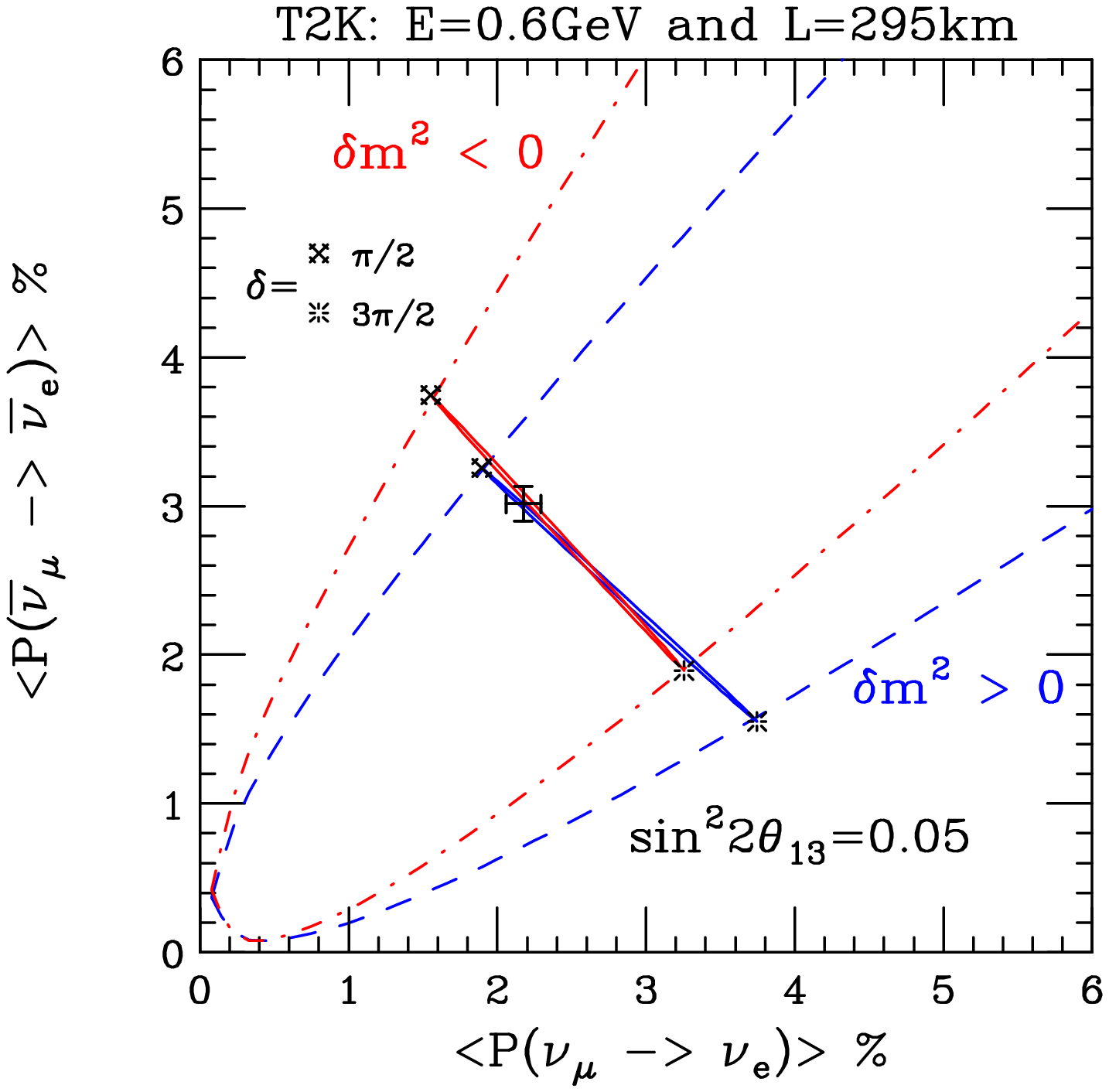}
\includegraphics[height=8.5cm]{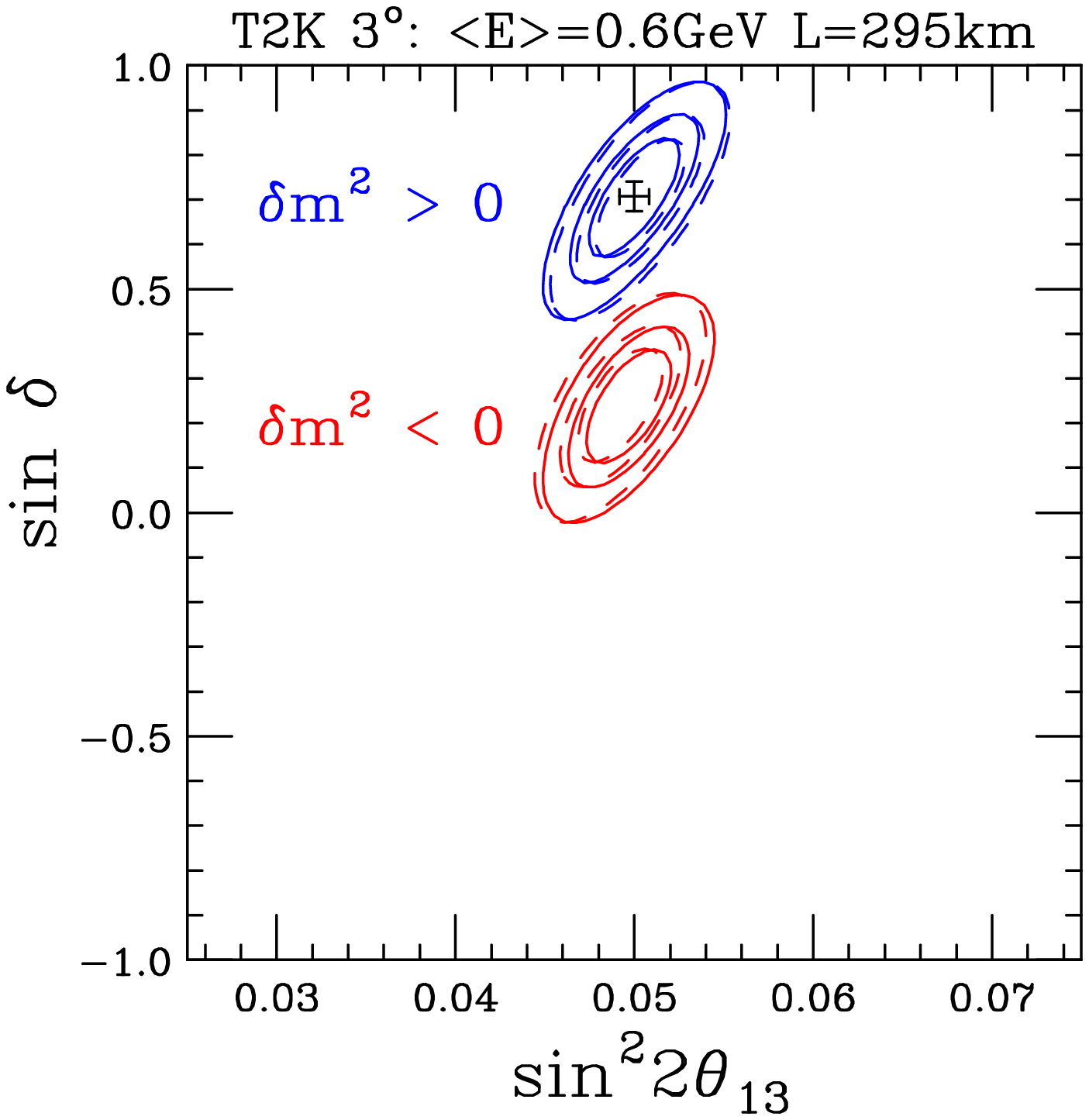}
\vskip .4cm
\includegraphics[height=8.5cm]{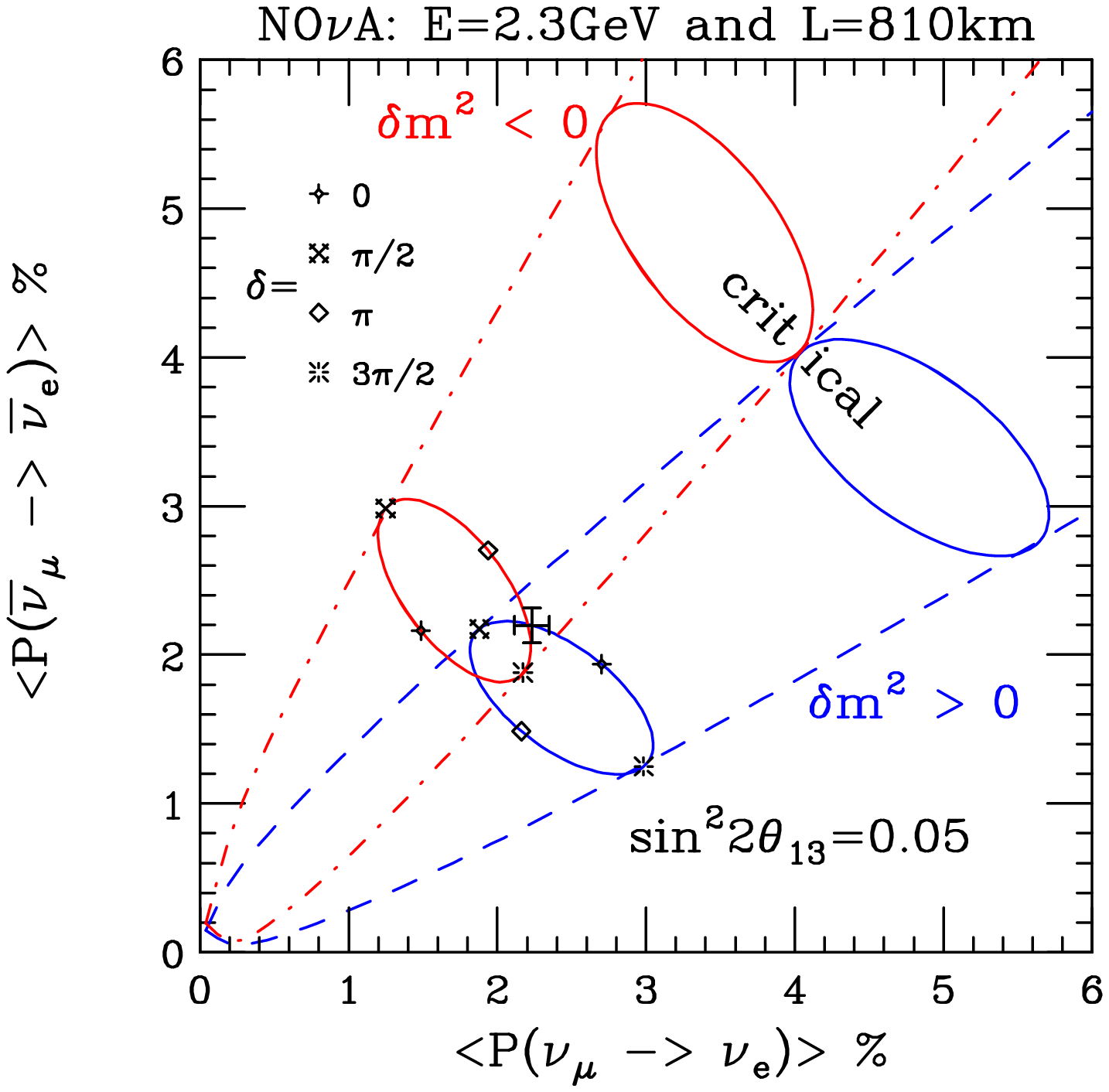}
\includegraphics[height=8.5cm]{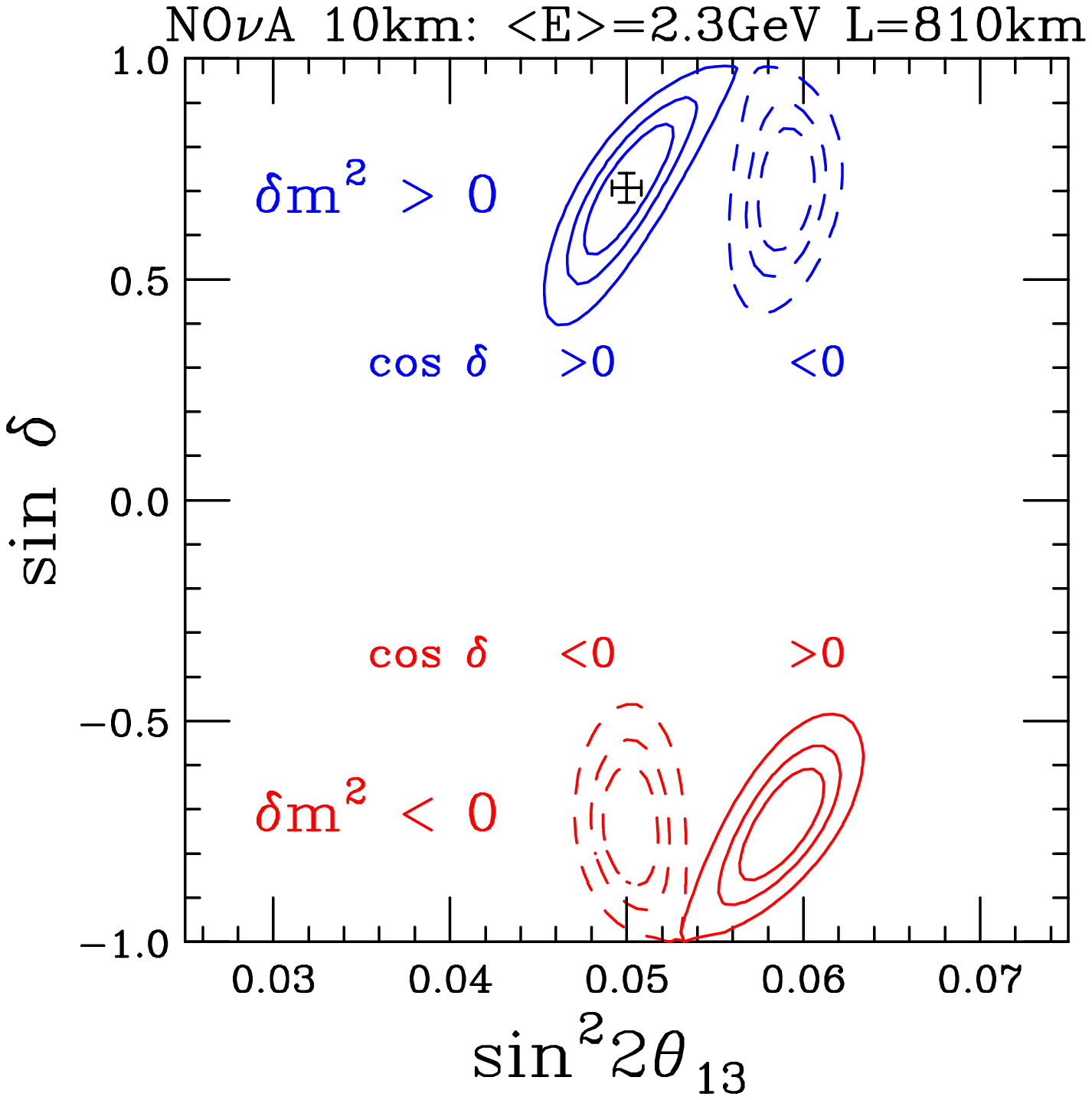}
\caption{(a) The bi-probability plot for T2K (top) NO$\nu$A (bottom),
  showing the allowed regions for the normal (dashed) and inverted
  (dot-dashed) hierarchies.  The cross is the assumed measured value
  of $(P,\bar{P})$.  The value of $\theta_{13}$ for which the two
  hierarchies separate is labeled as the critical value,
  $\theta_{13}^{crit}$.  (b) The four solutions for the assumed point
  in the $\sin^2 2 \theta_{13}$ vs. $\sin \delta$ plane.  Since T2K is
  near vacuum oscillation maximum all four solutions have
  approximately the same value of $\sin^2 2\theta_{13}$. But the
  solutions for the normal and inverted hierarchies occur at different
  values of $\sin \delta$. Since NO$\nu$A is above vacuum oscillation
  maximum, there are two distinct allowed values of $\sin^2
  2\theta_{13}$. Again the solutions for the same hierarchy have the
  same value of $\sin \delta$. But the solutions for the normal and
  inverted hierarchies occur at different values of $\sin \delta$.}
\label{fig:T2K-NOVA-biprob}
\end{figure}

However, this quantity is not directly observable as the actual
positions of the $\Delta P_{\nu \bar{\nu}}$ contours are modified by
matter effects, as shown in Fig.~\ref{fig:DeptaP-cont-t2k} for T2K and
Fig.~\ref{fig:DeptaP-cont-nova} for NO$\nu$A.  As expected, the
modifications due to the matter effect, which can be seen by the
asymmetry of contours with respect the line $\delta = \pi$, are
smaller for T2K than for NO$\nu$A.

In Fig.~\ref{fig:CPCpositions} we have show the location of the CP
conserving solutions of one hierarchy in the $(\sin^2 2\theta_{13},
\sin \delta)$ plane of the other hierarchy. These locations depend on
the matter effect for the experiment and thus differ for T2K and
NO$\nu$A. In order to claim the observation of CP violation, one must
be sufficiently away (with a required CL) not only from the line $\sin
\delta = 0$ but also from the dot-dashed (T2K) and dashed (NO$\nu$A)
curves shown in these plots unless the mass hierarchy is known.

For baseline $L < 1500$ km or so, one can expand $\Delta P_{\nu
  \bar{\nu}}(a) $ in matter as a Taylor series about the vacuum
$\Delta P_{\nu \bar{\nu}}(0) $ as follows,
\begin{equation} 
\Delta P_{\nu \bar{\nu}}(a)
= \Delta P_{\nu \bar{\nu}}(0)
\pm 4(aL) G(\Delta_{32}) 
[P_{\text{atm}}(0) +  \sqrt{ P_{\text{atm}}(0)P_{\text{sol}} }
\cos \Delta_{32} \cos\delta ],
\end{equation}
where $G(x) \equiv 1/x-\cot x$ and the + (-) sign is for the normal
(inverted) hierarchies.  This simple expression gives the relationship
of Figs.~\ref{fig:DeptaP-cont-t2k} and \ref{fig:DeptaP-cont-nova} to
Fig.~\ref{fig:DeptaP-cont-vac}.

\subsection{Bi-probability plots}
\label{sec:bi-probability-plots}

Of course, the neutrino and anti-neutrino probabilities are
correlated.  This correlation depends on the CP violating phase,
$\delta$, and the mass hierarchy.  Therefore it is useful to use
bi-probability diagrams.  In the bi-probability space spanned by
$P(\nu_\mu \to \nu_e)$ and $P(\bar\nu_\mu \to \bar\nu_e)$, for a given
$\theta_{13}$ (fixing all the mixing parameters except for $\delta$), 
energy and baseline, 
the variation of $\delta$ from 0 to $2\pi$ gives a
closed trajectory which is an ellipse\footnote{ We note that the
  trajectory is exactly elliptic as it was shown~\cite{Kimura:2002hb}
  that even in matter, oscillation probabilities take the form of $P =
  A\cos \delta + B \sin \delta + C$ where $A$, $B$ and $C$ are some
  constants which depend on mixing parameters as well as experimental
  parameters such as energy and baseline.}.  In
Figs.~\ref{fig:biprob1} and ~\ref{fig:biprob2}, we give the
bi-probability diagrams for both T2K and NO$\nu$A, respectively.
For T2K the separation between the hierarchies is small but
non-negligible whereas for NO$\nu$A the separation is much larger with
complete separation occurring for $\sin^2 2\theta_{13}=0.11$.  The
value of $\theta_{13}$ for which the two hierarchies separate is
called the critical value, $\theta_{13}^{crit}$.  For T2K and NO$\nu$A
the bi-probability diagrams are shown for two energies, one such that
$\Delta_{31} \approx \pi/2$ which is known as vacuum oscillation
maximum (VOM) and at an energy above VOM.  At VOM the ellipses in the
bi-probability plot are squashed to lines as the probabilities become
independent of $\cos \delta$ since $\cos \Delta_{31} = 0$.

\subsection{Parameter Degeneracy}
\label{sec:parameter-degeneracy}

Unfortunately the measurement of $P(\nu_\mu \to \nu_e)$ and
$P(\bar{\nu}_\mu \to \bar{\nu}_e)$ does not determine the values of
$\theta_{13}$ and $\delta$ uniquely due to the possibility of drawing
more than one ellipse through any $(P,\bar{P})$ point.  In general
four such ellipses can be drawn assuming $\sin ^2 \theta_{23}$ is
known uniquely. Eight if only $\sin^2 2 \theta_{23} (\neq 1)$ is
known.  Fig.~\ref{fig:T2K-NOVA-biprob} demonstrate this point.

The cross in the left hand panel of these two figure, are
the transition probabilities for the normal hierarchy with $\sin^2
2\theta_{13}=0.05$ and $\delta=\pi/4$ for both T2K and NO$\nu$A.  The
right hand panels show the four allowed solutions in the $\sin^2
2\theta_{13}$ v $\sin \delta$ plane that are consistent with this
point in the $(P,\bar{P})$ plane.
%
Clearly, the four solutions are related to one another: \\ \vskip .3cm
$\bullet$ 
  Within the same hierarchy, the intrinsic degeneracy
  \cite{Burguet-Castell:2001ez}.\\  The two solutions have approximately
  the same $\sin \delta$, but different signs for $\cos \delta$ i.e.
  $\delta$ and $\pi - \delta$.  But the values of $\theta_{13}$ differ
  by
$$
  \Delta \theta_{13} = \cos \delta \sin 2\theta_{12} \Delta_{21} \cot \Delta_{31} 
  $$
  At vacuum oscillation maximum, $\Delta_{31} =\pi/2$, these two
  different values of $\theta_{13}$ coincide, see
  Fig.~\ref{fig:T2K-NOVA-biprob}.\\ \vskip .3cm $\bullet$ Between the
  two hierarchies\cite{Minakata:2001qm}.\\ The allowed values for $\sin^2 2 \theta_{13}$ are
  the same for both hierarchies but the $\sin \delta$'s differ between
  the normal (NH) and inverted (IH) hierarchies as follows, $\sin
  \delta \vert_{NH} - \sin \delta \vert_{IH} =
  2\theta_{13}/\theta^{\text{crit}}_{13}$, see \cite{Mena:2004sa}.
  Where $\theta^{\text{crit}}_{13}$ is the largest value of
  $\theta_{13}$ for which the two regions in the bi-probability
  diagram overlap\footnote{The value of $\theta^{\text{crit}}_{13}$
    for any experiment is given by
\begin{eqnarray} 
\theta^{\text{crit}}_{13} & \approx & \frac{\pi^2}{8} ~\frac{\sin 2\theta_{12}}{\tan \theta_{23}} ~\frac{\Delta m^2_{21}}{\Delta m^2_{31}} 
\left( \frac{4\Delta_{31}^2/\pi^2}{1-\Delta_{31} \cot \Delta_{31}}\right) /(aL). 
\label{theta_crit}
\end{eqnarray}
}, see Fig.~\ref{fig:T2K-NOVA-biprob}.  For T2K and NO$\nu$A this
corresponds to
\begin{eqnarray}
\sin \delta \vert_{NH} - \sin \delta \vert_{IH} = 
\left\{  \begin{array}{lc}
0.47 \sqrt{\frac{\sin^2 2 \theta_{13}}{0.05} } & {\rm T2K} \\[0.3in]
1.42 \sqrt{\frac{\sin^2 2 \theta_{13}}{0.05} } & {\rm NO} \nu {\rm A} 
\end{array} \right.
\label{eqn:diffsindelta}
\end{eqnarray}

The coefficient is 3 times bigger for NO$\nu$A than T2K primarily
because the baseline for NO$\nu$A is approximately 3 times that of T2K
so that the matter effects are 3 times larger.  Notice that the matter
effects are smaller but not negligible for T2K, see
Fig.~\ref{fig:CPCpositions}.  Clearly the combination of T2K and
NO$\nu$A can be used to determine the hierarchy at least for large
values of $\theta_{13}$.

If $\sin^2 2 \theta_{23} \neq 1$ then there is a further
degeneracy~\cite{Fogli:1996pv} associated with whether
\begin{eqnarray}
\sin^2 \theta_{23} = \frac{ 1 \mp \sqrt{1-\sin^2 2\theta_{23}}}{2}.
\end{eqnarray}

The simultaneous presence of these three independent 2-fold
degeneracies leads up to 8-fold degeneracies~\cite{Barger:2001yr}.
Resolving all three of these degeneracies will be challenging but,
except for the hierarchy degeneracy, they do not affect much whether
or not one can claim the observation of leptonic CP violation.
  
For the mass hierarchy degeneracy this ambiguity can affect whether or
not one can claim CP violation especially for larger values of
$\theta_{13}$.  At larger values of $\theta_{13}$, NO$\nu$A alone may
be able to determine the hierarchy if $\text{sign}(\Delta m^2_{31})
\sin \delta$ is close to one. Otherwise a combination of T2K and
NO$\nu$A is our best option for determining the hierarchy with only
these two experiments.  Even if the hierarchy is not determined there
is still a substantial region of $(\theta_{13}, \delta)$ space where
not knowing the hierarchy does not preclude being able to claim the
observation of leptonic CP violation.  However there is a region in
both experiments where one hierarchy is CP violating and the other CP
conserving, fortunately these troublesome regions occur in different
locations for the two experiments.  This makes the combination of T2K
and NO$\nu$A a very powerful tool to observe leptonic for large
$\theta_{13}$.

\subsection{Beyond the First Oscillation Maximum:}
\label{sec:beyond-first}

To untangle all of the degeneracy associated with the neutrino mass
hierarchy, the quadrant of $\theta_{23}$ and the sign of $\cos
\delta_{CP}$ discussed in the previous subsection, it is probable that
experiments beyond the first atmospheric oscillation maximum will be
necessary.  First, let us discuss what happens in
vacuum~\cite{Marciano:2001tz} and then add matter effects.  In vacuum,
the atmospheric amplitude given by Eq.~(\ref{eq:patm-psol}) has the
same magnitude at each of the successive maxima whereas the solar
amplitude at the $n$-th peak is $(2n-1)$ times larger than at the
first peak as long as $\Delta_{21}$ is less than 1.  Thus, the
magnitude of the ratio of the solar to atmospheric amplitude grows as
$(2n-1)$.  Hence, the relative magnitude of the probabilities
associated with the atmospheric and solar $\delta m^2$ and their
interference in the total $\nu_\mu \rightarrow \nu_e$ transition
probability, the three terms in Eq.~(\ref{eq:pme}), changes at the
successive peaks.  At the n-th peak, the interference term, and hence
CP violation, grows as $(2n-1)$ compared to the first peak whereas the
term associated with purely the solar $\delta m^2$ grows as
$(2n-1)^2$.  This change in the relative magnitude of the three terms
can be exploited to untangle the degeneracies.

In vacuum, the transition probability $\nu_\mu \rightarrow \nu_e$ is
the same if one goes to the successive peaks by lowering the energy or
increasing the baseline or some combination of the two (as long as
$E/L$ is kept to be the same).  However, as we will see shortly, in
matter, the $\nu_\mu \to \nu_e$ transition probability depends on
exactly how one goes to the successive peaks.  E.g. the transition
probability at the second peak is different for lowering the energy by
a factor of three at the same baseline than increasing the baseline by
a factor of three at the same energy.  To understand this we need to
understand how matter effects change the atmospheric and solar
amplitudes which is most easily done for a constant matter density.
The solar amplitude changes little except at very large distances
whereas the atmospheric amplitude is substantially modified by matter
effects, see Eq.~(\ref{eq:pme_matter}).  The change in the atmospheric
amplitude depends on whether one goes to the successive peaks by
varying the baseline holding the energy fixed or varying the energy
and holding the baseline fixed.

For fixed energy and varying baseline, the modification of the
atmospheric amplitude can be understood as just a change in the mixing
angle and $\delta m^2$ which depend on the energy and matter density
but are independent of the baseline, see section \ref{sec:matter}.  So
the amplitude of the oscillation is the same at all successive peaks
and the position of $n$-th peak is just $(2n-1)$ times the distance to
the first peak.  See top left panel of Fig.~\ref{fig:pmatterE}.

For fixed baseline and varying energy, again one can think of the
modification as just changing the mixing angle and $\delta m^2$ but
these parameters are energy dependent and approach the vacuum values
as the energy gets smaller and smaller, see Fig.~\ref{fig:dmsqs}.
Again, see the discussion in section \ref{sec:matter}. 
So the matter effects are reduced by $(2n-1)$ in amplitude at the
$n$-th peak since the energy at the $n$-th peak is reduced by $(2n-1)$
compared to the first peak.
Therefore, matter effects become increasingly less important at the
successive peaks.  This can be see in the bottom left panel of
Fig.~\ref{fig:pmatterE}.


\begin{figure}[t]
\includegraphics[height=7.cm,width=0.45\textwidth]{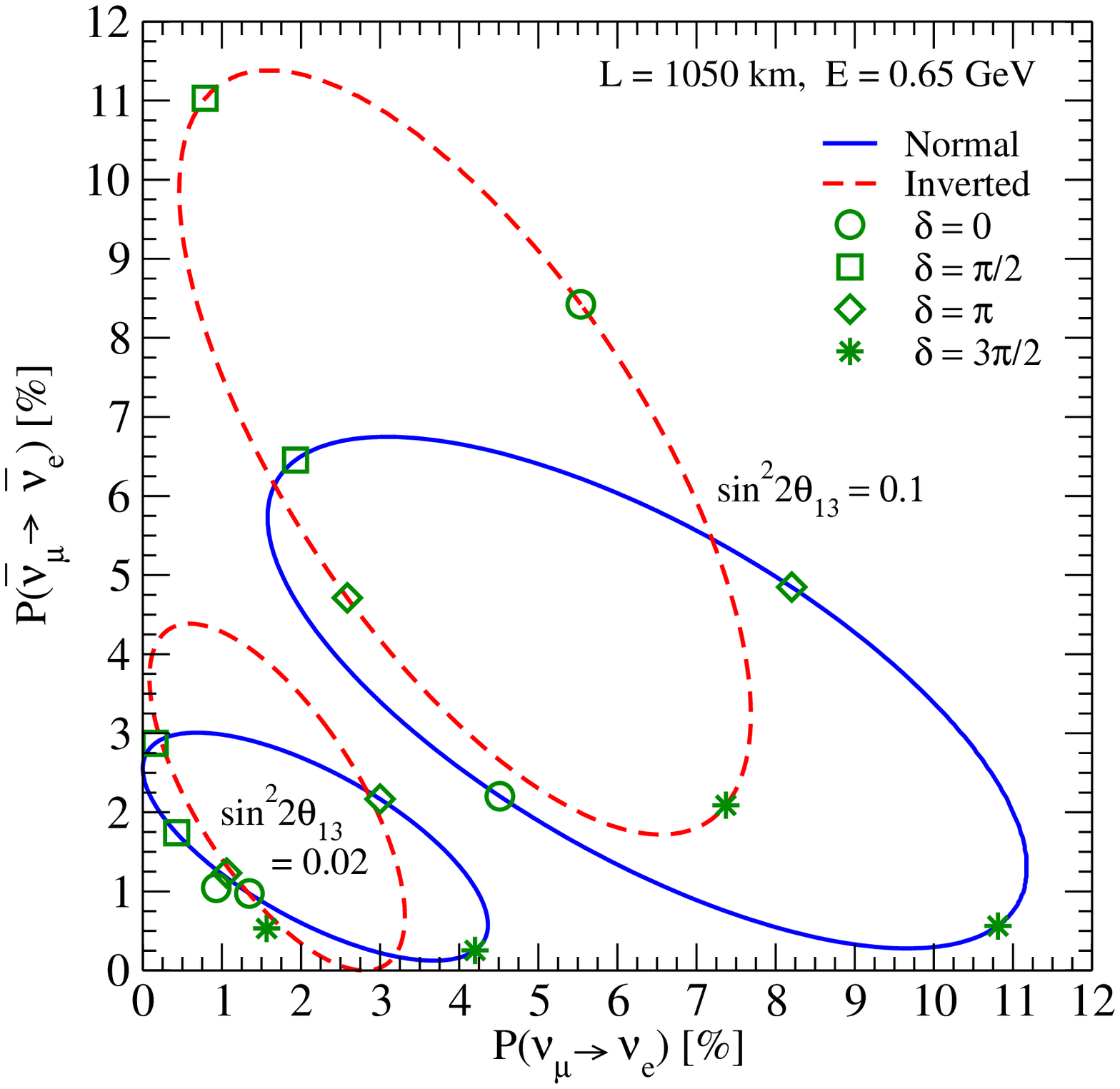}
\includegraphics[height=7.cm,width=0.45\textwidth]{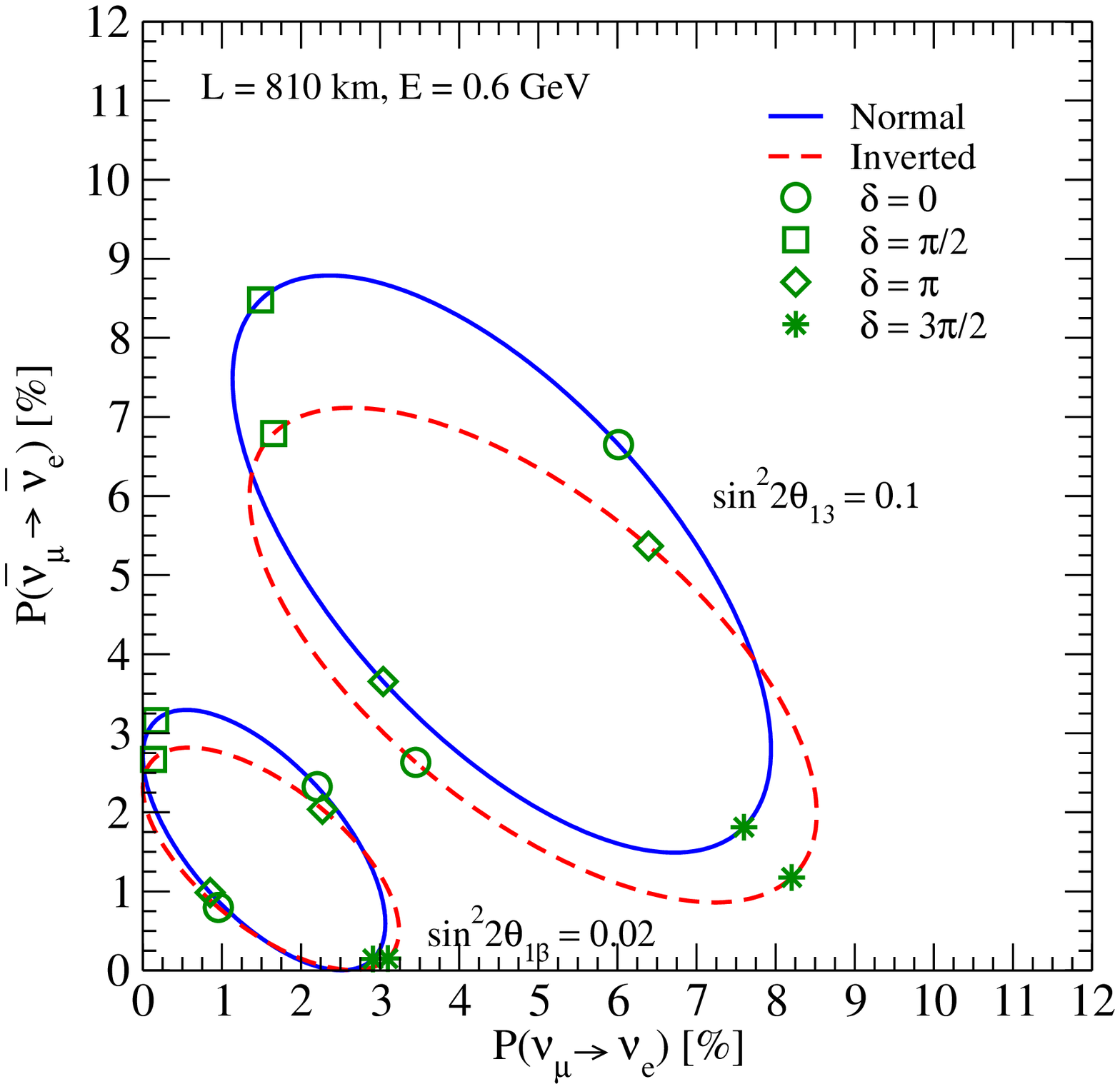}
\caption{Examples of the bi-probability $P(\nu_\mu \to  \nu_e)$ diagram 
corresponding to near the second oscillation maximum
including matter effects.}
\label{fig:biprob3}
\end{figure}

\begin{figure}[!h]
\includegraphics[height=7.cm,width=0.45\textwidth]{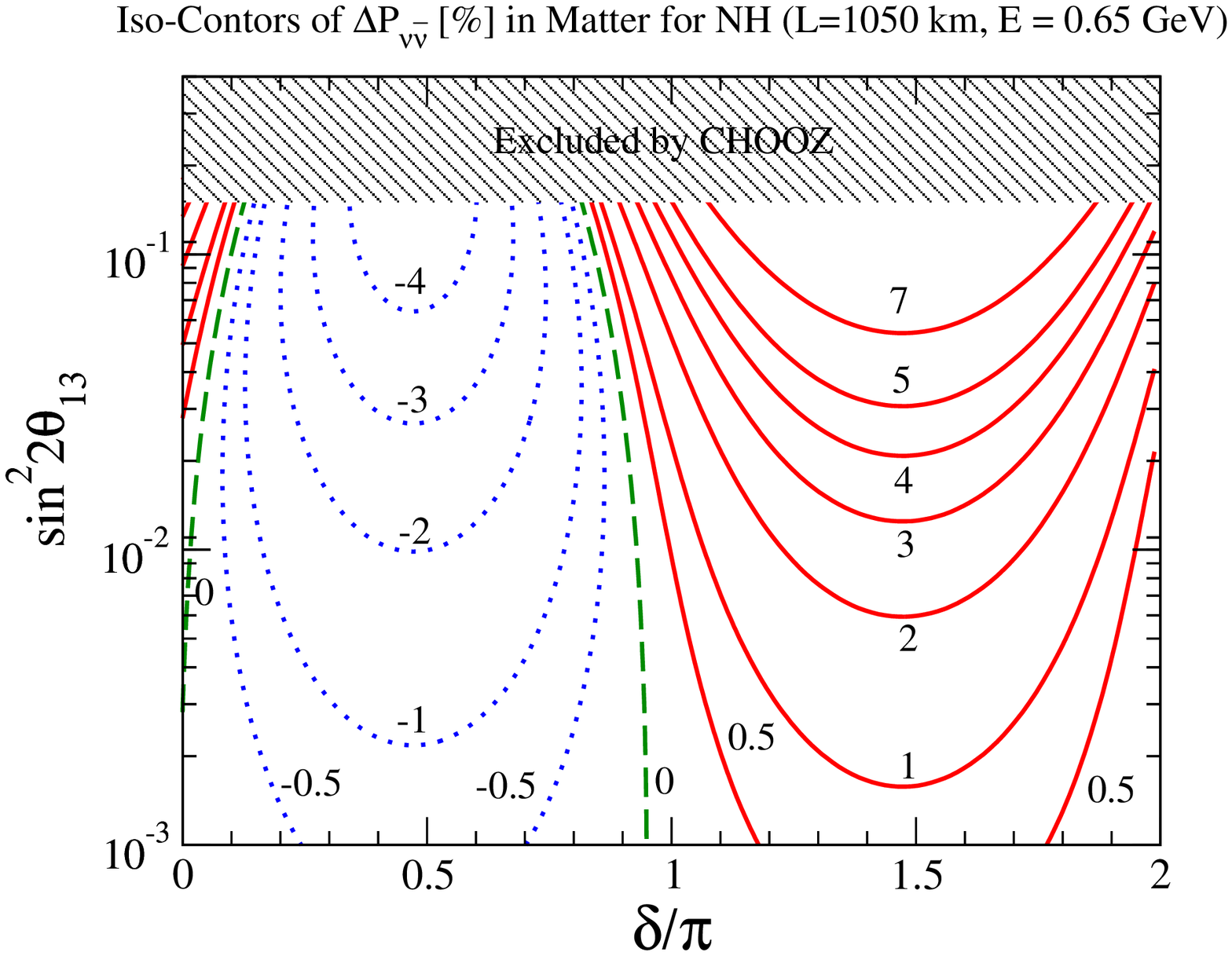}
\includegraphics[height=7.cm,width=0.45\textwidth]{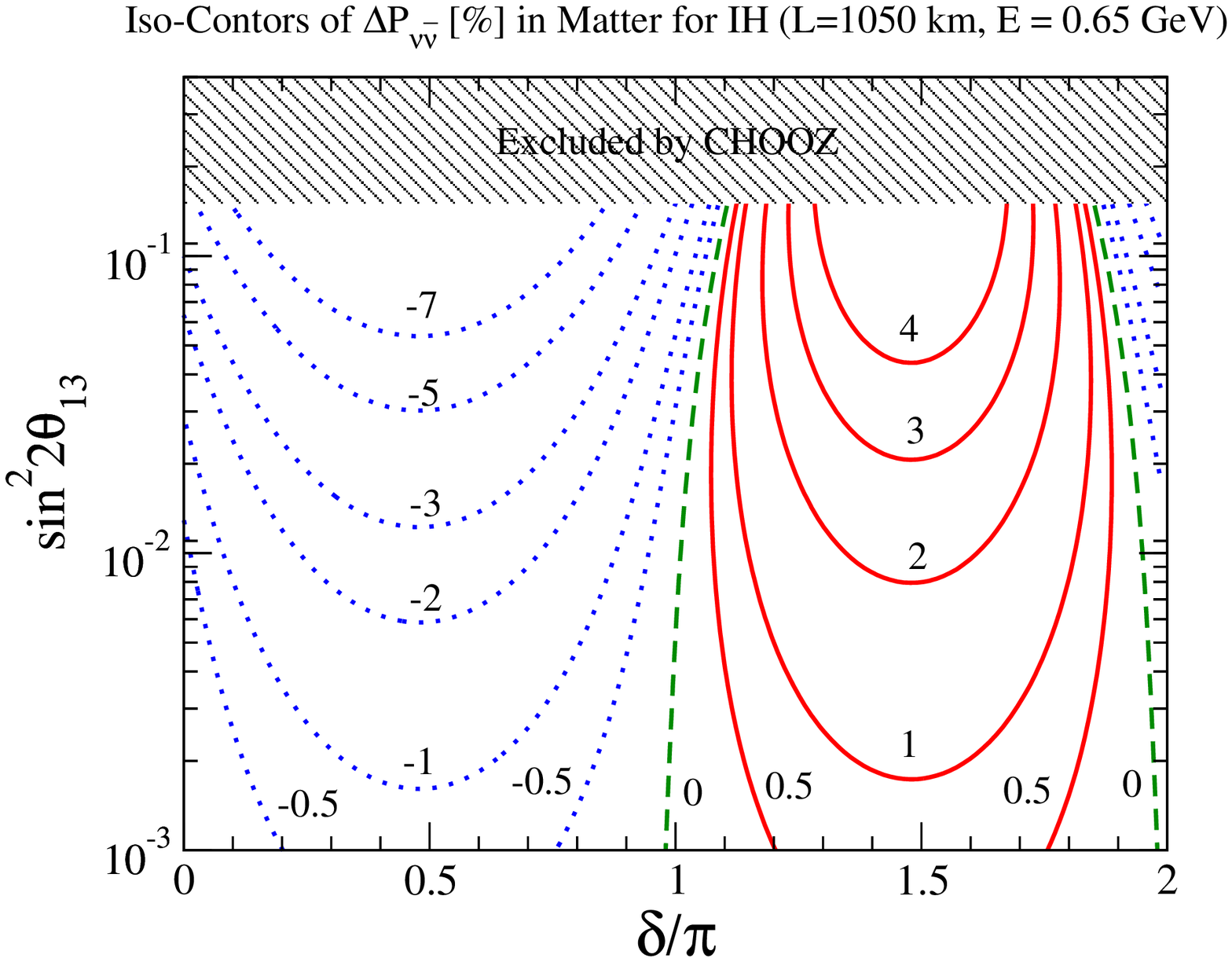}
\caption{Iso-contours of $\Delta P_{\nu \bar{\nu}} \equiv P(\nu_\mu
  \to \nu_e)-P(\bar{\nu}_\mu \to \bar{\nu}_e)$ in the $\delta$ and
  $\sin^2 2\theta_{13}$ for T2KK (L=1050 km) for the normal (left
  panel) and the inverted (right panel) mass hierarchy, with the
  matter effect included.  }
\label{fig:DeptaP-cont-t2kk}
\end{figure}

\begin{figure}[!h]
\includegraphics[height=7.cm,width=0.45\textwidth]{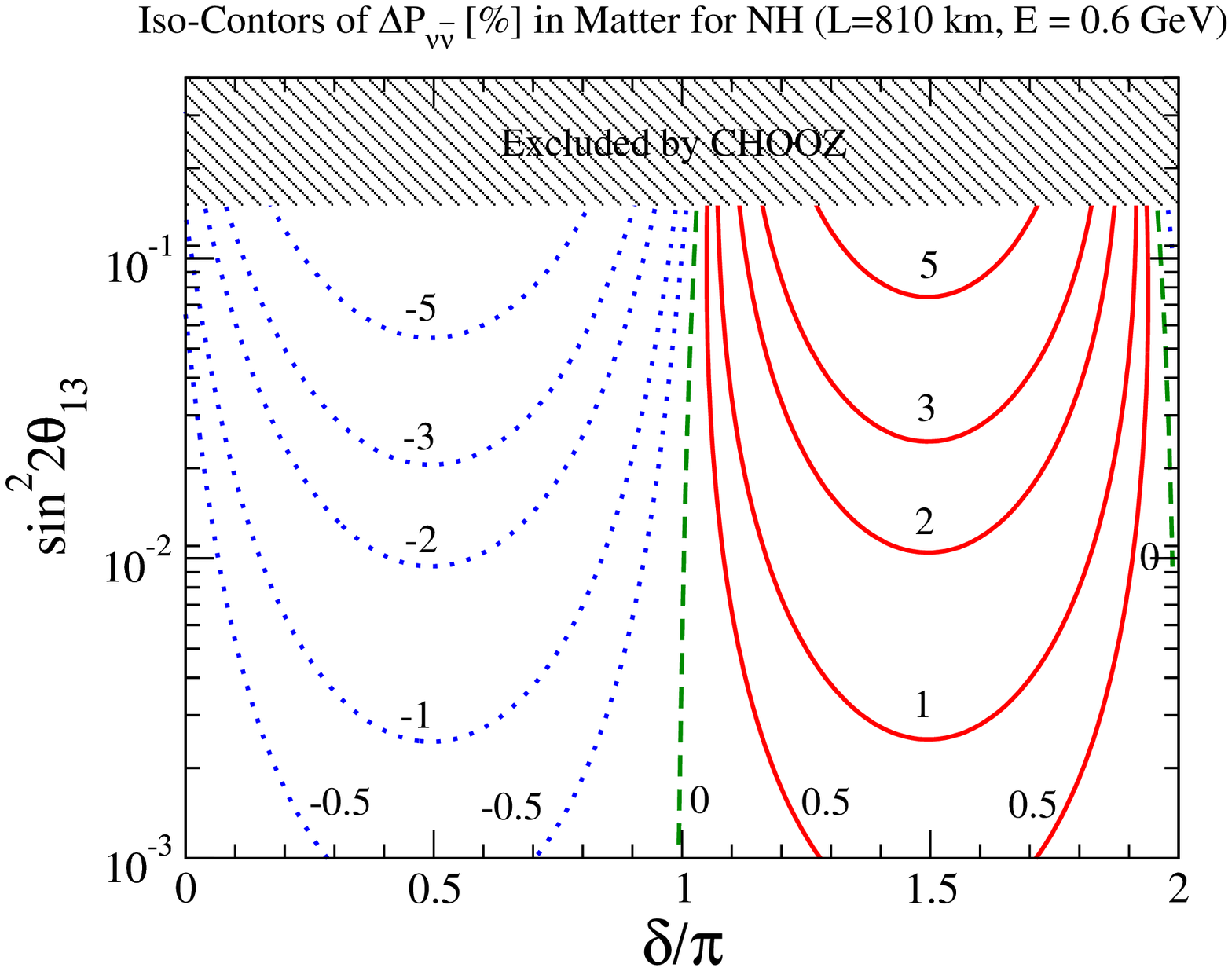}
\includegraphics[height=7.cm,width=0.45\textwidth]{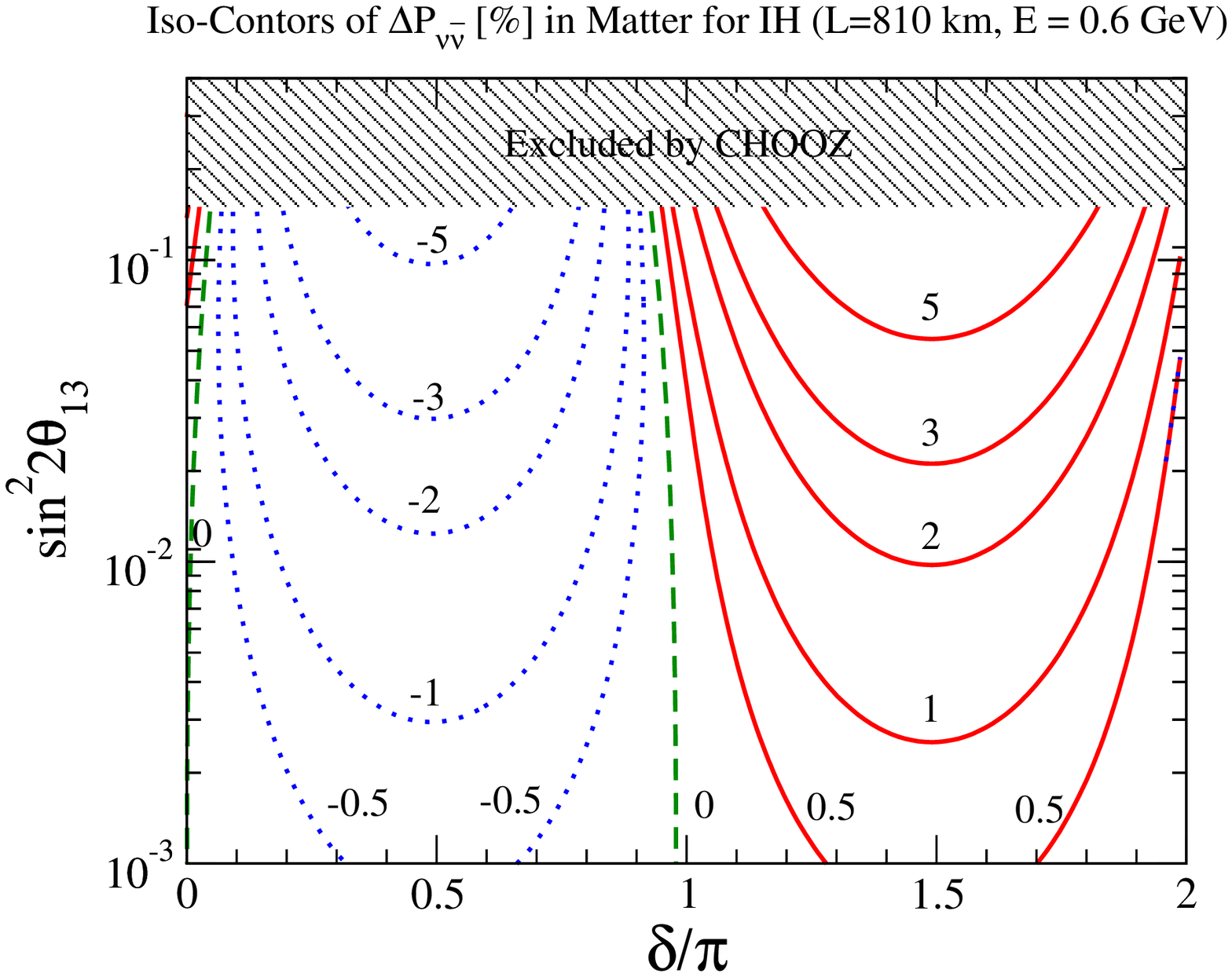}
\caption{Iso-contours of $\Delta P_{\nu \bar{\nu}} \equiv P(\nu_\mu
  \to \nu_e)-P(\bar{\nu}_\mu \to \bar{\nu}_e)$ in the $\delta$ and
  $\sin^2 2\theta_{13}$ for the NO$\nu$A baseline (L=810 km) but for 
the neutrino energy $E=0.6$ GeV for 
the normal (left panel) and the
  inverted (right panel) mass hierarchy, with the matter effect
  included.  }
\label{fig:DeptaP-cont-nova2}
\end{figure}

\subsection{Possible Experiments}

A possible extension of the T2K experiment, called
T2KK~\cite{Ishitsuka:2005qi,Kajita:2006bt} (see also
\cite{Hagiwara:2005pe,Hagiwara:2006vn}), consists of building two new
large detectors, one at the first oscillation peak at Kamioka, Japan
and the other in Korea, near the second oscillation peak.  Both
detectors would be at the same off-axis angle so they would see the
same neutrino energy spectrum but the Korean detector would be three
times further from the source than the Kamioka detector.  This is an
example of getting to the second peak by varying the baseline holding
the neutrino energy fixed.  In the left panel of
Fig.~\ref{fig:biprob3} we show the examples of bi-probability plots
corresponding to such an experimental set up, which should be compared
with the right panel of Fig. ~\ref{fig:biprob1}.  From these two
plots, we confirm that the oscillation probabilities as well as its
dependence on $\delta$ are significantly enhanced at Korea due to the
increase of the solar term.  In Fig.~\ref{fig:DeptaP-cont-t2kk} we
show the iso-contour of $\Delta P_{\nu \bar{\nu}}= P(\nu_\mu \to
\nu_e) - P(\bar{\nu}_\mu \to \bar{\nu}_e)$ in the $\delta-\sin^2
2\theta_{13}$ plane for the normal (left panel) and the inverted
(right panel) hierarchy, which should be compared with
Fig.~\ref{fig:DeptaP-cont-t2k}.  We observe that the qualitative
behaviour of these plots are similar but magnitude of $\Delta P_{\nu
  \bar{\nu}}$ for larger baseline is significantly greater.  By
studying the neutrino spectrum at both detectors one could untangle
all of the degeneracy~\cite{Kajita:2006bt}.

Another possibility is to build a second large off-axis detector near
the NO$\nu$A detector such that the energy of the neutrino beam is one
third of that seen by the NO$\nu$A detector~\cite{Barger:2007yw}.
Both detectors would be approximately at the same distance so this is
an example of getting to the second peak by varying the neutrino
energy and holding the baseline fixed.  At the second peak, CP
violation is three times larger than at the first peak and matter
effects are three times smaller.  Therefore, counting experiments at
both peaks could resolve the entanglement of the mass hierarchy and
CP-violation.

In the right panel of Fig.~\ref{fig:biprob3} we show an
example of bi-probability plot corresponding to this case which should
be compared with Fig.  ~\ref{fig:biprob2}.  We confirm that the effect
of the CP phase (i.e., the size of the ellipses) become larger and the
matter effects get smaller.  The corresponding iso-contour plots of
$\Delta P_{\nu \bar{\nu}}$, shown in Fig.~\ref{fig:DeptaP-cont-nova2},
also support this feature.

A third possibility is to build a large detector on-axis at a distance
of 1000 - 1500 km from the source and to measure the oscillated
neutrino spectrum over a wide neutrino energy covering the first,
second and possibly the third atmospheric oscillation
peaks~\cite{Diwan:2003bp,Barger:2007yw}.
This again is an example of a fixed baseline and varying the neutrino
energy.  The hierarchy would be determined by studying the first peak
and CP-violation from studies of the second peak.

\section{CP violation and lepton number violation: \nbb}
\label{sec:cp-violation-lepton}

As we have seen, neutrino oscillations are insensitive to the absolute
scale of neutrino masses. While the origin of the latter remains a
theoretical mystery, one expects on general grounds that neutrinos are
Majorana fermions, and that this accounts for their relative lightness
with respect to the other fundamental fermions.

The search for lepton number violating processes such as \nbb opens
the way to probe the basic nature - Dirac or Majorana - of neutrinos,
and also to probe the CP violation induced by so--called Majorana
phases~\cite{schechter:1980gr}. As already mentioned, the latter does
not show up in conventional neutrino oscillation experiments
~\cite{bilenky:1980cx,Schechter:1981gk,doi:1981yb} but can affect
neutrinoless double beta decays and electromagnetic properties of
neutrinos~\cite{Schechter:1982bd,schechter:1981hw,Wolfenstein:1981rk}
~\cite{pal:1982rm,kayser:1982br}.

The most direct mechanism engendering \nbb is the so-called ``mass
mechanism'' involving the exchange of massive Majorana neutrinos.  The
associated amplitude is proportional to
$$m_{ \beta\beta} = \sum_i K_{ei} m_i  K_{ei}~,$$
where the $K_{ei}$ are the first row in the lepton mixing matrix.  An
important feature of this amplitude is that it involves the
lepton-number-violating propagator in Eq.~\ref{eq:LNV} and, as a
result, none of the $K_{ei}$ factors appear with complex conjugation,
hence the effect of Majorana phases and the possibility of destructive
interference among amplitudes arising from different neutrino types.

Such destructive interferences may also take place without CP
violation. This is what happens for example in the case of a pure
Dirac neutrino, the exact cancellation coming in this case from the
phase present in Eq.~(\ref{eq:DECOMP}). Related concepts are those of
partial cancellations due to symmetry. One example is the case of a
Quasi-Dirac neutrino, made up of one active and one sterile neutrino,
almost degenerate in mass due to the nearly exact conservation of
standard lepton number~\cite{valle:1983yw}. Mass splittings between
the two neutrinos may be radiatively calculable in some gauge
models~\cite{valle:1983dk} and lead to active-sterile neutrino
oscillations. Depending on the size of the mass splitting, these
active-to-sterile conversions may affect primordial Big Bang
Nucleosynthesis~\cite{enqvist:1992qj,dolgov:2002wy}.  A different
concept is that of a pseudo-Dirac neutrino made up of two active
neutrinos, nearly degenerate due to the approximate conservation of
some non-standard combination of lepton
numbers~\cite{wolfenstein:1981kw}.

\begin{figure*}[!h]
  \centering
\includegraphics[width=6cm,height=4cm]{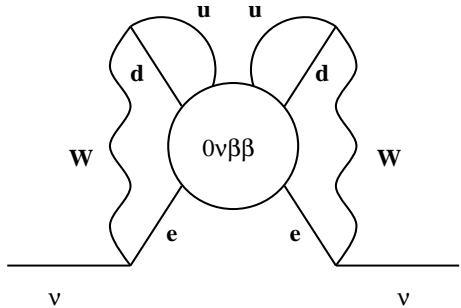}
  \caption{Neutrinoless double beta decay and Majorana mass are theoretically 
  equivalent~\cite{Schechter:1982bd}.}
 \label{fig:bbox}
\end{figure*}

The significance of neutrinoless double beta decay stems from the fact
that, in a gauge theory, {\sl irrespective of which is the mechanism}
that induces \nbb, it necessarily implies a Majorana neutrino
mass~\cite{Schechter:1982bd}, as illustrated in
Fig.~\ref{fig:bbox}~\footnote{Such black-box theorem does {\sl not}
  exist for the case of lepton flavour violation, which may proceed in
  the absence of neutrino mass~\cite{bernabeu:1987gr}.}.  This is
specially relevant given the fact that gauge theories bring in other
possible ways of inducing \nbb and it is conceivable that one of such
mechanisms might, perhaps, give the leading contribution.  Hence the
importance of searching for neutrinoless double beta decay.
Quantitative implications of the ``black-box'' argument are strongly
model-dependent, but the theorem itself holds in any ``natural'' gauge
theory.

Now that oscillations are experimentally confirmed we know that there
is a contribution to \nbb involving the exchange of light Majorana
neutrinos, the "mass-mechanism". The corresponding amplitude is
sensitive both to the absolute scale of neutrino mass as well as the
two Majorana CP phases that characterize the minimal 3-neutrino mixing
matrix~\cite{schechter:1980gr}, none of which can be probed in
oscillations.
\begin{figure}[!h]
 \centering
\includegraphics[clip,width=.45\linewidth,height=10cm]{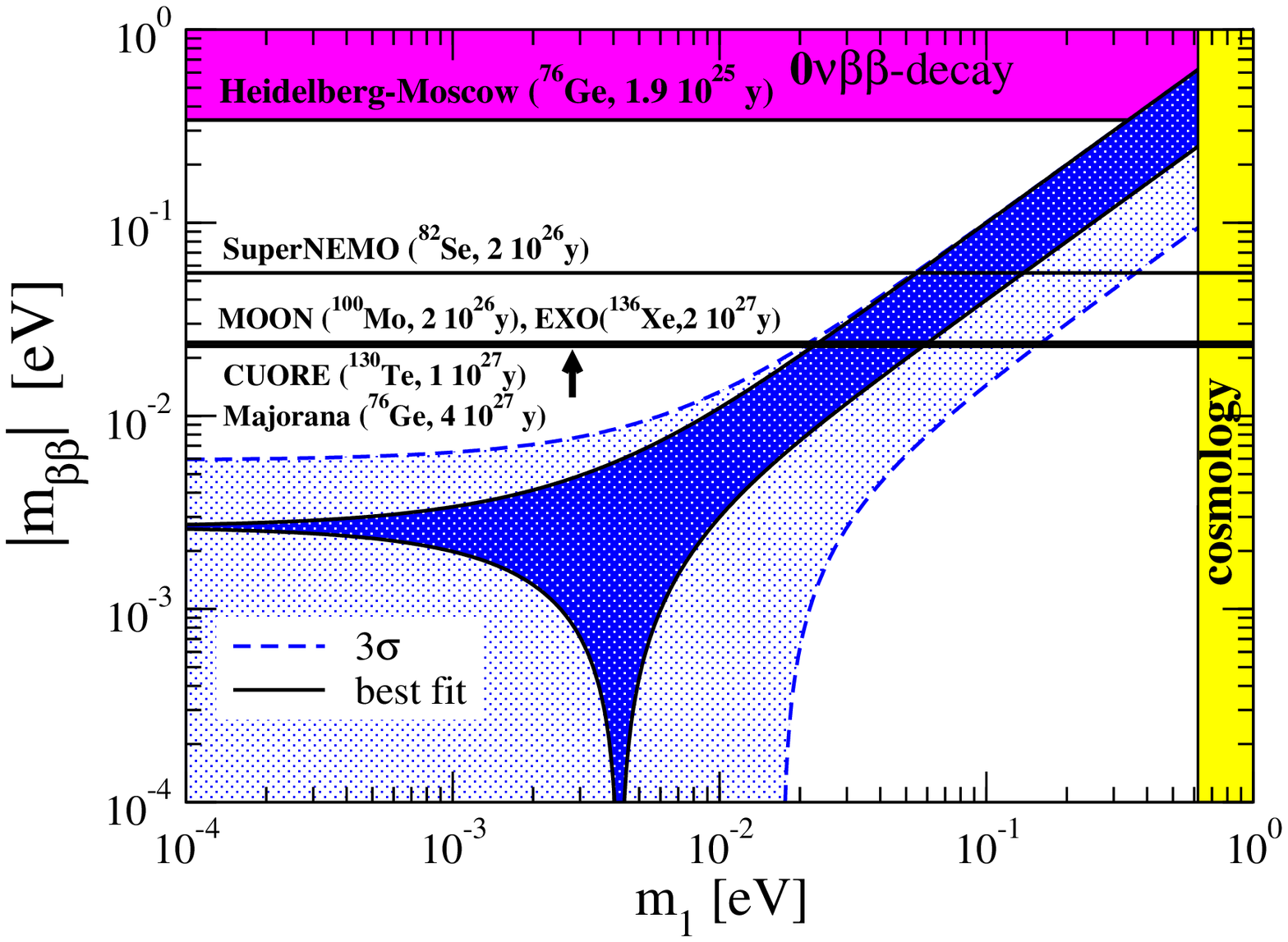}
\includegraphics[clip,width=.45\linewidth,height=10cm]{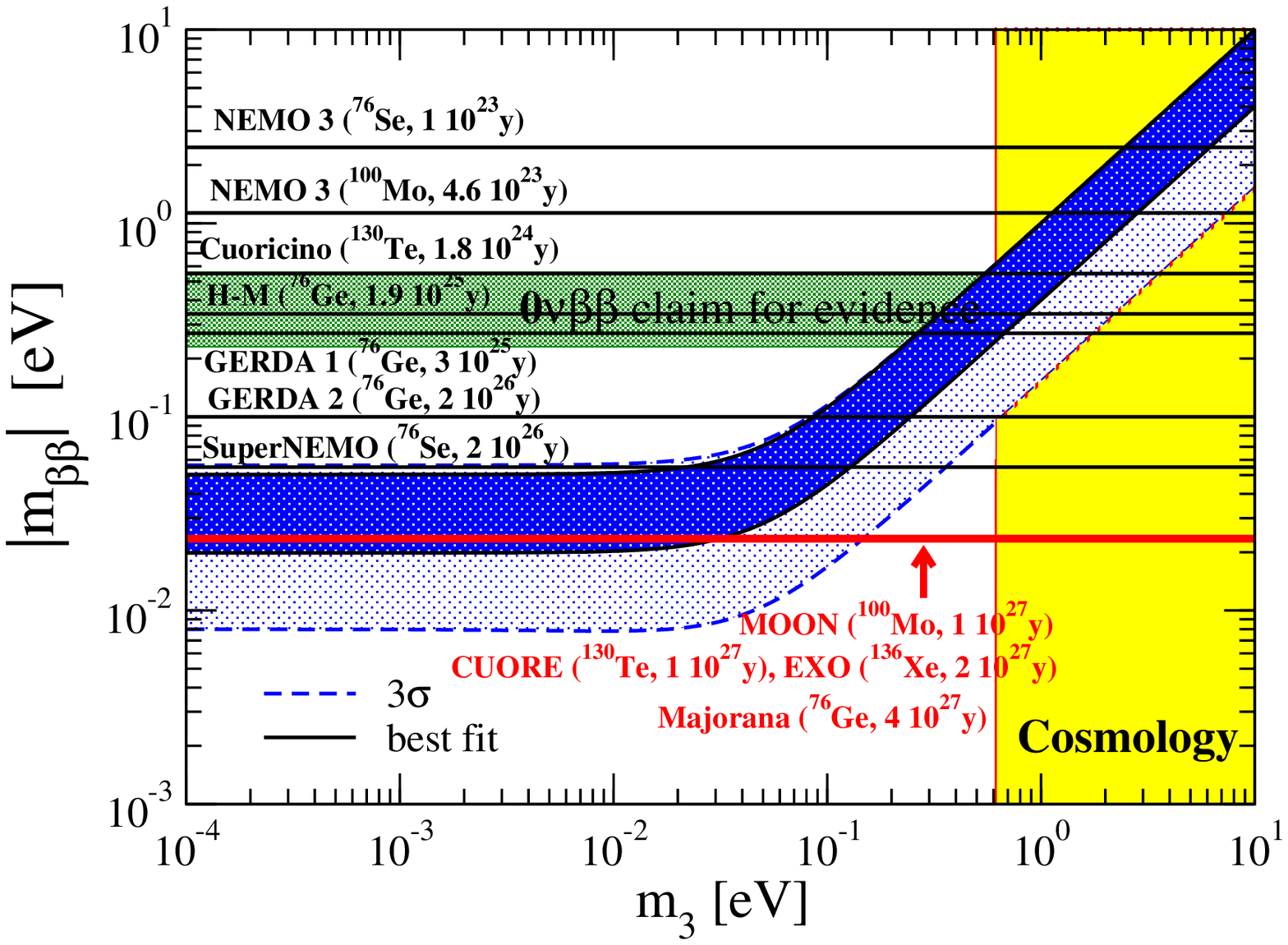}
 \caption{Neutrinoless double 
beta decay amplitude versus current oscillation data~\cite{Rodin:2006yk}.}
\label{fig:nbbfut}
\end{figure}

Fig.~\ref{fig:nbbfut} shows the estimated average mass parameter
characterizing the neutrino exchange contribution to \nbb versus the
lightest neutrino mass.  The calculation takes into account the
current neutrino oscillation parameters in \cite{Maltoni:2004ei} and
latest nuclear matrix elements of~\cite{Rodin:2006yk} and compares
with experimental sensitivities.

The most stringent lower bounds on the half-life of \nbb decay were
obtained in the Heidelberg-Moscow $^{76} \rm{Ge}$ \cite{Baudis:1999xd}
and CUORICINO $^{130} \rm{Te}$ \cite{Arnaboldi:2004qy} experiments:
\begin{equation}
  \label{eq:bb0n}
T^{0\nu}_{1/2}(^{76} \rm{Ge})\geq 1.9 \cdot 10^{25}\, years,~~~~
T^{0\nu}_{1/2}(^{130} \rm{Te})\geq 1.8 \cdot 10^{24}\, years.
\end{equation}
\label{eq:nbb}
Using recently calculated nuclear matrix elements with significantly
reduced theoretical uncertainties \cite{Rodin:2006yk} from these data
the following upper bounds for the effective Majorana mass $m_{
  \beta\beta} \equiv \meff$ can be inferred
\begin{eqnarray}
  \label{eq:nbb2}
|m_{\beta\beta}| \leq
0.34 \,~\rm{eV}\,~~~(\rm{Heidelberg-Moscow})\nonumber\\
 |m_{\beta\beta}| \leq
0.55 \,~\rm{eV}\,~~~(\rm{CUORICINO}).
\end{eqnarray} 
The Heidelberg group, which includes a few authors of the
Heidelberg-Moscow collaboration, recently claimed
\cite{Klapdor-Kleingrothaus:2004wj} evidence for the \nbb decay of
$^{76} \rm{Ge}$ with $ T^{0\nu}_{1/2} = (0.69-4.18)\cdot 10^{25}$
years at the $4.2\sigma$ confidence level.  Using the nuclear matrix
element obtained in Ref. \cite{Rodin:2006yk}, from this data one finds
for the effective Majorana mass the range $0.23~eV\le |m_{\beta\beta}|
\le 0.56~eV$.
Future experiments will extend the sensitivity of current \nbb
searches and provide an independent check of this claim and even go
further in sensitivity, as seen in the figure~\cite{dbd06}.

\begin{table}[h]
  \begin{center}
    \caption{Sensitivities of future $0\nu\beta\beta$-decay
      experiments to the effective Majorana neutrino mass calculated
      with the RQRPA nuclear matrix elements $M^{0\nu}(A,Z)$ of Ref.
      \protect\cite{Rodin:2006yk}.  For the axial coupling constant
      $g_A$ the value $g_A=1.25$ was assumed.  $T^{0\nu-exp}_{1/2}$ is
      the maximal half-life, which can be reached in the experiment
      and $m_{ \beta\beta} \equiv \meff$ is the corresponding upper
      limit of the effective Majorana neutrino mass.  }
    \label{tab.1}
\vglue .5cm
\begin{tabular}{lccccccc}
\hline\hline
Nucleus & Experiment & Source  & $T^{0\nu-exp}_{1/2}$ [yr] & Ref. & & $M^{0\nu}(A,Z)$ &
$|m_{\beta\beta}|$ [eV] \\ \hline
$^{76}Ge$ & GERDA(I)  & 15 kg of $^{enr}Ge$   & $3~ 10^{25}$ & \cite{Abt:2004yk} &  & 3.92  & $0.27$ \\ 
          & GERDA(II)  & 100 kg of $^{enr}Ge$ & $2~ 10^{26}$ & \cite{Abt:2004yk} &  & 3.92  & $0.10$ \\ 
          & Majorana & 0.5 t of $^{enr}Ge$ & $4~ 10^{27}$ & \cite{Aalseth:2005mn} &  & 3.92  & $0.023$ \\
$^{82}Se$ & SuperNEMO & 100 kg of $^{enr}Se$ & $2~ 10^{26}$ & \cite{Barabash:2004pu} &  & 3.49 & $0.055$ \\ 
$^{100}Mo$ & MOON  & 3.4 t of $^{nat}Mo$ & $1~ 10^{27}$ & \cite{elliott:2002xe} &  & 2.78 & $0.024$ \\ 
$^{116}Cd$ & CAMEO & 1 t of $CdWO_4$ crystals & $\approx 10^{26}$ & \cite{elliott:2002xe} &  & 2.42 & $0.085$ \\ 
$^{130}Te$ & CUORE & 750 kg of $TeO_2$  & $\approx 10^{27}$ & \cite{Giuliani:2005mv} &  & 2.95  & $0.023$ \\ 
$^{136}Xe$ & XMASS & 10 t of liq. Xe & $3~ 10^{26}$ & \cite{elliott:2002xe} &  & 1.97 & $0.062$ \\ 
           &       &                 &                   &          &  & 1.67 & $0.073$ \\
           & EXO   & 1 t $^{enr}Xe$  & $2~ 10^{27}$ & \cite{Danilov:2000pp} &  & 1.97 & $0.024$ \\ 
           &       &      &     &   &  & 1.67 & $0.028$ \\
\hline\hline
\end{tabular}
  \end{center}
\end{table}
The left (right) panel in Fig.~\ref{fig:nbbfut} corresponds to the
cases of normal (inverted) neutrino mass spectra. In these plots the
``diagonals'' correspond to the case of quasi-degenerate
neutrinos~\cite{babu:2002dz}, which give the largest \nbb amplitude.
In the normal hierarchy case there is in general no lower bound on 
the \nbb rate since there can be a destructive interference amongst
the neutrino amplitudes. In contrast,  the inverted neutrino mass
hierarchy implies a ``lower'' bound for the \nbb amplitude.

\begin{figure}[!h] \centering
\includegraphics[width=.47\linewidth,height=6cm]{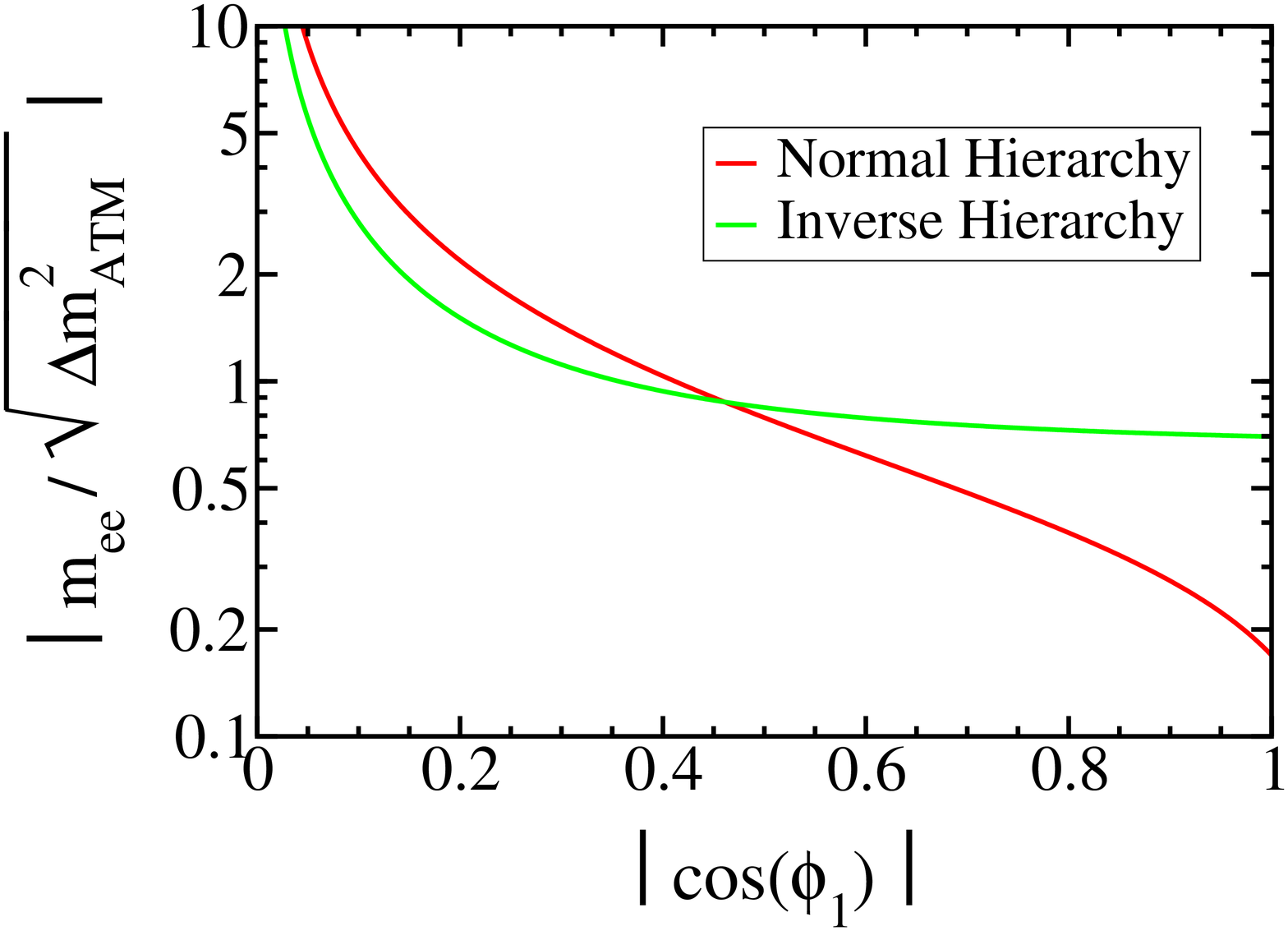}
\includegraphics[width=.47\linewidth,height=6cm]{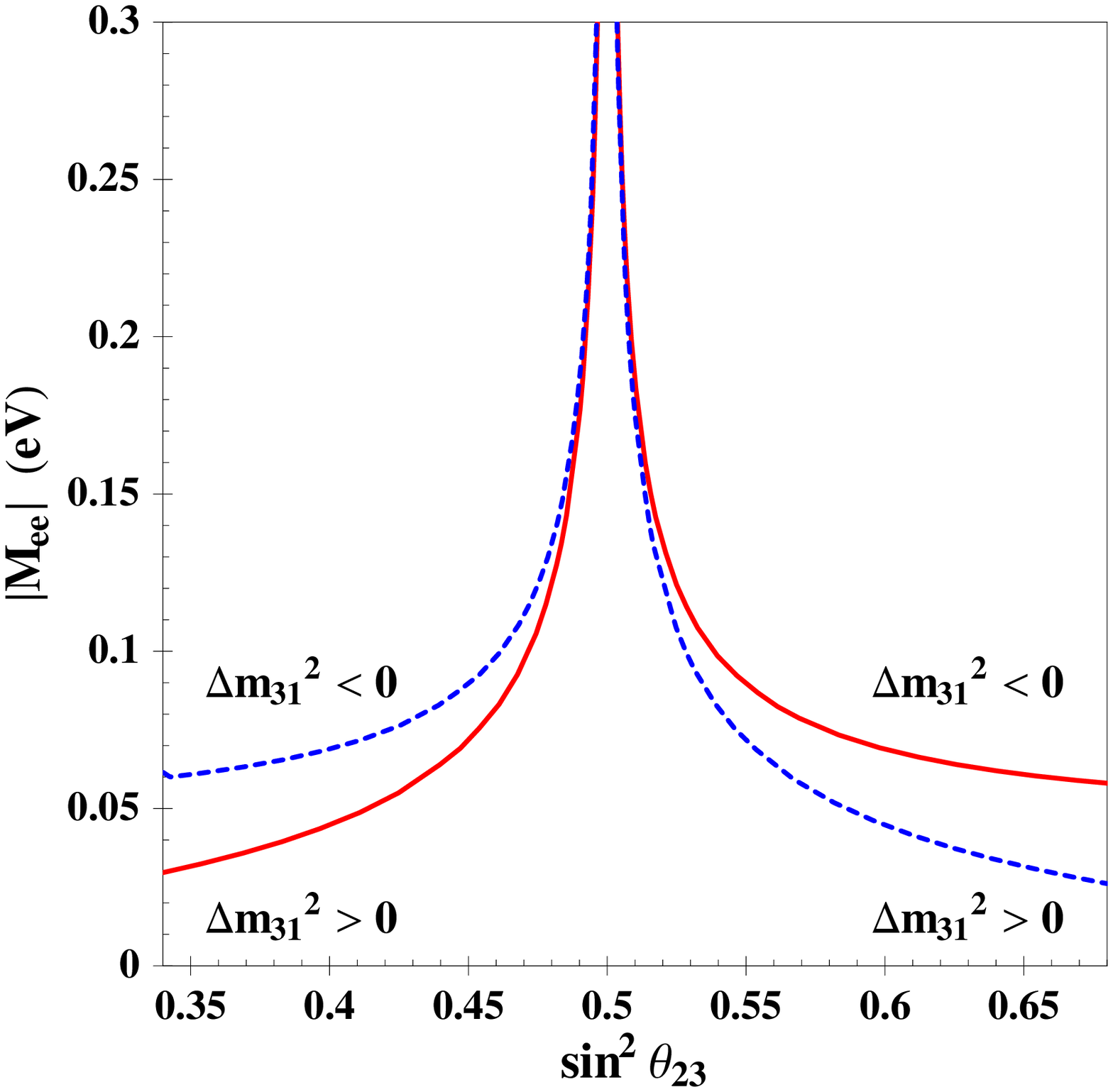}
\caption{\label{fig:bbn-a4} %
  Lower bound on neutrinoless double beta decay in the models of
  Refs.~\cite{Hirsch:2005mc} (left) and~\cite{Hirsch:2007kh} (right).
  In first model 
  the lower bound depends on the Majorana phase $\phi_1$, while in the
  second one has a strong dependence of the lower bound on the value
  of the atmospheric angle. }
\end{figure}

A normal hierarchy model with a lower bound on \nbb is given in
Ref.~\cite{Hirsch:2005mc}. An interesting feature is that the lower
bound obtained depends on the value of the Majorana violating phase
$\phi_1$, as indicated in Fig.~\ref{fig:bbn-a4}. Note that the lines
in dark (red) and grey (green) of the left panel correspond to normal
and inverse hierarchy, respectively.
An alternative model based on the $A_4$ flavour symmetry has been
suggested~\cite{Hirsch:2007kh} which implies a lower bound on the
neutrinoless double beta decay rate, corresponding to an effective
mass parameter $M_{ee} \gsim 0.03$ eV, as illustrated in the right
panel of the same figure.

\vskip .2cm

Complementary information on the absolute scale of neutrino mass is
expected to come from future beta decays searches such as in the
KATRIN experiment~\cite{Osipowicz:2001sq}. Altogether beta and \nbb
sensitivities should be confronted with information coming from
cosmology~\cite{Lesgourgues:2006nd,Hannestad:2006zg}.

\section{Beyond neutrino oscillations}
\label{sec:robustn-neutr-oscill}

How robust is the current oscillation picture of neutrino data?  For
example, how well do we know the astrophysics of the Sun, neutrino
propagation and/or neutrino detection to be confident that there are
no loop holes in our understanding of neutrinos?

\subsection{Solar magnetic fields}
\label{sec:magnetic-fields}

Here we give two examples of how solar magnetic fields can affect the
neutrino oscillation interpretation of solar neutrino data.

\subsubsection{Radiative zone magnetic fields and density noise}
\label{sec:rz-magnetic-fields}

The Sun can harbor magnetic fields in its radiative zone as well as in
its convective zone. It has been shown~\cite{Burgess:2003fj} that
magnetic fields deep within the solar radiative zone can produce
density fluctuations that could affect solar neutrino fluxes in an
important way~\cite{Burgess:2002we}.
However it has been shown that consistency with KamLAND reactor data
restores robustness of the determination of neutrino oscillation
parameters~\cite{Burgess:2003su}.

\subsubsection{Convective zone magnetic fields and spin flavour precession}
\label{sec:conv-zone-magn-1}

Convective zone magnetic fields can lead to spin flavour precession if
neutrinos have non-zero transition magnetic
moments~\cite{schechter:1981hw} that may affect neutrino propagation
both in {\it vacuo} and in matter~\cite{Lim:1987tk,Akhmedov:1988uk}.
A global analysis of spin-flavour precession (SFP) solutions to the
solar neutrino problem is characterized by three effective parameters:
$\Dms \equiv \Delta m^2$, the neutrino mixing angle $\theta_\Sol
\equiv \theta$ and the magnetic field parameter $\mu
B_\perp$~\cite{miranda:2000bi,miranda:2001hv}. For $\mu = 10^{-11}$
Bohr magneton, and an optimum self-consistent magneto-hydrodynamics
magnetic field profile with maximum strength $B_\perp \sim 80$ KGauss
in the convective zone~\cite{barranco:2002te} one finds an excellent
description of the solar neutrino data in terms of spin flavor
precession. However, after combining with data from the KamLAND
reactor experiment, one finds that such solutions to the solar
neutrino problem are ruled out, as they can not account for the
deficit and spectral distortion observed at KamLAND.

\subsubsection{Probing for neutrino magnetic moments}
\label{sec:prob-neutr-magn}

Even though by themselves Majorana neutrino transition moments can not
provide an acceptable explanation of the solar plus KamLAND data, they
can still be present at a sub-leading level and be studied using
existing solar and reactor neutrino data. They would contribute to the
neutrino--electron scattering cross section and hence alter the signal
observed in Super-Kamiokande and at Borexino.  In
ref.~\cite{Grimus:2002vb} constraints were placed on the neutrino
transition moments by using the solar neutrino data. It was found that
all transition elements can be bounded at the same time.  Furthermore,
improved reactor data play a complementary role to the solar neutrino
data in further improving the sensitivity, which is currently at $2
\times 10^{-10} \mu_B$ at the 90\% C.L. for the combined solar +
reactor data.  The upcoming Borexino experiment will improve the
bounds from today's data by roughly one order of magnitude, thanks to
the lower energy.

\subsection{Non-standard neutrino interactions}
\label{sec:non-stand-inter}

\begin{figure}[h!] \centering
    \includegraphics[height=3.7cm,width=.5\linewidth]{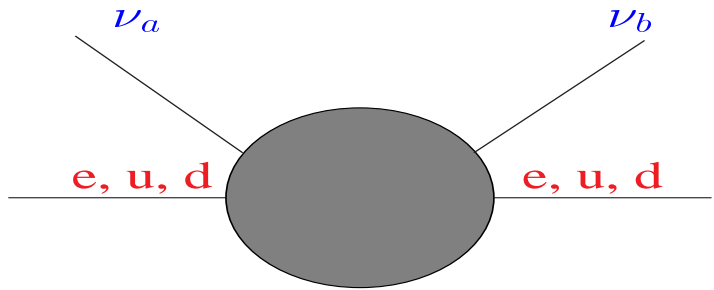}
    \caption{\label{fig:nuNSI} %
      Flavour-changing effective operator for non-standard neutrino interaction.}
\end{figure}

It is hard to find a neutrino mass generation scheme that does not
bring in dimension-6 non-standard neutrino interaction (NSI) terms.
Such sub-weak strength $\varepsilon G_F$ operators are illustrated in
Fig.~\ref{fig:nuNSI}.  They can be of two types: flavour-changing (FC)
and non-universal (NU) and may arise in several ways. One is from the
non-trivial structure of charged and neutral current weak interactions
characterizing the broad class of seesaw-type
models~\cite{schechter:1980gr}.
If present, these interactions would in general bring additional
sources of CP violation.  While the expected magnitude of the
non-standard interactions is rather model dependent, it may well fall
in the range that can be tested in the precision neutrino oscillation
studies discussed in Sec.~\ref{sec:cp-viol-neutr}.
The first manifestation non-standard interactions can have is through
the violation of unitarity in the lepton mixing
matrix~\cite{schechter:1980gr}.
The non-unitary piece of the lepton mixing matrix can be sizable in
inverse seesaw-type models~\cite{mohapatra:1986bd} and hence can be
phenomenologically
important~\cite{bernabeu:1987gr,branco:1989bn,rius:1990gk}.
With neutrino physics entering the precision age it becomes an
important challenge to scrutinize how good is the unitary
approximation of the lepton mixing matrix in future experiments, given
its theoretical fragility.
 
Relatively sizable NSI strengths may also be induced through the
exchange of new scalar bosons, as present in models with radiatively
induced neutrino masses. These typically contain new relatively light
scalar states such as Higgs bosons~\cite{zee:1980ai}, scalar
leptoquarks, etc.
NSI terms may also arise from renormalization group evolution in
supersymmetric unified models~\cite{hall:1986dx}.
                                        
Non-standard physics may affect neutrino production and detection
cross sections, as well as propagation properties.
In their presence, the Hamiltonian describing neutrino propagation
has, in addition to the standard oscillation part, another term
$H_\mathrm{NSI}$,
\begin{equation}
    H_\mathrm{NSI} = \pm \sqrt{2} G_F N_f
    \left( \begin{array}{cc}
        0 & \varepsilon \\ \varepsilon & \varepsilon'
    \end{array}\right) \,.
\end{equation}
Here $+(-)$ holds for neutrinos (anti-neutrinos) and $\varepsilon$ and
$\varepsilon'$ parametrize the NSI: $\sqrt{2} G_F N_f \varepsilon$ is
the forward scattering amplitude for the FC process $\nu_\mu + f \to
\nu_\tau + f$ and $\sqrt{2} G_F N_f \varepsilon'$ represents the
difference between $\nu_\mu + f$ and $\nu_\tau + f$ elastic forward
scattering. Here $N_f$ is the number density of the fermion $f$ along
the neutrino path.  
A remarkable fact is that, in the presence of non-standard
interactions resonant neutrino conversions can take place even in the
absence of neutrino masses~\cite{valle:1987gv}.\\

{\bf Solar neutrino oscillations}\\

\begin{figure}[h!] \centering
    \includegraphics[height=5.5cm,width=.7\linewidth]{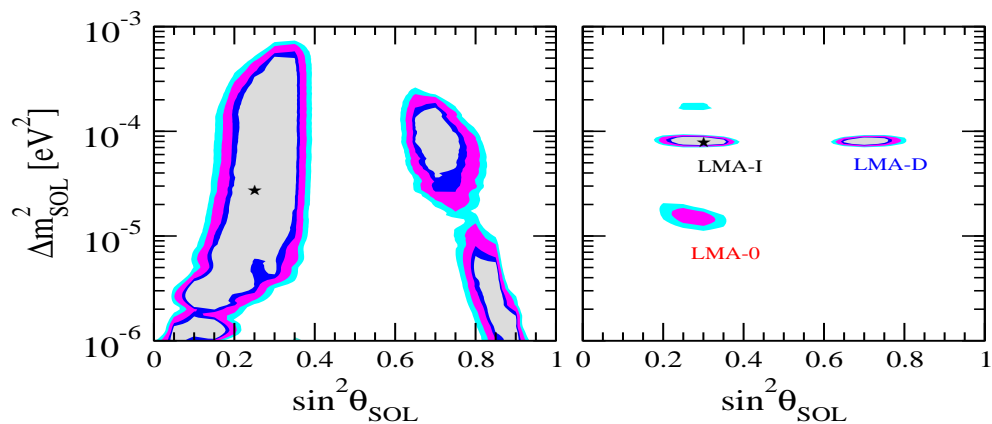}
    \caption{\label{fig:global-NSI-06} %
      New solar neutrino oscillation solutions in the presence of NSI~\cite{Miranda:2004nb}.}
\end{figure}

Although non-standard interaction effects may provide an alternative
way to account for the current solar neutrino
data~\cite{guzzo:2001mi}, consistency with KamLAND reactor neutrino
data requires them to be sub-leading~\cite{pakvasa:2003zv}.
However the oscillation interpretation of solar neutrino data is still
``fragile'' against the presence of non-standard interactions in the
$e-\tau$ sector, opening another degenerate solution in the second
octant of the solar mixing angle~\cite{Miranda:2004nb}.
Indeed, as seen in Fig.~\ref{fig:global-NSI-06}, in the presence of
NSI, there appear new solar neutrino oscillation solutions in addition
to the standard one. The two degenerate solutions are denoted LMA-I or
``normal'' and LMA-D ``degenerate or dark--side'' solution.
Although some ways of inducing the new LMA-D solution, for example via
NSI between neutrinos and down-type-quarks-only, are already in
conflict with the combination of current atmospheric data and
accelerator neutrino oscillation data of the CHARM experiment, the
general form of these solutions is not yet ruled out. Moreover,
further precision KamLAND reactor measurements will not resolve the
ambiguity in the determination of the solar neutrino mixing angle,
since they are expected to constrain mainly $\Dms$.

\begin{figure}[!h] \centering
    \includegraphics[height=7cm,width=.7\linewidth]{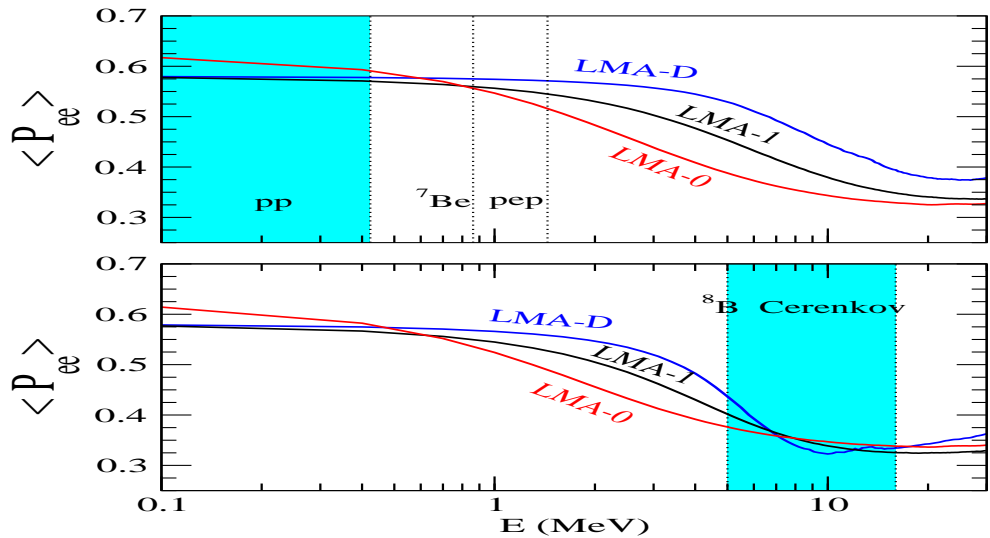}
    \caption{\label{fig:fut-NSI-solar} %
      Probing for NSI in future low energy solar neutrino
      experiments~\cite{Miranda:2004nb}.}
\end{figure}

However, as can be seen from Fig.~\ref{fig:fut-NSI-solar}, the
``normal'' LMA-I solutions have quite different predicted neutrino
survival probabilities for low energy neutrinos (left panel) and boron
neutrinos (right panel). Thus future precision low energy solar
neutrino experiments will help in lifting the degeneracy between the
``normal'' and the new degenerate solution.

On the other hand, even if very small, non-standard interaction
effects may play an important role in supernova astrophysics, opening
the possibility of new resonances that could take place in the
internal neutron-rich
regime~\cite{nunokawa:1996tg,nunokawa:1996ve,grasso:1998tt,Esteban-Pretel:2007yu}.\vskip 1cm

{\bf Atmospheric neutrino oscillations}\\

In contrast to the solar case, it has been
shown~\cite{fornengo:2001pm} that, within the 2--neutrino
approximation, the determination of atmospheric neutrino parameters
$\Dma$ and $\sin^2\theta_\Atm$ has been shown to be practically
unaffected by the presence of NSI on down-type quarks ($f=d$).  There
are however loop-holes in the three neutrino
case~\cite{Friedland:2004ah}. Let us also mention here that future
neutrino factories will have good potential for probing non-standard
neutrino in the mu-tau channel~\cite{huber:2001zw}.\\ [.2cm]

{\bf NSI-oscillation confusion theorem}\\

For the coming generation of experiments we note that even a small
residual non-standard interaction of neutrinos in the $e-\tau$ channel
characterized by a parameter $\epsilon_{e\tau}$ can have dramatic
consequences for the prospects of probing neutrino oscillations at
neutrino factories. For example, it has been shown~\cite{huber:2002bi}
that the presence of such NSI leads to a drastic loss in sensitivity
in the $\theta_{13}$ determination at a neutrino factory. It is
therefore important to improve the sensitivities on NSI, another
window of opportunity for neutrino physics in the precision age.
           
\section{Summary}
\label{sec:summary}

We have reviewed the basic mechanisms to generate neutrino mass,
analysing the corresponding structure of the lepton mixing matrix.
After an overview of the status of neutrino oscillation parameters as
determined from current data, we have focused on the next neutrino
oscillation experiments using accelerator neutrinos, and discussed the
prospects for probing the strength of CP violation in two of them, T2K
and NO$\nu$A. 
Intermediate term experiments using wide band beams and capable of
probing also the second Oscillation Maximum, such as T2KK, have also
been discussed
~\cite{Ishitsuka:2005qi,Hagiwara:2005pe,Hagiwara:2006vn,Kajita:2006bt}.

Farther in the future are experiments with novel neutrino beams, such
as neutrino factories and beta beams.  One of the advantages of the
neutrino beam coming from a muon storage ring is that the energy
spectrum is accurately known, in contrast with conventional muon
neutrino beam from pion decay, where the energy spectrum is not very
well determined.
It was suggested in ~\cite{Zucchelli:2002sa} that pure $\nu_e$ or $\bar
\nu_e$ beams can be obtained by using $\beta$ unstable isotopes.  The
idea is to produce first a huge number of $\beta$ unstable ions and
then accelerate them in a storage ring to some reference energy, and
let them to decay in the straight section of a storage ring in order
to get an intense $\nu_e$ or $\bar \nu_e$ beam.  The advantage of this
method is that the energy spectrum is precisely known (as in the case
of neutrino beam from muon decay) and the there is no contamination of
the other neutrino species unlike the case of neutrino beam from muon
decay.
None of these possibilities were considered here, but they are
discussed in a recent review by Geer and Zisman in this same
series~\cite{Geer:2006tb} for neutrino factories, and e.g., in
Ref.~\cite{Burguet-Castell:2005pa} for the case of beta beam.
Another extensive review is being prepared for the International
scoping study of a future Neutrino Factory and super-beam facility.
We have also briefly commented on the possibility of probing CP
violation effects induced by Majorana phases in
neutrinoless double beta decay.\\

{\bf Acknowledgements}\\

We thank Martin Hirsch, Filipe Joaquim, Stefano Morisi, Jorge Rom\~ao,
Joe Schechter and M. Tortola for useful comments. We thank Fedor
Simkovic for the updated results on nuclear matrix elements. This work
has been supported by Spanish grants FPA2005-01269 and
FPA2005-25348-E, by the European Commission RTN Contract
MRTN-CT-2004-503369 and ILIAS/N6 Contract RII3-CT-2004-506222, and
also by a Humboldt Research Award (JWFV).  Fermilab is operated under
DOE Contract No.~DE-AC02-76CH03000.


\renewcommand{\arraystretch}{1}

\end{document}